\documentclass[a4paper,12pt]{report}
\usepackage{amsmath,amssymb,amsfonts,graphicx,fancyhdr,calc,sectsty,epigraph,makeidx}
\usepackage[gregorian]{eukdate}
\usepackage[bookmarks=true,pdffitwindow=true,colorlinks=false,backref=page]{hyperref}
\usepackage{dbnsymb}
\makeindex

\newcommand{\sect}[1]{\setcounter{equation}{0}\section{#1}}

\def\fft#1#2{{\frac{#1}{#2}}}

\def\im{{{\rm i}}}

\def\half{{1 \over 2} } 
\def\be{\begin{equation}}
\def\ee{\end{equation}}
\def\ben{\begin{equation}}
\def\een{\end{equation}}
\def\bena{\begin{eqnarray}} 
\def\eena{\end{eqnarray}}
\def\bea{\begin{eqnarray}} 
\def\eea{\end{eqnarray}}
\def\ie{{\it i.e.\ }}
\def\eg{{\it e.g.\ }}
\def\Tr{{\rm Tr}}
\def\const{\rm constant}
\def\nn{\nonumber}
\def\sgn{{\rm sgn }}

\def\R{{{\mathbb R}}}
\def\bR{{{\mathbb R}}}
\def\bZ{{{\mathbb Z}}}

\def\Im{{{\frak{Im}}}}
\def\prd{Physical Review D}
\input amssym.def
\input amssym.tex

\begin{document}

\hfuzz=50pt
\begin{titlepage}
\begin{center}
\vspace*{1in}
{\bf {\LARGE \textsf{Geometric Aspects of Gauge and Spacetime Symmetries}}}
\par
\vspace{1in}
{\Large Steffen Christian Martin Gielen \\ Trinity College}
  \vspace{1in}

  {\large 
  A dissertation submitted to the\\
  University of Cambridge \\
  for the degree of \\
  Doctor of Philosophy}
  \\\vspace{1cm}
  \includegraphics[scale=0.1]{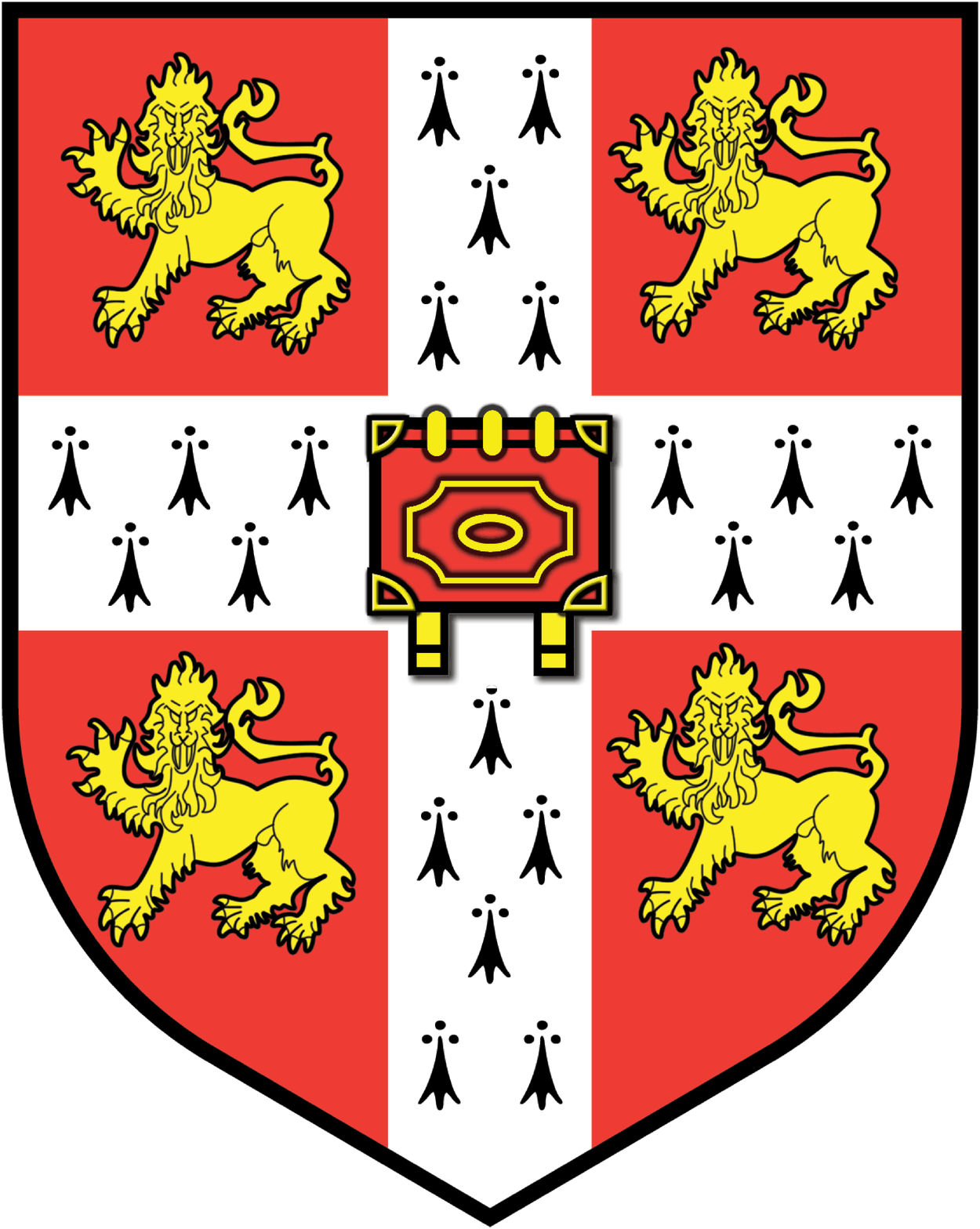}
  \\\vspace{1cm}
  {\large
  \monthname\;2011}
\end{center}
\end{titlepage}

\section*{Declaration}
\setcounter{page}{2}
This dissertation is my own work and contains nothing which is the outcome of work done in collaboration with others, except as specified in the text (section \ref{struktur}). It is not substantially the same as any other thesis that I have submitted or will submit for a degree at this or any other university.

This thesis is based on the research presented in the following refereed publications:
\begin{itemize}
\item[(1)] {\sc{Gary W. Gibbons and Steffen Gielen}}. {\em{The Petrov and Kaigorodov-Ozsv\'ath Solutions: Spacetime as a Group Manifold}}. Classical and Quantum Gravity {\bf 25} (2008), 165009. 
\item[(2)] {\sc{Gary W. Gibbons and Steffen Gielen}}. {\em{Deformed General Relativity and Torsion}}. Classical and Quantum Gravity {\bf 26} (2009), 135005.
\item[(3)] {\sc{Gary W. Gibbons, Steffen Gielen, C. N. Pope, and Neil Turok}}. {\em{Measures on Mixing Angles}}. Physical Review D{\bf 79} (2009), 013009.
\item[(4)] {\sc{Steffen Gielen}}. {\em{$CP$ Violation Makes Left-Right Symmetric Extensions With Non-Hermitian Mass Matrices Appear Unnatural}}. Physical Review D{\bf 81} (2010), 076003.
\item[(5)] {\sc{Steffen Gielen and Daniele Oriti}}. {\em{Classical general relativity as BF-Plebanski theory with linear constraints}}. Classical and Quantum Gravity {\bf 27} (2010), 185017.
\end{itemize}

These papers appear in the text as references \cite{petrovpaper}, \cite{dsrpaper}, \cite{cpviolation}, \cite{unnatural}, and \cite{bflinearc} respectively.
The results of section 5 of (3) were also separately published in
\begin{itemize}
\item[(6)] {\sc{Gary W. Gibbons, Steffen Gielen, C. N. Pope, and Neil Turok}}. {\em{Naturalness of CP Violation in the Standard Model}}. Physical Review Letters {\bf 102} (2009), 121802. 
\end{itemize}

\begin{flushright}
\vspace{15mm}
Steffen Gielen
\\\today
\end{flushright}

\newpage
\section*{Acknowledgments}
My first and foremost thanks go to my supervisor, Prof. Gary Gibbons, for teaching me so much about physics and mathematics, for many fruitful discussions and suggestions, for his encouragement and for his support of all of my plans over the last three years. 

I have also greatly enjoyed working with, and learning from, Dr. Daniele Oriti, Prof. C. N. Pope and Dr. Neil Turok on different projects that are part of this thesis. Invitations by Dr. Turok and Dr. Oriti that allowed me to spend an intensive week at Perimeter Institute in October 2008 and several months at AEI Golm in 2009/10 deserve special thanks.

I am grateful to Trinity College for all its support, both financial and otherwise, over the course of the last four years, without which I would not have been able to complete a PhD at Cambridge. My work has also been supported by an EPSRC studentship.

I would like to thank my friends in Cambridge, who have hopefully improved my understanding of physics and made my time more enjoyable, not least through countless hours of `foosball', and my friends in Germany who have kept in touch while I was away for so long, and gave support in difficult times.

Going back a bit further in time, I would not have ended up as a PhD student in Cambridge (and certainly not at Trinity) without the encouragement of Dipl.-Phys. Igor Pikovski and the strong support of Profs. Ingo Peschel and J\"urgen Bosse. The latter two together with Hermann Schulz (Hochschuldozent) were also influential in shaping my interest in theoretical physics early on.

Last but not least, I would like to thank my aunt Odina and uncle Uwe for their support, especially for the first year at Cambridge.

\newpage

\begin{center}
{\large {\bf Geometric Aspects of Gauge and Spacetime Symmetries}
\\{\sc Steffen Christian Martin Gielen}}
\end{center}

\section*{Summary}

We investigate several problems in relativity and particle physics where symmetries play a central role; in all cases geometric properties of Lie groups and their quotients are related to physical effects.

\

The first part is concerned with symmetries in gravity. We apply the theory of Lie group deformations to isometry groups of exact solutions in general relativity, relating the algebraic properties of these groups to physical properties of the spacetimes. We then make group deformation local, generalising deformed special relativity (DSR) by describing gravity as a gauge theory of the de Sitter group. We find that in our construction Minkowski space has a connection with torsion; physical effects of torsion seem to rule out the proposed framework as a viable theory. A third chapter discusses a formulation of gravity as a topological BF theory with added linear constraints that reduce the symmetries of the topological theory to those of general relativity. We discretise our constructions and compare to a similar construction by Plebanski which uses quadratic constraints.

\

In the second part we study $CP$ violation in the electroweak sector of the standard model and certain extensions of it. We quantify fine-tuning in the observed magnitude of $CP$ violation by determining a natural measure on the space of CKM matrices, a double quotient of $SU(3)$, introducing different possible choices and comparing their predictions for $CP$ violation. While one generically faces a fine-tuning problem, in the standard model the problem is removed by a measure that incorporates the observed quark masses, which suggests a close relation between a mass hierarchy and suppression of $CP$ violation. Going beyond the standard model by adding a left-right symmetry spoils the result, leaving us to conclude that such additional symmetries appear less natural.

\setcounter{tocdepth}{1}
\tableofcontents

\chapter{Introduction}
\setlength{\epigraphwidth}{10.2cm}
\epigraph{\em{``Ich kann es nun einmal nicht lassen, in diesem Drama von Mathematik und Physik -- die sich im Dunkeln befruchten, aber von Angesicht zu Angesicht so gerne einander verkennen und verleugnen -- die Rolle des (wie ich genugsam erfuhr, oft unerw\"unschten) {\rm Boten} zu spielen.'' \cite{weylgroup}}}{Hermann Weyl}

It is impossible to overemphasise the importance of symmetry as a fundamental concept in both mathematics and theoretical physics. 

In mathematics, the study of groups arose as an abstraction of ``transformation groups"\index{transformation group}, invertible maps of a given ``point field" into itself \cite{weylgroup}. Groups are the mathematical realisation of the concept of symmetry. Felix Klein's Erlangen Programme\index{Erlangen Programme} \cite{klein} aimed to describe any geometry by its transformation group; as we will detail shortly, the geometries considered in the Erlangen Programme are given by coset spaces of the form $G/H=\{gH,g\in G\}$, where $G$ is a continuous (Lie) group and $H$ a closed subgroup of $G$. Klein himself found the concept of groups most characteristic of $19^{{\rm th}}$ century mathematics \cite{weylgroup}. The study of continuous symmetries is, in  mathematical terms, the study of Lie groups and homogeneous spaces which nicely combines aspects of group theory and differential geometry. Beyond the direct study of groups, symmetry is an important concept in other parts of mathematics, such as the study of differential equations.

In physics, the underlying symmetries of any theory are arguably its most fundamental property. As an example, the first physical theory in the modern sense, Newtonian mechanics, fundamentally rests on assumptions about the symmetries of the physical world: A configuration of bodies in (gravitational) interaction will behave in the same way if it is rotated as a whole (space is isotropic); a common motion of all bodies at constant velocity is not observable (relativity of inertial frames). Consequently, the laws of Newtonian mechanics are invariant under the action of the semidirect product of the rotation group $SO(3)$ and commutative Galileian boosts, the {\bf Euclidean group}\index{Euclidean group} $E(3)$. When Einstein understood in 1905, in the wake of the Michelson-Morley experiment \cite{mmex}, that it was necessary to modify the concept of relativity of inertial frames, he did so by replacing the Euclidean group by the Lorentz group $SO(3,1)$ as the fundamental symmetry group. We will see in chapter \ref{defspacetime} that this process can be mathematically viewed as a Lie group (or algebra) deformation\index{Lie group/algebra deformation}.

The Galilean and Lorentz groups are examples of {\bf spacetime} symmetries; their primary action on space(time) itself  induces an action on all fields according to their tensorial structure. In general relativity, these global spacetime symmetries are made local, although global symmetries also appear in many exact solutions of physical interest, as we shall see in chapter \ref{defspacetime}. Both global and local spacetime symmetries will play a role in the first part of this thesis.

More generally, the equations defining a given theory (or the action, in the usual case that the equations of motion can be derived from an action principle) may also be invariant under a group of transformations of the physical variables which is not associated to a transformation of spacetime. A simple example in field theory is that of a complex scalar field $\phi$ with equations of motion invariant under a global rotation $\phi\rightarrow e^{\im\alpha}\phi$. Promoting such a global symmetry to a local symmetry leads to a {\bf gauge} theory\index{gauge theory}. We will discuss gauge symmetries in both gravity and particle physics in this thesis. 

Mathematically, the fundamental structures describing local symmetries are principal fibre bundles over a given manifold which is interpreted as spacetime, where the structure group becomes the ``gauge group" of the theory, and a connection on such a bundle is interpreted as a gauge potential. The remaining fields that make up the physical content of the theory then live in associated vector and tensor bundles. While this underlying structure suggests a very close relationship of all fundamental interactions in Nature, there are important differences between Yang-Mills theory, defined in terms of a principal bundle with compact gauge group, and general relativity, where in addition to the connection one has a frame field as a second structure which is crucial for the geometric interpretation of the theory.\footnote{A second difference is of course the non-compactness of the gauge group for Lorentzian signature, but this seems a less important difference to Yang-Mills theory in the basic structure of the theory -- one often considers Riemannian signature as well.} In chapter \ref{defgenrel}, we will study a formulation of general relativity as a Yang-Mills-like theory which tries to exploit this analogy; the usual Lorentz algebra valued connection and the frame field are combined into a connection on a bundle of the Poincar\'e group (or its deformation, the (anti-)de Sitter group). This study will elucidate some of the virtues as well as shortcomings of such a formulation.

Symmetry is our most important guiding principle towards discovering the laws of Nature. In particle physics, the electroweak model\index{electroweak model} \cite{electroweak, electroweak3, electroweak2} unifying the electromagnetic and weak interactions was discovered by postulating the fundamental symmetries and writing down the most general action invariant under them. A crucial step in completing the present picture of particle physics was the proof by 't Hooft and Veltman \cite{hooftvelt} that the mathematical structure of Yang-Mills theory is compatible with the framework of quantum field theory; namely, Yang-Mills theory is renormalisable\index{renormalisation}. By far most attempts to go beyond the standard model of particle physics, such as supersymmetric\index{supersymmetry} theories and GUTs, are based on postulating new symmetries. Part of our analysis of chapter \ref{natural} will be to discuss possible observable consequences of such an existence of additional symmetries beyond those of the electroweak model.

General relativity, however, has so far evaded a similar treatment in terms of conventional field theory formulated on a background Minkowski spacetime, and by now it seems unlikely that adding (super)symmetries will be sufficient for a consistent quantum-mechanical treatment, although there has recently been increasing interest in the possiblity that maximal ($N=8$) supergravity in four dimensions is perturbatively finite \cite{finite}. String theory\index{string theory} is one attempt to go beyond standard quantum field theory by replacing the notion of point particles by extended objects; symmetries (such as dualities) play a fundamental theory in string theory as well. Besides the symmetries that will be important for a consistent fundamental theory of quantum gravity, if we manage to construct (and understand) one, one may wonder whether at low energy some effects of quantum gravity\index{quantum gravity} will be manifest in a deformation of spacetime symmetries. This idea, which has been proposed on a purely phenomenological basis \cite{amelino} as well as a consequence of developments in non-perturbative quantum gravity \cite{quantsym}, will be taken up in chapter \ref{defgenrel}.

The notion of symmetries we have discussed so far refers to invariance of the action or equations of motion under certain group actions. From a canonical viewpoint, there is another important class of symmetries, namely those generated by constraints. To obtain the canonical formulation of a given theory, one performs a Legendre transformation, leading to a Hamiltonian
\ben
p_i(x)\equiv\frac{\delta S}{\delta \dot{q}^i(x)}\,,\quad \mathcal{H}\equiv \int d^{d-1}x\,\mathcal{H}(x)\equiv\int d^{d-1} x\left(p_i(x)\dot{q}^i(x) - \mathcal{L}(x)\right)\,,
\een
where $S=\int dt \int d^{d-1} x \,\mathcal{L}(x)$ is the action defining the theory, and a symplectic structure\index{symplectic structure} on the phase space parametrised by $(q^i,p_i)$,
\ben
\{q^i(x),p_j(y)\}=\delta^i_j\,\delta^{d-1}(x-y)\,,\quad\{q^i(x),q^j(y)\}=\{p_i(x),p_j(y)\}=0\,.
\een
The Hamiltonian density $\mathcal{H}(x)$ typically splits into a ``dynamical" part which is a generic function of the $p_i$ and $q^i$, and a ``constraint" part of the form $\lambda^\alpha f_\alpha(p_i,q^i)$; here $\lambda^\alpha$ are Lagrange multipliers, variables that appear in the original action whose canonical momentum is identically zero and that are conventionally not treated as belonging to the phase space. The functions $f_\alpha(p_i,q^i)$ then act as {\bf constraints}\index{constraints} imposed on the dynamical variables. A subset of these will be {\em first class}\index{constraints!first class}\footnote{We are here and in the following using the terminology introduced by Dirac \cite{diracform}.}, \ie their Poisson brackets vanish up to a linear combination of constraints. The Poisson brackets $\{f_{\alpha},\cdot\}$ of the first class constraints then determine vector fields on the phase space whose integral curves are interpreted as describing physically equivalent configurations. Although not necessarily associated with connections on fibre bundles, the transformations on the phase space generated by these vector fields are, by their physical interpretation, usually interpreted as ``gauge transformations" as well. We will see an instance of such a symmetry in the context of a topological field theory\index{topological field theory}, which is then constrained by a term that is added to the action, in chapter \ref{bflinear}.

Although our discussions of chapter \ref{bflinear} will remain classical, the motivation for the reformulation of general relativity that we will consider comes from the fact that the quantisation of topological field theories is relatively well understood. One might therefore hope that starting from a topological quantum theory which is then constrained may provide a different route to quantum general relativity. This also provides a bridge to the discussion of chapter \ref{defgenrel} in that both consider possible reformulations of general relativity at the classical level which are suggested by developments in non-perturbative approaches to quantising gravity.

In all of our discussions, an important distinction is that between discrete and continuous symmetries. The first systematic study of groups in physics was probably that of discrete groups describing the symmetries of crystals \cite{weylgroup}. We shall be mainly interested in continuous symmetries which are described by Lie groups. The spacetime and gauge symmetries described above are of this type. Their differentiable structure means that differential geometry can be used to study them. In particular, we will see that the group structure is encoded in a bracket operation on the tangent space to the group at the identity element.

We will also encounter discrete symmetries\index{discrete symmetries}, such as charge conjugation and parity in particle physics which are represented by the finite group $\mathbb{Z}_2$. These will not be studied mathematically, but their physical significance will be discussed.

In the spirit of Hermann Weyl's self-image as a ``messenger" between mathematics and physics, we will highlight different examples where a study of the differential geometry of (spacetime or gauge) symmetry groups leads to important physical consequences. Let us first introduce the most important aspects of the theory of Lie groups and homogeneous spaces that will be important throughout this thesis.

\sect{Lie Groups and Homogeneous Spaces}\index{Lie group}
\label{lie}

The study of symmetries is the study of groups. If we consider continuous symmetries, the corresponding groups have an additional differentiable structure: They are manifolds as well as groups\footnote{One also requires the group multiplication map $(x,y)\mapsto x\cdot y$ and the inversion map $x\mapsto x^{-1}$ to be differentiable.}. As simple examples one may think of the groups of isometries of $\mathbb{R}^n,\mathbb{C}^n$ or $\mathbb{H}^n$, denoted by $O(n)$, $U(n)$ and $Sp(n)$ respectively, or more generally, of groups of matrices with continuous parameters which are submanifolds of $\mathbb{R}^{n\times n},\mathbb{C}^{n\times n}$ or $\mathbb{H}^{n\times n}$. All examples of relevance in this thesis will be of this type.

Such groups are known as {\bf Lie groups}, named after the Norwegian mathematician Sophus Lie (1842-1899) who initiated their systematic study in the context of systems of differential equations. Lie himself seems to have referred to Lie groups as ``continuous groups" which describes appropriately what they are. The study of Lie groups combines aspects of differential geometry with group theory.

Two natural actions of a Lie group $G$ on itself are given by the left and right multiplication maps
\ben
L : G\times G \rightarrow G\,,\quad (h,g)\mapsto hg\,, \qquad R : G\times G \rightarrow G\,,\quad (h,g)\mapsto gh\,.
\een
By definition of a group these maps act simply transitively: Given $a,g\in G$ there is a unique element $h$ such that $L(h,g)=a$ and a unique element $h'$ such that $R(h',g)=a$. The pull-back and push-forward of these maps act on the tangent and cotangent bundle of $G$. 

Starting with a basis of the tangent space at the identity element $e$, one can use the push-forward of left translation to define a basis of vector fields on $G$, known as {\bf left-invariant vector fields}\index{left-invariant vector field}. These define a global frame field on $G$ and hence every Lie group is parallelisable. (Right-invariant vector fields can be defined analogously but often left invariance is preferred.) The tangent space at $e$, or alternatively the set of left-invariant vector fields on $G$, is known as the {\bf Lie algebra}\index{Lie algebra} $\frak{g}$. It inherits a {\bf Lie bracket}\index{Lie bracket} $[\cdot,\cdot]$ from the commutator of vector fields. For matrix Lie groups, the bracket is given by the commutator of matrices.

Lie's important results are summarised in the following theorem \cite{arvani}\index{Lie's theorems}:
\newtheorem{grund}{Theorem}[section]
\begin{grund} (Lie)
\\(1) For any Lie algebra $\frak{g}$ there is a Lie group $G$ (not necessarily unique) whose Lie algebra is $\frak{g}$.
\\(2) Let $G$ be a Lie group with Lie algebra $\frak{g}$. If $H$ is a Lie subgroup of $G$ with Lie algebra $\frak{h}$, then $\frak{h}$ is a subalgebra of $\frak{g}$. Conversely, for each Lie subalgebra $\frak{h}$ of $\frak{g}$, there exists a unique connected Lie subgroup $H$ of $G$ which has $\frak{h}$ as its Lie algebra. Furthermore, normal subgroups of $G$ correspond to ideals in $\frak{g}$.
\\(3) Let $G_1,G_2$ be Lie groups with corresponding Lie algebras $\frak{g}_1, \frak{g}_2$. Then if $\frak{g}_1$ and $\frak{g}_2$ are isomorphic as Lie algebras, then $G_1$ and $G_2$ are locally isomorphic. If the Lie groups $G_1,G_2$ are simply connected, then $G_1$ is isomorphic to $G_2$.
\end{grund}

Loosely speaking, all of the structure of the group $G$ is encoded in its Lie algebra. If one is not interested in the topological properties of $G$ and considers only connected and simply connected Lie groups, one can study the geometry of $G$ by studying $\frak{g}$ which is just a linear vector space together with a bracket structure.

For matrix Lie groups, it is usually most convenient to fix a basis of generators $M_a$ and to expand the {\bf Maurer-Cartan form}\index{Maurer-Cartan form} in terms of this basis:
\ben
g^{-1}\,dg=\sigma^a M_a
\label{mcform}
\een
determines a basis of left-invariant one-forms $\sigma^a$. Similarly, right-invariant forms can be obtained by computing $dg\, g^{-1}$ which is clearly invariant under $g\mapsto gh$. The basis dual to $\{\sigma^a\}$ is then a basis of left-invariant vector fields. By taking the exterior derivative of both sides of (\ref{mcform}) and using $d(g^{-1})=-g^{-1}\,dg\,g^{-1}$ for matrices, one can show that the forms $\sigma^a$ satisfy the Maurer-Cartan relations\index{Maurer-Cartan relations}
\ben
d\sigma^a = -\half {{C_b}^a}_c\, \sigma^b\wedge \sigma^c\,,
\een
where ${{C_a}^c}_b$ are the {\bf structure constants}\index{structure constants} of the Lie algebra defined by
\ben
[M_a,M_b]={{C_a}^c}_b M_c\,.
\label{structcon}
\een
One defines left-invariant tensors accordingly, \eg left-invariant metrics\index{left-invariant metric} as generated by symmetrised combinations of $\sigma^a$: Any metric of the form
\ben
g = h_{ab}\sigma^a\otimes \sigma^b
\een
for constant symmetric $h_{ab}$ is by definition left-invariant. One can phrase this by saying that there is a one-to-one correspondence between scalar products on the Lie algebra $\frak{g}$ (specified by $h_{ab}$) and left-invariant metrics on $G$ \cite{arvani}. These metrics are the ones compatible with the left action on $G$ on itself.

Lie groups are in some sense special cases of a more general class of manifolds that will be important in the following: In many situations of physical interest, one considers a manifold $X$ which a Lie group $G$ acts on transitively, but not necessarily simply transitively. That is, for any $x,x'\in X$, there is a (not necessarily unique) $g\in G$ with $g\cdot x=x'$. (Again we take this to be a left action, the case of a right action is analogous.) Then $X$ is called a {\bf homogeneous space}\index{homogeneous space}. For a given $x\in X$, the  subgroup $G_x=\{g\in G: g\cdot x=x\}$ leaving $x$ invariant is called the {\it stabiliser}, {\it little group} or {\it isotropy subgroup} of $x$. Then we have the following result \cite{arvani}
\newtheorem{homog}[grund]{Proposition}
\begin{homog} Let $G\times X\rightarrow X$ be a transitive action of a Lie group $G$ on a manifold $X$, and let $H=G_x$ be the isotropy subgroup of a point $x$. Then:
\\(a) The subgroup $H$ is a closed subgroup of $G$.
\\(b) The natural map $j:G/H\rightarrow X$ given by $j(gH)=g\cdot x$ is a diffeomorphism. In other words, the orbit $G\cdot x$ is diffeomorphic to $G/H$.
\\(c) The dimension of $G/H$ is $\dim G - \dim H$.
\end{homog}

Since we assume transitivity of the group action, the orbit $G\cdot x$ is the whole of $X$; one can then identify $X$ with the coset space $G/H$.\footnote{This result justifies dropping the distinction between the terms ``coset space" and ``homogeneous space" as we will do in the following.} The choice of $x$ in this identification is arbitrary and corresponds to a choice of origin in $X$; different choices for $x$ lead to groups $G_x$ related by conjugation. The natural left action of $G$ on $G/H$ is just induced by the group multiplication law.

In many applications, an additional assumption is satisfied which simplifies the construction of metrics on a homogeneous space: A given homogeneous space $G/H$ is called {\bf reductive}\index{reductive geometry} if there is a splitting of Lie algebras of the form
\ben
\frak{g} = \frak{h} \oplus \frak{x}
\label{reduk}
\een
invariant under the adjoint action of $H$. Then the subspace $\frak{x}$ of the Lie algebra $\frak{g}$ is isomorphic \cite{arvani} to the tangent space to $G/H$ at $H$ (or any other coset $gH$). It therefore defines a basis of left-invariant vector fields on $X$, and one can take the dual basis of a basis for $\frak{x}$ to construct left-invariant metrics on $X$. Reductive geometries will be of importance in chapter \ref{defgenrel} (with $G=SO(d,1)$ and $H=SO(d-1,1)$) and in chapters \ref{su3quot} and \ref{natural} (with $G=SU(3)$ and $H=U(1)^2$). We will encounter an example of a non-reductive geometry in section \ref{kaigorodovsect}.

Homogeneous spaces are the simplest generalisations of (flat) Euclidean geometry: In the $19^{{\rm th}}$ century, attempts to deduce Euclid's fifth postulate\index{Euclid's fifth postulate}, the parallel postulate \cite{euclid}, from the other four\footnote{About 30 such ``proofs" were shown to be unsatisfactory in the 1763 dissertation of A. Kaestner's student Georg Simon Kl\"ugel entitled ``Conatuum praecipuorum theoriam parallelarum demonstrandi recensio" (Critique of the foremost attempts to prove the theory of parallels) \cite[p.274]{kleinnichteuklid}.}, led to the discovery of elliptic and hyperbolic geometries based on the sphere $S^n\simeq SO(n+1)/SO(n)$ and hyperbolic space $H^n\simeq SO(n,1)/SO(n)$, the natural counterparts of Euclidean space $\mathbb{R}^n\simeq E(n)/SO(n)$. In Lorentzian geometry, the spaces of constant curvature, Minkowski, de Sitter and anti-de Sitter space, are also homogeneous spaces. Many other known exact solutions in general relativity are also homogeneous spaces; see \cite[chapter 12]{exact}. 

Homogeneous spaces appear whenever there are (gauge or spacetime) symmetry groups. Another example would be a Higgs potential spontaneously breaking a symmetry group $G$ down to a subgroup $H$; then the space of vacua would be $G/H$. 

\sect{Structure of this Thesis and Conventions}
\label{struktur}

The thesis consists of two main parts. 

The first part explores symmetries in relativity which are deformed or constrained. In chapter \ref{defspacetime} we introduce deformation theory of Lie algebras and investigate its application to global symmetries (isometries) of certain exact solutions to Einstein's equations in four dimensions. We will see how the group-theoretic viewpoint of spacetimes in terms of their isometry groups relates to physical properties such as causality violation. Section \ref{petrovsect} contains calculations and results taken from \cite{petrovpaper} and are the result of a collaboration with Gary Gibbons, except subsection \ref{causalsect} which is new. We omit those parts of \cite{petrovpaper} which are of less relevance to the discussion of symmetry groups. In chapter \ref{defgenrel} we turn to a study of local symmetries in relativity, namely those of the tangent space. We investigate how Cartan geometry, a generalisation of Riemannian geometry where the flat tangent space is replaced by an arbitrary homogeneous space, can be used to generalise the proposed deformed special relativity (DSR) to a ``deformed general relativity" in the same way that (pseudo-)Riemannian geometry describes general relativity as an extension of special relativity. We analyse the physical predictions of a particular example supposedly describing Minkowski space, and note that it involves a connection with torsion potentially leading to disastrous predictions of charge nonconservation. This analysis serves as another example of the interplay between an algebraic viewpoint on (infinitesimal) symmetries and physical effects. The material presented in chapter \ref{defgenrel} is mainly taken from \cite{dsrpaper} which again resulted from collaboration with Gary Gibbons. In chapter \ref{bflinear} we study yet another possible reformulation of classical general relativity as a topological BF theory with constraints. Such a formulation has been considered before and is of relevance in covariant approaches to quantising gravity, but we give a new version of it where the constraints are linear, removing some of the ambiguity in their solutions. We see how a very large symmetry on the phase space of the theory -- the local equivalence of all solutions to the equations of motion -- can be constrained to give general relativity, a theory with local degrees of freedom. The results detailed in chapter \ref{bflinear}, published in \cite{bflinearc}, were obtained in collaboration with Daniele Oriti.

The second part is concerned with gauge symmetries in particle physics and their influence on $CP$ violation in the electroweak sector. We aim to give estimates for the likelihood of the observed magnitude of $CP$ violation, measured by the Jarlskog invariant $J$, by constructing a natural measure on the space of Cabibbo-Kobayashi-Maskawa (CKM) matrices.
This space is a double quotient of $SU(3)$ and does not have a clearly preferred measure. After introducing different choices for the measure in section \ref{su3quot}, motivated first only by geometric and then also by physical considerations -- we will incorporate the observed quark masses into the analysis as well -- in chapter \ref{natural} we make a detailed study of their physical predictions. We find that once the quark masses are assumed, predictions for $CP$ violation agree well with observation. We then provide another example of how modified symmetries affect physical predictions by assuming an extended left-right symmetry given by a second $SU(2)$ gauge group. This will influence our predictions on $CP$ violation and lead to completely different results. The analysis for the standard model was done in collaboration with Gary Gibbons, Chris Pope and Neil Turok in \cite{cpviolation}. Much of the calculations presented in sections \ref{jstats} and \ref{gaussian} are taken from this paper; I have omitted some parts of \cite{cpviolation} where my original contribution has been limited. The calculations in subsection \ref{nonherm} are my own and were published in \cite{unnatural}. 

We close with a brief summary and outline possible directions for future research.

\

Our conventions for general relativity are those of Hawking \& Ellis; the signature of a Lorentzian metric is mainly positive; Greek letters are used for spacetime and Latin letters for internal indices. The curvature of a connection is always $F=dA+A\wedge A$ without any coupling constants as customary in some of the particle physics literature. In discussions of $CP$ violation in particle physics we follow the book of Jarlskog \cite{jarlskog}.

\part{Deformed and Constrained Symmetries in Gravity}

\chapter{Deformations of Spacetime Symmetries}
\label{defspacetime}
\sect{Overview}

The archetypal examples of groups describing spacetime symmetries in physics are the Galilean and Lorentz groups of classical mechanics and special relativity. The crucial insight of Einstein in 1905 was the necessity of replacing the Galilean group by the Lorentz group in order to have an invariant speed of light $c$; as we will see shortly, this process is also the archetypal example of a {\bf Lie algebra deformation}. Historically, the development of a mathematical theory of deformation and contraction of Lie algebras was motivated by advances in theoretical physics, but it is at least conceivable that the replacement of the Galilean group by the Lorentz group had been suggested by mathematicians working on Lie group theory, instead of physicists like Einstein.

Since physical theories are to some extent always idealised approximations of a much more complex Nature, the process of deformation of a symmetry group allows one to make statements about the stability of a given theory\index{stability of theories}. This stability point of view was explored in detail in \cite{vilela}. The passage from non-relativistic to special-relativistic mechanics can be understood as a deformation of an ``unstable" to a ``robust stable" theory \cite{vilela}. The point here is that the zero commutators of Galilean boosts constitute ``fine-tuning'' and any perturbation of the commutation relations leads to noncommuting boosts. On the other hand, given noncommuting boosts, a slight perturbation of the coefficient appearing in the commutator can be undone by redefining the algebra generators, hence one has ``robustness" in the theory.

In \cite{vilela}, the possible central extension of the Poisson algebra of the phase-space coordinates of Hamiltonian classical mechanics, leading to quantum mechanics, was interpreted as a similar instability of the original algebra. It was then argued that although the resulting Heisenberg algebra admits a deformation, it can be undone by a nonlinear redefinition of the position coordinate. 

In this chapter, we shall focus on Lie algebra deformations\index{Lie group/algebra deformation}, perturbations of the structure constants which can be expanded in a power series. We are maintaining the structure of a Lie algebra, \ie of a vector space, and redefinitions of the generators will always be linear transformations. One could consider more general kinds of deformations, where the Lie algebra structure is lost, either by allowing commutators of two generators to be nonlinear in the generators or by allowing ``structure functions".

While the existence or non-existence of deformations of a given Lie algebra is a mathematically well-posed question, the physical interpretation of its answer is less clear in general. Both the generators and the deformation parameters of a given Lie algebra may have different possible interpretations. For example, while the Poincar\'e group admits a deformation to the de Sitter or anti de Sitter groups, it was argued in \cite{vilela} that the deformation parameter $R$, defining a length scale, may be taken to be infinity if one is only interested in local kinematics, which of course is what happens in general relativity. It is also clear that while linear transformations can formally be applied to the generators in the Lie algebra, it may not appear physically meaningful to take linear combinations of, for instance, translation and rotation generators in the Poincar\'e algebra. This point was stressed in \cite{okon}, who also pointed out that the interpretation of position operators appearing in a ``kinematical algebra" acting on single- or multi-particle quantum states is physically rather unclear, since position is not additive in a way that momentum and angular momentum are.

The classic cases of the de Sitter group and its contractions were studied in detail in \cite{bacry}, as we will review shortly. One could argue that the mathematical theory behind deformations seems to have predictive power in terms of extensions or generalisations of fundamental theories, but these are only {\it a posteriori} predictions \cite{okon}. There seem to be two obvious classes of new applications for the concept of deformations of spacetime symmetries:

\begin{itemize}

\item Situations where the relevant symmetry group is a subgroup of the Poincar\'e group, or more generally a group of lower dimension than the Poincar\'e group. One example for the former is the proposed ``very special relativity" \cite{cohenglashow}, where the fundamental symmetry group is taken to be ISIM$(n-2)$, consisting of the maximal subgroup of the Lorentz group leaving a null direction invariant plus translations. The possibility of deforming this group to obtain ``general very special relativity" was discussed in \cite{finsler}; a physically acceptable one-parameter deformation was found, similar to the case of the Poincar\'e group which can be deformed into the (anti-)de Sitter group. Interestingly enough, this deformed group turns out to be the symmetry group of an asymptotically AdS spacetime playing a role in the construction of nonrelativistic hydrodynamics in the context of AdS/CFT\index{AdS/CFT}: The ``Schr\"odinger spacetime" \cite{adscft}\index{Schr\"odinger spacetime}
\ben
ds^2 = r^2\left(-2\,du\,dv-r^{2\nu}du^2+d\vec{x}^2\right)+\frac{dr^2}{r^2}
\een 
can be viewed as a left-invariant metric on a group manifold which is a subgroup of the deformed group ${\rm DISIM}_b(n-2)$\cite[app. B]{maxwellsim}.

On a less fundamental level, given a homogeneous space $G/H$ with isometry group $G$, \eg as a given solution of general relativity, one may always ask what deformations its isometry group admits, and try to construct spaces with the deformed symmetry. The two examples we give in this chapter describe this kind of situation. In principle, this could be used to facilitate the search for new solutions of general relativity, although one is restricted to homogeneous spaces, and curvature invariants are constant. One may still obtain spaces with interesting causal properties, as exemplified by the BTZ black hole \cite{steif} (obtained from anti-de Sitter space by periodic identifications), and by the closed timelike curves\index{closed timelike curves} that we will see occur in the Petrov and Kaigorodov-Ozsv\'ath solutions.

\item Alternatively, one may assume that the symmetry group is larger than the Poincar\'e group. In particular, the formalism can be straightforwardly extended to supergroups \cite{superfarrill}. It was found in \cite{superfarrill} that the Killing superalgebra of the M2-brane in eleven dimensions admits a deformation, which leads to the conjecture that a solution with the deformed Killing superalgebra exists, possibly having an interpretation as a perturbation of the M2-brane. Unfortunately, the search for such a solution does not seem to have been successful\footnote{Jos\'e Figueroa-O'Farrill, private communication (May 2009)}. 

Another example is Peter West's conjecture of $E_{11}$ as being the relevant symmetry group for M-theory\index{M-theory} \cite{west}, where one could look for possible deformations.

\end{itemize}

As outlined in section \ref{lie}, the general setup for our considerations is a manifold $X$ which a group of transformations $G$ acts on transitively, which may then be identified with $G/H$ where $H$ is a subgroup of $G$.

When using deformations to look for interesting manifolds ``close to" a given one, one considers group deformations of the group $G$ and then identifies the appropriate stabiliser $H$ in the deformed group. It simplifies the interpretation of the manifold one obtains if this stabiliser can be viewed as a deformation of the stabiliser subgroup of the original undeformed group, but in general this need not be the case \cite{finsler}.

By definition, the deformation theory reviewed here can only make statements about Lie algebras. As explained in section \ref{lie}, Lie's theorems allow us to use the terms ``Lie algebra" and ``Lie group" interchangeably, meaning by ``Lie group" the unique connected, simply connected group specified by a given algebra. In particular, terms such as ``Poincar\'e group" or ``de Sitter group" refer to connected components of the identity. 

We review the mathematical theory of Lie algebra deformations in section \ref{deftheory}, and then give two examples for pairs of solutions of general relativity related by such deformations: The Petrov and Kaigorodov-Ozsv\'ath solutions are discussed in section \ref{petrovsect}. We shall see that they are both geodesically complete, and try to relate their causal properties. We then turn to the Kaigorodov solution in section \ref{kaigorodovsect}, and find it can be deformed into a solution known as a ``Lobatchevski plane gravitational wave".

\sect{Deformation Theory}
\label{deftheory}

The theory of deformation of Lie algebras was reviewed in \cite{levy}, where it was connected to the perhaps more familiar operation of Lie algebra (Inonu-Wigner \cite{inonu}) contraction. A Lie algebra $\frak{g}$\index{Lie group/algebra deformation} is deformed by redefining the Lie brackets as a power series in a parameter $t$
\ben
f_t(a,b)=[a,b]+tF_1(a,b)+t^2 F_2(a,b)+\ldots,\quad a,b\in\frak{g}\,,
\een
where the series is required to converge in some neighbourhood of the origin. The Jacobi identity\index{Jacobi identity} then leads to integrability conditions on the functions $F_i$ at each order in $t$:
\ben
\sum_{\mathcal{P}(a,b,c)}\sum_{\mu+\nu=n}F_{\mu}(F_{\nu}(a,b),c)+F_{\nu}(F_{\mu}(a,b),c)\stackrel{!}{=}0\,,
\label{intcond}
\een
where the sum is over all cyclic permutations and $n=0,1,2,\ldots$ 

Choosing $n=0$ gives back the Jacobi identity for $[\cdot,\cdot]$, and at linear order ($n=1$) one gets
\ben
\sum_{\mathcal{P}(a,b,c)}F_1([a,b],c)+[F_1(a,b),c]\stackrel{!}{=}0\,.
\een
A deformation that only corresponds to a change of basis in the algebra $\frak{g}$ will be regarded as trivial; these are of the form
\ben
f(a,b)=S[S^{-1}a,S^{-1}b]\,,\quad S\in GL(n)\,,
\een
and expanding $S$ in a power series around the identity gives the general form of trivial deformations at each order in $t$. 

It is perhaps more common to express these operations in terms of {\it structure constants}\index{structure constants} ${{C_a}^c}_b$, as defined in (\ref{structcon}) for a given basis. Denoting the structure constants of the deformed algebra, with respect to the same basis, by ${{\hat{C}_a}\,^c}_{b}$, the Jacobi identity for the deformed algebra can be written as
\ben
{{\hat{C}_d}\,^e}_{[a}{{\hat{C}_b}\,^d}_{c]}=0\,.
\label{jacobi}
\een
One can then contemplate $\R^{n^2(n-1)/2}$ (where $n$ is the dimension of $\frak{g}$) parametrised by coordinates ${{\hat{C}_a}\,^c}_b$ antisymmetric in the lower indices, and the submanifold of $\R^{n^2(n-1)/2}$ defined by the Jacobi identity. A deformation of the given algebra $\frak{g}$ describes a smooth curve in this submanifold which can be expanded in a power series in a parameter $t$:
\ben
{{\hat{C}_a}\,^c}_b (t)={{C_a}^c}_b+t{{A_a}^c}_b+t^2{{B_a}^c}_b+\ldots
\een
At linear order in $t$ the Jacobi identity (\ref{jacobi}) holds if
\ben
{{C_d}^e}_{[a}{{A_b}^d}_{c]}+{{A_d}^e}_{[a}{{C_b}^d}_{c]}=0\,.
\label{jaclin}
\een
Then a linear deformation only gives rises to a deformation if the requirement (\ref{jacobi}) can be satisfied at each order in $t$.

Under a change of basis in $\frak{g}$, expressed as a matrix $S$, the structure constants will change according to
\ben
{{\hat{C}_a}\,^b}_c (t)={S^b}_e{{C_d}^e}_f{(S^{-1})^d}_a{(S^{-1})^f}_c\,.
\een
Expanding ${S^a}_b(t)=\delta^a_b+t{M^a}_b+\ldots$, this means that to first order a trivial deformation can be written as
\ben
{{A_a}^b}_c={M^b}_e{{C_a}^e}_c-{{C_e}^b}_c{M^e}_a-{{C_a}^b}_e{M^e}_c\,.
\label{nontriv}
\een

One can rephrase these conditions in the language of differential forms by choosing a basis $\{\lambda^a\}$ of left-invariant 1-forms for the original algebra satisfying the Maurer-Cartan relations\index{Maurer-Cartan relations} $d\lambda^a=-\frac{1}{2}{{C_b}^a}_c\lambda^b\wedge\lambda^c$, and also defining $A^a=\frac{1}{2}{{A_b}^a}_c\lambda^b\wedge\lambda^c$ to be a vector-valued 2-form and ${C^a}_b={{C_c}^a}_b\lambda^c$ to be a matrix-valued 1-form \cite{finsler}. Here ${{C_b}^a}_c$ are the structure constants of the original algebra and ${{A_b}^a}_c$ is their infinitesimal deformation. Then at linear order, the Jacobi identity is
\ben
DA:=dA+C\wedge A=0\,,
\een
and a deformation that can be written as $A=DS$ for some $S={S^a}_b\lambda^b$ is regarded as trivial. Hence non-trivial infinitesimal deformations is in one-to-one correspondence with the cohomology group $H^2(\frak{g},\frak{g})$ \cite{farrill}. At higher order one has to consider the higher cohomology\index{cohomology} groups. In general, if a non-trivial deformation has been found infinitesimally (\ie at linear order), it does not necessarily extend to a deformation satisfying the full set of conditions (\ref{intcond}); obstructions to ``integrability" of infinitesimal deformations are given by the cohomology group $H^3(\frak{g},\frak{g})$ \cite{farrill}. In practice, once all infinitesimal deformations have been found, one can verify the full Jacobi identity, and if it is violated, try to add higher order terms to the deformation.

Using cohomology theory, one can determine all general deformations of a given algebra, as was done for the Galilean algebra, with and without central extension, in \cite{farrill}. We will focus on deformations at linear order in the following.

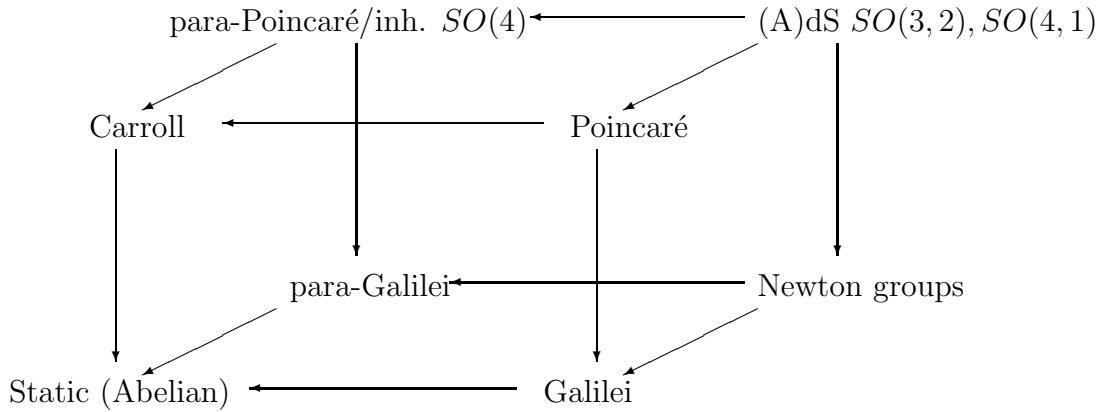
\begin{figure}[htp]
\centering
\begin{picture}(300,180)
\put(250,160){(A)dS $SO(3,2), SO(4,1)$}\put(245,165){\vector(-1,0){80}}\put(30,160){para-Poincar\'e/inh. $SO(4)$}
\put(280,155){\vector(0,-1){80}}\put(245,65){\vector(-1,0){110}}\put(100,155){\vector(0,-1){80}}
\put(250,60){Newton groups}\put(75,60){para-Galilei}
\put(250,155){\vector(-2,-1){50}}\put(70,155){\vector(-2,-1){50}}\put(250,55){\vector(-2,-1){50}}\put(70,55){\vector(-2,-1){50}}
\put(180,120){Poincar\'e}\put(0,120){Carroll}\put(190,115){\vector(0,-1){80}}\put(10,115){\vector(0,-1){80}}
\put(170,20){Galilei}\put(-30,20){Static (Abelian)}\put(160,25){\vector(-1,0){100}}\put(170,125){\vector(-1,0){120}}
\end{picture}
\caption[Possible contractions and deformations of kinematical groups.]{{\small Possible contractions and deformations of kinematical groups, according to \cite{bacry}. Arrows denote possible Inonu-Wigner \cite{inonu} contractions.}}
\label{contract}
\end{figure}

Applying this formalism, the authors of \cite{bacry} classified all possible kinematical groups\index{kinematical group} in four dimensions, by which they meant classifying ten-dimensional Lie algebras consisting of ``translations", ``rotations", and ``inertial transformations". This splitting meant singling out three generators (``rotations") and fixing their commutators with the other generators to assure isotropy of space. They further assumed that parity and time-reversal (defined on the generators) leave the Lie brackets invariant, and that the subgroup of inertial transformations is non-compact. The classification one then obtains consists of eight types of Lie algebras, all of which may be obtained by either contracting the de Sitter algebra $\frak{so}(4,1)$ or anti-de Sitter algebra $\frak{so}(3,2)$, or alternatively as deformations of the Abelian algebra, as shown in figure \ref{contract}. The possible contractions correspond to different physical limits such as vanishing cosmological constant, infinite speed of light or zero speed of light, as explained in detail in \cite{bacry}. For instance, the Newton groups describe the symmetries of a non-relativistic cosmological model with nonzero cosmological constant; this was discussed in detail in \cite{patricot}.

Looking at figure \ref{contract}, it could be argued that the introduction of noncommuting translations, the most direct effect of a curved spacetime, is very natural from the group deformation viewpoint as the de Sitter and anti-de Sitter groups arise as deformations of the Poincar\'e group.

For completeness, we mention that in order to obtain the algebra of quantum mechanics from a deformation of that of classical mechanics, as discussed in \cite{vilela}, one either deforms the Poisson algebra of functions on phase-space, replacing the Poisson bracket by a Moyal bracket\index{Moyal bracket}
\ben
\{f,g\}_M=\{f,g\}-\frac{\hbar^2}{4\cdot 3!}\sum_{{{i_1,i_2,i_3}\atop{j_1,j_2,j_3}}}\omega^{i_1 j_1}\omega^{i_2 j_2}\omega^{i_3 j_3}\partial_{i_1 i_2 i_3}(f)\partial_{j_1 j_2 j_3}(g)+\ldots\,,
\een
or alternatively considers the phase-space coordinates as elements of an Abelian Lie algebra which is replaced by the Heisenberg algebra. In the first case one has an infinite-dimensional algebra, in the second case one performs a central extension together with a deformation. The very illuminating discussion in \cite{vilela} used purely mathematical arguments as well as physical intuition to establish the stability of the Heisenberg algebra, considered as a test for the ``robustness" of quantum mechanics. We stress again that this is a different notion of deformation, since one is now considering non-linear transformations of the generators.

The maximal algebra one could consider in this discussion is the algebra of the Poincar\'e group, together with ``spacetime coordinates" and an ``identity" (a central element) which appear as operators in quantum mechanics. In four dimensions, this is a 15-dimensional group which is naturally deformed into $\frak{so}(6-t,t)$, where the original central element ceases to commute with the other generators in general. This maximally extended case is most obviously fraught with interpretational difficulties, as clearly illustrated in \cite{okon}. Formally, the deformation process naturally leads to non-commuting spacetime coordinates \cite{vilela}. This proposal goes back to Snyder\index{Snyder's non-commuting coordinates} \cite{snyder}, whose algebra is however not the same as the one proposed in \cite{vilela}. The idea of ``spacetime being non-commutative" is very interesting physically, however since it is separate from the deformation theory of spacetime symmetries reviewed in this chapter, it will be discussed in more detail in chapter \ref{defgenrel}.

\sect{The Petrov and Kaigorodov-Ozsv\'ath Solutions}
\label{petrovsect}

Among the wide range of known exact solutions to Einstein's field equations (the most comprehensive source is \cite{exact}), there are some with interesting symmetry properties. In this section we concentrate on the particularly simple case of a simply transitive group of motions, such that the stabiliser is trivial and there is a one-to-one correspondence between spacetime points and elements of the group of motions. Therefore the spacetime is not only a manifold but also a Lie group. One example, and in fact the only example among vacuum solutions without a cosmological constant, is provided by the Petrov solution (\ref{petrov}). An analogous example with negative cosmological constant is provided by the Kaigorodov-Ozsv\'ath solution (\ref{lambda}).

The Petrov solution has a physical interpretation as the exterior solution of an infinite rigidly rotating cylinder. To gain an understanding of the physical properties of the Kaigorodov-Ozsv\'ath solution, we compute the stress-energy tensor of the boundary theory in the context of the AdS/CFT correspondence from an expansion near the conformal boundary. We then give a matrix representation of the isometry group of the Petrov solution and its algebra and use it to construct left-invariant one-forms on the group which could be used to compute left-invariant metrics.

We apply the theory of Lie algebra deformations to relate the Petrov solution to the Kaigorodov-Ozsv\'ath solution, and discuss the physical interpretation of this result. It is well known that the Petrov solution contains closed timelike curves (CTCs), so an obvious question is whether the Kaigorodov-Ozsv\'ath solution exhibits similarly exotic properties. Since in the context of possible causality violation by CTCs a central question is whether these can be created by some process in a spacetime which did not exhibit CTCs initially, we give a brief discussion of the possible appearance of CTCs by spinning up a rotating cylinder. This elementary example supports Hawking's chronology protection conjecture which asserts that the appearance of CTCs is forbidden by the laws of physics \cite{hawking92}.

In a more detailed analysis of the physical properties of the Petrov and Kaigorodov-Ozsv\'ath solutions, we show that they are geodesically complete. We discuss the global causality properties of the two spaces; while the Petrov solution is totally vicious, due to a theorem of Carter, a similar analysis for the Kaigorodov-Ozsv\'ath solution remains inconclusive. The computation of the stress-energy tensor in AdS/CFT however suggests causal pathologies in the Kaigorodov-Ozsv\'ath solution as well.

\subsection{Vacuum Solutions With Simply-Transitive Groups of Motions}

\subsubsection*{$\Lambda=0$ - The Petrov Solution}
The Petrov solution\index{Petrov solution} is introduced in \cite{exact} in the following theorem: {\it The only vacuum
solution of Einstein's equations admitting a simply-transitive
four-dimensional maximal group of motions is given by}
\ben
k^2
ds^2=dr^2+e^{-2r}dz^2+e^r(\cos\sqrt{3}r(d\phi^2-dt^2)-2\sin\sqrt{3}r\,d\phi\,dt)\,,
\label{petrov}
\een
where $k$ is an arbitrary constant, which shall be set equal to
one, and we have relabelled the coordinates compared to
\cite{exact}. The solution was first given in \cite{petrov} and also discussed in \cite{debever}. It describes a hyperbolic plane $H^2$ (the $(r,z)$-plane) with a timelike two-plane attached to each point.

The isometry group is generated by the Killing vector fields\index{Killing vector}
\ben
T\equiv\frac{\partial}{\partial t}\,,\;\Phi\equiv\frac{\partial}{\partial {\phi}}\,,\;Z\equiv\frac{\partial}{\partial z}\,,\;R\equiv\frac{\partial}{\partial r}+z\frac{\partial}{\partial z}+\half(\sqrt{3}t-\phi)\frac{\partial}{\partial {\phi}}-\half(t+\sqrt{3}\phi)\frac{\partial}{\partial t}\,,
\label{killing}
\een
which satisfy the algebra
\ben
[R,T]=\half
T-\frac{\sqrt{3}}{2}\Phi\,,\quad [R,\Phi]=\half\Phi+\frac{\sqrt{3}}{2}T\,,\quad[R,Z]=-Z\,.
\label{algebra}
\een
The isometry group contains three-dimensional subgroups of Bianchi
types I and ${\rm VII}_h$ \index{Bianchi classification}acting on timelike hypersurfaces, and the
solution (\ref{petrov}) is Petrov type I \index{Petrov classification}\cite{exact}. The first three Killing vectors obviously generate translations while the action of the one-parameter subgroup generated by $R$ on spacetime is given by the integral curves of $R$, solutions of
\ben
\frac{dx^{\mu}(\lambda)}{d\lambda}=R^{\mu}(x(\lambda))\,.
\een
These integral curves have the form
\bea
x^{\mu}(\lambda)& = &\left(r_0+\lambda\,,z_0e^{\lambda}\,,\phi_0e^{-\frac{\lambda}{2}}\cos\frac{\sqrt{3}}{2}\lambda+t_0e^{-\frac{\lambda}{2}}\sin\frac{\sqrt{3}}{2}\lambda\,,\right.\nn
\\& & \left.-\phi_0e^{-\frac{\lambda}{2}}\sin\frac{\sqrt{3}}{2}\lambda+t_0e^{-\frac{\lambda}{2}}\cos\frac{\sqrt{3}}{2}\lambda\right)\,,
\label{integralcurves}
\eea
where we label the coordinates by $x^{\mu}=(r,z,\phi,t)$.

The metric components $g_{\phi\phi}$ and $g_{tt}$ become zero at certain values\footnote{these values of $r$ can be shifted by a coordinate transformation $\phi\rightarrow\alpha\phi+\beta t,\;t\rightarrow-\beta \phi+\alpha t$ and hence have no coordinate-independent significance} of $r$, but as the determinant of the metric in (\ref{petrov}) is always $-1$, it
is possible to extend the coordinates to infinite ranges and the coordinates ($r,z,\phi,t$) define a global chart. The manifold is also time-orientable\index{time-orientable}, as the vector field $t^{\mu}=(0,0,\sin\frac{\sqrt{3}}{2}r,\cos\frac{\sqrt{3}}{2}r)$ defines a global arrow of time, though this amounts to $\frac{\partial}{\partial t}$ being future- as well as past-directed at some points (as well as spacelike at others).

Bonnor \cite{bonnor} pointed out that the solution can be viewed as a special case of the exterior part of a Lanczos-van Stockum solution \cite{lanczos, vanstockum} describing an infinite cylinder\index{cylinder, rotating} of rigidly rotating dust. Since this allows a physical interpretation of (\ref{petrov}), let us give the general solution for
an infinite rigidly rotating dust cylinder, which in Weyl-Papapetrou form is given by \cite{tipler74}
\ben
ds^2=H(\rho)(d\rho^2+d\tilde{z}^2)+L(\rho)d\chi^2+2M(\rho)\,d\chi\,d\tau-F(\rho)d\tau^2,
\label{vanstockum}
\een
where $H,\;L,\;M,\;F$ are functions of the radial variable $\rho$
containing two parameters $a$ and $R$, interpreted as the angular
velocity and radius of the cylinder respectively.

The high energy case $aR>\frac{1}{2}$, which contains closed
timelike curves\index{closed timelike curves} (CTCs), is of interest here, and with the choices
$R=\sqrt{e}$ and $aR=1$ the exterior solution is given by
\bena
& & H(\rho)=\frac{1}{\rho^2}\,,\quad L(\rho)=-2\sqrt{\frac{e}{3}}\rho\sin\left(\sqrt{3}\log\frac{\rho}{\sqrt{e}}\right)\,,
\\& & M(\rho)=\frac{2}{\sqrt{3}}\rho\sin\left(\frac{\pi}{3}+\sqrt{3}\log\frac{\rho}{\sqrt{e}}\right)\,,\quad F(\rho)=\frac{2}{\sqrt{3e}}\rho\sin\left(\frac{\pi}{3}-\sqrt{3}\log\frac{\rho}{\sqrt{e}}\right)\,,\nn
\eena
where $\rho$ is a radial coordinate in the exterior of the cylinder and so is restricted to 
\ben
\rho\ge\sqrt{e}\,, 
\label{rhobig}
\een
and $\chi$ is an angular coordinate and periodically identified with period $2\pi$; $\tau$ and $\tilde{z}$ are unconstrained. 
Applying the coordinate transformations
\ben
\tilde z=\sqrt{e}\,z\,,\;\rho=\sqrt{e}\,e^r\,,\;\chi=\frac{1}{\sqrt[4]{3}\sqrt{2e}}\left(\sqrt{2+\sqrt{3}}z+\sqrt{2-\sqrt{3}}t\right)\,,\tau=\frac{1}{\sqrt[4]{3}}(z-t)
\een
to the line element (\ref{vanstockum}) indeed gives back (\ref{petrov}). Hence if we adopt the interpretation of (\ref{petrov}) as describing the exterior of a spinning cylinder, we restrict the coordinates to $r\ge 0$ and identify $\sqrt{2+\sqrt{3}}z+\sqrt{2-\sqrt{3}}t$ with $\sqrt{2+\sqrt{3}}z+\sqrt{2-\sqrt{3}}t+2\pi\sqrt[4]{3}\sqrt{2e}$.

The general Lanczos-van Stockum solution has three linearly independent Killing vectors $\frac{\partial}{\partial {\chi}},\;\frac{\partial}{\partial {\tau}}$ and $\frac{\partial}{\partial z}$ but a fourth Killing vector is only present in the special case $aR=1$ because the algebraic invariants of the Riemann tensor are independent of $\rho$ just in this case \cite{bonnor80}.

\subsubsection*{$\Lambda<0$ - The Kaigorodov-Ozsv\'ath Solution}
A solution of the vacuum Einstein equations with {\it negative} cosmological constant which has a simply-transitive four-dimensional group of motions was first given by Kaigorodov \cite{kaigorodov} and rediscovered by Ozsv\'ath \cite{ozsvath}\index{Kaigorodov-Ozsv\'ath solution}. It has the line element
\ben
ds^2=-\frac{3}{\Lambda}dr^2+e^{-2r}(dz^2+2\,dt\,d\phi)+e^{4r}d\phi^2-2\sqrt{2}e^r\, dz\,d\phi\,.
\label{lambda}
\een
This solution was also given in \cite{exact}. It is Petrov type III\index{Petrov classification} and the metric asymptotically (as $r\rightarrow -\infty$) approaches that of anti-de Sitter space. Obvious Killing vectors\index{Killing vector} are 
\ben
Z\equiv\frac{\partial}{\partial z}\,,\;\Phi\equiv\frac{\partial}{\partial {\phi}}\,,\;T\equiv\frac{\partial}{\partial t}\,,
\een
and the metric (\ref{lambda}) has a further isometry
\ben
r\rightarrow r+\lambda,\;z\rightarrow e^{\lambda}z,\;\phi\rightarrow e^{-2\lambda}\phi,\;t\rightarrow e^{4\lambda}t
\een
which is generated by the fourth Killing vector $R\equiv\frac{\partial}{\partial r}+z\frac{\partial}{\partial z}-2\phi\frac{\partial}{\partial {\phi}}+4 t\frac{\partial}{\partial t}$. The Killing vector fields satisfy the algebra
\ben
[R,Z]=-Z\,,\;[R,\Phi]=2\Phi\,,\;[R,T]=-4 T\,.
\label{algnew}
\een
No analogous solution for a positive cosmological constant exists \cite{ozsvath, exact}. Because of the similarity to (\ref{algebra}) one would expect (\ref{algnew}) to arise as a deformation of (\ref{algebra}). Physically both algebras describe the isometries of vacuum solutions of Einstein's equations, one with $\Lambda<0$ and one with $\Lambda=0$, and hence one might expect one of them to arise as some limit of the other.

This spacetime is also time-orientable\index{time-orientable}, as the vector field $t^{\mu}=(0,-\frac{1}{\sqrt{2}}e^{3r},-1,1)$ defines a global arrow of time.

We can express (\ref{lambda}) in coordinates corresponding to Poincar\'e coordinates on AdS (with $\Lambda=-3$)
\ben
ds^2=\frac{d\rho^2+dz^2+d\phi^2-dt^2}{\rho^2}-2\rho\,dz(dt+d\phi)+\half\rho^4 (dt+d\phi)^2\,.
\label{poincare}
\een

The limit $\rho\rightarrow 0$ in Poincar\'e coordinates corresponds to the timelike boundary $\mathcal{I}$ of anti-de Sitter space. After setting $\tilde\rho=\rho^2$ the line element is
\ben
ds^2=\frac{d\tilde{\rho}^2}{4\tilde{\rho}^2}+\frac{1}{\tilde\rho}\left(dz^2+d\phi^2-dt^2-2\tilde{\rho}^{3/2}\,dz(dt+d\phi)+\half\tilde{\rho}^3 (dt+d\phi)^2\right)\,,
\label{poincare2}
\een
and (\ref{poincare2}) is an expansion of the form
\ben
ds^2=\frac{d\tilde{\rho}^2}{4\tilde{\rho}^2}+\frac{1}{\tilde\rho}g_{\mu\nu}dx^{\mu} dx^{\nu}\,,\quad g_{\mu\nu}(x,\tilde{\rho})=g_{\mu\nu}^{(0)}(x)+g_{\mu\nu}^{(2)}(x)\tilde{\rho}+g_{\mu\nu}^{(3)}(x)\tilde{\rho}^{3/2}+\ldots
\label{stresst}
\een
as given in \cite{anomaly}. The coefficient $g^{(3)}=-dz\otimes(dt+d\phi)-(dt+d\phi)\otimes dz$ encodes the stress energy tensor of the boundary dual theory in the context of the AdS/CFT\index{AdS/CFT} correspondence \cite{subra,anomaly}. It does not satisfy even the null energy condition on the three-dimensional conformal boundary, since $g_{\mu\nu}^{(3)}n^{\mu} n^{\nu}=-2$ for the null vector $n=\frac{\partial}{\partial t}+\frac{\partial}{\partial z}$. This presumably reflects causal pathologies of the bulk spacetime that we will try to understand by more conventional means shortly.

A general analysis of stationary cylindrically symmetric Einstein spaces was done in \cite{maccallum}. These authors assume the Lewis form\index{Lewis form} of the metric, where a cross term $dz\,d\phi$ would be absent. We did not find it possible
to bring (\ref{lambda}) to the Lewis form. The theorem by Papapetrou \cite{papapetrou} that any solution with two commuting Killing vectors (one timelike, one spacelike with periodic orbits) can be written in the Lewis form, only applies to solutions of the vacuum Einstein equations without cosmological term. Hence one cannot make a connection with spaces of the Lewis form as was possible for the Petrov solution.

\subsection{Left-Invariant Forms}

Since the action of the elements of the group manifold on itself has been given in (\ref{integralcurves}), we can write down a matrix representation of this group of motions, with a general element given by
\ben
g=\left(\begin{matrix}1&0&0&0&r \cr 0&e^{r}&0&0&z \cr
  0&0&e^{-\frac{r}{2}}\cos(\frac{\sqrt{3}}{2}r)&e^{-\frac{r}{2}}\sin(\frac{\sqrt{3}}{2}r)&\phi
  \cr
  0&0&-e^{-\frac{r}{2}}\sin(\frac{\sqrt{3}}{2}r)&e^{-\frac{r}{2}}\cos(\frac{\sqrt{3}}{2}r)&t
  \cr 0&0&0&0&1\end{matrix}\right)\,.
\label{paramet}
\een
The group is generated by
\ben
Z=\left(\begin{matrix}0&0&0&0&0 \cr 0&0&0&0&1 \cr 0&0&0&0&0 \cr 0&0&0&0&0
  \cr 0&0&0&0&0\end{matrix}\right)\,,\;\Phi=\left(\begin{matrix}0&0&0&0&0 \cr 0&0&0&0&0 \cr 0&0&0&0&1 \cr 0&0&0&0&0
  \cr 0&0&0&0&0\end{matrix}\right)\,,
\een
\ben
T=\left(\begin{matrix}0&0&0&0&0 \cr 0&0&0&0&0 \cr 0&0&0&0&0 \cr 0&0&0&0&1
  \cr 0&0&0&0&0\end{matrix}\right)\,,R=\left(\begin{matrix}0&0&0&0&1 \cr 0&1&0&0&0 \cr 0&0&-\frac{1}{2}&\frac{\sqrt{3}}{2}&0 \cr 0&0&-\frac{\sqrt{3}}{2}&-\frac{1}{2}&0
  \cr 0&0&0&0&0\end{matrix}\right)\,,
\een
so that $g=e^{zZ+\phi\Phi+tT}e^{rR}$ and $(r,z,\phi,t)$ are coordinates on the group. The generators satisfy the algebra
\ben
[R,T]=-\frac{1}{2}T+\frac{\sqrt{3}}{2}\Phi\,,\quad
  [R,\Phi]=-\frac{1}{2}\Phi-\frac{\sqrt{3}}{2}T\,,\quad [R,Z]=Z\,.
\label{genalg}
\een
This differs from the Killing algebra (\ref{algebra}) by the usual overall minus sign coming from the fact that right-invariant vector fields generate left actions and vice versa. 

The matrix representation (\ref{paramet}) gives the group multiplication law
\ben
(r,z,\phi,t)\cdot(r',z',\phi',t')=\left(r+r',z+e^r z',\phi+e^{-\frac{r}{2}}(t's+\phi'c),t+e^{-\frac{r}{2}}(t'c-\phi's)\right)\,,
\label{multlaw}
\een
where $s\equiv\sin\frac{\sqrt{3}}{2}r,\;c\equiv\cos\frac{\sqrt{3}}{2}r$. The Maurer-Cartan form\index{Maurer-Cartan form} is
\ben
g^{-1}dg=e^{-rR}(Z\,dz+\Phi\, d\phi+T\,dt)e^{rR}+R\,dr\equiv R\,\lambda^1+ Z\,\lambda^2+\Phi\,\lambda^3+T\,\lambda^4\,,
\label{maurer}
\een
which gives the desired basis of left-invariant one-forms\index{left-invariant one-form}:
\bena
& \lambda^1=dr\,,\;\lambda^2=e^{-r}\,dz\,,\;\lambda^3=e^{\frac{r}{2}}\left(\cos\left(\frac{\sqrt{3}}{2}r\right)\,d\phi-\sin\left(\frac{\sqrt{3}}{2}r\right)\,dt\right)\,,\nn
\\& \lambda^4=e^{\frac{r}{2}}\left(\sin\left(\frac{\sqrt{3}}{2}r\right)\,d\phi+\cos\left(\frac{\sqrt{3}}{2}r\right)\,dt\right)\,.\label{1forms}
\eena
We obtain a left-invariant metric\index{left-invariant metric} on the group
\bea
ds^2 & = & \eta_{ab}\lambda^{a}\otimes\lambda^{b}\nn
\\& = & dr^2+e^{-2r}dz^2+e^{r}\left(\cos(\sqrt{3}r)(d\phi^2-dt^2)-2\sin(\sqrt{3}r)d\phi\,dt\right)
\eea
with $\eta={\rm diag}(1,1,1,-1)$, which is the same as (\ref{petrov}) and shows that our chosen coordinates agree with the initial Petrov coordinates. We see how to recover a metric on a group manifold; note that one could obtain this metric by just starting from the algebra (\ref{algebra}).

\subsection{Deformations of the Petrov Killing Algebra}
We examine possible infinitesimal deformations of the four-dimensional Lie algebra. Using {\sc Mathematica} for the computations, equations
(\ref{jaclin}) give the following conditions on linear deformations\index{Lie group/algebra deformation}:
\bena
0 & = & {{A_z}^r}_{\phi}+\sqrt{3}{{A_t}^r}_z={{A_z}^r}_t-\sqrt{3}{{A_{\phi}}^r}_z={{A_{\phi}}^r}\nn
_t\,;\nn
\\0 & = & 2{{A_{\phi}}^r}_r+{{A_{\phi}}^z}_z+\sqrt{3}{{A_t}^z}_z=-\sqrt{3}{{A_{\phi}}^z}_z+2{{A_t}^r}_r+{{A_t}^z}_z={{A_{\phi}}^z}_t={{A_{\phi}}^r}_t\,;\nn
\\0 & = & {{A_r}^r}_z+2{{A_{\phi}}^{\phi}}_z+\sqrt{3}{{A_z}^t}_{\phi}-\sqrt{3}{{A_t}^{\phi}}_z=\sqrt{3}{{A_r}^r}_z+2{{A_z}^{\phi}}_t-\sqrt{3}{{A_z}^t}_t-\sqrt{3}{{A_{\phi}}^{\phi}}_z\nn
\\& = &\sqrt{3}{{A_r}^r}_{\phi}+{{A_t}^{\phi}}_{\phi}-\sqrt{3}{{A_{\phi}}^t}_t-{{A_t}^r}_r=\sqrt{3}{{A_z}^r}_{\phi}-{{A_t}^r}_z\,;\nn
\\0 & = &\sqrt{3}{{A_r}^r}_z-\sqrt{3}{{A_z}^{\phi}}_{\phi}-2{{A_z}^t}_{\phi}-\sqrt{3}{{A_t}^t}_z={{A_r}^r}_z-\sqrt{3}{{A_z}^{\phi}}_t+2{{A_t}^t}_z+\sqrt{3}{{A_{\phi}}^t}_z\nn
\\& = &{{A_r}^r}_{\phi}-\sqrt{3}{{A_{\phi}}^{\phi}}_t+\sqrt{3}{{A_t}^r}_r-{{A_t}^t}_{\phi}={{A_z}^r}_{\phi}+\sqrt{3}{{A_t}^r}_z\,.
\label{constr}
\eena
These constraints for ${{A_a}^c}_b$ reduce the number of free parameters from 24 to twelve. We list the most general deformation parameters satisfying (\ref{constr}) in a table:
\begin{table}[htp]
\caption{{\small Infinitesimal deformations of the Petrov Killing algebra.}}
\begin{center}
\begin{tabular}{c|c|c|c|c}
& $c=r$&$c=z$&$c=\phi$&$c=t$
\\\hline
${{A_r}^c}_z$&$2C$&$x_1$&$x_2$&$x_3$
\\\hline
${{A_r}^c}_{\phi}$&$-\sqrt{3}A-B$&$x_4$&$x_5$&$x_6$
\\\hline
${{A_r}^c}_t$&$-A+\sqrt{3}B$&$x_7$&$x_8$&$x_9$
\\\hline
${{A_z}^c}_{\phi}$&0&$2B$&$C$&$\sqrt{3}C$
\\\hline
${{A_z}^c}_t$&0&$2A$&$-\sqrt{3}C$&$C$
\\\hline
${{A_{\phi}}^c}_t$&0&0&$-\sqrt{3}B-A$&$B-\sqrt{3}A$
\end{tabular}
\end{center}
\end{table}
\\The parameters $x_1,x_2,\ldots,x_9,A,B$ and $C$ can be arbitrary real constants. We need to investigate which of these correspond to trivial deformations. The conditions (\ref{nontriv}) mean that trivial deformations can be written as
\ben
{{A_r}^r}_z=-{M^r}_z\,,\;{{A_r}^z}_z=-{M^r}_r\,,\;{{A_r}^{\phi}}_z=-\frac{3}{2}{M^{\phi}}_z+\frac{\sqrt{3}}{2}{M^t}_z\,,\;{{A_r}^t}_z=-\frac{\sqrt{3}}{2}{M^{\phi}}_z-\frac{3}{2}{M^t}_z\,.
\een
The parameters $C,x_1,x_2$ and $x_3$ correspond to trivial deformations and can be set to zero by a change of basis. Furthermore,
\bea
& & {{A_r}^r}_{\phi}=\frac{1}{2}{M^r}_{\phi}+\frac{\sqrt{3}}{2}{M^r}_t\,,\;{{A_r}^r}_t=-\frac{\sqrt{3}}{2}{M^r}_{\phi}+\frac{1}{2}{M^r}_t\,,\nn
\\& & {{A_z}^z}_{\phi}=-{M^r}_{\phi}\,,\;{{A_z}^z}_t=-{M^r}_t
\eea
etc., so that we can set $A=B=0$,
\bea
& & {{A_r}^z}_{\phi}=\frac{3}{2}{M^z}_{\phi}+\frac{\sqrt{3}}{2}{M^z}_t\,,\;{{A_r}^{\phi}}_{\phi}=-\frac{1}{2}{M^r}_r+\frac{\sqrt{3}}{2}{M^{\phi}}_t+\frac{\sqrt{3}}{2}{M^t}_{\phi}\,,\nn
\\& & {{A_r}^t}_{\phi}=-\frac{\sqrt{3}}{2}{M^r}_r-\frac{\sqrt{3}}{2}{M^{\phi}}_{\phi}+\frac{\sqrt{3}}{2}{M^t}_t\,,
\eea
so that we can set $x_4=x_5=x_6=0$,
\bea
& & {{A_r}^z}_t=\frac{3}{2}{M^z}_t-\frac{\sqrt{3}}{2}{M^z}_{\phi}\,,\;{{A_r}^{\phi}}_t=\frac{\sqrt{3}}{2}{M^r}_r-\frac{\sqrt{3}}{2}{M^{\phi}}_{\phi}+\frac{\sqrt{3}}{2}{M^t}_t\,,\nn
\\& & {{A_r}^t}_t=-\frac{1}{2}{M^r}_r-\frac{\sqrt{3}}{2}{M^{\phi}}_t-\frac{\sqrt{3}}{2}{M^t}_{\phi}\,,
\eea
so that we can set $x_7=0$, but must treat $x_8$ and $x_9$ as nontrivial perturbations. After a relabelling of these parameters, the modified Lie algebra is now
\ben
[R,T]=-aT-b\Phi\,,\quad  [R,\Phi]=-\frac{1}{2}\Phi-\frac{\sqrt{3}}{2}T\,,\quad [R,Z]=Z\,.
\een

These relations satisfy the full Jacobi identity and so the linear deformation indeed defines a deformation of the Lie algebra. We may modify the matrix representation by setting
\ben
R=\left(\begin{matrix}0&0&0&0&1 \cr 0&1&0&0&0 \cr 0&0&-\frac{1}{2}&-b&0 \cr 0&0&-\frac{\sqrt{3}}{2}&-a&0
  \cr 0&0&0&0&0\end{matrix}\right)\,.
\een

In the case where $a\neq\frac{1}{2}$, one can always find a linear transformation of the basis vectors $\Phi$
and $T$ such that the algebra takes the more symmetric form
\ben
[R,T]=a'T+b'\Phi\,,\quad  [R,\Phi]=a'\Phi\pm b'T\,,\quad [R,Z]=Z\,,
\een
with $\pm$ depending on the value of $b$ in the original deformation. This means that there
are three distinct cases: 

{\bf First case: Positive sign.} Then a matrix representation of
$R$ is
\ben
R=\left(\begin{matrix}0&0&0&0&1 \cr 0&1&0&0&0 \cr 0&0&a'&b'&0 \cr 0&0&b'&a'&0
  \cr 0&0&0&0&0\end{matrix}\right)
\een
and a general group element looks like
\ben
g=e^{zZ+\phi\Phi+tT}e^{rR}=\left(\begin{matrix}1&0&0&0&r \cr 0&e^{r}&0&0&z \cr
  0&0&e^{a' r}\cosh(b' r)&e^{a' r}\sinh(b' r)&\phi
  \cr
  0&0&e^{a' r}\sinh(b' r)&e^{a' r}\cosh(b' r)&t
  \cr 0&0&0&0&1\end{matrix}\right)\,.
\label{deformedgroup}
\een
The Maurer-Cartan form\index{Maurer-Cartan form} is
\bea
g^{-1}dg & = & Z\,e^{-r}\,dz+\Phi \,e^{-ar}(\cosh(b'r)\,d\phi-\sinh(b'r)\,dt)\nn
\\& & +T \,e^{-ar}(\cosh(b'r)\,dt-\sinh(b'r)\,d\phi)+R\, dr
\eea
and we can read off the left-invariant forms.

The special case $a'=1,\;b'=3$ gives the algebra of the Kaigorodov-Ozsv\'ath solution (\ref{lambda}), as can be seen by setting $\Phi'=\Phi-T$ and $T'=\Phi+T$, which amounts to
\ben
[R,\Phi]=\Phi+3T\,,\;[R,T]=3\Phi+T\quad\Rightarrow\;[R,\Phi']=-2\Phi'\,,\;[R,T']=4T'\,,
\een
which is just the sign-reversed version of (\ref{algnew}). One can recover the metric (\ref{lambda}) by choosing the symmetric matrix
\ben
h_{ab}=\left(\begin{matrix}-\frac{3}{\Lambda}&0&0&0\cr 0&1&-\sqrt{2}&\sqrt{2} \cr 0&-\sqrt{2}&-1&-1\cr 0&\sqrt{2}&-1&3\end{matrix}\right)
\label{newmatrix}
\een
and computing the left-invariant metric\index{left-invariant metric}
\ben
h_{ab}\lambda^a\otimes\lambda^b=-\frac{3}{\Lambda}dr^2+e^{-2r}dz^2-2e^{-2r}(d\phi^2-dt^2)+e^{4r}(d\phi-dt)^2-2\sqrt{2}e^r dz(d\phi-dt)\,,
\een
which after the coordinate transformations $\phi-t=\phi'$ and $-\phi-t=t'$ reduces to (\ref{lambda}). Since the matrix (\ref{newmatrix}) has one negative and three positive eigenvalues, there exists a (vierbein) basis of left-invariant one-forms $\sigma^a$ such that $\eta_{ab}\sigma^a\otimes\sigma^b$ gives the metric (\ref{lambda}).

{\bf Second case: Negative sign.}
\ben
R=\left(\begin{matrix}0&0&0&0&1 \cr 0&1&0&0&0 \cr 0&0&a'&b'&0 \cr 0&0&-b'&a'&0
  \cr 0&0&0&0&0\end{matrix}\right)\,,\;
g=e^{zZ+\phi \Phi+tT}e^{rR}=\left(\begin{matrix}1&0&0&0&r \cr 0&e^{r}&0&0&z \cr
  0&0&e^{a' r}\cos(b' r)&e^{a' r}\sin(b' r)&\phi
  \cr
  0&0&-e^{a' r}\sin(b' r)&e^{a' r}\cos(b' r)&t
  \cr 0&0&0&0&1\end{matrix}\right)
\een
The Maurer-Cartan form\index{Maurer-Cartan form} is
\ben
g^{-1}dg=Z\,e^{-r}dz+\Phi \,e^{-ar}(\cos(b'r)d\phi-\sin(b'r)dt)+T \,e^{-ar}(\cos(b'r)dt+\sin(b'r)d\phi)+R\, dr
\een
and a left-invariant metric\index{left-invariant metric} will be given by
\ben
ds^2=\eta_{ab}\lambda^a\otimes\lambda^b=dr^2+e^{-2r}dz^2+e^{-2a'r}(\cos(2b'r)(d\phi^2-dt^2)-2\sin(2b'r)d\phi\,dt)\,.
\label{linv}
\een
The original Petrov algebra is of course the special case $a'=-\half,\;b'=\frac{\sqrt{3}}{2}$.

For the metric (\ref{linv}) the Ricci tensor has non-vanishing components
\bena
& R_{rr}=-1-2a'^2+2b'^2\,,\;R_{zz}=-(1+2a')e^{-2r}\,,\nn
\\& R_{\phi\phi}=-(1+2a')e^{-2a'r}(a'\cos(2b'r)+b'\sin(2b'r))\,,\nn
\\& R_{\phi t}=(1+2a')e^{-2a'r}(a'\sin(2b'r)-b'\cos(2b'r))\,,\nn
\\& R_{tt}=(1+2a')e^{-2a'r}(a'\cos(2b'r)+b'\sin(2b'r))\,.
\eena
The manifold is an Einstein manifold only if $a'=-\half$ and $b'=\pm\frac{\sqrt{3}}{2}$ ($\Lambda=0$, Petrov solution) or $a'=1$ and $b'=0$ ($\Lambda=-3$, anti-de Sitter space). In the general case the energy-momentum tensor defined by $T_{\mu\nu}=\frac{1}{8\pi G}G_{\mu\nu}$ does not satisfy the weak energy condition\index{weak energy condition}; without loss of generality assume $\sin(\sqrt{3}r)=0$ and $\cos(\sqrt{3}r)=1$ and choose a timelike vector $t^{\mu}=(0,0,t_3,t_4)$ ($t_4^2\ge t_3^2$), then
\ben
G_{\mu\nu}t^{\mu} t^{\nu}=(1+a'+a'^2-b'^2)(t_3^2-t_4^2)-2(1+2a')b't_3t_4
\een
can be made arbitrarily negative by letting $t_3,t_4\rightarrow\pm\infty$ while keeping $t_4^2-t_3^2$ small and positive, unless $a'=-\half$ and $b'^2\ge \frac{3}{4}$. In the case $a'=-\half$ and $b'^2\ge \frac{3}{4}$, the Einstein tensor can be written as
\ben
G_{\mu\nu}=-\lambda g_{\mu\nu}+2\lambda u_{\mu} u_{\nu},\quad\lambda\equiv b'^2-\frac{3}{4}\ge 0,\quad u_{\mu}=(1,0,0,0)\,,\;u_{\mu} u^{\mu}=1\,.
\een
Note that $u_{\mu}$ is spacelike. This tensor satisfies the dominant energy condition\index{dominant energy condition} as $2t^r v^r-t^a v_a\ge 0$ for any timelike and future-directed $t,v$. These statements are independent of the choice of the arrow of time, \ie hold for both $t_4<0$ or $t_4>0$.

{\bf Third case: $a=\frac{1}{2}$}. Let us introduce a new parameter $c$, so that
\ben
[R,T]=-\frac{1}{2}T-\frac{2}{\sqrt{3}}c^2 \Phi\,,\quad
  [R,\Phi]=-\frac{1}{2}\Phi-\frac{\sqrt{3}}{2}T\,,\quad [R,Z]=Z
\een
for positive $b$ which gives
\ben
R=\left(\begin{matrix}0&0&0&0&1 \cr 0&1&0&0&0 \cr 0&0&-\frac{1}{2}&-\frac{2c^2}{\sqrt{3}}&0 \cr 0&0&-\frac{\sqrt{3}}{2}&-\frac{1}{2}&0
  \cr 0&0&0&0&0\end{matrix}\right)
\een
and a general group element looks like
\ben
g=e^{zZ+\phi\Phi+tT}e^{rR}=\left(\begin{matrix}1&0&0&0&r \cr 0&e^{r}&0&0&z \cr
  0&0&e^{-\frac{r}{2}}\cosh(c r)&-\frac{2c}{\sqrt{3}}e^{-\frac{ r}{2}}\sinh(c r)&\phi
  \cr
  0&0&-\frac{\sqrt{3}}{2c}e^{-\frac{ r}{2}}\sinh(c r)&e^{-\frac{ r}{2}}\cosh(c r)&t
  \cr 0&0&0&0&1\end{matrix}\right)\,.
\een
The Maurer-Cartan form \index{Maurer-Cartan form}is
\bea
g^{-1}dg & = & R\,dr+Z\,e^{-r}\,dz+\Phi\left(e^{\frac{r}{2}}\cosh(cr)\,d\phi+\frac{2c}{\sqrt{3}}e^{\frac{r}{2}}\sinh(cr)\,dt\right)\nn
\\& & +\,T\left(e^{\frac{r}{2}}\cosh(cr)\,dt+\frac{\sqrt{3}}{2c}e^{\frac{r}{2}}\sinh(cr)\,d\phi\right)\,.
\eea
In the limit $c\rightarrow 0$ or $b\rightarrow 0$ this becomes
\ben
g^{-1}dg=R\,dr+Z\,e^{-r}\,dz+\Phi\,e^{\frac{r}{2}}\,d\phi+T\left(e^{\frac{r}{2}}\,dt+\frac{\sqrt{3}}{2}e^{\frac{r}{2}}r\,d\phi\right)\,.
\een
The only remaining case, namely negative $b$, is
\ben
[R,T]=-\frac{1}{2}T+\frac{2}{\sqrt{3}}c^2 \Phi\,,\quad
  [R,\Phi]=-\frac{1}{2}\Phi-\frac{\sqrt{3}}{2}T\,,\quad [R,Z]=-Z\,,
\een
which turns the hyperbolic into trigonometric functions. 

If $c\neq 0$ one can rescale the coordinate $t$ (for instance) and recover the same left-invariant forms as before, hence this does not give anything new. In the case $c=0$ a left-invariant metric\index{left-invariant metric} is given by
\ben
ds^2=dr^2+e^{-2r}dz^2+e^r\left(\left(1-\frac{3}{4}r^2\right)d\phi^2-\sqrt{3}r\,d\phi\,dt-dt^2\right)\,.
\een
The Einstein tensor for this metric can be written as
\ben
G_{\mu\nu}=\frac{3}{16} g_{\mu\nu}+\frac{3}{8}\left(\mbox{diag}\left(-2,e^{-2r},2e^r,0\right)\right)_{\mu\nu}\,.
\een
This does not satisfy the weak energy condition\index{weak energy condition} as $G_{\mu\nu}t^{\mu} t^{\nu}=-\frac{3}{16}(3+e^r)<0$ for the timelike vector $t^{\mu}=(1,0,0,1)$.

Of course, for all three possible cases considered, one could take any constant symmetric matrix with one negative and three positive eigenvalues instead of $\eta_{ab}$ to construct a left-invariant metric; this gives a 10-parameter family (subject to constraints) of metrics whose Killing algebras contain one of the deformations of the four-dimensional Petrov Killing algebra as a subalgebra. One of these parameters can be absorbed into an overall scale, and any cross terms $dr\,dz$, $dr\,d\phi$ or $dr\,dt$ can be absorbed into a redefinition of $z,\phi$ and $t$. One is left with a 6-parameter family of metrics, by far most of which will not appear to have a physical interpretation. We have chosen to just give examples; more efficient methods of constructing general metrics with a given isometry group are certainly known.

\subsection{Spinning Cylinders}
In the case of an infinite rigidly rotating dust cylinder\index{cylinder, rotating} one could imagine trying to speed up this cylinder by shooting in particles with some angular momentum which enter the interior region on causal curves and increase the angular velocity $a$, so as to reach and surpass the critical value $aR=\half$ above which CTCs appear. We will show that this is not possible.

The interior part of the general van Stockum solution
\ben
ds^2=H(\rho)(d\rho^2+dz^2)+L(\rho)d\phi^2+2M(\rho)d\phi\, dt-F(\rho)dt^2\,,
\een
describing the region $\rho<R$ is given by \cite{tipler74}
\ben
H=\exp(-a^2 \rho^2)\,,\quad L=\rho^2(1-a^2 \rho^2)\,,\quad M=a\rho^2\,,\quad F=1\,.
\een
As there are closed timelike curves for $\rho>\frac{1}{a}$ we require $aR\le 1$. The exterior solution is for $a<\frac{1}{2}$, from now on setting $R=1$ for simplicity which is no loss of generality,
\ben
H=e^{-a^2}\rho^{-2a^2}\,,\;L=\frac{\rho\sinh(3\epsilon+\theta)}{2\sinh 2\epsilon \cosh\epsilon}\,,\;M=\frac{\rho\sinh(\epsilon+\theta)}{\sinh 2\epsilon}\,,\;F=\frac{\rho\sinh(\epsilon-\theta)}{\sinh\epsilon}
\een
where $\theta(\rho)=\sqrt{1-4a^2}\log\rho$ and $\epsilon={\rm Artanh }\sqrt{1-4a^2}$. Note that always $-FL-M^2=-\rho^2$ and so the metric has the right signature for all $\rho$ (this of course is also true as $a\rightarrow\half$). The point-particle Lagrangian\index{point particle} is (a dot denotes differentiation with respect to an affine parameter $\lambda$)
\ben
\mathcal{L}=g_{\mu\nu}\dot{x}^{\mu}\dot{x}^{\nu}=H(\dot{\rho}^2+\dot{z}^2)+L\dot{\phi}^2+2M\dot{\phi}\dot{t}-F\dot{t}^2\,,
\een
and since the Lagrangian does not depend on $z,\phi$ and $t$ there are three conserved quantities associated with geodesics:
\ben
P\equiv H\dot{z}\,,\quad J\equiv L\dot{\phi}+M\dot{t}\,,\quad E\equiv F\dot{t}-M\dot{\phi}\,.
\een
The Lagrangian for timelike or null geodesics becomes
\ben
\mathcal{L}=H\dot{\rho}^2+\frac{P^2}{H}+\frac{1}{\rho^2}\left(FJ^2-2MEJ-LE^2\right)=-\mu^2\le 0
\label{lagrarsch}
\een
and we obtain the radial equation
\bea
\left(\frac{d\rho}{d\lambda}\right)^2 & = & \frac{L}{H\rho^2}\left(-\frac{\rho^2
  \mu^2}{L}-\frac{\rho^2P^2}{HL}+E^2+\frac{2M}{L}EJ-\frac{F}{L}J^2\right)\nn
\\& = &\frac{L}{H\rho^2}(E-V_{eff}^+(\rho))(E-V_{eff}^-(\rho))\,,
\eea
where we have introduced an effective potential\index{effective potential}
\ben
V_{eff}^{\pm}(\rho)=\frac{M(\rho)}{L(\rho)}J\pm \rho\sqrt{\frac{1}{L(\rho)}\left(\mu^2+\frac{P^2}{H(\rho)}+\frac{J^2}{L(\rho)}\right)}\,.
\een
This is well-defined for all $\rho$ as $H,\;L,\;M$ all remain positive for all $\rho$. A particle falling in on a geodesic can enter the cylinder if 
\ben
E>V_{eff}^+(1)=\frac{a}{1-a^2}J+ \sqrt{\frac{1}{1-a^2}\left(\mu^2+P^2 e^{a^2}+\frac{J^2}{1-a^2}\right)}\ge\frac{a+1}{1-a^2}J=\frac{1}{1-a}J\,.
\een
Any particle entering the cylinder on a geodesic must have $\frac{J}{E}<1-a.$ In the limit $a\rightarrow 0$ the conserved quantities $J$ and $E$ clearly describe angular momentum and energy per mass. We can identify the ratio $\frac{J}{E}$ with the angular velocity of an infalling particle at $R=1$.

If we are considering accelerated observers, equation (\ref{lagrarsch}) still holds, but $P,\;E$ and $J$ will no longer be conserved quantities. However, only the local values of these quantities at $R=1$ will decide about whether or not a particle will be able to enter the interior region of the cylinder.

This means that the above considerations also hold for accelerated observers and as any particle entering the cylinder must have $\frac{J}{E}<\half$ for $a=\half$, one cannot speed up the cylinder beyond $a=\half$ using particles on timelike or null curves.

\subsection{Geodesic Completeness}
We ask whether the Petrov spacetime, with the radial coordinate $r$ extended to take arbitrary values, is
geodesically complete\index{geodesic completeness}, \ie whether all timelike and null geodesics can be extended to
infinite values of the affine parameter. First we give an example that this need not be possible on a group manifold: 
Remove the null hyperplane $z=t$ from Minkowski space and consider the half-space $z>t$, denoted by $M^-$. It is clearly geodesically incomplete. Null translations and boosts
\ben
(z,t)\rightarrow  (z+c,t+c)\,;\quad (t+z,t-z)\rightarrow \left(\lambda(t+z),\frac{1}{\lambda}(t-z)\right)
\een
act on $M^-$, and together with translations $(x,y)\rightarrow(x+a,y+b)$ they form a four-dimensional group which acts simply-transitively on $M^-$. For instance, the point $(x,y,z,t)=(0,0,1,0)$ is, by a null translation and a successive boost, taken to
\ben
(0,0,1,0)\rightarrow (0,0,1+c,c)\rightarrow \left(0,0,\lambda c+\half-\frac{1}{2\lambda},\lambda c+\half+\frac{1}{2\lambda}\right)\,.
\een
There is a one-one correspondence between points in $M^-$ and group parameters $(a,b,\lambda,c)$, where $\lambda>0$. The space $M^-$ can be identified with the group $G\times \bR^2$, where $G$ is the unique two-dimensional non-Abelian Lie group.

As a second example, introduced in a slightly different context in \cite{polyakov}, consider the dilatation group generated by translations and dilatations\index{dilatation}
\ben
x^{\mu}\rightarrow x^{\mu}+c^{\mu}\,,\quad x^{\mu}\rightarrow \rho x^{\mu}\,,
\een
where we denote the generators by $P_a$ and $D$ respectively. Parametrising the group elements by\footnote{The coordinate $\lambda$ only covers dilatations with $\rho>0$, so that one may obtain geodesic completeness by adding the disconnected component with $\rho<0$. This suggests that geodesic completeness of groups is a topological question that cannot be decided at the level of the Lie algebra.} $g=e^{x^a P_a}e^{\lambda D}$, the Maurer-Cartan form\index{Maurer-Cartan form} is
\ben
g^{-1}dg=e^{-\lambda}dx^a\,P_a+d\lambda\,D
\een
and hence a left-invariant metric\index{left-invariant metric} is
\ben
ds^2=d\lambda^2+e^{-2\lambda}\eta_{\mu\nu}dx^{\mu} dx^{\nu}=\frac{1}{\rho^2}(d\rho^2+\eta_{\mu\nu}dx^{\mu} dx^{\nu})\,,
\een
where $\rho=e^{\lambda}$. This is the metric of anti-de Sitter space\index{anti-de Sitter space} in five dimensions in Poincar\'e coordinates, which is geodesically incomplete as these coordinates cover only a patch of the full spacetime. Hence geodesic completeness is a non-trivial property of a group manifold.

To show geodesic completeness of the Petrov solution, we need to show that no geodesic reaches infinity for finite values of the affine parameter. Consider the Lagrangian
\ben
\mathcal{L}=g_{\mu\nu}\dot{x}^{\mu}\dot{x}^{\nu}\,,
\een
which for the metric (\ref{petrov}) is
\ben
\mathcal{L}=\dot{r}^2+e^{-2r}\dot{z}^2+e^r\left(\cos\sqrt{3}r(\dot{\phi}^2-\dot{t}^2)-2\,\dot{\phi}\,\dot{t}\,\sin\sqrt{3}r\right)\,.
\een
Evidently, from the Euler-Lagrange equations\index{Euler-Lagrange equations}, there are three conserved
quantities because the Lagrangian does not depend on $z,\phi$ or $t$
explicitly, the conjugate momenta
\ben
P\equiv e^{-2r}\dot{z}\,,\;j\equiv
e^r(\dot{\phi}\cos\sqrt{3}r-\dot{t}\sin\sqrt{3}r)\,,\;E\equiv e^r(\dot{\phi}\sin\sqrt{3}r+\dot{t}\cos\sqrt{3}r)\,.
\een
The Lagrangian now takes the form
\ben
\mathcal{L}=\dot{r}^2+e^{2r}P^2+e^{-r}\left(\cos\sqrt{3}r(j^2-E^2)+2Ej\sin\sqrt{3}r\right)\,,
\een
and since the Lagrangian is a conserved quantity in geodesic motion the equation
\ben
\left(\frac{dr}{d\lambda}\right)^2=e^{-r}(E^2\cos\sqrt{3}r-2Ej\sin\sqrt{3}r-j^2\cos\sqrt{3}r)-e^{2r}P^2-\mu^2
\label{radial}
\een
is satisfied by any geodesic, where $\mu^2$ is positive for timelike,
zero for null and negative for spacelike geodesics. For timelike geodesics ($\mu^2>0$), right-hand side of (\ref{radial}) becomes negative for large
$r$, so that $r$ is bounded and we may extend geodesics
infinitely. For null geodesics, $\dot{r}$ is bounded\footnote{If we allow $r$ to take negative values, then for $E\neq 0$ or $j\neq 0$ the right-hand side becomes oscillatory for large negative $r$, taking positive as well as negative values. Hence both timelike and null geodesics are bounded from below in $r$. For $E=j=0$ the right-hand side is either constant zero or always negative.}. So for any finite values of the affine parameter,
$\dot{r}$ remains finite and so do $\dot{z},\dot{\phi}$ and
$\dot{t}$. Hence the Petrov spacetime is geodesically complete. It will be incomplete if we interpret it as the exterior solution of a rotating cylinder and cut off the region described by the Petrov solution at some value of $r$.

By very similar arguments we can show geodesic completeness of the Kaigorodov-Ozsv\'ath solution with line element (\ref{poincare}). In this case, the Lagrangian is
\ben
\mathcal{L}=\frac{1}{\rho^2}\left(\dot{\rho}^2+\dot{z}^2+\dot{\phi}^2-\dot{t}^2\right)-2\rho\,\dot{z}(\dot{t}+\dot{\phi})+\half\rho^4(\dot{t}+\dot{\phi})^2
\een
and the conserved quantities are
\ben
P\equiv \frac{1}{\rho^2}\dot{z}-\rho(\dot{t}+\dot{\phi}),\;j\equiv
\frac{1}{\rho^2}\dot{\phi}-\rho\dot{z}+\half\rho^4(\dot{t}+\dot{\phi}),\;E\equiv \frac{1}{\rho^2}\dot{t}+\rho\dot{z}-\half\rho^4(\dot{t}+\dot{\phi})
\een
so that the radial equation is
\ben
\left(\frac{d\rho}{d\lambda}\right)^2=-\rho^4(P^2+j^2-E^2)-2\rho^7 P(j+E)-\half\rho^{10}(j+E)^2-\rho^2 \mu^2\,.
\label{radial2}
\een
The right-hand side becomes negative if $\mu^2>0$ for both $\rho\rightarrow\infty$ and $\rho\rightarrow 0$, hence all timelike geodesics are bounded in $\rho$. For $\mu=0$ the right-hand side is either constant zero or becomes negative for large $r$, so null geodesics are bounded from above in $\rho$. For very small $\rho$, $\dot{\rho}$ goes to zero, so as before $\dot{\rho}$ is bounded for null geodesics. This shows geodesic completeness. Note that in the case of anti-de Sitter space\index{anti-de Sitter space} (\ref{radial2}) would be 
\ben
\left(\frac{d\rho}{d\lambda}\right)^2=-\rho^4(P^2+j^2-E^2)-\rho^2 \mu^2
\een
and depending on the magnitudes of $E$, $j$ and $P$ the right-hand side blows up as $\rho\rightarrow\infty$ for some geodesics, which can reach $\rho=\infty$ in finite affine parameter distance.

\subsection{Causal Properties}
\label{causalsect}
It was observed by Tipler \cite{tipler74} that the Petrov solution contains CTCs\index{closed timelike curves}. As shown in \cite{ulanovskii}, its causal behaviour is actually even more pathological: The Petrov solution is {\it totally vicious}\index{totally vicious} (for a summary of causality conditions see \cite{causalreview}), \ie for any two points $a$ and $b$ one has $a<b$ and $b<a$, where ``$<$" is the usual causal relation ``can be joined by a timelike curve". The proof of this statement proceeds by giving the future and past light cones at the identity element of the group manifold, and the observation that the adjoint action of elements of the form $\exp(\lambda R)$ rotates the cone around, such that for certain values of $\lambda$ a future-directed timelike tangent vector is mapped to a past-directed timelike tangent vector.

One might hope that a similar statement could be made for the Kaigorodov-Ozsv\'ath solution, but this turns out not to be case, as is apparent from the deformed group element (\ref{deformedgroup}) or the commutation relations (\ref{algnew}). The action of $R$ on the plane generated by $\Phi$ and $T$ is now just a scaling of the generators, which amounts to an opening and closing of the light cones, but no rotation. The assumptions of Theorem 2 in \cite{ulanovskii} are not satisfied.

Apart from the results in \cite{ulanovskii} which refer to homogeneous spacetimes, we may try to apply a criterion for the causality properties of spacetimes with an Abelian isometry group acting transitively on timelike surfaces which was given by Carter in his 1967 PhD thesis \cite{carterthm}. For the Kaigorodov-Ozsv\'ath solution, this Abelian isometry group is generated by $\frac{\partial}{\partial z},\;\frac{\partial}{\partial {\phi}}$ and $\frac{\partial}{\partial t}$ which act on surfaces $\{r=\const\}$. Carter's result is that a sufficient condition for such a spacetime to be totally vicious is the nonexistence of a one-form in the dual of the Abelian Killing algebra which is everywhere timelike or null. In the present case, such a one-form is of the form $\omega = A\,dz+B\,d\phi+C\,dt$, with $A,\;B$ and $C$ constants. Conversely, if there is such a one-form which is everywhere timelike, the spacetime is {\em virtuous}, \ie causally well-behaved. In the present case, one finds that the norm of such an $\omega$ is 
\ben
\tilde{g}(\omega,\omega)=e^{2r}\left(A^2+2\sqrt{2}A\,C\,e^{3r}+C(2B+Ce^{6r})\right)\,,
\label{this}
\een
where $\tilde{g}$ is the contravariant version of (\ref{lambda}). In order to make (\ref{this}) non-positive for all values of $r$, one has to choose $A=C=0$ to be left with the null one-form $d\phi$. Put differently, the subgroup generated by $\frac{\partial}{\partial z}$ and $\frac{\partial}{\partial t}$ acts on null surfaces $\{r=\phi=\const\}$; were these spacelike it would follow that the Kaigorodov-Ozsv\'ath solution is causal (cf. Proposition 9 in \cite{cartercomm}). 

Carter's condition for causality violation was used by Tipler in \cite{tipler74}\footnote{whose wording is slightly unfortunate, since he states the relevant condition on $\omega$ as being ``everywhere timelike"} to show the pathological behaviour of the Petrov solution. Here the situation is less clear; we cannot show directly that in our example spacetimes related by Lie algebra deformation have similar causal properties. The strongest indication we have for causal pathologies of the Kaigorodov-Ozsv\'ath spacetime comes from the computation of the boundary stress-energy tensor in AdS/CFT in (\ref{stresst}).

\sect{The Kaigorodov Solution and Lobatchevski Plane Gravitational Waves}
\label{kaigorodovsect}
As a different example of a homogeneous space which appears as an exact solution in general relativity, we consider the Kaigorodov solution, which has a well-known interpretation as the AdS analogue of homogeneous pp-waves\index{pp-wave} on a Minkowski background \cite{podolsky}. It has received recent interest in the context of string theory\index{string theory} as the near-horizon limit of an M2 brane \cite{popem2,patricot2}. By using Lie algebra deformations we again try to find a spacetime which has a physical interpretation ``close to" the Kaigorodov solution. An obvious candidate would be a member of the family of ``Lobatchevski plane gravitational waves" discussed by Siklos in \cite{siklos} as spacetimes admitting a Killing spinor. The expectation that the particular solution considered by Siklos which has a five-dimensional isometry group arises as a deformation of the Kaigorodov solution will be confirmed. Since the isometry group is now five-dimensional and not four-dimensional as in the previous section, one has to consider a quotient $G/H$. The larger symmetry will place constraints on the general form of a left-invariant metric.

\subsection{The Kaigorodov Solution}
The {\bf Kaigorodov} solution (of Petrov type N\index{Petrov classification}) \cite{kaigorodov} is another example of a
homogeneous space-time which solves the vacuum Einstein equations with a
cosmological constant $\Lambda<0$. The form given in \cite{exact} is\index{Kaigorodov solution}
\ben
ds^2=-\frac{12}{\Lambda}dz^2+10k\,e^{2z}dx^2+e^{-4z}dy^2-10u\,e^z
dz\,dx-2e^z\,du\,dx\,.
\label{kaigo}
\een
The five-dimensional Killing algebra\index{Killing vector} of (\ref{kaigo}) is spanned by
\bea
& & Y\equiv \frac{\partial}{\partial y}\,, \quad X\equiv \frac{\partial}{\partial x}\,,\quad U\equiv e^{-5z}\frac{\partial}{\partial u}\,,\nn
\\& & Z\equiv\frac{\partial}{\partial z}+2y\frac{\partial}{\partial y}-x\frac{\partial}{\partial x}\,,\quad W\equiv ye^{-5z}\frac{\partial}{\partial u}+x\frac{\partial}{\partial y}\,.
\eea
The stabiliser of the origin is the one-parameter subgroup generated by $W$, which we call $H$ to be consistent with previous notation. The transformations generated by $Z$ are
\ben
z\rightarrow z+\lambda,\quad y\rightarrow ye^{2\lambda},\quad x\rightarrow xe^{-\lambda}\,,
\een
while $W$ generates
\ben
y\rightarrow y+\lambda x,\quad u\rightarrow u+ye^{-5z}\lambda+\frac{1}{2}xe^{-5z}\lambda^2\,.
\een
The Killing vectors have non-zero commutators
\bea
& & [W,Y]=-U\,,\;[W,X]=-Y\,,\;[Z,U]=-5U\,,\nn
\\& &[Z,Y]=-2Y\,,\;[Z,X]=X\,,\;[W,Z]=3W\,;
\eea
$Y,X$ and $U$ span a three-dimensional Abelian subalgebra. The
generators of the kinematical group will have the sign-reversed algebra
\bea
& & [W,Y]=U\,,\;[W,X]=Y\,,\;[Z,U]=5U\,,\nn
\\& &[Z,Y]=2Y\,,\;[Z,X]=-X\,,\;[W,Z]=-3W\,.
\label{kaigkilling}
\eea
We denote the isometry group by $\mathcal{K}$ \footnote{The $\mathcal{K}$ stands for Kaigorodov or Killing, according to taste}; it has three-dimensional subgroups of Bianchi types I, II and VI$_{h}$\index{Bianchi classification}. The spacetime is naturally represented as a coset space
\ben
\mathcal{M}=\mathcal{K}/H\,.
\een
Represented as matrices, the group $\mathcal{K}$ is generated by
\bena
& X=\left(\begin{matrix}0 & 0 & 0 & 0 & 0 \cr 0 & 0 & 0 & 0 & 0 \cr 0 & 0 & 0 & 0 & 0 \cr 0 & 0 & 0 & 0 & 1 \cr 0 & 0 & 0 & 0 & 0\end{matrix}\right)\,,\quad Y=\left(\begin{matrix}0 & 0 & 0 & 0 & 0 \cr 0 & 0 & 0 & 0 & 0 \cr 0 & 0 & 0 & 0 & 1 \cr 0 & 0 & 0 & 0 & 0 \cr 0 & 0 & 0 & 0 & 0\end{matrix}\right)\,,\quad Z=\left(\begin{matrix}5 & 0 & 0 & 0 & 0 \cr 0 & 0 & 0 & 0 & 1 \cr 0 & 0 & 2 & 0 & 0 \cr 0 & 0 & 0 & -1 & 0 \cr 0 & 0 & 0 & 0 & 0\end{matrix}\right)\,,\nn
\\& U=\left(\begin{matrix}0 & 0 & 0 & 0 & -1 \cr 0 & 0 & 0 & 0 & 0 \cr 0 & 0 & 0 & 0 & 0 \cr 0 & 0 & 0 & 0 & 0 \cr 0 & 0 & 0 & 0 & 0\end{matrix}\right)\,,\quad W=\left(\begin{matrix}0 & 0 & -1 & 0 & 0 \cr 0 & 0 & 0 & 0 & 0 \cr 0 & 0 & 0 & 1 & 0 \cr 0 & 0 & 0 & 0 & 0 \cr 0 & 0 & 0 & 0 & 0\end{matrix}\right)\,.
\eena
Any element of the group can be parametrised by
\ben
g = e^{ x X + y Y + u e^{5z} U} e^{z Z} e^{w W}\,,
\een
such that $(x,y,z,u)$ are coordinates on the coset space $\mathcal{M}$. The Maurer-Cartan form\index{Maurer-Cartan form}
\ben
\omega_{{\rm MC}}\equiv g^{-1} dg =e_a \lambda^a\,,
\een
where $e_a$ are the group generators, gives a basis of left-invariant one-forms on $\mathcal{K}$:
\bena
& \lambda^1= e^z\,dx\,,\quad \lambda^2=e^{-2z}\,dy-w\,e^z\,dx\,,\quad \lambda^3=dz\,,\nn
\\& \lambda^4=du-w\,e^{-2z}\,dy+\frac{1}{2}w^2\,e^z\,dx+5u\,dz\,,\quad \lambda^5=dw+3 w\,dz\,.
\eena
As usual, one can now construct a left-invariant metric\index{left-invariant metric} on the group by
\ben
ds^2 = G_{ab}\lambda^a\lambda^b\,,
\een
where $G_{ab}$ is constant and non-degenerate. Any such metric will depend on the coordinate $w$ on $W$, since the Lie algebra splitting
\ben
\frak{k} = \frak{m} \oplus \frak{h}
\een
is not reductive: $[W,Z]=3W$ where one would require $[\frak{h},\frak{m}]\subset\frak{m}$. (Recall the definition of a reductive geometry\index{reductive geometry} given in (\ref{reduk})).

The original Kaigorodov metric can be reconstructed by taking
\bena
ds^2 & = & 10k(\lambda^1)^2 + (\lambda^2)^2 -\frac{12}{\Lambda}(\lambda^3)^2 - 2 \lambda^1\lambda^4 + (\lambda^5)^2
\\& = & 10 k\,e^{2z}dx^2 + e^{-4z}\,dy^2 - \frac{12}{\Lambda}dz^2 -2e^z\,dx(du+5\,u\,dz) + (dw + 3 w\,dz)^2\nn
\eena
and reducing to the coset by dropping the last term. 

We summarise the most important physical properties of this solution, as discussed in more detail in \cite{podolsky}: The Kaigorodov spacetime has a curvature singularity at $z=\infty$,
and $z=-\infty$ represents null and spacelike infinity; the metric
approaches the AdS metric asymptotically as $z\rightarrow -\infty$. The patch covered by the coordinates $(x,y,z,u)$ can be extended by changing coordinates to $q=e^{2z}$, and extending to negative $q$. The enlarged spacetime splits
into two disjoint regions $q>0$ and $q<0$ which cannot mutually
communicate. Then $X=\frac{\partial}{\partial x}$ is a timelike Killing vector for
$q<0$, and the enlarged spacetime is stationary in this region. This means that gravitational radiation\index{gravitational radiation} can exist in a stationary spacetime.

\subsection{Deformations of the Lie Algebra}\index{Lie group/algebra deformation}
In order to find infinitesimal deformations of the Killing algebra (\ref{kaigkilling}), we again consider the Jacobi identity at linear order (\ref{jaclin}), again using {\sc Mathematica} for explicit computations. Imposing these constraints on ${{A_a}^c}_b$, one is left with 20 free parameters, as shown in table \ref{deformkai}.
\begin{table}[h]
\caption{{\small Infinitesimal deformations of the Kaigorodov Killing algebra.}}
\label{deformkai}
\begin{center}
\begin{tabular}{c|c|c|c|c|c}
& $c=x$&$c=y$&$c=z$&$c=u$&$c=w$
\\\hline
${{A_x}^c}_y$& $(2\epsilon+\eta)/6$ & $(2\delta+5\tau)/3$ & 0 & $-\gamma/4$ & 0
\\\hline
${{A_x}^c}_z$& $-\omega+2P$ & $3\alpha+\beta$ & $-(\delta+\tau)/3$ & $c_2$ & $\gamma$
\\\hline
${{A_x}^c}_u$& $\sigma/3$ & $\eta/2$ & 0 & $5/3(\delta+\tau)$ & 0
\\\hline
${{A_x}^c}_w$& $(c_9-2c_8-5c_7)/5$ & $c_1$ & $(2\epsilon+\eta)/6$ & $\alpha$ & $\delta$
\\\hline
${{A_y}^c}_z$& $(6c_9-7c_8-15c_7)/5$ & $-\omega+P$ & $(2\epsilon+\eta)/3$ & $\beta$ & $\tau$
\\\hline
${{A_y}^c}_u$& 0 & $-2\sigma/3$ & 0 & $(5\epsilon+4\eta)/3$ & 0
\\\hline
${{A_y}^c}_w$& $m$ & $c_7$ & $\sigma/3$ & $c_5$ & $\varepsilon$
\\\hline
${{A_z}^c}_u$& $-6m$ & $c_8$ & $-5\sigma/3$ & $\omega$ & $\eta$
\\\hline
${{A_z}^c}_w$& $c_3$ & $c_4$ & $(3c_9-c_8)/5$ & $c_6$ & $P$
\\\hline
${{A_u}^c}_w$& 0 & $-m$ & 0 & $c_9$ & $\sigma$
\end{tabular}
\end{center}
\end{table}

Again it turns out that almost all of the possible deformations should be considered trivial (\ie correspond to a change of basis in the Lie algebra). Only a one-parameter deformation, which can be given by $\omega$, cannot be obtained by a basis transformation. This infinitesimal deformation satisfies the full Jacobi identity\index{Jacobi identity}, and we obtain the new algebra
\bea
& & [W,Y]=U\,,\;[W,X]=Y\,,\;[Z,U]=(5+\alpha)U\,,\nn
\\& &[Z,Y]=(2+\alpha)Y\,,\;[Z,X]=(\alpha-1)X\,,\;[W,Z]=-3W\,.
\label{defalg}
\eea
One would perhaps have expected that at least one possible deformation would be to deform the commutation relations of $Z$ to arbitrary numbers:
\ben
[Z,U]=aU\,,\;[Z,Y]=bY\,,\;[Z,X]=cX\,,\;[W,Z]=dW\,.
\een
The one-parameter deformation we have obtained is the most general deformation of this type: By a rescaling of $Z$ one can always keep one of the four numbers $a,b,c,d$ fixed. Then the Jacobi identity puts two constraints on changing the coefficients in the three other commutators, leaving a single deformation parameter.

Put differently, one is free to replace the deformed algebra (\ref{defalg}) by any one-parameter deformation of the action of $Z$ which satisfies the Jacobi identity, e.g.
\bea
& & [W,Y]=U\,,\;[W,X]=Y\,,\;[Z,U]=5U\,,\nn
\\& &[Z,Y]=(2+\alpha)Y\,,\;[Z,X]=(2\alpha-1)X\,,\;[W,Z]=(\alpha-3)W\,.
\eea
We will use this last form in the following. In the matrix representation we only have to modify the matrix for $Z$:
\ben
Z_{\alpha}=\left(\begin{matrix}5&0&0&0&0 \cr 0&0&0&0&1 \cr 0&0&2+\alpha&0&0 \cr 0&0&0&2\alpha-1&0
  \cr 0&0&0&0&0\end{matrix}\right)\,.
\een

\subsection{Construction of the Manifold}
We are trying to construct a metric on the quotient space
\ben
\mathcal{M}'=\mathcal{K}'/H\,,
\een
where the stabiliser $H$ is again generated by $W$. We use the same parametrisation as before:
\ben
g = e^{ x X + y Y + u e^{5z} U} e^{z Z_{\alpha}} e^{w W}
\een
By computing the Maurer-Cartan form\index{Maurer-Cartan form} one finds the left-invariant forms are now
\bena
& \lambda^1= e^{(1-2\alpha)z}\,dx\,,\quad \lambda^2=e^{-(2+\alpha)z}\,dy-w\,e^{(1-2\alpha)z}\,dx\,,\quad \lambda^3=dz\,,
\\& \lambda^4=du-w\,e^{-(2+\alpha)z}\,dy+\frac{1}{2}w^2\,e^{(1-2\alpha)z}\,dx+5u\,dz\,,\quad \lambda^5=dw+(3-\alpha) w\,dz\,.\nn
\eena
Again, it is not possible to find a left-invariant metric\index{left-invariant metric}
\ben
ds^2 = G_{ab}\lambda^a\lambda^b
\een
which is independent of $w$. We can, however write any such metric in the form
\ben
ds^2 = G_{55}(dm + A_{\mu}dx^{\mu})^2 + \tilde{G}_{\mu\nu}dx^{\mu}dx^{\nu}\,,
\een
where $x^{\mu}=(x,y,z,u)$, and demand that $\tilde{G}_{\mu\nu}$, in order to be a metric on the coset space, be independent of $w$. This reduces the number of free parameters in $G$ from 15 to nine, and the coset metric is (setting $G_{55}$, which determines the overall scale, to $-1$)
\ben
ds^2 = p_3 e^{z(2-4\alpha)}\,dx^2 + p_1 e^{-2z(2+\alpha)}\,dy^2+p_2 dz^2 - 2 e^{z(1-2\alpha)}\,dx(p_1 du + (5\,p_1\,u-p_4)dz)\,,
\een
where we have defined
\ben
p_1:=G_{25}^2-G_{22}G_{55}\,,\;p_2=G_{35}^2-G_{33}G_{55}\,,\;p_3=G_{15}^2-G_{11}G_{55}\,,\;p_4=G_{15}G_{35}-G_{13}G_{55}\,.
\een
It is clear that by shifting $u$ by a constant one can always achieve that $p_4=0$. This means the most general metric on the deformed coset space is
\ben
ds^2 = p_3 e^{z(2-4\alpha)}\,dx^2 + p_1 e^{-2z(2+\alpha)}\,dy^2+p_2 dz^2 - 2 p_1\,e^{z(1-2\alpha)}\,dx(du + 5u\,dz)\,.
\label{newmetric}
\een
One discovers that the determinant of (\ref{newmetric}) is
\ben
\det g = -e^{-2(1+3\alpha)z}p_1^3 p_2\,,
\een
and one finds that for Lorentzian signature $(-+++)$ we must have $p_1>0$ and $p_2>0$. One can verify that the metric, as it should be, has five Killing vectors\index{Killing vector}
\bea
& & Y\equiv \frac{\partial}{\partial y}\,, \quad X\equiv \frac{\partial}{\partial x}\,,\quad U\equiv e^{-5z}\frac{\partial}{\partial u}\,,\nn
\\& & Z_{\alpha}\equiv\frac{\partial}{\partial z}+(2+\alpha)y\frac{\partial}{\partial y}+(2\alpha-1)x\frac{\partial}{\partial x}\,,\quad W\equiv ye^{-5z}\frac{\partial}{\partial u}+x\frac{\partial}{\partial y}\,,
\eea
which satisfy the algebra
\bena
& [W,Y]=-U\,,\;[W,X]=-Y\,,\;[Z_{\alpha},U]=-5U\,,
\cr & [Z_{\alpha},Y]=-(2+\alpha)Y\,,\;[Z_{\alpha},X]=(1-2\alpha)X\,,\;[W,Z_{\alpha}]=(3-\alpha)W\,.
\eena
One finds that the Ricci tensor of (\ref{newmetric}) is, in the coordinate basis,
\ben
R_{\mu\nu}=-\frac{3(2+\alpha)^2}{p_2}g_{\mu\nu}-\frac{5e^{(2-4\alpha)z}p_3\alpha(\alpha-3)}{p_2}{\delta_{\mu}}^x{\delta_{\nu}}^x\,.
\een
There is a cosmological term with
\ben
\Lambda = -\frac{3(2+\alpha)^2}{p_2}\,,
\een
consistent with $p_2=-12/\Lambda$ for the original Kaigorodov solution, and a traceless second term which describes null dust. In an orthonormal frame\index{orthonormal frame}, given by
\bena
E^0 & = & du + e^{z(1-2\alpha)}\left(-\frac{p_3}{2p_1}+\frac{25p_1u^2}{2p_2}+\frac{p_1}{2}\right)dx\,,\nn
\\E^1 & = & du + e^{z(1-2\alpha)}\left(-\frac{p_3}{2p_1}+\frac{25p_1u^2}{2p_2}-\frac{p_1}{2}\right)dx\,,\nn
\\E^2 & = & \sqrt{p_2}\,dz - \frac{5p_1}{\sqrt{p_2}}\,u\,e^{z(1-2\alpha)}dx\,,\nn
\\E^3 & = & \sqrt{p_1}e^{-z(2+\alpha)}dy\,,
\eena
the nonvanishing components of $T^{ab}$ are
\ben
T^{00}=T^{01}=T^{10}=T^{11}=\frac{5p_3\alpha(3-\alpha)}{p_2 p_1^2}\,.
\label{tcomp}
\een
The number (\ref{tcomp}) is chosen to be positive so that $T^{ab}$ describes an electromagnetic field; then depending on the value of $\alpha$ the sign of $p_3$ has to be chosen appropriately.

Using deformations of the isometry group of the Kaigorodov solution we have constructed a spacetime which belongs to a more general class of ``Lobatchevski plane gravitational waves\index{gravitational radiation}" discussed in \cite{siklos}. To bring the metric (\ref{newmetric}) into the standard form we apply the coordinate transformation
\ben
q=\frac{\sqrt{p_2}}{2+\alpha}e^{(2+\alpha)z}\,,\quad \tilde{y}=\sqrt{p_1}y\,,\quad \tilde{u}=-p_1e^{5z}u\,.
\een
In these coordinates the metric becomes
\ben
ds^2=\frac{p_2}{(2+\alpha)^2q^2}\left[dq^2+d\tilde{y}^2+2dx\,d\tilde{u}+H(q)dx^2\right]\,,
\een
with
\ben
H(q)=p_3 p_2^{\frac{\alpha-3}{2+\alpha}}\left((2+\alpha)q\right)^{\frac{6-2\alpha}{2+\alpha}}\,.
\een
In the notation of \cite{exact}, we have the metric form,
\ben
ds^2=\frac{3}{|\Lambda|q^2}\left(dq^2 + d\tilde{y}^2 + 2dx\,d\tilde{u}\pm q^{2k}dx^2\right)
\label{siklosform}
\een
with $k=\frac{3-\alpha}{2+\alpha}$ and the sign corresponding to the sign of $p_3$. The condition for positive energy given in \cite{exact} is equivalent to
\ben
p_3(3-\alpha)\alpha> 0\,,
\een
which is consistent with the remarks below (\ref{tcomp}). The case $k=-1$, corresponding to Defrise's pure radiation solution \cite{defrise} with enhanced symmetry, is the only value of $k$ which does not correspond to any choice of $\alpha$. Metrics of the form (\ref{siklosform}) are conformal to pp-waves and a more detailed physical interpretation is given in \cite{siklos}.

Similar to the previous case of the Petrov solution, using a deformation of the Killing algebra of a given spacetime we have constructed a solution with similar physical properties, namely in this case a pp-wave on an AdS background which now describes gravitational radiation as well as a null Maxwell field.

\section{Summary}

We have seen how the mathematical theory of Lie algebra deformations can relate different physical situations, in this case exact solutions of general relativity. The Petrov and Kaigorodov-Ozsv\'ath solutions, the unique vacuum solutions without and with cosmological constant with simply-transitive four-dimensional isometry groups, were shown to be related by Lie algebra deformations; they were shown to be both geodesically complete. While the Petrov solution displays pathological causal behaviour as first observed by Tipler \cite{tipler74} (for the more general class of Lanczos-van Stockum solutions), we were not able to determine the causal properties of the Kaigorodov-Ozsv\'ath solution. There is however strong indication that the latter shows causal pathologies since we found an unphysical energy-momentum tensor on the boundary in the context of AdS/CFT\index{AdS/CFT}. 
We have to leave a more concrete study of the relation of these two properties to further work. Another interesting question that has been left open is whether the Kaigorodov-Ozsv\'ath solution can be understood as a certain limit of an exterior solution of a matter distribution, perhaps representing a rotating cylinder as well.\footnote{A class of rotating cylinder solutions with negative cosmological constant has been studied in \cite{godelext}, who discuss the appearance of CTCs as well, but leave out certain limiting null cases. Our study strongly suggests that the Kaigorodov-Ozsv\'ath solution, if related to the results of \cite{godelext}, would be a limiting case not considered in \cite{godelext}.} More generally its physical interpretation which is so far largely unknown deserves further study. 

In a second calculation, we analysed deformations of the Killing algebra of the Kaigorodov spacetime and saw that one is naturally led to a solution including an electromagnetic field. Although we have not constructed new exact solutions here, we have shown how this more algebraic approach to exact solutions of general relativity can give new insights on physical properties.

\chapter{Deformed General Relativity and Torsion}
\label{defgenrel}
\epigraph{\em{``Seit die Mathematiker \"uber die Relativit\"atstheorie hergefallen sind, verstehe ich sie selbst nicht mehr.'' \cite{flatau}}}{Albert Einstein}
\def\vol{\rm Vol}
\def\herm{\rm Herm}
\def\dS{\mathcal{D}}
\def\dSG{SO(d,1)}
\def\L{SO(d-1,1)}
\def\tr{{\rm tr}}
\def\P{\mathcal{P}}
\def\M{\mathcal{M}}
\def\ir{{\rm irr}}

\section{Introduction}

In the previous chapter, we have discussed exact solutions of general relativity which are homogeneous spaces $G/H$. Since these are special cases with more symmetry than one would expect in a generic physical situation, it is natural to investigate a more general geometric description in which the symmetry of a given homogeneous space is only present infinitesimally. 

The passage from special to general relativity can be understood as making the symmetries of a homogeneous space local: General relativity is formulated in terms of (pseudo-)Riemannian geometry, which makes the symmetries of Euclidean space, which can be thought of as a homogeneous space $E(p,q)/SO(p,q)$ \footnote{For $p$ spacelike and $q$ timelike dimensions. We will set $p=d-1$ and $q=1$ in the following.}, local, and can thus be formulated on arbitrary manifolds. Physically, this is the incorporation of Einstein's equivalence principle\index{equivalence principle}: At at a given point, any spacetime looks to an inertial observer like Minkowski space. In the vier-/vielbein formalism of general relativity (originally due to Cartan as well as the Cartan connections we will use in this chapter), one chooses a basis of the tangent space at each point, and thus a ``moving frame" of vector fields, such that the metric in this frame is constant and diagonal\index{orthonormal frame}. The frame field then encodes the metric information of the manifold.

In this chapter, we shall investigate a generalised formulation of general relativity in terms of {\bf Cartan geometry}\index{Cartan geometry} which incorporates the symmetries of an arbitrary given homogeneous space $G/H$ on an infinitesimal level, thus providing an extension of both Felix Klein's Erlangen Programme\index{Erlangen Programme} \cite{klein} and (pseudo-)Riemannian geometry, as summarised in the following picture that we essentially take from \cite{sharpe}:
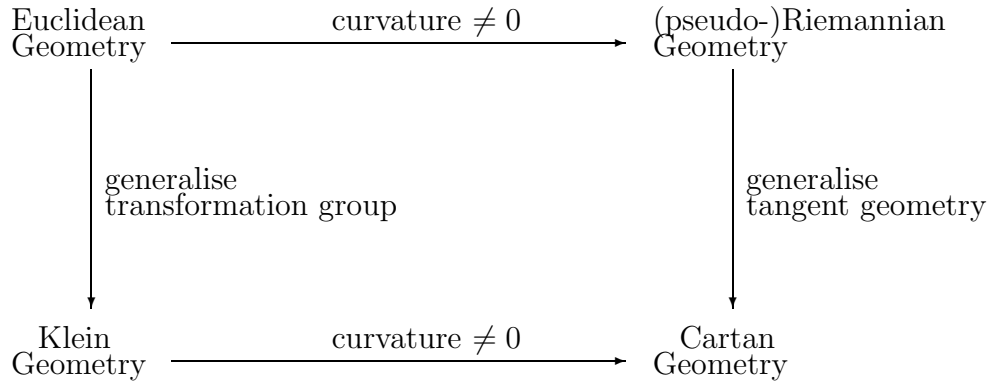
\begin{figure}[htp]
\centering
\begin{picture}(300,150)
\put(0,140){Euclidean}\put(0,130){Geometry}\put(60,135){\vector(1,0){170}}\put(120,140){curvature $\neq 0$}\put(240,140){(pseudo-)Riemannian}\put(240,130){Geometry}
\put(30,125){\vector(0,-1){90}}\put(270,125){\vector(0,-1){90}}
\put(35,80){generalise}\put(35,70){transformation group}
\put(275,80){generalise}\put(275,70){tangent geometry}
\put(10,20){Klein}\put(0,10){Geometry}\put(60,15){\vector(1,0){170}}\put(120,20){curvature $\neq 0$}\put(250,20){Cartan}\put(240,10){Geometry}
\end{picture}
\caption{{\small Cartan geometry extends both Klein geometry and (pseudo-)Riemannian geometry.}}
\end{figure}

In general relativity, if an orthonormal frame has been found, one can still use general local Lorentz transformations to obtain a different frame with the same properties, and so in this formulation general relativity appears as a gauge theory. The description of general relativity is in terms of an $\frak{h}$-valued Ehresmann connection (with $H=SO(d-1,1)$, the Lorentz group), together with a soldering form (the frame field) determining how the tangent spaces at different points are ``soldered" to the spacetime manifold. The gauge group $H$ acts on the tangent space which is viewed as a vector space $\R^{d-1,1}$.

We will use a similar gauge-theoretic formulation, but of a more general character: Instead of viewing Minkowski space as $\R^{d-1,1}$, we will view it as a homogeneous space $E(d-1,1)/SO(d-1,1)$. A description in terms of Cartan geometry will allow us to {\em deform} general relativity by replacing $E(d-1,1)$ by its deformation, the de Sitter group \footnote{The anti-de Sitter group $SO(d-1,2)$ could be used equally well, and in fact is used in \cite{stellewest}.} $SO(d,1)$. We effectively try to construct a theory of gravity built on tangent de Sitter spaces which replace the usual (co-)tangent spaces in general relativity. As we saw in the previous chapter, from the viewpoint of Lie group deformations this process is a very natural extension of the standard framework. In addition to the speed of light $c$, a second scale $\kappa$ with dimensions of momentum is introduced on a fundamental level. It is tempting to speculate that this scale could be related to quantum gravity, although the basic idea of ``deforming" the kinematic algebra has existed at least since Heisenberg, who thought in terms of atomic rather than quantum-gravitational scales.

The motivation for investigating this possible generalisation of the usual formulation of Einstein-Cartan theory, or general relativity in first order formulation, comes from the recently proposed framework of deformed, or doubly, special relativity (DSR), where one effectively proposes a momentum space that is acted on by the de Sitter, not the Poincar\'e group, rather in the spirit of Snyder's proposal\index{Snyder's non-commuting coordinates} \cite{snyder}. It seems natural to try to construct a general geometric framework which incorporates this symmetry on the infinitesimal level. We will use a formulation of Einstein-Cartan theory given by Stelle and West, and interpret it in the concrete example of Minkowski space. By doing this construction, we hope to shed light on the physical interpretation of DSR, which seems rather unclear at present. In particular, while the noncommuting translations appearing in the de Sitter group are commonly interpreted as noncommuting ``spacetime coordinates", and one passes to a noncommutative geometry description to account for this, our discussion will stay within conventional differential geometry. We will see that the description of Minkowski space in this theory includes a connection with torsion\index{torsion} instead of a fundamentally noncommutative structure.

In order to confront this mathematical framework with physical reality, we note that such a connection with torsion may lead to the usual ambiguities in minimal coupling. We give estimates concerning observable effects and note that observable violations of charge conservation induced by torsion should happen on a time scale of $10^3$ s, which seems to rule out these modifications as a serious theory. Our considerations show, however, that the noncommutativity of translations in the Snyder algebra need not correspond to noncommutative spacetime in the usual sense.

To set the scene, we start with the Einstein-Hilbert action\index{Einstein-Hilbert action} in four dimensions, without cosmological constant, written in first order (Palatini) form:
\ben
S_{{\rm EH}} = \frac{1}{16\pi G}\int \epsilon_{abcd} \left(e^a\wedge e^b \wedge R^{cd}\right)\,,
\label{palatini}
\een
where $R^{cd}$ is the curvature two-form of an $\frak{so}(3,1)$-valued connection one-form ${\omega^a}_b$. One discovers that variation with respect to the vierbein field $e^a$ gives the usual vacuum field equation ${\rm Ric}_a=0$ (where ${\rm Ric}_a$ is the Ricci one-form), while variation with respect to ${\omega^a}_b$ gives the condition of vanishing torsion. Thus, this formulation has the advantage over the second order form that the condition of a torsion-free connection does not have to be put in by hand. On the other hand, if one now adds a matter action to the Einstein-Hilbert action, both equations now potentially have a non-zero right-hand side. While the functional derivative of the matter action with respect to the vierbein determines an energy-momentum tensor, one will also have nonvanishing torsion if
\ben
\frac{\delta S_{{\rm mat}}}{\delta {\omega^a}_b}\neq 0.
\een
This is the case for matter with internal spin, such as spinor fields. This generalisation of general relativity which incorporates the possibility of torsion is known as {\bf Einstein-Cartan theory}\index{Einstein-Cartan theory}. Since torsion does not propagate, \ie the field equation contains the full torsion two-form instead of contractions of it (as for curvature in Einstein's equations for $d>3$), it seems difficult to rule out such a theory experimentally, since one would only measure torsion inside a matter distribution with spin.

The basic observation is now the following: Combine the connection ${\omega^a}_b$ and the vierbein $e^a$ into a single connection, taking values in the algebra of the Poincar\'e group, 
\ben
A=\left(\begin{matrix}
&  & & \cr & {\omega^a}_b & & \frac{1}{L}e^i \cr & & & \cr  & 0 & & 0
\end{matrix}\right)\,,
\label{poinconn}
\een
where $L$ is a length introduced to get dimensions right, but which is completely arbitrary at this stage. Then consider the Yang-Mills-like action (with now $d=4$)\index{MacDowell-Mansouri formulation}
\ben
S=-\frac{L^2}{32\pi G}\int \epsilon_{abcd} \left(F^{ab} \wedge F^{cd}\right)= -\frac{L^2}{32\pi G}\int d^4 x \, \frac{1}{4}\epsilon_{abcd}\epsilon^{\mu\nu\rho\tau}F_{\mu\nu}^{ab}F_{\rho\tau}^{cd}\,,
\label{ymaction}
\een
involving the curvature $F^{ab}$ of the connection ${A^a}_b$ (Latin indices here run from 1 to 4, and so one projects $F^{ab}$ to its $\frak{so}(3,1)$ part in this action). Writing out $F^{ab}$ in terms of the curvature $R^{ab}$ of ${\omega^a}_b$ and the vierbein field $e^i$, one discovers that the action (\ref{ymaction}) contains a topological Gauss-Bonnet term plus the Einstein-Hilbert action (\ref{palatini}). Thus, the field equations of Einstein-Cartan theory are recovered.

The idea that general relativity may be interpreted as a gauge theory with the Poincar\'e group as gauge group is of course far from new. In the usual textbook discussions, the significance of a gauge theory\index{gauge theory} is expressed as the promotion of a global symmetry in a given theory to a local one by the introduction of an additional field which appears in gauge-covariant derivatives. This viewpoint was probably first elaborated in detail by Utiyama \cite{utiyama}, who showed that the promotion of a global to a local symmetry directly leads to a gauge field, valued in the Lie algebra\index{Lie algebra} of the symmetry group under consideration, with the usual transformation properties, and thus geometrically to the well known formulation in terms of connections on principal bundles (see e.g. \cite{nakahara}). Utiyama also considered the Lorentz group as an example and was led to general relativity, but only by making several assumptions, such as the existence of a Riemannian metric and a local Lorentz (vierbein) frame and the symmetricity of the affine connection in its lower indices, without further motivation. It was then realised by Kibble \cite{kibble} that the more natural gauge group to take is the Poincar\'e group, so that the vierbein appears naturally as part of the connection. Kibble showed that one is led to general relativity in the first order formalism, \ie to Einstein-Cartan theory, and so to a connection with torsion when matter with internal spin, such as fermionic matter, is present. The appearance of torsion when fermions are coupled to gravity had earlier been noted by Weyl \cite{weyl50}.\footnote{Rather amusingly, Weyl wondered in \cite{weyl50} whether the ``gauge transformations" of electromagnetism should not be called ``phase transformations" instead -- the term ``gauge transformation"\index{gauge transformation} had of course been introduced by Weyl himself \cite{spacetimematter} in the context of local conformal transformations (changes in ``gauge").}

The result of MacDowell and Mansouri \cite{mmgravity} is essentially that by considering a connection taking values in the algebra $\frak{so}(4,1)$ of the de Sitter group, one obtains a cosmological term in addition to the Einstein-Hilbert action (\ref{palatini}). (MacDowell and Mansouri also extended the calculations to supergravity, which we will not discuss further.) We will see that the result is not just a mathematical trick, but has a geometric interpretation in terms of a Cartan connection.

The rest of this chapter is organised as follows: We discuss the idea of a curved momentum space in section \ref{intro} and give a brief introduction into the ideas of {\bf deformed (doubly) special relativity} (DSR) most relevant to the following discussion in section \ref{dsr}. In section \ref{gauge} we outline how Einstein-Cartan theory can be formulated as a gauge theory of gravity with the de Sitter group $\dSG$ as gauge group; this theory includes a gauge field that plays a crucial role in what follows. In this section we essentially re-derive the results of Stelle and West, using a different set of coordinates which we find more closely related to the DSR literature. This section uses the language of gauge theory familiar to physicists; we then give a more mathematical account of the geometry of Cartan connections in section \ref{app}.

We then return to a concrete example: To justify our claim that this geometric framework can be used to generalise the ideas of DSR, we show in section \ref{synth} how, if spacetime is taken to be Minkowski space, the simplest non-trivial choice of zero section leads to a connection with torsion, providing a geometric interpretation for the noncommuting ``coordinates" appearing in the Snyder algebra. We close with a discussion of our results and their possible physical implications, which show that the theory, at least in its given form, is not physically viable. We conclude that there may be different physical interpretations of algebraic commutation relations such as those used in DSR.

Since the two most obvious extensions of general relativity are admitting either connections with torsion or non-metric connections, we briefly discuss the theory of a torsion-free non-metric connection, known as symmetric affine theory, in a final section. We will see that it does not fit as well into a description by Cartan geometry as the case highlighted in this chapter.

\sect{Curved Momentum Space}
\label{intro}

It is commonly assumed that quantum gravity\index{quantum gravity} sets a fundamental length scale, the Planck scale \cite{planck}, which can not be resolved by any physical experiment. Different approaches to quantum gravity, such as string theory\index{string theory} or loop quantum gravity, incorporate such a scale. This leads to the idea that some kind of ``space discreteness" should be apparent even in a low-energy ``effective" theory.

The idea of putting quantum mechanics on a discrete lattice\footnote{with spacing equal to the Compton wavelength of the proton, $l_c\approx 1.3$ fm} seems to have been first considered by Heisenberg in the spring of 1930 \cite{heisenberg}, in an attempt to remove the divergence in the electron self-energy. Because the absence of continuous spacetime symmetries leads to violations of energy and momentum conservation, this approach was not pursued further, but later in the same year Heisenberg considered modifying the commutation relations involving position operators instead \cite{heisenberg}. 

A fundamental length scale is absent in special relativity, where two observers will in general not agree on lengths or energies they measure. Hence the usual ideas of Poincar\'e invariance must be modified in some way. Snyder observed \cite{snyder} that this could be done by deforming the Poincar\'e algebra into the de Sitter algebra, \ie considering the isometry group of a (momentum) space of constant curvature. As we have seen in chapter \ref{defspacetime}, if one maintains the structure of a Lie algebra and considers deformations\index{Lie group/algebra deformation} of the Poincar\'e algebra, replacing it by the de Sitter (or anti-de Sitter) algebra is the unique way of implementing a modified kinematic framework\index{kinematical group} \cite{bacry}.

A $d$-dimensional de Sitter momentum space\index{de Sitter space} with curvature radius $\kappa$ is defined as the submanifold of a $(d+1)$-dimensional flat space with metric signature ($d,1$) by
\ben
(P^1)^2 + (P^2)^2 + \ldots + (P^{d-1})^2 - (P^d)^2 + (P^{d+1})^2 = \kappa^2\,,
\label{ads}
\een
where $\kappa$ has dimensions of mass. Its isometry group is generated by the algebra
\bena
[M_{ab},M_{cd}] =\eta_{ac}M_{bd}+\eta_{bd}M_{ac}-\eta_{bc}M_{ad}-\eta_{ad}M_{bc}\,, \nn
\\ \left[ X_{a},M_{bc} \right]=\eta_{ac}X_b-\eta_{ab}X_c\,, \quad [X_a,X_b]=\frac{1}{\kappa^2}M_{ab}\,.
\label{commutators}
\eena
Here $M_{ab}$ correspond to a Lorentz subalgebra of the de Sitter algebra, while $X_a\equiv\frac{1}{\kappa}M_{d+1,a}$ are interpreted as (noncommuting) translations. These translations are then interpreted as corresponding to coordinates on spacetime; Snyder thought of operators acting on a Hilbert space. Since the operators $X_1,\ldots,X_{d-1}$ correspond to compact rotations in the $(d+1)$-dimensional space, their spectrum is discrete. In this way, one obtains ``quantised spacetime"\index{Snyder's non-commuting coordinates}, while maintaining Lorentz covariance.

It should be stressed at this point that we are interested in the full isometry group acting on momentum space, \ie the Poincar\'e group in the case of a Minkowski space. Clearly, an equation such as (\ref{ads}) defines a scale $\kappa$ already in the $(d+1)$-dimensional Minkowski space; this scale would however not be invariant under translations of this space. In other words, there is no natural length scale in Minkowski space, whereas there is one in de Sitter space, namely the scale $\kappa$ defining the manifold. There has been some confusion in the literature (see \eg \cite{schuller}) about the way in which Lorentz invariance is deformed in the Snyder algebra and the more recent ideas of DSR that we will review shortly. It should be clear from the commutation relations (\ref{commutators}) that the Lorentz subgroup generated by $M_{ab}$ is {\em not} deformed in any way.

One can give explicit expressions for the algebra elements by choosing coordinates on de Sitter space (\ref{ads}). The choice made by Snyder is taking Beltrami coordinates
\ben
p^1=\kappa\frac{P^1}{P^{d+1}}\,,\;p^2=\kappa\frac{P^2}{P^{d+1}}\,,\ldots\,,\;p^d=\kappa\frac{P^d}{P^{d+1}}\,,
\een
whence one has $(P^{d+1})^2=\kappa^4/(\kappa^2+\eta_{ab}p^a p^b)$ to satisfy (\ref{ads}), and $\eta_{ab}p^a p^b \ge -\kappa^2$, corresponding to an apparent maximal mass if $p^a$ were interpreted as Cartesian coordinates on a Minkowski momentum space. (Up to this point one could in principle have chosen anti-de Sitter instead of de Sitter space. Then this inequality becomes $\eta_{ab}p^a p^b \le \kappa^2$, which perhaps seems less motivated physically.) A necessary sign choice means that these coordinates only cover half of de Sitter space. In these coordinates, the translation generators
\ben
X_a=\frac{1}{\kappa}\left(P^{d+1}\frac{\partial}{\partial P^a}-P_a\frac{\partial}{\partial P^{d+1}}\right)=\frac{\partial}{\partial p^a}+\frac{1}{\kappa^2}p_a p^b\frac{\partial}{\partial p^b}
\een
generate ``displacements" in de Sitter space. (In this notation, indices are raised and lowered with $\eta_{ab}$, the $d$-dimensional Minkowski metric, so that $p_a=\eta_{ab}p^b$.)

The motivation behind these ideas was to cure the infinities of quantum field theory, which evidently arise from allowing arbitrary high momenta (or short distances). In a somewhat similar spirit, Gol'fand suggested \cite{golfand} to define quantum field theory on a momentum space of constant curvature, using Beltrami coordinates as momentum variables. This makes the volume of the corresponding Riemannian space finite and so presumably leads to convergent loop integrals in the Euclideanised theory. The consequences for standard quantum field theory were further explored in \cite{kada,golfand2}.

 Gol'fand only assumed that $\kappa\gg m$ for all elementary particles; thinking of quantum gravity, one would perhaps identify $\kappa$ with the Planck scale, whereas the original authors seem to have thought of the Fermi (\ie weak interaction) scale.

 The induced metric on de Sitter space in terms of the coordinates $p^a$ is
\ben
g_{nr}=\frac{\kappa^2}{\kappa^2+p\cdot p}\left(\eta_{nr}-\frac{p_n p_r}{\kappa^2+p\cdot p}\right)\,,
\label{metric}
\een
where $p\cdot p\equiv \eta_{cd}p^c p^d$. The metric (\ref{metric}) becomes singular when $p \cdot p\rightarrow -\kappa^2$, and negative definite when extended to what Gol'fand calls the exterior region $p \cdot p < -\kappa^2$. In four dimensions, 
\ben
\det g = -\kappa^{10}(\kappa^2+p\cdot p)^{-5}\,,
\een
and the volume element is $d^4 p\,\kappa^5(\kappa^2+p\cdot p)^{-5/2}$.

 In Gol'fand's approach (assuming $d=4$ of course), the standard Feynman rules were modified by replacing the addition of momenta $p$ and $k$ at a vertex by
\ben
(p(+)k)^a=\frac{\kappa}{\kappa^2 - p\cdot k}\left(p^a\sqrt{\kappa^2+k\cdot k}+k^a\left(\kappa-\frac{p\cdot k}{\kappa+\sqrt{\kappa^2+k\cdot k}}\right)\right)\,,
\een
which corresponds to a translation by $k$ of the vector $p$. (Again $p\cdot k \equiv \eta_{ab}p^a k^b$, etc.) It was also noted that spinors now transform under ``displacements" as well, which is made more explicit in \cite{kada} and \cite{golfand2}. As is well known, five-dimensional Dirac spinors still have four components and the matrix $\gamma^5$ appears in the Dirac Lagrangian, hence there is no chirality\index{chirality}. This alone seems to imply that the original Gol'fand proposal cannot be used for an appropriate model of the known particles.

Gol'fand's approach is very different from more recent approaches to quantum field theory on noncommutative spaces (see \eg \cite{nekrasov}) in that the field theory is defined on a momentum space which is curved, but neither position nor momentum space are noncommutative in the usual sense. 

\sect{Deformed Special Relativity}
\label{dsr}

The idea that the classical picture of Minkowski spacetime should be modified at small length scales or high energies was re-investigated in more recent times, motivated by the apparent existence of particles in ultra high energy cosmic rays whose energies could not be explained within special relativity \cite{experiment}. The proposed framework of deformed special relativity (DSR)\index{deformed special relativity (DSR)} \cite{amelino} modifies the Poincar\'e algebra, introducing an energy scale $\kappa$ into the theory, in addition to the speed of light $c$. This leads to a quantum ($\kappa$-)deformation of the Poincar\'e algebra \cite{majid}, with the parameter $\kappa$ associated with the newly introduced scale.

It was soon realised \cite{kowalski} that this deformed algebra is the algebra of the isometry group of de Sitter space, and that the symmetries of DSR could hence be obtained by identifying momentum space with de Sitter space, identifying $X_a$ as the generators of translations on this space. The constructions of DSR thus appear to be a resurrection of Snyder's and Gol'fand's ideas. We take this observation as the defining property of DSR, and will seek to describe a framework in which momentum space, or rather the (co-)tangent space in general relativity, is replaced by an ``internal" de Sitter space. We will see that this can best be done using Cartan geometry.

When discussing DSR as a modification of special relativity\index{special relativity}, we take the view that special relativity is defined as a kinematic framework with preferred inertial systems, related to one another by (proper) Lorentz transformations. That is, one has a flat spacetime on which there exist certain preferred coordinate systems, those in which the metric is diagonal with entries $\pm 1$. From this point of view, the choice of coordinates on the internal de Sitter space plays quite an important role if one is looking for a ``deformation" of special relativity including an energy scale $\kappa$. Such a deformation can only arise if the chosen coordinate system reduces to Cartesian coordinates on Minkowski space as $\kappa\rightarrow\infty$. The choice of coordinates is obviously not unique.

The generators of the algebra will take different explicit forms when different coordinate systems (on four-dimensional de Sitter space\index{de Sitter space}) are chosen. In \cite{kowalski} ``natural coordinates" are defined by, in the notation of section \ref{intro}, \footnote{Capital Latin indices such as $I$ and $J$ used in this section only run over spatial coordinates (from 1 to 3).}
\ben
g=\exp\left[p^I (M_{I4}+X_I)\right] \exp\left[p^4 X_4\right]\mathcal{O}\,,
\een
where $\mathcal{O}=(0,0,0,0,\kappa)$ is taken to be the origin of de Sitter space in five-dimensional Minkowski space, and $M_{I5}$ and $M_{45}$ correspond to translations in space and time. The coordinates one obtains are related to the five-dimensional coordinates by
\ben
P^I=p^I e^{\frac{p^4}{\kappa}}\,,\quad P^4=\kappa \sinh\left(\frac{p^4}{\kappa}\right)+\frac{\vec{p}^2}{2\kappa}e^{\frac{p^4}{\kappa}}\,,\quad P^5=\kappa \cosh\left(\frac{p^4}{\kappa}\right)-\frac{\vec{p}^2}{2\kappa}e^{\frac{p^4}{\kappa}}\,.
\een
Again, these cover only half of de Sitter space where $P^4+P^5>0$. The metric in these ``flat" coordinates is
\ben
ds^2=-(dp^4)^2+e^{\frac{2p_4}{\kappa}}\delta_{IJ}dp^I \, dp^J\,.
\een
Slices of constant $p_4$ are flat; to an observer using these coordinates the spacetime appears as expanding exponentially. An illuminating discussion of different coordinate systems and kinematics on de Sitter space is given in \cite{special}.

The Magueijo-Smolin model \cite{msmodel} corresponds to the following choice of coordinates:
\ben
p^1=\kappa\frac{P^1}{P^5-P^4}\,,\;p^2=\kappa\frac{P^2}{P^5-P^4}\,,\;p^3=\kappa\frac{P^3}{P^5-P^4}\,,\;p^4=\kappa\frac{P^4}{P^5-P^4}\,,
\een
The generators of boosts in de Sitter space take the form
\ben
K^I\equiv p^I\frac{\partial}{\partial p^4}+p^4\frac{\partial}{\partial p^I}+\frac{1}{\kappa}p^I p^J\frac{\partial}{\partial p^J}\,,
\een
and translations (not considered by the authors of \cite{msmodel}) would take the form
\ben
X_I=\frac{p^4+\kappa}{\kappa}\frac{\partial}{\partial p^I}+\frac{1}{\kappa^2} p_I p^b\frac{\partial}{\partial p^b}\,,\quad X_4=\frac{1}{\kappa}p^b\frac{\partial}{\partial p^b}+\frac{p^4+\kappa}{\kappa}\frac{\partial}{\partial p^4}\,.
\een
This choice of coordinates is somewhat peculiar as $p^4$ takes a special role, as is also apparent from the modified dispersion relations presented in \cite{msmodel}. The quantity
\ben
||p||^2=\frac{\eta_{ab}p^a p^b}{(1+\frac{1}{\kappa}p^4)^2}
\een
is invariant under boosts and rotations in de Sitter space, as would $\eta_{ab}p^a p^b$ be in Beltrami coordinates.

Each DSR model corresponds to a choice of coordinates on de Sitter space, such that all expressions reproduce the expressions for special-relativistic Minkowski coordinates as $\kappa\rightarrow\infty$. What Magueijo and Smolin call a ``$U$ map" is essentially a coordinate transformation from Beltrami coordinates to a different set of coordinates, which becomes the identity as $\kappa\rightarrow\infty$. In the remaining sections we shall use Beltrami coordinates. Note that this means we always have $p\cdot p\ge -\kappa^2$.

\sect{A de Sitter Gauge Theory of Gravity}
\label{gauge}

The most direct implementation of the ideas discussed so far into a framework describing more general spacetimes is replacing the cotangent (or tangent) bundle usually taken as phase space by a general symplectic manifold $\{\P,\omega\}$, which can be locally viewed as a product $U \times \dS$ of a subset $U\subset \M$ of spacetime $\M$ with de Sitter space $\dS$. We want to retain the differentiable structure of a manifold, which we do not assume to be present in a full theory of quantum gravity\index{quantum gravity}. We also assume that the structure of momentum space is fixed and in particular does not depend on matter fields, as suggested in \cite{moffat}.

If phase space is described as such a manifold, with a choice of origin in the ``tangent" de Sitter space at each point, the appropriate mathematical language is that of fibre bundles. The theory of connections on fibre bundles of this type, called {\it homogeneous bundles}\index{homogeneous bundles} in \cite{russians}, was developed by \'Elie Cartan (\eg in \cite{cartan}). Adopting this framework means there is now an $\frak{so}(d,1)$ connection, instead of an $\frak{so}(d-1,1)$ connection, defining parallel transport on spacetime.

It was noted by MacDowell and Mansouri \cite{mmgravity} that gravity with a cosmological term in four dimensions could be obtained from a theory of such an $\frak{so}(d,1)$ connection by projecting it onto its $\frak{so}(d-1,1)$ part in the action. A more elaborate description in terms of Einstein-Cartan theory was then given by Stelle and West \cite{stellewest}. Their analysis included the gauge field needed to identify the fibres at different spacetime points, which will be crucial for the interpretation of the theory. The mathematical side of MacDowell-Mansouri gravity as a theory of a Cartan connection is nicely illustrated in \cite{wise}; we follow this article as well as the more computationally based presentation of \cite{stellewest}, who use the language of non-linear realisations. An overview over the mathematics of Cartan connections is given in \cite{sharpe}. 

For clarity we first describe the framework in a language more common to physicists; a more mathematical account of Cartan connections on homogeneous bundles is given in the following section \ref{app}. 

The usual description of general relativity as a gauge theory of the Lorentz group, known as vier-/vielbein formalism, was reviewed at the beginning of this chapter. Since the tangent bundle is in our description replaced by a homogeneous bundle with a curved ``tangent" space, one has to effectively use a ``double vielbein" formalism, in which spacetime vectors are mapped to vectors in the tangent space to the internal (curved) space by a soldering form (vielbein). The picture we have in mind is that of a de Sitter space rolled along the manifold. One then needs to introduce a new field which specifies the point of tangency, expressed in a given coordinate system on the internal space, at each spacetime point. We denote it by $p^a(x)$. This corresponds mathematically to a necessary choice of zero section (see below), and physically to a gauge field. Picking a point of tangency at each spacetime point breaks the gauge group $SO(d,1)$ down to the Lorentz subgroup $SO(d-1,1)$ leaving the point of tangency invariant.

In more general terms \cite{wise2}, one has a principal bundle over a base manifold $\mathcal{M}$ with gauge group $G$. Locally, a connection on this bundle can be viewed as a one-form taking values in the Lie algebra $\frak{g}$. One then considers the projection of the connection to its $\frak{g}/\frak{h}$ part and demands that at any point in the manifold, this be an isomorphism between the tangent space at this point and the algebra $\frak{g}/\frak{h}$. In other words, the $\frak{g}/\frak{h}$ part ${e_{\mu}}^i$, written as a matrix, has to have full rank. This of course means that the dimension of $\frak{g}/\frak{h}$ must be equal to the dimension of the manifold. The gauge field $p^a(x)$ now comes in because the projection of the connection to its $\frak{g}/\frak{h}$ part is not canonical; for a given subgroup $H\subset G$ viewed as the stabiliser of a particular point in $G/H$, one could equally well take a conjugate group $gHg^{-1} \simeq H$ stabilising a different point in $G/H$. This will become more explicit shortly when we consider the MacDowell-Mansouri action.

The procedure does not depend on the choice of gauge group $G$, and would be equally possible if the gauge group were chosen to be the Poincar\'e group $E(d-1,1)$; such a formulation for gravity, including copuling to matter, is indeed discussed in \cite{grignard}. In that case, the homogeneous space $G/H$ would be Minkowski space, viewed as an affine space. As the value of the gauge field $p^a(x)$ at a particular spacetime point picks an origin in this affine space, it may then be identified with the vector space $\bR^{d-1,1}$: At a given point in Minkowski space, the tangent space to this point can be identified with Minkowski space itself. Although all calculations we will present go through in the case of $E(d-1,1)$ as gauge group (resulting in different formulae of course), the formalism may seem a bit redundant since there is no real necessity for the second vielbein field. In the more general case of a homogeneous space $G/H$ which acts as a ``tangent space'' instead of a flat vector space, it is however necessary.

We consider a theory with gauge group $SO(d,1)$, so that the connection $A$ takes values in the Lie algebra\index{Lie algebra} $\frak{so}(d,1)$. It can be split as (again introducing a length $l$ on dimensional grounds)
\ben
A=\left(\begin{matrix}
&  & & \cr & {\omega^a}_b & & \frac{1}{l}e^i \cr & & & \cr  & -\frac{1}{l}e_i & & 0
\end{matrix}\right)\,,
\een
so that ${\omega^a}_b$ acts as the usual $\frak{so}(d-1,1)$-valued connection of general relativity and $e^i$ as a vielbein one-form. We have simultaneously unified the usual connection and the vielbein, and replaced the (flat) tangent space by a curved ``internal" space, such that the de Sitter group and not the Poincar\'e group now appears as a gauge group. (Lorentz) indices on ${\omega^a}_b$ and $e^i$ are raised and lowered using $\eta^{ab}$. 

A gauge transformation\index{gauge transformation}, \ie a local transformation $g(x)$ taking values in the de Sitter group, can be split as $g(x)=s(x)\Lambda(x)$, where $s(x)$ changes the zero section, \ie changes the local identification of points of tangency at each spacetime point, and $\Lambda(x)$ is a local Lorentz transformation in the vielbein formalism of general relativity which does not mix the ${\omega^a}_b$ and $e^i$ parts of the connection. In this notation, the connection transforms under a gauge transformation as
\ben
A(x)\rightarrow A'(x)=\Lambda^{-1}(x)s^{-1} (x) A(x) s(x)\Lambda(x) + \Lambda^{-1}(x)s^{-1} (x) ds(x)\Lambda(x) + \Lambda^{-1}(x)d\Lambda(x)\,.
\label{gaugetransf}
\een

One can use this equation to relate the connection $A_0$ corresponding to the trivial zero section, where the point of tangency is the origin of the internal space at each spacetime point, $p^a(x)\equiv (0,0,0,0)$, to a connection corresponding to any given zero section. The physical significance of this is the following. Assume we have fixed $p^a(x)\equiv (0,0,0,0)$. Then an action can be defined from the curvature of the connection $A$ (here $R$ is the curvature of the $\frak{so}(d-1,1)$ part of $A$),
\ben
F=dA+A\wedge A=\left(\begin{matrix}
& & & \cr & {R^a}_b - \frac{1}{l^2}(e^a \wedge e_b) & & \frac{1}{l}T^i\equiv\frac{1}{l}(de^i + {\omega^i}_j \wedge e^j) \cr \cr & -\frac{1}{l}T_i & & 0
\end{matrix}\right)\,.
\een
In four dimensions, the MacDowell-Mansouri action\index{MacDowell-Mansouri formulation} \cite{mmgravity, wise} is
\ben
S=-\frac{3}{32\pi G\Lambda}\int \epsilon_{abcd} \left(F^{ab} \wedge F^{cd}\right)= -\frac{3}{32\pi G\Lambda}\int d^4 x \, \frac{1}{4}\epsilon_{abcd}\epsilon^{\mu\nu\rho\tau}F_{\mu\nu}^{ab}F_{\rho\tau}^{cd}\,,
\label{akshn}
\een
where the Latin indices run from 1 to 4, and so one projects $F$ to its $\frak{so}(3,1)$ part in this action. 

Apart from a topological Gauss-Bonnet term, the action (\ref{akshn}) is equivalent to the Einstein-Hilbert action\index{Einstein-Hilbert action} with a cosmological term
\ben
S = \frac{3}{16\pi G\Lambda}\frac{1}{l^2}\int \epsilon_{abcd} \left(e^a\wedge e^b \wedge R^{cd} - \frac{1}{2 l^2}e^a\wedge e^b \wedge e^c \wedge e^d \right)\,,
\label{einsthilb}
\een
where we have to identify
\ben
\Lambda=\frac{3}{l^2}
\een
as the cosmological constant.\footnote{Note that in the formulation in terms of the Poincar\'e group, with connection (\ref{poinconn}), no cosmological constant appears and the length scale $L$ remains undetermined.}

In order to define the projection of $F$ in the action (\ref{akshn}), one has used a splitting 
\ben
\frak{so}(d,1) \simeq \frak{so}(d,1)/\frak{so}(d-1,1) \oplus \frak{so}(d-1,1)\,,
\label{split1}
\een
 which depends on the gauge field since the subgroup $SO(d-1,1)$ leaving a given point in de Sitter space invariant depends on the choice of this point. 

When the action (\ref{akshn}) is coupled to matter, the $\frak{so}(d,1)/\frak{so}(d-1,1)$ part $e^a$ of the connection appears in a volume element in the matter Lagrangian. By varying the action one obtains the field equations of Einstein-Cartan theory with a cosmological constant $\Lambda=3/l^2$. The length scale $l$, which is so far arbitrary, can be chosen to reproduce the $\Lambda$ of the observed universe, which means it must be chosen to be very large (the ``cosmological constant problem"\index{cosmological constant problem}). By the field equations, one can determine for a given matter distribution a connection $A_0$ consisting of an $\frak{so}(d-1,1)$ connection $({\omega^a}_b)_0$ and a vielbein $e^i_0$. 

The MacDowell-Mansouri action reproducing Einstein-Cartan theory with a cosmological constant includes a gauge choice. We can hence view it as the gauge-fixed version of a more general theory. Since (\ref{gaugetransf}) determines how the connection transforms under a gauge transformation, we can generalise a given solution of Einstein-Cartan theory to an arbitrary gauge choice. The extension of the theory to arbitrary configurations of the gauge field, and hence arbitrary choices of tangency points of the internal space to spacetime, is what we call {\bf Einstein-Cartan-Stelle-West theory}\index{Einstein-Cartan-Stelle-West theory}. Any solution of Einstein-Cartan theory, in particular any (torsion-free) solution of general relativity, gives rise to more general solutions of Einstein-Cartan-Stelle-West theory via (\ref{gaugetransf}). We will later see that one can construct an $\frak{so}(d-1,1)$ connection with torsion from a torsion-free one.

In (\ref{gaugetransf}), $s(x)$ takes values in the de Sitter group, more precisely in the subgroup generated by ``translations" which leaves no point of de Sitter space invariant. The correspondence between Beltrami coordinates $p^a(x)$ on de Sitter space and such group elements is given explicitly by
\ben
s(p(x))=\exp\left[ \frac{p^i(x)}{\sqrt{-p(x)\cdot p(x)}}\,{\rm Artanh}\left(\frac{\sqrt{-p(x)\cdot p(x)}}{\kappa}\right)\kappa\,X_i \right]\,.
\label{param}
\een 
Then the group element $s(p(x))$ maps $(0,0,0,0)$ to $(p^1(x),p^2(x),p^3(x),p^4(x))$ in Beltrami coordinates. A different choice of coordinates in the internal de Sitter space would correspond to a different parametrisation of the elements of the subgroup of translations of the de Sitter group.

Inserting (\ref{param}) into (\ref{gaugetransf}) and setting $\Lambda(x)\equiv e$, we obtain
\bena
\omega^{ab}(p(x)) & = & \frac{p^a e_0^b}{l \kappa \gamma(p)} + \left(1 - \frac{1}{\gamma(p)}\right)\frac{p^a dp^b + \omega_0^{ca}p^b p_c}{p\cdot p} + \frac{1}{2}\omega_0^{ab} - (a \leftrightarrow b)\,,\nn
\\ e^i(p(x)) & = & \frac{l \kappa}{p\cdot p+\kappa^2}\left(p^i \frac{p_c dp^c }{p\cdot p}(1-\gamma(p)) +dp^i \gamma(p) + ({\omega^i}_b)_0 p^b \gamma(p)\right)\nn
\\& & +p^i e_0^a p_a\frac{1 + \frac{\kappa^2}{p\cdot p}(1-\gamma(p))}{p\cdot p+\kappa^2} + \frac{e_0^i}{\gamma(p)}\,, 
\label{explicit}
\eena
where
\ben
\gamma(p)\equiv\sqrt{\frac{p\cdot p +\kappa^2}{\kappa^2}}=1+\frac{p\cdot p}{2\kappa^2}+\ldots\,
\een

Because $p\cdot p \ge -\kappa^2$ in Beltrami coordinates, the square root is always real. In the limit $p\cdot p\rightarrow 0$, our parametrisation is the same as that used in \cite{stellewest}, and we recover their results
\bena
 \omega^{ab}(p(x)) & = & \left(\half\omega^{ab}_0+\frac{1}{l \kappa} p^a e_0^b + \frac{1}{2\kappa^2} \left(p^a dp^b + \omega_0^{ca}p^b p_c \right)\right) - (a \leftrightarrow b)\,,\nn
\\ e^i(p(x))& = & e_0^i + \frac{l}{\kappa}\left(-\frac{1}{2\kappa^2}p^i p_c dp^c +dp^i + \omega^{ib}_0 p_b \right)+\frac{1}{2\kappa^2} p^i e_0^a p_a \,. 
\eena
Near $p=0$, we have
\ben
\omega^{ab}(p(x)) = \omega_0^{ab} + O\left(\frac{p}{\kappa}\right)\,, \quad e^i(p(x)) = e_0^i + \frac{l}{\kappa}dp^i + O\left(\frac{p}{\kappa}\right)\,. 
\label{smallp}
\een

As mentioned above, the $\frak{so}(d,1)/\frak{so}(d-1,1)$ part of the connection $A$ acts as a vielbein and maps vectors in the tangent space at a point $x$ in spacetime to vectors in the tangent space at $p(x)$ in the internal de Sitter space, given in components with respect to an orthonormal basis\index{orthonormal frame} at $p(x)$. In order to give their components in the coordinate-induced basis $\{\frac{\partial}{\partial p^a}\}$, we need another vielbein, which can be obtained from (\ref{explicit}) by setting $\omega_0=e_0=0$ (corresponding to spacetime being de Sitter space with cosmological constant $\Lambda$) and $p^a(x)=\frac{\kappa}{l}x^a$, as in \cite{stellewest}. We obtain
\ben
{l_n}^a(p(x))=\kappa^2\frac{{\delta_n}^a (p\cdot p) \gamma(p) - p^a p_n (\gamma(p)-1)}{(p \cdot p)(p\cdot p+\kappa^2)}\,,
\label{vierbein}
\een
where $n$ is a coordinate index in the internal space and $a$ denotes a Lorentz index, as before. This vielbein is of course independent of the underlying spacetime.

Parallel transport can be defined for the $\frak{so}(d,1)$ connection using the notion of development, which generalises the usual covariant derivative. One introduces a development\index{development} operator \cite{stellewest}
\ben
D = d - \half\omega^{ab}M_{ab} - (e\cdot V)\,,
\een
where the second term is the usual infinitesimal relative rotation of tangent spaces at different spacetime points, and the last term compensates for the change of point of tangency and hence generates maps from the tangent space at one point of de Sitter space to the tangent space at a different point of de Sitter space. Again one should think of an internal space rolled along spacetime \cite{wise}. 

In components, in our conventions we have
\ben
{(\omega^{ab}M_{ab})^c}_d=-2{\omega^c}_d\,,
\een
and the combination $e^a V_a$ acts on Lorentz indices as an element of $\frak{so}(d-1,1)$, representing the map from one tangent space to another in the respective bases. We use the result obtained by \cite{stellewest} using the techniques of non-linear realisations, namely that when expressed as an $\frak{so}(d-1,1)$ matrix,
\ben
l(e\cdot V) = \kappa\,s(p)^{-1}(e^a X_a )s(p)-s(p)^{-1}\left[s(p+\delta p)-s(p)\right]\,,
\een
where $s(p)$ is defined according to (\ref{param}) and $\delta p$ is determined from the equation
\ben
{\left[s(p+\delta p)\right]^a}_5={\left[(1+e^b X_b \kappa)s(p)\right]^a}_5
\een
where only terms linear in $e^a$ are kept in $\delta p$. An explicit calculation shows that
\ben
\delta p^a = \frac{p^a}{\kappa}(\eta_{bc}e^b p^c)+e^a\kappa\,,
\een
and hence near $p=0$, we have $\delta p^a = \kappa e^a$, as expected. We find that $(e\cdot V)$ has components
\ben
{(e\cdot V)^b}_c=\frac{\kappa(e^b p_c - e_c p^b)(1-\gamma(p))}{l (p\cdot p)}\,.
\label{development}
\een

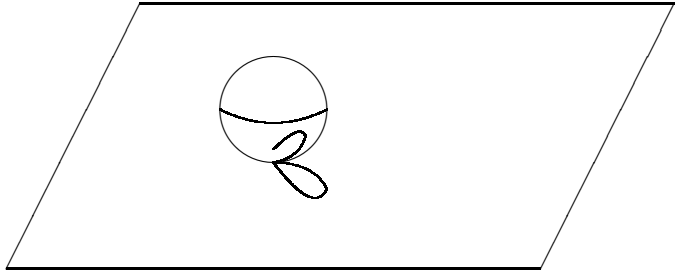
\begin{figure}[h]
\begin{center}
\begin{picture}(300,120)
\put(0,0){\line(1,0){200}}\put(0,0){\line(1,2){50}}
\put(50,100){\line(1,0){200}}\put(200,0){\line(1,2){50}}
\put(100,60){\circle{40}}\bezier{313}(80,60)(100,50)(120,60)
\bezier{125}(100,40)(115,40)(120,30)\bezier{578}(100,40)(115,20)(120,30)
\bezier{125}(100,40)(110,42)(112,50)\bezier{578}(100,45)(110,55)(112,50)
\end{picture}
\end{center}
\caption[Rolling an internal de Sitter space along a closed path in Minkowski space.]{{\small When the curved internal space is rolled along Minkowski space, a path in spacetime corresponds to a path in the internal space. Because of the curvature of the internal space, a closed path in Minkowski space does not correspond to a closed path in the internal space, which is manifest as torsion. (We have drawn Riemannian spaces, a plane and a sphere, for clarity of presentation.)}}
\label{fig}
\end{figure}

One then has a notion of holonomy, mapping closed loops in spacetime into the internal space by development. In particular, if one develops the field $p(x)$ describing the point of tangency around an infinitesimal closed loop at $x_0$, the developed value will in general differ from the original value at $x_0$ \cite{stellewest}:
\ben
\Delta p^a (x_0) \propto {T_{\mu\nu}}^i(x_0){l^a}_i(p(x_0))\oint x^{\mu} dx^{\nu}\,,
\een
where ${l^a}_i(p(x))$ is the inverse of the vielbein (\ref{vierbein}) and ${T_{\mu\nu}}^i$ are the components of the torsion tensor $T=de+\omega\wedge e$. This is because by specifying $p^a(x)$, one locally identifies the ``internal'' de Sitter spaces, thinking of a tangent de Sitter space rolled along the manifold. In a generic situation, where the manifold $\mathcal{M}$ is not identical to the internal space, such an identification is, strictly speaking, only possible at a point. To understand the geometry, one may picture a sphere being rolled along a plane, as in figure \ref{fig}. In general, when the sphere is rolled around and returns to its initial position on the plane, its point of tangency will not be the original one. This is just the situation for the Lorentzian analogues, Minkowski space and de Sitter space, as we will discuss shortly.

The central result we will try to justify in the following is that, starting from Minkowski spacetime, if we assume the internal de Sitter space is rolled along Minkowski space in a non-trivial way, we obtain a connection with torsion\index{torsion}. In our interpretation, this is the only way that ``coordinates" can act as translations on momentum space, as one normally assumes when associating the Snyder algebra with a noncommutative spacetime\index{Snyder's non-commuting coordinates}.

\sect{Cartan Connections on Homogeneous Bundles}
\label{app}

This more mathematical introduction into Cartan connections on homogeneous bundles relies mainly on \cite{wise}, but mentions some additional points which are of importance to our discussion of Einstein-Cartan-Stelle-West theory.

The tangent bundle of a manifold needs to be replaced by a fibre bundle whose fibres are homogeneous spaces $\dS\equiv\dSG/SO(d-1,1)$. This can be achieved by starting with a principal bundle $P(\M,\dSG)$, and considering the associated bundle $\P=E(\M,\dS,\dSG,P)=P \times_{\dSG} \dS$ (taken as phase space); it can be identified with $P/SO(d-1,1)$ by the map
\ben
\nu: \P \rightarrow P/SO(d-1,1), \quad [u,a\cdot SO(d-1,1)] \mapsto u a \cdot SO(d-1,1).
\een
Then the structure group $\dSG$ is reducible to $ SO(d-1,1)$ if the associated bundle $\P$ admits a cross section $\sigma: \M \rightarrow \P$ \cite{kobayashi}; furthermore, there is a one-to-one correspondence between reductions of the structure group and cross sections. This cross section, called a {\it zero section}\index{zero section} in \cite{petti}, corresponds to a choice of origin in the momentum space attached to each point. In physicist's terms, the de Sitter group is spontaneously broken down to the Lorentz group by the choice of points of tangency in the tangent de Sitter spaces at each spacetime point.

The bundle reduction depends on the choice of zero section, or rather, its local representation in coordinates as a function $\M \supset U \rightarrow U \times \dS$. This is because the embedding of $ SO(d-1,1)$ into $\dSG$ is not canonical, as the stabilisers of different points in $\dS$ are isomorphic but related by conjugation. In other words, the mappings appearing in the exact sequence\index{exact sequence}
\ben
{\bf 0} \rightarrow  SO(d-1,1) \rightarrow \dSG \rightarrow \dSG /  SO(d-1,1) \rightarrow {\bf 0}
\een
are not canonically chosen (cf. the discussion for the affine group in \cite{petti}).

It is of course possible to choose canonical coordinates such that the function representing the zero section is just $x\mapsto (x,[e])\equiv (x, SO(d-1,1))\in \M \times \dS$. However, in general we want to locally identify the fibres at nearby base space points, adopting the viewpoint that there is a single tangent $\dS$ space which is ``rolled along" the manifold. Then we need to retain the general coordinate freedom. (This point is missing in the discussion of \cite{wise}.) An exact identification is only possible when the connection is flat. Let us assume that coordinates on $\P$ have been fixed, and that it is the zero section, and hence the identification of the fibres, that is varied\footnote{one is free to choose an ``active" or ``passive" viewpoint here}. After a choice of zero section, there is still a local gauge freedom corresponding to the stabiliser $SO(d-1,1)$. We express a given section as $s(x)$, where $\sigma(x)=(x,s(x))\in \M \times \dS$ in our coordinates. The section that corresponds to $s_0(x)\equiv [e]$ will be called ``trivial".

An $\frak{so}(d,1)$-valued Ehresmann connection $\mathbf{A}$ in $P$ is in general not reducible to an $\frak{so}(d-1,1)$-valued connection in the reduced $ SO(d-1,1)$ bundle $P_R(\M, SO(d-1,1))$. It can, however, be pulled back using the inclusion
\ben
\iota_x:  SO(d-1,1) \rightarrow \dSG, \quad  \Lambda \mapsto s(x)\Lambda s(x)^{-1}
\een
to a Cartan connection $\mathbf{A}_C$ on the reduced bundle\footnote{We assume here that the necessary condition $\ker\mathbf{A} \cap (\iota_x)_*(TP_R(\M, SO(d-1,1)))=\{0\}$ (see \cite{sharpe}) is satisfied.}. Of course reducing the connection to an $\frak{so}(d-1,1)$-valued connection and pulling it back to a Cartan connection are very different operations, since in the latter case one wants the $\frak{so}(d,1)/\frak{so}(d-1,1)$ part of the pulled-back connection to act as a soldering form, so in particular to be non-singular. We obtain a bundle sequence (cf. \cite{wise})
\begin{center}
\begin{picture}(250,110)
\put(-60,90){$P_R(\M, SO(d-1,1))$}\put(50,93){\vector(1,0){25}}\put(80,90){$P(\M,\dSG)$}\put(160,93){\vector(1,0){25}}\put(190,90){$P/\iota_x( SO(d-1,1))\simeq \P$}\put(285,100){{\small $\nu_x^{-1}$}}
\put(35,80){\vector(1,-1){60}}\put(110,80){\vector(0,-1){60}}\put(185,80){\vector(-1,-1){60}}\put(105,5){$\M$}
\end{picture}
\end{center}
The reduced bundle $P_R(\M, SO(d-1,1))$ is mapped into $P(\M,\dSG)$ by
\bea
p \mapsto [p,e] &= &\{(p\Lambda^{-1},s(x)\Lambda s(x)^{-1})|\Lambda \in  SO(d-1,1)\}\nn
\\& & \in P_R(\M, SO(d-1,1)) \times_{\iota_x( SO(d-1,1))} \dSG\,.
\eea
The connection one-form $A$ on $\M$, induced by the connection $\mathbf{A}$ on $P$, depends on a choice of section $\tau:\M\rightarrow P(\M,\dSG)$. If the zero section $\sigma$ is fixed, one can choose an arbitrary (local) section $\tau_R:\M\rightarrow P_R(\M, SO(d-1,1))$ to obtain a section $\tau$; in local coordinates,
\bea
& & \sigma(x)=(x,s(x))\,,\; \tau_R(x)=(x,\Lambda(x))\nn
\\& &\longrightarrow\; \tau(x)=(x,s(x)\Lambda(x) s(x)^{-1}\cdot s(x))=(x,s(x)\Lambda(x))\,.
\eea
For practical computations, it is often useful to first consider the trivial section. The induced connection corresponding to this section, denoted by $A_0(x)$, is related to the connection for a general section by
\ben
A(\tau(x))=\Lambda^{-1}(x)s^{-1} (x) A_0(x) s(x)\Lambda(x) + \Lambda^{-1}(x)s^{-1} (x) ds(x)\Lambda(x) + \Lambda^{-1}(x)d\Lambda(x)\,.
\label{ageneral}
\een
Once the zero section $s(x)$ has been fixed, there is still the freedom of $ SO(d-1,1)$ transformations, corresponding to different choices of $\Lambda(x)$ in (\ref{ageneral}). These are the standard local Lorentz transformations in the vielbein formalism of general relativity.

The choice of zero section induces a local splitting of the $\frak{so}(d,1)$ connection, according to
\ben
\frak{so}(d,1) \simeq \frak{so}(d,1)/\frak{so}(d-1,1) \oplus \frak{so}(d-1,1)\,.
\label{splitting}
\een
This splitting is invariant under the adjoint action of $SO(d-1,1)$, thus the different parts of the connection will not mix under $SO(d-1,1)$ transformations. We are considering a reductive geometry\index{reductive geometry} (cf. (\ref{reduk})); recall that the existence of such a splitting
\ben
\frak{g}=\frak{x}\oplus\frak{h}
\label{splitting2}
\een
invariant under the adjoint action of $H$, implies that we may identify $\frak{g}/\frak{h}$ with the subspace $\frak{x}$ of $\frak{g}$. In the present case, this is of course the subspace spanned by the generators $M_{d+1,a}$ of $SO(d,1)$. For the de Sitter group, just as for the Poincar\'e and anti de Sitter groups, one can say more: The Lie algebra $\frak{so}(d,1)$ is equipped with a $\bZ_2$ grading\index{$\bZ_2$ grading}; the subspaces in (\ref{splitting2}) satisfy
\ben
[\frak{h},\frak{h}]\subset\frak{h}\,,\quad [\frak{h},\frak{x}]\subset\frak{x}\,,\quad [\frak{x},\frak{x}]\subset\frak{h}
\een
and the homogeneous space $G/H$ is a {\bf symmetric space}\index{symmetric space} \cite{wise2}.

The fact that de Sitter space\index{de Sitter space} viewed as a homogeneous space is a symmetric space is directly responsible for the splitting of the curvature $F$ of the Cartan connection $A$ into curvature and torsion of the $\frak{so}(d-1,1)$ connection $\omega$ that we have observed above \cite{wise2}. A general Cartan connection $A=\omega+e$, where $\omega$ takes values in $\frak{h}$ and $e$ takes values in $\frak{x}$, has curvature
\ben
F = R + \frac{1}{2}[e,e] + d_{\omega}e\,,
\een
where $[e,e]$ denotes taking the wedge product on the form part and the commutator on the Lie algebra part, \ie
\ben
[e,e]_{\mu\nu}^i={{C_j}^i}_k\, \left(e^j\wedge e^k\right)_{\mu\nu} = {{C_j}^i}_k\, \left(e^j_{\mu}e^k_{\nu}-e^k_{\mu}e^j_{\nu}\right)
\een
if ${{C_j}^i}_k$ are the structure constants\index{structure constants} as in (\ref{structcon}). The middle term generically takes values in $\frak{h}$ and $\frak{x}$, so that one has a splitting
\ben
F_{\frak{h}} = R + \frac{1}{2}[e,e]_{\frak{h}}\,,\quad  F_{\frak{x}} = \frac{1}{2}[e,e]_{\frak{x}} + d_{\omega}e\,.
\een
For a symmetric space, however, $[e,e]_{\frak{x}}$ vanishes \cite{wise2}, and we obtain the result used in the construction of the MacDowell-Mansouri action (\ref{akshn}).

Because we assume $A$ to be a Cartan connection, the $\frak{so}(d,1)/\frak{so}(d-1,1)$ part acts as a soldering form, corresponding to the standard vielbein of general relativity; in particular, ${e_{\mu}}^i$ is an invertible matrix. The soldering form maps vectors in the tangent space $T_x \M$ at a point $x$ in spacetime to vectors in the tangent space $T_{p(x)} \dS$ at $p(x)$ in the internal de Sitter space, given in components with respect to an orthonormal basis at $p(x)$. The vielbein that maps between the components of a vector in the orthonormal basis and the coordinate-induced basis is given in (\ref{vierbein}).

\sect{Synthesis}
\label{synth}

The notion of development\index{development} along curves in spacetime is central to the interpretation of Einstein-Cartan-Stelle-West theory, because it allows ``spacetime coordinates" to act as translations in the internal de Sitter space. The situation described by DSR, where noncommuting translations on a curved momentum space are interpreted as noncommuting spacetime coordinates, here corresponds to a Minkowski spacetime with an internal de Sitter space rolled along this Minkowski space. The gauge field $p^a(x)$ specifies the points of tangency of the internal space at each spacetime point, and we have chosen Beltrami coordinates on de Sitter space which look like Cartesian coordinates on Minkowski space near the ``origin" of de Sitter space. Since the internal space has a natural scale $\kappa$ and we needed to introduce a natural scale $l$ in spacetime, we choose the gauge field to be
\ben
p^a(x)=\frac{\kappa}{l}x^a
\label{choice}
\een
in a vicinity of the origin of spacetime which is now taken to be Minkowski space, where $x^a$ are the standard Minkowski coordinates such that the connection vanishes in general relativity. In general a closed path in spacetime will not correspond to a closed path traced out on the internal space, hence such an identification is only local and, strictly speaking, only valid the origin of Minkowski spacetime. On dimensional grounds, the effects of torsion\index{torsion} scale as $\frac{x}{l}$ or $\frac{p}{\kappa}$. For (\ref{choice}) to be well-defined, we must guarantee that $x\cdot x \ge -l^2$, so $l$ should be large in Planck units. We will comment on the significance of the scale $l$ at the end of this section.

It should perhaps be emphasised that the gauge field $p^a(x)$ does not represent physical momentum, but determines the point of tangency of the internal space we have introduced which is to some extent arbitrary. Tangent vectors to the original spacetime can be mapped to tangent vectors to the internal space via the vielbein. The physical interpretation of motion in an internal ``momentum" space which is related to motion in spacetime seems obscure, but if coordinates are to act as translations in the internal space, the two must be connected in some way. In this sense, we are constructing the minimal non-trivial gauge field which leads to observable effects, and an alternative interpretation of noncommuting generators $X_a$ in the Snyder algebra. 

In our interpretation, different points in the internal de Sitter space do not represent different values for physical four-momentum. Hence we avoid problems with the physical interpretation of DSR, such as the ``spectator problem" of noncommutative momentum addition and the ``soccer ball problem" of how to describe extended objects, given that any momentum would appear bounded by the relation $p\cdot p\ge -\kappa^2$. In our framework, tangent vectors representing a particle's (or extended body's) velocity remain vectors and as such live in an unbounded space with commutative addition.

As explained before, we can use equations (\ref{explicit}) to obtain the connection components $\omega$ and $e$ corresponding to this choice of our gauge field; we set $\omega_0=0$ and $({e_{\mu}}^a)_0={\delta_{\mu}}^a$ and substitute (\ref{choice}) to get
\bena
{\omega_{\mu}}^{ab} & = & \left(x^a {\delta_{\mu}}^b - x^b {\delta_{\mu}}^a \right)\frac{x\cdot x+l^2(\gamma(x)-1)}{l^2 (x\cdot x) \gamma(x)}\,,
\label{connection}
\\{e_{\mu}}^i & = & \frac{1}{(x\cdot x)(x\cdot x + l^2)}\Big(x^i x_{\mu}(x\cdot x-2 l^2(\gamma(x)-1))+{\delta_{\mu}}^i (x\cdot x) 2 l^2 \gamma(x)\Big) \nn
\eena
and
\bena
\partial_{\nu}{e_{\mu}}^i-\partial_{\mu}{e_{\nu}}^i & = & \left(x_{\nu}{\delta_{\mu}}^i-x_{\mu}{\delta_{\nu}}^i\right)\frac{l^2(2l^2(\gamma(x)-1)-3(x\cdot x))-(x\cdot x)^2}{(x\cdot x)(x\cdot x+l^2)^2}\,,\nn
\\{\omega_{\nu}}^{ib}e_{\mu b}-{\omega_{\mu}}^{ib}e_{\nu b} & = & \frac{\left(2 l^2+x\cdot x\right)\left(x\cdot x+ l^2(\gamma(x)-1)\right)}{(x\cdot x)(x\cdot x+l^2)l^2\gamma(x)}\left(x_{\nu}{\delta_{\mu}}^i-x_{\mu}{\delta_{\nu}}^i\right)\,,
\eena
where now
\ben
\gamma(x)\equiv\sqrt{\frac{x\cdot x + l^2}{l^2}}\,,
\een
which gives a non-zero torsion\index{torsion}
\ben
{T_{\mu\nu}}^i=\left(x_{\nu}{\delta_{\mu}}^i-x_{\mu}{\delta_{\nu}}^i\right)\frac{1}{l^2 \sqrt{\frac{x\cdot x}{l^2}+1}}\,.
\een
Interestingly enough, for the choice of zero section (\ref{choice}) the scale $\kappa$ drops out of all expressions. Expressed in coordinates on the internal space, one has
\ben
{T_{\mu\nu}}^i=\left(p_{\nu}{\delta_{\mu}}^i-p_{\mu}{\delta_{\nu}}^i\right)\frac{1}{l \kappa \sqrt{\frac{p\cdot p}{\kappa^2}+1}}\,.
\een
The quantity ${T_{\mu\nu}}^i$ will be multiplied by an infinitesimal closed loop $\oint x^{\mu} dx^{\nu}$ to give the difference in the value $p(x)$ caused by development along this loop. In momentum coordinates, this is equal to $\frac{l}{\kappa}\oint p^{\mu} dp^{\nu}$, and the effect of going around the developed curve in the internal space is (near $x=0$ or $p=0$) proportional to $\kappa^{-2}$, just as was suggested by (\ref{commutators}). 

Expressing Minkowski space in the usual coordinates, together with the (local) identification $p^a(x)=\frac{\kappa}{l}x^a$, in this framework gives a connection with torsion. Developing a closed curve in spacetime in the internal space will give a curve that does not close in general, which is the effect of noncommuting translations in the internal space. 

The reader may wonder how the ``deformation" of the Minkowski solution described here is manifest in a metric. We can define a metric by the usual expression
\ben
g_{\mu\nu}=e_{\mu}^a e_{\nu}^b \eta_{ab}\,.
\een
This metric would not determine the connection, but could be used to define distances in the spacetime in the usual way. Then, from (\ref{connection}), we get
\ben
g_{\mu\nu}=\eta_{\mu\nu}\,\frac{4}{1+\frac{x \cdot x}{l^2}}+x_{\mu}x_{\nu}\frac{(x\cdot x)}{\left((x\cdot x)+l^2\right)^2}\,.
\een
It should be stressed that the connection on spacetime is {\it not} the Levi-Civita connection of this metric, as we are working in a first order formulation where metric and connection are independent. There is a factor of 4 because of a term in (\ref{smallp}) which does not necessarily go to zero as $p\rightarrow 0$. With the identification (\ref{choice}), the soldering form always gets a contribution
\ben
{e_{\mu}}^i(x) = ({e_{\mu}}^i)_0 + {\delta_{\mu}}^i + O\left(\frac{x}{l}\right)\,.
\een
The limit $\kappa\rightarrow\infty$ is now identified with the limit $l\rightarrow\infty$, in which we recover the (rescaled) Minkowski metric. 

In deriving the expressions (\ref{connection}) we started with Minkowski space, which clearly solves the field equations of the Einstein-Cartan theory for an energy-momentum tensor cancelling the cosmological constant term, and vanishing internal spin. In changing the zero section, we then performed a $\dSG$ gauge transformation, under which the curvature $F$ transformed as
\ben
F(s(x))=s^{-1} (x) F(x) s(x).
\een
This is a general $\dSG$ rotation which mixes up the $\frak{so}(d-1,1)$ and $\frak{so}(d,1)/\frak{so}(d-1,1)$ parts of the connection and the curvature. Hence, the resulting connection will no longer solve the original field equations, but the field equations for an energy-momentum tensor which has also undergone a $\dSG$ transformation. This mixes up the energy-momentum and internal spin parts, combining them into an element of the Lie algebra $\frak{so}(d,1)$, the interpretation of which seems obscure at least.

A comment is in order with regard to physical units\index{physical units}. In addition to the energy scale $\kappa$, which is perhaps naturally identified with the Planck scale, the identification of lengths with momenta, necessary in the framework presented here, requires the choice of a unit of length $l$ which is not necessarily connected to the scale $\kappa$. It may well be that it is instead the cosmological constant which sets this length scale, leading to an astronomical scale instead of a sub-atomic one. And indeed, some more recent approaches to quantum gravity (\eg \cite{friedel}) use the product ${G\Lambda}$ as a dimensionless parameter in a perturbative expansion. A fixed positive $\Lambda$ also seems to be required in non-perturbative approaches to quantum gravity \cite{quantsym}. Then the cosmological constant may play the role of a fundamental parameter in quantum gravity.

\sect{Summary and Discussion}

It has been argued that the algebra of DSR describes the symmetries of a semiclassical limit of (a generic theory of) quantum gravity\index{quantum gravity} (see \eg \cite{quantsym}). If this claim is taken seriously, one has to give an interpretation of the noncommuting translations appearing in the algebra, and usually they are supposed to represent a spacetime with a fundamentally noncommutative structure \cite{madore}. Alternatively, one may view the apparent noncommutativity as an artefact of the finite resolution of lengths \cite{oriti}. However, there are fundamental difficulties in associating these noncommuting operators directly with coordinates on spacetime, as position is not {\it additive} in a way that momentum and angular momentum are \cite{okon}. Furthermore, as also pointed out in \cite{okon}, a proposed noncommutativity of spacetime of the form (\ref{commutators}), proportional to angular momentum or boost generators, and hence vanishing at a given ``origin", seems deeply at odds with any idea of (even Galilean) relativity. This would also be an obvious criticism of the framework presented in this chapter, when taken as a theory that is supposed to describe the real world.

What we have shown here, is that using the framework of Einstein-Cartan-Stelle-West theory, one reaches a different conclusion from the usual one: The noncommutativity of translations on a momentum space of constant curvature is interpreted as torsion of a connection that solves the equations of Einstein-Cartan theory with a modified energy-momentum tensor that mixes with the spin tensor. If one takes this seriously, one is led to conclude that there is an effect of torsion induced by quantum gravity, whose effects would however only become measurable over distances comparable to $l$, a length scale presumably associated with the cosmological constant.

No such effect appears in de Sitter space with an appropriate cosmological constant, or indeed any vacuum solution of the theory. Vacuum solutions are then just described by the Poincar\'e algebra, and hence undeformed special relativity. It is curious that the curvature in the internal ``momentum'' space is reflected in torsion when Minkowski spacetime is considered as spacetime, which would however seem rather unnatural as it is not a vacuum solution any more. This result does however seem to be the most natural interpretation of the DSR algebra in a differential-geometric framework capable of describing more general spacetimes.

Any non-zero energy-momentum tensor, however, will lead to a connection having torsion\index{torsion}. In theories such as Einstein-Cartan theory, this leads to well-known problems when trying to couple the gravitational field to Maxwell fields, for instance, as there is no unambiguous procedure of minimal coupling\index{minimal coupling}. This is because the statement that the exterior derivative is independent of the choice of connection,
\ben
(dA)_{\mu\nu}\propto\partial_{[\mu}A_{\nu]}=\nabla_{[\mu}A_{\nu]}\,,
\een
is true precisely when torsion vanishes. Using an $\frak{so}(d-1,1)$ connection, this is apparent from
\ben
d(e^i A_i)=\nabla(e^i A_i)=A_i \nabla e^i - e^i \wedge\nabla A_i = - e^i \wedge \nabla A_i + A_i T^i
\een
where $\nabla e^i=de^i+{\omega^i}_j\wedge e^j$ etc. One has two different candidates for the field strength $F$, namely $e^i\wedge \nabla A_i$ and $d(e^i A_i)$, with possibly observable differences between these choices, although it could be argued that $F=dA$ is the only meaningful choice because it preserves gauge invariance \cite{benn}.

In the framework of Einstein-Cartan-Stelle-West theory, gauge fields should be coupled to gravity via development\index{development}, \ie replacing $F=dA$ by $F=DA$. We compute from (\ref{explicit}) and (\ref{development}) that development can be expressed in terms of $\omega_0$ and $e_0$ by
\ben
D= d + \omega - (e\cdot V) = d + \omega_0 + 2(p \otimes_A e_0)\kappa\frac{(\gamma(p)-1)}{l(p\cdot p)}=: d + \omega_{{\rm eff}}\,,
\een
where $\otimes_A$ is an antisymmetrised tensor product, $(U\otimes_A V)^{ab} \equiv U^{[a}V^{b]} \equiv \half(U^a V^b - U^b V^a)$. Parallel transport is effectively described by the connection $\omega_{{\rm eff}}$, whose torsion is in general non-zero. One can give an explicit formula for the torsion which is however rather complicated and does not seem to give much insight; to linear order in $p^i$, one has
\ben
T^i = \frac{l}{\kappa}({R^i}_b)_0 p^b - \frac{1}{2 l\kappa}e^i_0 \wedge (e^0_j p^j) + \frac{1}{2\kappa^2}\left(p^i e_{j0}\wedge dp^j - p_j e_0^i \wedge dp^j\right)+O(p^2).
\een
If we assume a universal relation of internal momenta and spacetime lengths of the form $p\sim\frac{\kappa}{l}x$, the second and third terms seem to give contributions of order $x/l^2$. The first term is proportional to the local curvature of $\omega_0$, ${R^i}_b = d{\omega^i}_b + {\omega^i}_j\wedge{\omega^j}_b$, contracted with $x^b$. Note that it is the Riemann tensor, not the Ricci tensor, that appears, so that propagating degrees of freedom of the gravitational field are included. This first term should in realistic situations, even in vacuum, give the dominant contribution.

Assuming that minimal coupling is achieved through the development operator $D$, or equivalently by using the effective connection which has torsion, one would couple vector or matter fields (using $D\psi$ for spinors) to torsion, breaking gauge invariance. Such an effect of course leads to the absence of charge conservation, and this should be experimentally observable in the presence of a non-trivial gravitational field, \ie in regions where spacetime is not exactly de Sitter. Let us recall that in standard tensor calculus one uses the identity
\ben
[\nabla_{\mu},\nabla_{\nu}]M_{\lambda\rho}={R_{\mu\nu\lambda}}^{\sigma}M_{\sigma\rho}+{R_{\mu\nu\rho}}^{\sigma}M_{\sigma\lambda}-{{T_{\mu}}^{\sigma}}_{\nu}\nabla_{\sigma}M_{\lambda\rho}
\een
which gives for an antisymmetric $M_{\lambda\rho}$ when contracted
\ben
[\nabla^{\lambda},\nabla^{\rho}]M_{\lambda\rho}=-g^{\mu\lambda}g^{\nu\rho}{{T_{\mu}}^{\sigma}}_{\nu}\nabla_{\sigma}M_{\lambda\rho},
\een
to establish that the right-hand side of Maxwell's equation $\nabla^{\lambda}F_{\lambda\rho}=4\pi J_{\rho}$ satisfies a continuity equation in the absence of torsion. With torsion present, one has then for any region $R$
\ben
\int_{\partial R}d^3x \;\sqrt{h}\;n^{\lambda} J_{\lambda} = \frac{1}{4\pi}\int_R d^4x\; \sqrt{g}\;\left(-g^{\mu\lambda}g^{\nu\rho}{{T_{\mu}}^{\sigma}}_{\nu}\nabla_{\sigma}F_{\lambda\rho}\right).
\een
Effects become important when the size of the region $R$ is comparable to the length scale of torsion.

As an example consider the Schwarzschild\index{Schwarzschild} solution, which has Kretschmann scalar
\ben
R_{abcd}R^{abcd} \sim \frac{r_S^2}{r^6},
\een
so roughly $R_{abcd}\sim r_S r^{-3}$. Assuming that the origin of the $x$ coordinate system corresponds to the centre of the Earth, we would, on the surface of the Earth, measure a torsion of order $r_S R^{-2}$, where $R$ is the radius of the Earth. With $r_S\sim 10^{-2}$ m and $R^2 \sim 10^{13}\,{\rm m}^2$, the length scale for effects of torsion would be about $10^{11}$ m. The other two contributions, given that $l\sim 10^{26}$ m, would be much smaller. Although this crude estimate suggests that effects will be very small, even tiny violations of charge conservation should have been observed experimentally. For a discussion of experimental tests of charge conservation and possible extensions of Maxwell theory in Minkowski space, see \cite{exptest}. Processes such as electron decay on a length scale of $10^{11}$ m, or a time scale of $10^{3}$ s, can clearly be ruled out. 

The example presented here shows that the correct physical interpretation of purely algebraic relations, such as the commutators of the Snyder algebra, may not be the seemingly obvious one. We conclude that the physical motivation for assuming spacetime is ``noncommutative" may not be as clear as often assumed. 

\sect{Gauge Invariance Broken?}

When discussing the issue of possible consequences of broken gauge invariance, we must bear several points in mind.

The idea that an asymmetry between the proton and electron charges could have interesting astrophysical consequences goes back to Lyttleton and Bondi \cite{bondi}, who argued that a charge difference, and hence a net charge of the hydrogen atom, of $10^{-18}$ elementary charges, might explain the observed expansion of the universe by electrostatic repulsion. This idea was proposed in connection with Hoyle's ideas of a universe in a steady state, which required continuous production of material via a ``creation field"\index{creation field} \cite{hoyle}, and a modification of Maxwell's equations was proposed to accommodate charge nonconservation. From Hoyle's perspective, however, the steady state model was incompatible with expansion of the universe by electrostatic repulsion, and should lead to electrostatic attraction instead \cite{hoyle2}. 

There seems to be no need for the electron and proton charges to be of equal magnitude to maintain gauge invariance. However, if the universe as a whole is not neutral, but it is homogeneous, gauge invariance must be broken. Hence the two issues are closely related. Modern laboratory experiments\index{experiment} \cite{laboratorium} give a bound of $10^{-21}$ elementary charges on the difference of electron and proton charge; astrophysical considerations give bounds of $10^{-26}$ elementary charges using the isotropy of the cosmic microwave background \cite{cmbbounds}, or $10^{-29}$ elementary charges by considering cosmic rays \cite{raybounds}. Recently, an interesting proposal to measure net charges of atoms and neutrons, sensitive to $10^{-28}$ elementary charges, was put forward \cite{expproposal}.

From a theoretical viewpoint, if gauge invariance is broken, it is natural to assume a nonvanishing photon mass. One then considers Einstein-Proca theory\index{Einstein-Proca theory}, an outline of which can be found in \cite{hejna}. The photon may also be charged. Here, experimental bounds on the charge are $10^{-29}$ elementary charges using pulsars \cite{pulsar}, and possibly $10^{-35}$ elementary charges from CMB isotropy \cite{cmbbounds}\index{cosmic microwave background (CMB)}. 

Experimental bounds on violations of gauge invariance in electrodynamics are very tight, and hence any theory predicting torsion which is coupled to electromagnetism faces severe problems when confronted by experiment. In the framework of Einstein-Cartan-Stelle-West theory, it is possible to maintain gauge invariance by choosing $F=dA$, but using the development operator is the most natural choice.

\sect{Symmetric Affine Theory}

If Einstein-Cartan theory is considered as the extension of general relativity which allows for torsion, there is an analogous extension which allows for a non-metric connection. This theory can be formulated in terms of a torsion-free $\frak{gl}(n,\bR)$ connection and is known as {\bf symmetric affine theory}\index{symmetric affine theory}. It is equivalent to standard general relativity with a massive vector field, known as (nonlinear) Einstein-Proca theory \cite{hejna}.

One could attempt to embed this theory into a theory of a connection taking values in the algebra of the affine group $\frak{a}(n,\bR)$\footnote{For a comprehensive review of general theories of this type, see \cite{hehl}.},
\ben
A=\left(\begin{matrix}
&  & & \cr & {\omega^a}_b & & \frac{1}{l}e^i \cr & & & \cr  & 0 & & 0
\end{matrix}\right)\,,
\een
where now ${\omega^a}_b$ is not constrained by $\omega^{ab}=-\omega^{ba}$. Geometrically, this means that the connection does not preserve the lengths of vectors under parallel transport.

The corresponding curvature of $A$ ($R$ is the curvature of the $\frak{gl}(n,\bR)$ part),
\ben
F=dA+A\wedge A=\left(\begin{matrix}
& & & \cr & {R^a}_b & & \frac{1}{l}T^i\equiv\frac{1}{l}(de^i + {\omega^i}_j \wedge e^j) \cr \cr & 0 & & 0
\end{matrix}\right)\,,
\een
would then be constrained by demanding that $T^i\equiv 0$. This seems rather unnatural from the perspective of Cartan geometry. Furthermore, the length scale $l$ is now completely arbitrary as it does not appear in the $\frak{gl}(n,\bR)$ part of the curvature any more, just as for a connection taking values in the Poincar\'e algebra.

One proceeds by considering Lagrangians that only depend on the Ricci tensor, which is a one-form obtained by contracting the components of the Riemann curvature, written in the basis of one-forms given by the vielbein $e^i$:
\ben
{\rm Ric}_a = {\rm Ric}_{ia}e^i,\quad {\rm Ric}_{ia}={{R_{ji}}^j}_a,
\een 
where the curvature two-form is
\ben
{R^a}_b = \frac{1}{2}{{R_{ij}}^a}_b e^i\wedge e^j.
\een
One then splits the Ricci tensor into symmetric and antisymmetric part, symmetrising over a component (with respect to the given basis) index and a $\frak{gl}(n,\bR)$ index. The antisymmetric part can be interpreted as a spacetime two-form
\ben
i_{e_a}\left({R^a}_b\wedge e^b\right),
\een
where $i_{e_a}$ is interior multiplication with the vector ${e_a}$, defined by being dual to the one-forms ${e^b}$:
\ben
e^b(e_a)={\delta^b}_a.
\een
No such construction is possible for the symmetric part, which is normally more relevant in concrete constructions. The splitting itself seems depend on the choice of basis.

\chapter{General Relativity From a Constrained Topological Theory}
\label{bflinear}
\epigraph{\em{``What, then, is time? If no one ask me, I know; if I wish to explain to him who asks, I know not.'' \cite{augustine}}}{St. Augustine}
\sect{Overview}

In this chapter, we shall investigate another possible formulation of general relativity in four dimensions, in terms of a constrained topological theory. Here the term ``topological" means that the theory has a very large symmetry: All of its solutions are locally gauge equivalent, and there are no local degrees of freedom. Globally, not all solutions are necessarily gauge equivalent if there are topological obstructions. Starting from such a theory, we will use constraints that restrict the possible gauge transformations and hence allow for a less trivial theory, namely one with local degrees of freedom that are those of general relativity.

The topological theory\index{topological field theory} we consider is usually known as BF theory\index{BF theory}, deriving its name from the action
\ben
S = \int B^{ab} \wedge F_{ab}[A]\,,
\label{BFakshn}
\een
where $F$ is the curvature of a $G$-connection $A$ and $B$ is a $\frak{g}$-valued two-form (in our case $G$ will be a rotation group). The equations of motion are
\ben
F^{ab}=0\,,\quad \nabla B^{ab}=0\,,
\een
where $\nabla$ is a covariant exterior derivative involving the connection $A$. All flat connections are locally gauge equivalent, and the (local) equivalence of all solutions to the second equation can be seen \cite{baez} by noting that the transformation
\ben
B^{ab}\mapsto B^{ab}+\nabla \eta^{ab}\,,
\een
where $\eta^{ab}$ is a $\frak{g}$-valued one-form, is a symmetry of the action (\ref{BFakshn}). This transformation is generated by the equation $F^{ab}=0$, in a canonical sense, and should be viewed as a generalised gauge transformation. It is a symmetry on the phase space of the theory. Locally, any solution to $\nabla B^{ab}=0$ is of the form $\nabla\eta^{ab}$, so any solution is gauge equivalent to $B^{ab}=0$ in this generalised sense.

The motivation for considering BF theory is that general relativity in three dimensions is just of this form: The first order formulation can be written as\index{Einstein-Hilbert action}
\ben
S = \int \epsilon_{abc}\left(e^a\wedge R^{bc}[\omega]\right)\,,
\een
and hence general relativity in three dimensions is topological. Since the quantisation of this system is relatively well understood (see \cite{witten} for Witten's\index{Witten} formulation as a Chern-Simons theory, and the book \cite{carlip} for an overview over different routes to quantisation\index{quantum gravity}), and seems to have many conceptual aspects such as diffeomorphism and Lorentz invariance in common with the higher-dimensional case, one might hope to gain understanding of higher-dimensional general relativity from studying BF theory. On the other hand, there is of course a fundamental difference between a system with an infinite number of (local) degrees of freedom and one with only (usually a finite number of) global degrees of freedom, which makes it rather unclear whether this approach will be fruitful.

It has been shown \cite{bfformul} that in any number of dimensions $d\ge 4$ general relativity can be formulated as a constrained BF theory, where one adds to the BF action constraints quadratic in the field $B^{ab}$, restricting it to be a wedge product of one-forms $e^a$ that can be interpreted as a frame field\index{orthonormal frame}. In four dimensions, where one has to enforce
\ben
B^{ab}={\epsilon^{ab}}_{cd}e^c\wedge e^d
\een
to reproduce the first order GR action (\ref{palatini}), this formulation (in slightly different complex form) was first studied by Plebanski \cite{plebanski}. This is also the situation most studied in the spin foam approach to quantum gravity\index{quantum gravity!spin foam approach}, where one tries to quantise general relativity by quantising a BF-type theory and then implementing constraints.

In this chapter, we shall investigate a formulation which is in the spirit of the Plebanski formulation, but involving only linear constraints, of the type used recently as a new idea in the spin foam approach to quantum gravity. We identify both the continuum version of the linear simplicity constraints used in the quantum discrete context and a linear version of the quadratic volume constraints that are necessary to complete the reduction from the topological theory to gravity. We illustrate and discuss also the discrete counterpart of the same continuum linear constraints. Moreover, we show under which additional conditions the discrete volume constraints follow from the simplicity constraints, thus playing the role of secondary constraints. 

In the general context of the thesis, we see how a very large symmetry present in the original action (\ref{BFakshn}) can be constrained appropriately so as to give a theory with rather different physics -- one with local degrees of freedom. This symmetry acts neither on spacetime nor on fibre bundles, but on a symplectic manifold, namely the phase space of a dynamical system.

\sect{Actions for Gravity}
It should not be surprising that the equations of general relativity can be derived from several different action principles, leading to equivalent classical theories (in the case of pure gravity, at least). The statement that given equations of motion arise from different actions is the functional equivalent of saying that there are many functions with the same stationary points. Apart from constructing different actions as functionals of the same variables, one can also express the same physical content in terms of different variables (by performing a Legendre transformation, or otherwise), leading to even more possibilities for putting the same physics into a different mathematical form.

This introduces several possible sources of ambiguity into quantisation; Feynman's path integral approach to quantum mechanics is based on the fact that off-shell configurations, \ie those that do not solve the equations of motion and are not stationary points of the action, contribute to transition amplitudes as well. Hence different functionals of the same variables, while giving the same equations of motion, may give different quantum theories. For formulations in terms of different variables it is even less clear whether the resulting quantum theories are related or not. One may see this as a problem or as a virtue; when a classical theory such as general relativity seems to resist conventional quantisation techniques, one may hope that one of the possible classical reformulations offers more promise for quantisation. The most obvious example of this is Ashtekar new variables\index{Ashtekar new variables} \cite{ashtekar} as a possible new route to quantising general relativity in four dimensions. In three dimensions, a number of consistent quantisation schemes for general relativity are known, leading to (at least apparently) inequivalent theories \cite{carlip}.

But already classically, in the case of general relativity equivalent formulations for pure gravity can give non-equivalent theories when coupled to matter. We encountered an instance of this in chapter \ref{defgenrel} when discussing Einstein-Cartan theory; in contrast to conventional general relativity, where torsion is set to zero by {\em fiat}, in Einstein-Cartan theory matter can act as a source of torsion. This is a non-equivalence on the level of equations of motion which can therefore be tested more easily by experiment. In this chapter we shall consider a reformulation of pure general relativity only.

\begin{figure}[htp]
\centering
\begin{picture}(350,250)
\put(0,0){\line(1,0){80}}\put(0,0){\line(0,1){50}}\put(0,50){\line(1,0){80}}\put(80,0){\line(0,1){50}}
\put(15,34){Hilbert-}\put(15,22){Palatini}\put(15,10){$\mathcal{L}(e^{\mu}_a,\omega^{ab}_{\mu})$}
\put(80,25){\vector(1,0){100}}\put(82,13){$\frac{\delta S}{\delta \omega}=0\Rightarrow\omega=\omega(e)$}
\put(180,0){\line(1,0){80}}\put(180,0){\line(0,1){50}}\put(180,50){\line(1,0){80}}\put(260,0){\line(0,1){50}}
\put(195,34){Einstein-}\put(195,22){Hilbert}\put(205,10){$\mathcal{L}(e^{\mu}_a)$}
\put(220,50){\vector(0,1){40}}\put(220,90){\vector(0,-1){40}}
\put(180,90){\line(1,0){80}}\put(180,90){\line(0,1){50}}\put(180,140){\line(1,0){80}}\put(260,90){\line(0,1){50}}
\put(230,72){Legendre}\put(230,60){transform}
\put(205,124){ADM}\put(190,112){Hamiltonian}\put(195,100){$\mathcal{H}(e^i_a,\pi_j^b)$}
\put(20,90){\line(1,0){80}}\put(20,90){\line(0,1){50}}\put(20,140){\line(1,0){80}}\put(100,90){\line(0,1){50}}
\put(20,124){Hilbert-Palatini}\put(30,112){Hamiltonian}\put(35,100){$\mathcal{H}(\omega^{ab}_i,\pi^i_{ab})$}
\put(60,50){\vector(0,1){40}}\put(60,90){\vector(0,-1){40}}
\put(70,72){Legendre}\put(70,60){transform}
\put(100,115){\vector(1,0){80}}\put(103,103){solve 2nd class}\put(103,91){constraints}
\put(0,180){\line(1,0){80}}\put(0,180){\line(0,1){50}}\put(0,230){\line(1,0){80}}\put(80,180){\line(0,1){50}}
\put(6,214){BF-Plebanski}\put(11,202){(real form)}\put(6,190){$\mathcal{L}(\omega^{ab}_{\mu},B^{ab}_{\mu\nu},\Xi)$}
\put(10,50){\vector(0,1){130}}\put(10,180){\vector(0,-1){130}}
\put(15,155){$\frac{\delta S}{\delta \Xi}=0\;\Rightarrow\;B=B(e)$}
\put(80,205){\vector(1,0){110}}\put(85,193){$\frac{\delta S}{\delta B}=0,\frac{\delta S}{\delta \Xi}=0 \Rightarrow $}\put(85,178){$B=B(\omega),\Xi=\Xi(\omega)$}
\put(82,210){$SO(3,1)\rightarrow SO(3,\mathbb{C})$}
\put(190,180){\line(1,0){80}}\put(190,180){\line(0,1){50}}\put(190,230){\line(1,0){80}}\put(270,180){\line(0,1){50}}
\put(210,214){CDJ}\put(205,202){Lagrangian}\put(205,190){$\mathcal{L}(\eta,\omega^{ab}_{\mu})$}
\put(300,120){\line(1,0){80}}\put(300,120){\line(0,1){50}}\put(300,170){\line(1,0){80}}\put(380,120){\line(0,1){50}}
\put(310,154){Ashtekar}\put(310,142){Hamiltonian}\put(310,130){$\mathcal{H}(E_{ai},A^{ai})$}
\put(260,125){\vector(1,0){40}}\put(265,110){complex canonical}\put(265,97){transform}
\put(300,170){\vector(-1,1){30}}\put(280,190){Legendre}\put(292,178){transform}
\end{picture}
\caption[Lagrangians and Hamiltonians for general relativity in four dimensions.]{{\small Lagrangians and Hamiltonians for general relativity in four dimensions. (Essentially taken from \cite{peldan}.)}}
\label{akshns}
\end{figure}
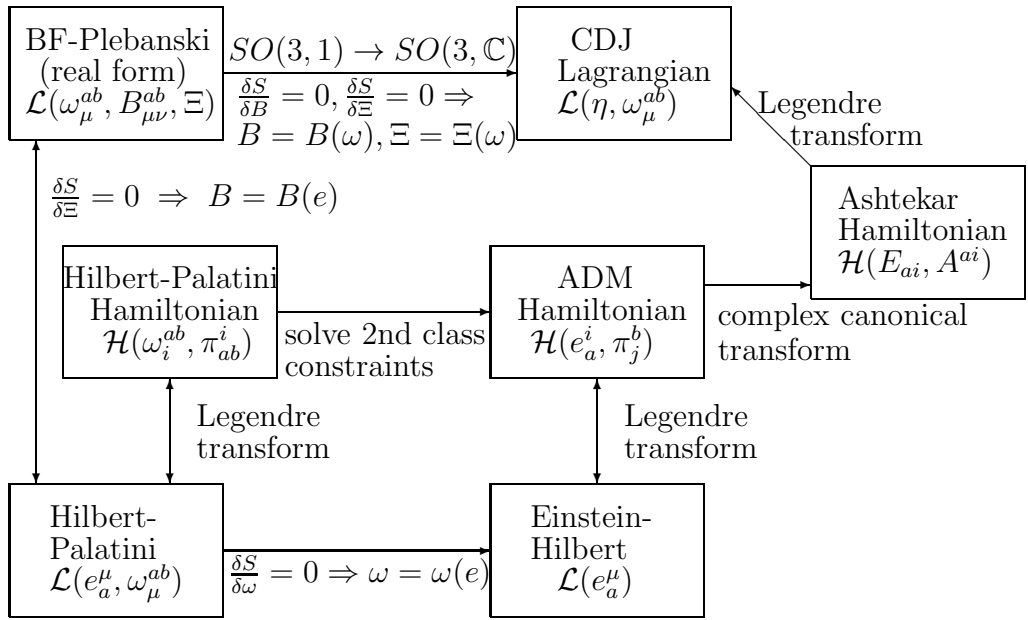

In figure \ref{akshns}, essentially taken from the review \cite{peldan}, we show a few possible action principles for general relativity and their interrelations.\footnote{Figure \ref{akshns} is by no means meant to be a complete display of all known actions for general relativity; for instance, an interesting variant of the (complex) Plebanski formulation has recently been explored by Eyo Eyo Ita III (Ph.D. thesis, University of Cambridge (2010)).} The most commonly used ones are the Einstein-Hilbert action \cite{hilbert}, the Palatini first order formulation, and its modification proposed by Holst \cite{holst}. This last one is of special interest because it is the classical, covariant starting point for the canonical quantisation leading to loop quantum gravity\index{quantum gravity!loop} \cite{rovelli}. Even if not the only possible useful one \cite{leeartem}, a particularly popular action in covariant approaches to quantising gravity \cite{libro}, like the spin foam \cite{SF} and group field theory approach \cite{iogft}, is the formulation as a constrained BF (or Plebanski \cite{plebanski, kirill}) theory. Here one starts from topological BF theory \cite{horowitzBF} in four spacetime dimensions, and adds suitable constraints\index{constraints} on the two-form $B$ variables of the theory such that, on solutions of these constraints, the action reduces to the Palatini or Holst action for general relativity. We will summarise the idea behind this formulation in the following. In the original Plebanski formulation the constraints on the $B$ variables are quadratic, and so are the discrete constraints that are then implemented in the spin foam models based on a simplicial discretisation. On the other hand, the most recent developments in the spin foam and group field theory approach to quantum gravity are based on a linear set of discrete constraints, which can be shown to be slightly stronger, in the restrictions they impose on the original BF configurations, than some of the original discrete quadratic constraints. Once more, we will detail this construction in the following. 

Here we investigate whether a formulation in terms of linear constraints is also possible in the classical continuum theory, and what it implies. We will see that the replacement of the so-called diagonal and cross-simplicity constraints with linear constraints  at the continuum level is relatively straightforward, after one has introduced new variables $n_a$ forming a basis of three-forms at each point. One then needs additional constraints corresponding to the volume constraints. We will see that one can linearise these constraints too. We then give a discrete version of these linear volume constraints (thinking of spacetime which is discretised in a triangulation), which bears a striking resemblance to the so-called ``edge simplicity" constraints of \cite{biancajimmy}. We note that only certain linear combinations of the volume constraints one would naively write down are necessary to constrain the bivectors $\Sigma^{AB}(\triangle)$ sufficiently.
Similarly to the quadratic case, we will also see that when linear diagonal and off-diagonal constraints hold everywhere in a 4-simplex\index{4-simplex}, and one also imposes ``closure" constraints on both bivectors $\Sigma^{ab}(\triangle)$ (associated to triangles) and normals $n_a(\tetrahedron)$ (associated to tetrahedra), the sufficient set of linear combinations of the linear volume constraints follows. This additional four-dimensional closure constraint on the normal vectors has, to the best of our knowledge, not been considered or implemented as an additional condition in the spin foam literature yet, although it does appear in some first order formulation of Regge calculus \cite{caselle}, and it plays also a role in the discrete analysis of \cite{FreidelConradysemiclassical}.

\sect{Constraining BF Theory}
\label{introBF}
Let us briefly review what is known at the classical continuum and discrete level, concerning the Plebanski formulation of classical gravity. We limit our considerations to the covariant, Lagrangian context, and to a very small subset of the available results, those which have been already of direct relevance for quantum gravity model building\index{quantum gravity}, especially in the spin foam context. For recent results in the canonical Hamiltonian setting, both continuum and discrete, see \cite{henneaux, biancajimmy, zapata}.

In this chapter we consider the Einstein-Hilbert-Palatini-Holst \cite{hilbert,holst} Lagrangian\index{Einstein-Hilbert action} (without cosmological constant), a modification of the action (\ref{palatini}) used earlier,
\ben
S_{{\rm EHPH}} = \frac{1}{8\pi G}\int_{\mathcal{R}\times\mathbb{R}}\left(\half\,\epsilon_{abcd}\,e^a\wedge e^b\wedge R^{cd}[\omega]+\frac{1}{\gamma}\,e^a\wedge e^b\wedge R_{ab}[\omega]\right)\,,
\label{einsthilbBF}
\een
where spacetime is assumed to be of the form $\mathcal{R}\times\mathbb{R}$ so that a (3+1) splitting can be performed, $\omega^{ab}$ is a $G$-connection one-form (the gauge group $G$ is $SO(3,1)$ or $SO(4)$, or an appropriate cover), $R^{ab}$ its curvature, and $e^a$ is an $\bR^4$-valued one-form representing an orthonormal frame. The term involving $\gamma$, known as the Holst term, is not relevant classically; it is, up to a total derivative\footnote{For a discussion of the total derivative term see \cite{mercuri}.}, proportional to $T^a\wedge T_a$, and hence does not modify the classical equation of motion $T^a=0$. When (\ref{einsthilbBF}) is coupled to matter, the Holst term leads to a rescaling of the coupling of the matter spin density to torsion.

Though not relevant classically, the Holst term is of fundamental importance in loop quantum gravity (LQG), where $\gamma$ is known as the Barbero-Immirzi parameter, and more generally for any canonical formulation of gravity, as it modifies the symplectic structure of the theory.

If one introduces a $\frak{g}$-valued two-form,
\ben
B^{ab}=\frac{1}{8\pi G}\left(\half{\epsilon^{ab}}_{cd}\,e^c\wedge e^d+\frac{1}{\gamma}e^a\wedge e^b\right),
\label{bfield}
\een
then the action (\ref{einsthilbBF}) becomes\footnote{We use indices from the beginning of the Greek alphabet such as $\alpha$ to denote abstract indices not associated with certain transformation groups.}
\ben
S = \int B^{ab}\wedge R_{ab}[\omega]+\lambda^{\alpha}C_{\alpha}[B]\,,
\een
\ie it takes the form of a topological BF theory with additional constraints\index{constraints} $C_{\alpha}$ which enforce that $B^{ab}$ is indeed of the form (\ref{bfield}), and that are enforced by means of Lagrange multipliers $\lambda_\alpha$. 

As said, BF theory without constraints is topological. Its equations of motion imply that $\omega^{ab}$ is flat and the covariant exterior derivative of $B^{ab}$ vanishes. Having no local degrees of freedom, the quantisation of such a theory is therefore rather simple and quite well understood. Inspired by this classical formulation, the main issue when trying to construct a quantum theory related to quantum gravity, in four dimensions, is then the correct implementation of appropriate constraints that lead to (\ref{bfield}) for some set of one-forms $e^a$, either at the level of quantum states or in a path integral formulation. Indeed, the bulk of the work in the spin foam approach \cite{eprlong, freidelkrasnov, consistent, SF} (as well as in the group field theory formalism \cite{danieleAristide, iogft, danieleGFtsimpl}), in recent years, has been devoted to this task.
These constraints are also the subject of this chapter.

To simplify the following calculations, we introduce another two-form field $\Sigma^{ab}$,
\ben
\Sigma^{ab}\equiv\frac{1}{1-s\gamma^2}\left(B^{ab}-\frac{\gamma}{2}{\epsilon^{ab}}_{cd}B^{cd}\right) \label{redefB}\,,
\een
where $s$ is the spacetime signature, $s=-1$ for $G=SO(3,1)$ and $s=+1$ for $G=SO(4)$ (and we assume $\gamma^2 \neq s$)\footnote{For uniformity of the discussion, we shall in the following talk about ``time'' and use the label $0$ even when the gauge group is $SO(4)$ and the signature Riemannian.}. This is a linear redefinition which simplifies the constraint (\ref{bfield}),
\ben
\Sigma^{ab} = \frac{1}{8\pi\gamma G}\,e^a\wedge e^b\,,
\label{sigma}
\een
but leads to more terms in the action. The translation of all calculations from one set of variables to the other is usually straightforward.

The traditional way to enforce the restriction (\ref{sigma}), the one matching the original classical Plebanski formulation of gravity, was to add quadratic {\bf simplicity constraints}\index{constraints!simplicity}\footnote{``Simplicity'' because a two-form that can be written as a wedge product of one-forms is called simple.} to the action \cite{freidelpleb,reisenclass},
\ben
\epsilon_{abcd}\Sigma^{ab}_{\mu\nu}\Sigma^{cd}_{\rho\sigma}=V\epsilon_{\mu\nu\rho\sigma}\,,
\label{quadconst}
\een
where $V$ can be expressed in terms of $\Sigma^{ab}$ by contracting (\ref{quadconst}) with $\epsilon^{\mu\nu\rho\sigma}$, to give: $V=\frac{s}{24} \epsilon^{\mu\nu\rho\sigma}\epsilon_{abcd}\Sigma^{ab}_{\mu\nu}\Sigma^{cd}_{\rho\sigma}$. This is itself a reformulation of the original Plebanski constraint, which would read:
\ben
\epsilon^{\mu\nu\rho\sigma}\Sigma^{ab}_{\mu\nu}\Sigma^{cd}_{\rho\sigma}=V\epsilon^{abcd}\,,
\label{quadconstB}
\een
and is equivalent to the first under assumption that $V\neq 0$ everywhere. 
The version (\ref{quadconst}) has the advantage of permitting a much simpler discretisation and thus a more straightforward implementation within the spin foam formalism. 
Under the same assumption $V\neq 0$, there are the following four classes of solutions to (\ref{quadconst}):
\ben
\mbox{either }\;\Sigma^{ab}=\pm e^a\wedge e^b\quad\mbox{or }\;\Sigma^{ab}=\pm\half{\epsilon^{ab}}_{cd}E^c\wedge E^d
\label{ambiguity}
\een
for some set of one-forms $E^a$ or $e^a$ (in the following, we reserve $e^a$ for one-forms satisfying the first relation, viewed as encoding the metric; since we consider pure gravity, the factor $8\pi\gamma G$ can obviously be introduced by rescaling). One would like to select only the first class of solutions $\Sigma^{ab}=+e^a\wedge e^b$, which, when substituted in the BF action, gives the Holst action (\ref{einsthilbBF}). Classically, this is not a severe problem. As shown in \cite{reisenclass}, non-degenerate initial data of a solution of the form $\Sigma^{ab}=+e^a\wedge e^b$ generically remain within the same branch of solutions. The situation in the quantum theory, where one necessarily has contributions from all branches, is less clear. 

More troublesome, if $V=0$, the field $\Sigma^{ab}$ does not permit a straightforward geometric interpretation at all. Since in the region of the phase space where $V=0$, the theory is less constrained, and hence has more degrees of freedom, these non-geometric configurations should be expected to be dominating in a path integral \cite{reisenclass}, unless measure factors are such that this is avoided. 

Spin foam models are usually defined in a piecewise flat context, and spin foam amplitudes are defined for given simplicial complexes \cite{SF}. Therefore one is interested in identifying a discrete version of the above constraints that could be imposed at the level of each complex. The version  (\ref{quadconstB}) of the simplicity constraints admits only a rather involved discrete counterpart \cite{freidelpleb} and, upon quantisation, leads to the Reisenberger model \cite{freidelpleb,reisenberger}, which has so far received only limited attention. 

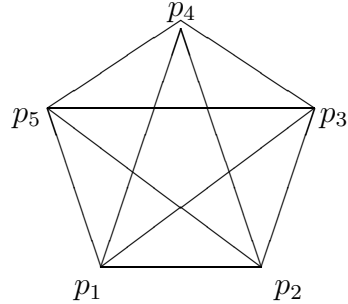
\begin{figure}[htp]
\centering
\begin{picture}(100,115)
\put(20,10){\line(1,0){60}}\put(10,0){$p_1$}\put(85,0){$p_2$}
\put(20,10){\line(-1,3){20}}\put(80,10){\line(1,3){20}}
\put(0,70){\line(3,2){50}}\put(100,70){\line(-3,2){50}}
\put(-13,65){$p_5$}\put(102,65){$p_3$}\put(45,105){$p_4$}
\put(0,70){\line(1,0){100}}
\put(20,10){\line(1,3){30}}\put(80,10){\line(-1,3){30}}
\put(20,10){\line(4,3){80}}\put(80,10){\line(-4,3){80}}
\end{picture}
\caption{{\small A 4-simplex is generated by five points.}}
\label{simpl}
\end{figure}

The discrete analogue of the constraints (\ref{quadconst}) led instead \cite{freidelpleb,reisenclass} to the construction of the Barrett-Crane model \cite{barrettcrane}, in the case in which the Barbero-Immirzi parameter is excluded from the original action ($\gamma\rightarrow\infty$). The construction is initially limited to a single 4-simplex, the convex hull of five points in $\mathbb{R}^4$ ($\mathbb{R}^{1,3}$, in the Lorentzian case) with the topology of a 4-ball, whose boundary is triangulated by the five tetrahedra identified by the five independent subsets of four such points, while subsets of three points  identify the four triangles belonging to each of these five tetrahedra, each of the triangles being shared by a pair of tetrahedra (see figure \ref{simpl}).
One then associates a Lie algebra element (bivector) $\Sigma^{ab}_\triangle\in\mathfrak{so}(4)\simeq \wedge^2\mathbb{R}^4$ (similarly in the Lorentzian case) to each triangle $\triangle$ in a given triangulation by integrating the two-form $\Sigma^{ab}$ over $\triangle$. The task is then to constrain appropriately these Lie algebra variables (or their quantum counterpart) following the continuum treatment.

It is useful to split the set of continuum equations (\ref{quadconst}) into two sets. Out of the 21 equations (\ref{quadconst}), one first identifies and imposes those 18 which have zero on the right-hand side (the ``diagonal'' and ``off-diagonal'' simplicity constraints), 
\ben
\epsilon_{abcd}\Sigma^{ab}_{\mu\nu}\Sigma^{cd}_{\mu\nu}=\epsilon_{abcd}\Sigma^{ab}_{\mu\nu}\Sigma^{cd}_{\mu\rho}=0 \quad \forall \, \mu,\nu,\rho\mbox{ (no sum over }\mu,\nu)\,.
\label{diagonal}
\een

This corresponds to the case if one or two of the indices of the two fields $\Sigma$ coincide. At the discrete level, this translates into two triangles on which the same fields are discretised which either coincide or at least share a single edge, and thus belong to the same tetrahedron. Thus all bivectors $\Sigma^{ab}_\triangle$ are required to satisfy 
$$\epsilon_{abcd} \Sigma_\triangle^{ab}\Sigma_\triangle^{cd}=0 \quad \text{(diagonal simplicity constraint)}$$ and $$\epsilon_{abcd}\,\Sigma_\triangle^{ab}\,\Sigma_{\triangle'}^{cd}\,=\,0 \quad \text{ for all} \quad \triangle,\triangle' \quad \text{sharing an edge (cross-simplicity constraint)}. $$ These two sets of equations can be imposed at the level of each tetrahedron in the 4-simplex.

The remaining three equations (the ``volume'' constraints) are equivalent to the requirement that:
\ben
\epsilon_{abcd}\Sigma^{ab}_{01}\Sigma^{cd}_{23}=-\epsilon_{abcd}\Sigma^{ab}_{02}\Sigma^{cd}_{13}=\epsilon_{abcd}\Sigma^{ab}_{03}\Sigma^{cd}_{12}\,\propto\, V(\Sigma)\,,
\label{quadvol}
\een
and can be imposed at the discrete level as the requirement that, for each 4-simplex:
\ben
\epsilon_{abcd}\,\Sigma_\triangle^{ab}\,\Sigma_{\triangle'}^{cd}\,=\,V \quad \text{ for all} \quad \triangle,\triangle' \quad \textbf{not} \quad \text{sharing an edge (volume constraints)}
\een
where $V$ is defined by the above equation, and is interpreted, on the solutions of the constraints, as the volume of the 4-simplex.

An additional condition on the bivectors is usually considered, namely the ``closure'' constraint, which states that the sum of four bivectors corresponding to the faces of one tetrahedron is zero:
\ben
\sum_{\triangle\subset\tetrahedron} \Sigma_{\triangle}^{ab}\,=\,0\, . \label{closure}
\een
This constraint can be understood in two ways. 
One can either view it as the condition that the triangles described by the variables $\Sigma^{ab}_{\triangle}$ close to form a tetrahedron \cite{quanttetra}, or as a consequence of the equations of motion. In a topologically trivial region such as the interior of a tetrahedron, a flat connection can be set to zero by a gauge transformation. Then using Stokes' theorem,\index{Stokes' theorem} the integral over the equation $d\Sigma^{ab}=0$ can be written as $\int_{\tetrahedron}\Sigma^{ab}=0$, which is the closure constraint. The canonical counterpart of this condition is the so-called Gauss constraint\index{constraints!Gauss}, which generates local gauge (rotation) transformations and is to be imposed on the quantum states of the theory.

The same picture appears in three spacetime dimensions, where there are no simplicity constraints and one directly deals with a $\mathfrak{su}(2)$ connection one-form $\omega^{a}$ and an $\mathfrak{su}(2)$-valued (using $\mathfrak{su}(2)\simeq \mathbb{R}^3$) one-form $e^a$. Here the equation $de^a=0$ is integrated over a (spacetime) triangle to give a closure constraint. The vectors associated to the edges of the triangle add up to zero, and thus have a consistent geometric interpretation as edge vectors in $\mathbb{R}^3$. In this sense, an $n$-form with vanishing exterior derivative and appropriate internal indices can be given a geometric interpretation as describing $n$-simplices closing up to form an $(n+1)$-simplex. We shall encounter another instance of this statement later on.

The closure constraint, being linear in the $\Sigma$'s and local in each tetrahedron, is obviously easier to impose at the discrete level, and in the quantum theory, than the volume constraints. Thus it is a useful fact that it can indeed be imposed instead of them. More precisely, it can be shown  \cite{consistent} that the volume constraints in each 4-simplex are implied if one has enforced the diagonal and cross-diagonal simplicity constraints, {\it plus the closure conditions} everywhere, \ie in all the tetrahedra of the 4-simplex (in general, \ie for non-degenerate 4-simplices, involving tetrahedra belonging to different \lq\lq time slices\rq\rq). From a canonical perspective, this observation is usually phrased as an interpretation of the volume constraints as ``secondary constraints" required to guarantee conservation of the other constraints (including the Gauss (closure) constraint) under time evolution. 

After a period of investigations, several potentially worrying issues have been put forward regarding the Barrett-Crane model \cite{alescirovelli, BCdegenerate} (for a more recent analysis of the geometry of the Barrett-Crane model, see \cite{danieleAristide}), and have given impetus to the development of alternative spin foam models \cite{eprlong,freidelkrasnov}. These models are known to have nice semiclassical properties \cite{nottingham}, and, importantly, generalise the spin foam setting to include the Barbero-Immirzi parameter at the quantum level (for an early attempt, see \cite{danieleeteraImmirzi}), and thanks to this allow for a more direct contact with the canonical loop quantum gravity. Their study is still somewhat preliminary, but the above properties make them promising candidates for a quantum theory related to gravity. One of the central features of the new models is the replacement of the quadratic simplicity constraints (\ref{diagonal}) by linear constraints of the form
\ben
n_a(\tetrahedron)\Sigma^{ab}(\triangle)=0 \quad \forall\triangle\subset\tetrahedron\,,
\label{linear}
\een
where $n_a$ is the normal associated to the tetrahedron $\tetrahedron$ and $\triangle$ is any of the faces of $\tetrahedron$. 

It can be shown that these are lightly stronger than the discrete diagonal and off-diagonal quadratic simplicity constraints, and remove some of the discrete ambiguity in the solution for $\Sigma^{ab}$: out of the classes of solutions (\ref{ambiguity}), one can restrict to (the discrete version of) $\Sigma^{ab}=\pm e^a\wedge e^b$ only. For a geometric analysis of these conditions in the discrete setting, see \cite{FreidelConradysemiclassical, eprlong, freidelkrasnov}, and for a proof that the same discrete conditions can also lead to the Barrett-Crane model, see \cite{danieleAristide}.

\sect{Linear Constraints for BF-Plebanski Theory}
\label{constraints}
The purpose of this chapter is to investigate whether a formulation in terms of linear constraints is also possible in the classical continuum theory, and what it implies.

Let us work backwards, at first. Assume that the two-form field $\Sigma^{ab}$ is of the form $\Sigma^{ab}=E^a\wedge E^b$, and that the ``frame field'' (not necessarily associated with a metric) $E^a$ is non-degenerate, \ie that the matrix $(E^a_\mu)$ is invertible. It follows that
\ben
E^\rho_a \Sigma^{ab}_{\mu\nu}=\delta_\mu^\rho E_\nu^b - \delta_\nu^\rho E_\mu^b\,.
\label{start}
\een
In order to make a connection to the discrete setting it is more convenient to work with exterior powers of the cotangent bundle only ($n$-forms can be integrated over $n$-dimensional submanifolds). Hence we multiply (\ref{start}) by $\epsilon_{\rho\sigma\tau\upsilon}$ and insert the relation $\epsilon_{\rho\sigma\tau\upsilon}E^\rho_a=\left(\det E^\mu_a\right)\epsilon_{adef}E^d_\sigma E^e_\tau E^f_\upsilon$, which is true for invertible matrices, obtaining
\ben
\epsilon_{adef}E^d_\sigma E^e_\tau E^f_\upsilon \Sigma^{ab}_{\mu\nu} = (\det E_\mu^a) \left(\epsilon_{\mu\sigma\tau\upsilon}E_\nu^b-\epsilon_{\nu\sigma\tau\upsilon}E^b_\mu\right)\,.
\label{linconst}
\een
One can define the three-form $n_{a\sigma\tau\upsilon}\equiv n_{a[\sigma\tau\upsilon]}\equiv \epsilon_{adef}E^d_{\sigma}E^e_{\tau}E^f_{\upsilon}$, so that (\ref{linconst}) take the form
\ben
n_{a\sigma\tau\upsilon} \Sigma^{ab}_{\mu\nu} = (\det E_\mu^a) \left(\epsilon_{\mu\sigma\tau\upsilon}E_\nu^b-\epsilon_{\nu\sigma\tau\upsilon}E^b_\mu\right)\,.
\label{betterform}
\een
$n_{a\sigma\tau\upsilon}$ can be interpreted as a 3D volume form for the submanifold parametrised by $(x^\sigma,x^\tau,x^\upsilon)$ embedded in 4D spacetime, whose internal index gives the normal to this submanifold. If $E^a$ are a basis of one-forms at each spacetime point, then $n_a$ are a basis of three-forms at each spacetime point, and so one can choose to work either with one or the other set of variables. Clearly $E^a$ can be reconstructed from $n_a$:
\ben
\frac{1}{6}\epsilon^{\rho\sigma\tau\upsilon}n_{a\sigma\tau\upsilon}=s(\det E_\mu^a)E^\rho_a=s\sqrt[3]{\det\left(\frac{1}{6}\epsilon^{\nu\sigma\tau\upsilon}n_{b\sigma\tau\upsilon}\right)}E^\rho_a\,.
\label{reconsta}
\een
This means that the set of variables $n_a(x)$ define a co-tetrad frame at any point of the spacetime manifold (for the discrete analogue of the above, see \cite{FreidelConradysemiclassical}).

\subsection{Linearised Diagonal and Off-Diagonal Constraints}

So far we have just rewritten the equation we want to obtain for $\Sigma^{ab}$. Let us now consider the implications of imposing (\ref{betterform}) as constraints, where we restrict to those with zero right-hand side, \ie those for which $\{\mu,\nu\}\subset\{\sigma,\tau,\upsilon\}$. These are half of the equations (\ref{betterform}). This will identify the continuum analogue of the linear simplicity constraints.

\newtheorem{kleem}{Claim}[section]
\begin{kleem} 
For a basis $n_{a}$ of three-forms, the general solution to
\ben
n_{a\sigma\tau\upsilon}\Sigma^{ab}_{\mu\nu}=0\quad\forall\{\mu,\nu\}\subset\{\sigma,\tau,\upsilon\}
\label{zerorighthandside}
\een
is
\ben
\Sigma^{ab}_{\mu\nu}=G_{\mu\nu} E^{[a}_{\mu} E^{b]}_{\nu}\,,
\label{solforb}
\een
where $E^a_\mu$ is defined in terms of $n_{a\sigma\tau\upsilon}$ as in (\ref{reconsta}), and so in particular non-degenerate, and $G_{\mu\nu}=G_{\nu\mu}$ and $G_{\mu\mu}=0$. (Obviously, as the variables $\Sigma^{ab}$ and $E^a$, the ``coefficients'' $G_{\mu\nu}$ are spacetime dependent.)
\end{kleem}

{\bf Proof.} First note that we can rewrite (\ref{zerorighthandside}) as
\ben
\epsilon_{adef}E^d_{\sigma}E^e_{\tau}E^f_{\upsilon} \Sigma^{ab}_{\mu\nu} = 0
\label{zero2}
\een
with $E^a_\mu$ defined by (\ref{reconsta}). Then $E^a_\mu$ by assumption defines a basis in the cotangent space, so that
\ben
\Sigma^{ab}_{\mu\nu}=G^{\chi\xi}_{\mu\nu}E^a_{\chi}E^b_{\xi}
\een
for some coefficients $G^{\chi\xi}_{\mu\nu}$ with $G^{\chi\xi}_{\mu\nu}\equiv G^{[\chi\xi]}_{[\mu\nu]}$. Substituting this into (\ref{zero2}), we get
\ben
0\stackrel{!}{=}\epsilon_{adef}E^a_{\chi} E^d_{\sigma}E^e_{\tau}E^f_{\upsilon} G^{\chi\xi}_{\mu\nu}E^b_{\xi} = \epsilon_{\chi\sigma\tau\upsilon}\det(E^a_{\mu})E^b_{\xi}G^{\chi\xi}_{\mu\nu}\,,
\een
and since $\det (E^a_{\mu})\neq 0$ and $E^b_{\xi}$ form a basis of (the internal) $\mathbb{R}^4$, this implies that
\ben
\epsilon_{\chi\sigma\tau\upsilon}G^{\chi\xi}_{\mu\nu} = 0\quad\forall\{\mu,\nu\}\subset\{\sigma,\tau,\upsilon\}\,.
\een
It follows that $G^{\chi\xi}_{\mu\nu}=0$ unless $\{\chi,\xi\}=\{\mu,\nu\}$ and so $G^{\chi\xi}_{\mu\nu}\equiv\delta^{[\chi}_{\mu}\delta^{\xi]}_{\nu}G_{\mu\nu}$.
\begin{flushright}
$\Box$
\end{flushright}

By a linear redefinition $e^a_\mu = \lambda_\mu E^a_\mu$ one might try to set some of the $G_{\mu\nu}$ to a given value (usually $\pm 1$, but one might prefer $\pm\frac{1}{8\pi\gamma G}$), but it is clear that one needs two conditions on the $G_{\mu\nu}$ for this to be possible.

In the discrete context, one sets $n_a(\tetrahedron)=(1,0,0,0)$ for each $\tetrahedron$ by a gauge transformation\index{gauge transformation}.\footnote{This seems to involve an implicit assumption, namely that there is a non-degenerate normal to each tetrahedron, as well.} One could use some of the gauge freedom here to restrict the form of $n_a$: This amounts to finding a convenient parametrisation for the coset space $GL(4)/SO(3,1)$. Let us make the (usual) assumption that the normal to hypersurfaces $\{t=\const\}$ is indeed timelike. Then one can use the boost part of $SO(3,1)$ to set $n_{a123}=(C,0,0,0)$. The remaining $SO(3)$ subgroup can then be used to make the ($3\times 3$) matrix $n_{i0\sigma\tau}$, where $i\in\{1,2,3\}$, upper diagonal, so that one has the form
\ben
n_{a\sigma\tau\upsilon}\sim\begin{pmatrix} * & * & * & * \cr 0 & * & * & * \cr 0 & 0 & * & * \cr 0 & 0 & 0 & * \end{pmatrix}\,.
\een
Clearly, when this form of $n_{a\sigma\tau\upsilon}$ is assumed, integrating the three-form $n_a$ over a region where $t$ is constant gives a vector in $\mathbb{R}^4$ that only has a time component. Its magnitude specifies the three-dimensional volume of such a region.

\subsection{Linearised Volume Constraints}
As in the quadratic case, further constraints, in addition to the linear simplicity constraints (\ref{zerorighthandside}), are needed to complete the identification $\Sigma^{ab}=\pm e^a \wedge e^b$.

First of all, one can show the following.

\newtheorem{claim2}[kleem]{Claim}\label{claim2}
\begin{claim2} Under the assumption that all $G_{\mu\nu}$ are non-zero, the necessary and sufficient conditions for the existence of a linear redefinition $e^a_\mu=\lambda_\mu E^a_\mu$, such that either $\Sigma^{ab}=c e^a\wedge e^b$ or $\Sigma^{ab}=-c e^a\wedge e^b$, where $c$ is a given positive number, are
\ben
G_{12}G_{03}=G_{01}G_{23}=G_{13}G_{02}(\neq 0)\,.
\label{gcondition}
\een
\end{claim2}

{\bf Proof.} Set $c=1$; the extension to arbitrary $c$ amounts to a further rescaling by $\sqrt{c}$. Then the required redefinition is possible if and only if there exist $\lambda_0,\ldots,\lambda_3$, such that either $G_{\mu\nu}=\lambda_\mu\lambda_\nu$ for all $\mu\neq \nu$ or $G_{\mu\nu}=-\lambda_\mu\lambda_\nu$ for all $\mu\neq \nu$. Clearly (\ref{gcondition}) are necessary. They are also sufficient: Take
\ben
\lambda_1=\sqrt{\left|\frac{G_{12}G_{13}}{G_{23}}\right|},\quad\lambda_2=\sgn\left(\frac{G_{12}G_{13}}{G_{23}}\right)\frac{G_{12}}{\lambda_1},\quad\lambda_3=\sgn\left(\frac{G_{12}G_{13}}{G_{23}}\right)\frac{G_{13}}{\lambda_1}\,,
\een
which solves the equations for $G_{12},G_{13}$ and $G_{23}$ with $\sgn\left(\frac{G_{12}G_{13}}{G_{23}}\right)$ specifying the overall sign. The remaining three equations for $G_{01},G_{02}$ and $G_{03}$ are then solved by the two relations (\ref{gcondition}) and 
\ben
\lambda_0=\sgn\left(\frac{G_{12}G_{13}}{G_{23}}\right)\frac{G_{01}}{\lambda_1}\,.
\een
\begin{flushright}
$\Box$
\end{flushright}

The assumption $G_{\mu\nu}\neq 0$ is necessary: One solution to (\ref{gcondition}) is $G_{12}=G_{23}=G_{13}=0$ with the other $G_{\mu\nu}$ non-zero, which cannot be expressed as $G_{\mu\nu}=\pm\lambda_\mu\lambda_\nu$.

Further constraints, in addition to the linear simplicity constraints (\ref{zerorighthandside}), are needed to complete the identification $\Sigma^{ab}=\pm e^a \wedge e^b$. One possibility is to use the quadratic volume constraints (\ref{gcondition}). Take the three volume constraints (\ref{quadvol}),
\ben
\epsilon_{abcd}\Sigma^{ab}_{01}\Sigma^{cd}_{23}=-\epsilon_{abcd}\Sigma^{ab}_{02}\Sigma^{cd}_{13}=\epsilon_{abcd}\Sigma^{ab}_{03}\Sigma^{cd}_{12}\,,
\een
and substitute the solution $\Sigma^{ab}_{\mu\nu}=G_{\mu\nu} E^{[a}_{\mu} E^{b]}_{\nu}$ of (\ref{zerorighthandside}). This gives precisely (\ref{gcondition}). The non-degeneracy assumption needed for (\ref{gcondition}) is then the usual one, namely $V\neq 0$ in (\ref{quadconst}).

This shows that imposing the linear version of the diagonal and off-diagonal simplicity constraints (\ref{zerorighthandside}) together with the quadratic volume constraints (\ref{quadvol}) and a non-degeneracy assumption on $\Sigma^{ab}$ implies that
\ben
\Sigma^{ab}=\pm c e^a \wedge e^b
\label{solucion}
\een
for some set of one-forms $e^a$, where $c>0$ can be chosen at will. Thus, linearising the diagonal and off-diagonal simplicity constraints means that two of the four types of solutions for $\Sigma^{ab}$ are removed, but on the other hand one needs to introduce a basis of three-forms $n_a$ at each spacetime point, which is put in as an additional variable. One also still has to assume $V\neq 0$. 

There is also generically no evolution of initial data with $V\neq 0$ into a degenerate $\Sigma^{ab}$ with $V=0$ and a non-geometric interpretation (this is part of the discussion of \cite{reisenclass}). The geometry of the spacetime manifold is specified by $e^a$ and not by $E^a$ which is only used to determine normals in the constraints.

Alternatively, one might prefer to use a linear version of the volume constraints as well. Consider the original equation (\ref{betterform})
\ben
n_{a\sigma\tau\upsilon} \Sigma^{ab}_{\mu\nu} = (\det E_\mu^a) \left(\epsilon_{\mu\sigma\tau\upsilon}E_\nu^b-\epsilon_{\nu\sigma\tau\upsilon}E^b_\mu\right)\,,
\een
which was equivalent to $\Sigma^{ab}=E^a\wedge E^b$ for a basis of one-forms $E^a$. So far we only considered one half of these equations, namely those with $\{\mu,\nu\}\subset\{\sigma,\tau,\upsilon\}$. The other half have the form
\ben
n_{a\nu\tau\upsilon} \Sigma^{ab}_{\mu\nu} = (\det E_\mu^a) \epsilon_{\mu\nu\tau\upsilon}E_\nu^b\,,\quad\mbox{no sum over }\nu\,,
\een
with $\epsilon_{\mu\nu\tau\upsilon}\neq 0$. One way to read these equations is as the requirement on the left-hand side to be totally antisymmetric in $(\mu,\tau,\upsilon)$:
\ben
n_{a\nu\tau\upsilon} \Sigma^{ab}_{\mu\nu} = n_{a\nu\upsilon\mu} \Sigma^{ab}_{\tau\nu} = n_{a\nu\mu\tau} \Sigma^{ab}_{\upsilon\nu}\,,\quad\mbox{no sum over }\nu\,.
\label{linearvol}
\een

We could again try to turn the argument around and impose (\ref{linearvol}) as constraints on a $\frak{g}$-valued two-form $\Sigma^{ab}$ together with the linear simplicity constraints (\ref{zerorighthandside}). Substituting the solution $\Sigma^{ab}_{\mu\nu}=G_{\mu\nu}E^{[a}_\mu E^{b]}_\nu$ of the linear simplicity constraints into (\ref{linearvol}) gives (after diving by a non-zero factor $\half \det(E_{\mu}^a)$)
\ben
\epsilon_{\mu\nu \tau\upsilon}G_{\mu\nu}E_{\nu}^b = \epsilon_{\tau\nu\upsilon\mu}G_{\tau\nu}E_\nu^b = \epsilon_{\upsilon\nu\mu\tau}G_{\upsilon\nu}E_\nu^b\,.
\label{multipleofe}
\een
For $\epsilon_{\mu\nu \tau\upsilon}\neq 0$ this would imply $G_{\mu\nu}=G_{\tau\nu}=G_{\upsilon\nu}$. By Claim \ref{claim2}, imposing (\ref{linearvol}) for one fixed $\nu$, say $\nu=0$, is generically not sufficient: If we know that $G_{01}=G_{02}=G_{03}\neq 0$, we still have the condition
\ben
G_{12}=G_{13}=G_{23}\,,
\een
so that one needs more conditions of the form (\ref{linearvol}). These will then imply that all $G_{\mu\nu}$ are equal, $G_{\mu\nu}=c'$ for some $c'$ that could be positive, negative, or zero. One can absorb $|c'|$ by an overall redefinition, so that one has
\ben
\Sigma_{ab}=\pm c\, e^a\wedge e^b\,,
\een
for any chosen $c$, as before. Note that here it is possible, if $c'=0$ at a point, that all $e^a$ are zero this point and so $\Sigma^{ab}=0$ as well. While this is a very degenerate geometry, it is still a geometry. 

While the conditions (\ref{linearvol}), imposed for all values of $\nu$, are therefore sufficient to complete the identification of the two-form field $\Sigma^{ab}$ as $\pm c e^a\wedge e^b$, note that (\ref{linearvol}) is a massively redundant set of constraints: In order to obtain at most five relations on the coefficents $G_{\mu\nu}$ (two relations (\ref{gcondition}) if all $G_{\mu\nu}$ are nonzero), we are imposing {\em eight vector} equations! We have not exploited the fact that (\ref{multipleofe}) is a multiple of one of the vectors $E_\nu^b$, which are by assumption linearly independent. We could add several of the equations (\ref{linearvol}) for different $\nu$, instead of considering all equations for different $\nu$ separately. Let us try to impose
\ben
\sum_\nu \sum_{\{\mu,\upsilon\}\not\in\{\nu,\tau\}} n_{a\nu\tau\upsilon} \Sigma^{ab}_{\mu\nu} = 0,\quad \tau\in\{0,1,2,3\}\mbox{ fixed}.
\label{sumconst}
\een
Again substituting the solution $\Sigma^{ab}_{\mu\nu}=G_{\mu\nu}E^{[a}_\mu E^{b]}_\nu$ of the linear simplicity constraints into (\ref{sumconst}), we obtain
\ben
\half\det(E_\mu^a) \sum_{\nu} \sum_{\{\mu,\upsilon\}\not\in\{\nu,\tau\}} \epsilon_{\mu\nu\tau\upsilon}G_{\mu\nu}E_\nu^b = 0,\quad \tau\mbox{ fixed},
\een
which implies, by linear independence of the $E_\nu^b$, that indeed $G_{\mu\nu}=G_{\upsilon\nu}$ for all $\tau\not\in\{\mu,\nu,\upsilon\}$. It is then sufficient to impose the constraint (\ref{sumconst}) for three different choices of $\tau$, say $\tau=0,1,2$, so that we only need three vector equations instead of eight. 

By absorbing the constant $G_{\mu\nu}=c'$ (all $G_{\mu\nu}$ are equal) we rescale all $E^a$ by the same factor to obtain the variables $e^a$ that will have the physical intepretation of frame fields\index{orthonormal frame} encoding the metric geometry of spacetime. While in the case of quadratic volume constraints the one-forms $E^a$, or alternatively the three-forms $n^a$, only specified the normals to submanifolds $\{x^\mu=\const\}$, for linear volume constraints they can be directly interpreted, up to a position-dependent normalisation, as specifying an orthonormal basis in the cotangent space.

Note that this implies that one can assume a convenient normalisation for the one-forms $E^a$. Instead of just assuming non-degeneracy $\det(E^a_\mu)\neq 0$, one could fix $\det (E^a_\mu)=1$. This is no restriction of the physical content of the theory as the $E^a$, for both linear and quadratic volume constraints, only have a geometric interpretation after rescaling. One could then interpret $E^a_\mu$ as a map into $SL(4,\mathbb{R})$. For linear volume constraints, the relation between the normalised one-forms $E^a$ and the variables $e^a$ that are interpreted as frame fields is a single function on spacetime which may be viewed as a ``gauge"\index{gauge} in the sense of Weyl \cite{spacetimematter}.

In contrast to the case of the quadratic volume constraint, no non-degeneracy assumption on the two-form $\Sigma^{ab}$ is needed to enforce simplicity. One might get $\Sigma^{ab}=0$ in some region as a solution to the constraints, in which case the action for this region will be zero. This is analogous to a metric with vanishing determinant in general relativity and, in contrast to the requirement $V\neq 0$ outlined above, not an additional issue. Notice, however, that one still has to assume that the tetrad field $E^a$ and, equivalently, the co-tetrad field $n_a$ are non-degenerate, in order for the simplicity and volume constraints to imply (\ref{solucion}). Failing this, one gets solutions of the constraints that admit no proper geometric interpretation.

In the end, writing the action for BF theory in terms of $\Sigma^{ab}$,
\ben
S = \int B^{ab}\wedge R_{ab} = \int \Sigma^{ab}\wedge R_{ab} + \frac{\gamma}{2}{\epsilon^{ab}}_{cd} \Sigma^{cd}\wedge R_{ab}\,,
\een
we substitute (\ref{solucion}) into this action, which gives (setting $c=\frac{1}{8\pi\gamma G}$)
\ben
S = \frac{1}{8\pi G}\int_{\mathcal{R}\times\mathbb{R}}\sigma(x)\left(\half\epsilon_{abcd}e^a\wedge e^b\wedge R^{cd}+\frac{1}{\gamma}e^a\wedge e^b\wedge R_{ab}\right)\,.
\een

One is left with a field $\sigma(x)$ that can take the values $\pm 1$, but in the classical theory one may again argue that if $\sigma=1$ everywhere on an initial hypersurface, there will be no evolution into $\sigma=-1$. What we obtain is first order general relativity where one uses $(\det e)=\pm|\det e|$ instead of $|\det e|$ as a volume element in the action. If $\sigma$ is continuous as classical fields usually are assumed to be, this differs from the action with $|\det e|$ by an overall sign at most.

To summarise, we have identified both a linear version of the quadratic simplicity constraints and a linear version of the (quadratic) volume constraints in the continuum, which can be used to reduce topological BF theory to 4D gravity in the continuum. We have found also that both linear versions are slightly stronger (\ie more restrictive) than the corresponding quadratic constraints, so that the resulting constrained theory is likely to be closer to gravity at the quantum level than the one in which quadratic constraints are implemented. We now discuss the discrete counterpart of the constraints found above.

\subsection{Discrete Linear Constraints and their Relations}
The discrete analogue of (\ref{zerorighthandside}) is just the linear constraint used in \cite{eprlong,freidelkrasnov}, as desired:
\ben
n_a(\tetrahedron)\Sigma^{ab}(\triangle)=0\quad\forall\triangle\subset\tetrahedron\,.
\een

One could write down also a discrete version of (\ref{linearvol}), obtained in the natural way, demanding that within the same 4-simplex
\ben
n_{a}(\tetrahedron) \Sigma^{ab}(\triangle') = n_{a}(\tetrahedron') \Sigma^{ab}(\triangle'') 
\label{discvol}
\een
whenever $\triangle'\not\subset\tetrahedron$ and $\triangle''\not\subset\tetrahedron'$ and the edge shared by $\triangle'$ and $\tetrahedron$ is the same as that shared by $\tetrahedron'$ and $\triangle''$. 

In the following we adopt the notation of \cite{consistent}, where the tetrahedra in a given 4-simplex are labelled by ${\bf A,B,C,D,E}$, so that triangles are represented by ${\bf AB, AC}$, etc., and edges by combinations ${\bf ABC, ABD}$, etc. The orientation of the triangles and tetrahedra in (\ref{discvol}) is then fixed by the signs of the permutations of the letters,
\ben
n_{a}(\tetrahedron_{\bf A}) \Sigma^{ab}(\triangle_{\bf BC}) = -n_{a}(\tetrahedron_{\bf B}) \Sigma^{ab}(\triangle_{\bf AC}) = n_{a}(\tetrahedron_{\bf C}) \Sigma^{ab}(\triangle_{\bf AB}),\quad\mbox{etc.}
\label{linvol}
\een

In analogy to the continuum case, it will be sufficient to impose, instead of the full set of conditions (\ref{linvol}), certain linear combinations of (\ref{linvol}) to complete the geometric interpretation of the bivectors $\Sigma^{ab}(\triangle)$. The discrete analogue of the three continuum equations (\ref{sumconst}), where the index $e$ was kept fixed, is to pick one of the tetrahedra and add those six equations out of (\ref{linvol}) which involve triangles belonging to this tetrahedron. Starting with ${\bf A}$, we impose the constraint
\ben
\sum_{\{i,j\}\not\ni A}n_{a}(\tetrahedron_{\bf i})\Sigma^{ab}(\triangle_{\bf Aj})=0,
\label{discsumc}
\een
and the equivalent conditions for the tetrahedra ${\bf B}$ to ${\bf E}$, thereby needing to satisfy only five instead of 20 volume constraints.

The above discrete formulation of the linearised volume constraints resembles strongly the edge simplicity constraints studied, in a canonical setting, in \cite{biancajimmy}, and it imposes indeed the same restriction on the discrete data. However, it does not match exactly any of the various expressions given for these edge simplicity constraints in \cite{biancajimmy}. The correspondence between the two, therefore, deserves to be studied in more detail, given also that edge simplicity constraints have been shown to be crucial for the kinematical phase space of BF theory (and of loop gravity) to reduce to that of discrete gravity, in accordance with what we find here in a covariant setting.

In spin foam models such as \cite{eprlong,freidelkrasnov}, as we mentioned earlier, only the diagonal and off-diagonal simplicity constraints, but no quadratic volume constraints (\ref{quadvol}) are imposed. This is because in the discrete setting, one can use the closure constraint (\ref{closure}), imposed in all the tetrahedra in a 4-simplex, to relate the (quadratic) simplicity constraints to the volume constraints, so that if the former are imposed everywhere the latter follows. Since the quadratic simplicity constraints follow from the linear ones, as can be easily checked, this argument is still valid if one uses linear simplicity constraints.

One might hope that the sufficient set of linear volume constraints (\ref{discsumc}) would also follow from the linear simplicity constraints and the closure constraints. This is almost the case, but not quite. In fact, one more constraint should be added to simplicity and closure imposed in the five tetrahedra in the 4-simplex. This is a ``4D closure'' constraint of the form
\ben
n_a(\tetrahedron_{\bf A})+n_a(\tetrahedron_{\bf B})+n_a(\tetrahedron_{\bf C})+n_a(\tetrahedron_{\bf D})+n_a(\tetrahedron_{\bf E})=0\,,
\label{nclosure}
\een
where $\tetrahedron_i$ are the (appropriately oriented) tetrahedra of a given 4-simplex. 

Just as for the usual closure constraint (\ref{closure}), there are two ways to understand why such a constraint must be imposed. Recall that if one demands the triangles described by discrete variables $\Sigma^{ab}_{\triangle}$ close to form a tetrahedron, they have to satisfy (\ref{closure}). Alternatively, one can start with the continuum field equation $\nabla_{[\mu}^{(\omega)}\Sigma^{ab}_{\nu\rho]}=0$, where $\nabla^{(\omega)}$ is the covariant derivative for the connection $\omega^{ab}$, set the (flat) connection to zero by a gauge transformation, and integrate this over an infinitesimal 3-ball (whose triangulation is a tetrahedron).

The new constraint (\ref{nclosure}) seems to be the analogous statement that tetrahedra close up to form a 4-simplex. By Hodge duality $\wedge^1\mathbb{R}^4\simeq\wedge^3\mathbb{R}^4$ and any internal covector $n_a(\tetrahedron)$ can be mapped to a three-form; unlike for two-forms, any three-form can be written as $E^1\wedge E^2\wedge E^3$ for some $e^\alpha$. Demanding that the tetrahedra described by these three-forms form a closed surface is then (\ref{nclosure}). Thus the simplicial geometric reasoning goes through also for this new constraint. In terms of the equations of motion of the theory, on the other hand, the only argument for the need of this constraint is the following. If $\nabla_{[\mu}^{(\omega)}\Sigma^{ab}_{\nu\rho]}=0$ and we assume that $\Sigma^{ab}=\pm e^a\wedge e^b$, then it follows that $\nabla_{[\mu}^{(\omega)}n^a_{\nu\rho\sigma]}=0$ as well. Integrating this equation (with the connection again set to zero) over a 4-ball (which can be thought of as our 4-simplex) whose boundary is a 3-sphere, triangulated by tetrahedra, then leads to (\ref{nclosure}). We then however have to assume simplicity of $\Sigma^{ab}$. A more direct derivation of (\ref{nclosure}) from the equations of motion would be desirable.

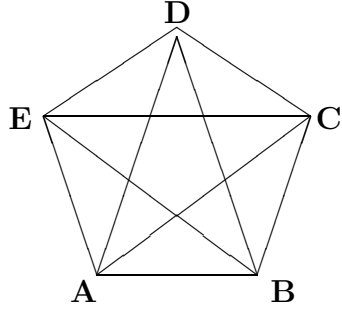
\begin{figure}[htp]
\centering
\begin{picture}(100,115)
\put(20,10){\line(1,0){60}}\put(10,0){{\bf A}}\put(85,0){{\bf B}}
\put(20,10){\line(-1,3){20}}\put(80,10){\line(1,3){20}}
\put(0,70){\line(3,2){50}}\put(100,70){\line(-3,2){50}}
\put(-13,65){{\bf E}}\put(102,65){{\bf C}}\put(45,105){{\bf D}}
\put(0,70){\line(1,0){100}}
\put(20,10){\line(1,3){30}}\put(80,10){\line(-1,3){30}}
\put(20,10){\line(4,3){80}}\put(80,10){\line(-4,3){80}}
\end{picture}
\caption[Dual picture of a 4-simplex.]{{\small Dual picture of a 4-simplex; points represent tetrahedra, lines represent triangles shared by adjacent tetrahedra.}}
\label{simplex}
\end{figure}
The role of this constraint, anyway, is the following. Consider the closure constraint
\ben
\Sigma^{ab}(\triangle_{\bf AB})+\Sigma^{ab}(\triangle_{\bf AC})+\Sigma^{ab}(\triangle_{\bf AD})+\Sigma^{ab}(\triangle_{\bf AE})=0.
\label{aclosure}
\een
Contracting with $n_a(\tetrahedron_{\bf B})$ gives, using the linear simplicity constraint $n_a(\tetrahedron_{\bf B})\Sigma^{ab}(\triangle_{\bf AB})=0$,
\ben
n_a(\tetrahedron_{\bf B})\Sigma^{ab}(\triangle_{\bf AC})+n_a(\tetrahedron_{\bf B})\Sigma^{ab}(\triangle_{\bf AD})+n_a(\tetrahedron_{\bf B})\Sigma^{ab}(\triangle_{\bf AE})=0.
\label{clocon1}
\een
Alternatively, one may start with the 4d closure constraint and contract with $\Sigma^{AB}(\triangle_{\bf AB})$ to get, again using the linear simplicity constraints,
\ben
n_a(\tetrahedron_{\bf C})\Sigma^{ab}(\triangle_{\bf AB})+n_a(\tetrahedron_{\bf D})\Sigma^{ab}(\triangle_{\bf AB})+n_a(\tetrahedron_{\bf E})\Sigma^{ab}(\triangle_{\bf AB})=0.
\label{clocon2}
\een
In total one obtains 20 + 10 = 30 equations of this kind that can be used to express some of the combinations $n_a(\tetrahedron)\Sigma^{ab}(\triangle)$ in terms of others. Substituting the resulting expressions into the five discrete volume constraints (\ref{discsumc}) one finds that the equations (\ref{discsumc}) indeed follow from the relations (\ref{clocon1}) and (\ref{clocon2}). We have seen in the continuum that the summed constraints (\ref{sumconst}) are sufficient to identify $\Sigma^{AB}=\pm e^A\wedge e^B$, and hence we find that in the discrete case the situation is analogous to the case of quadratic constraints in that a sufficient set of volume constraints can be viewed as secondary.

To see more clearly what happens in both our construction and in the case of quadratic constraints analyzed in \cite{consistent}, note that in our linear case one could use the 3d and 4d closure constraints to express the variables $n_a(\tetrahedron_{\bf E})$ and $\Sigma^{ab}(\triangle_{\bf AE}), \Sigma^{ab}(\triangle_{\bf BE})$, $\Sigma^{ab}(\triangle_{\bf CE}), \Sigma^{ab}(\triangle_{\bf DE})$ in terms of the others. Taking the linear simplicity constraints into account, one is then left with twelve independent combinations $n_a(\tetrahedron)\Sigma^{ab}(\triangle)$, just as in the continuum. In the continuum, we saw that one can impose the three additional constraints (\ref{sumconst}) on the twelve contractions $n_{a\mu\tau\upsilon}\Sigma^{ab}_{\mu\nu}$ to complete the identification $\Sigma^{ab}=\pm e^a\wedge e^b$. In the discrete case, one has the following three additional conditions coming from linear cross-simplicity constraints:
\bea
0&=&n_a(\tetrahedron_{\bf E})\Sigma^{ab}(\triangle_{\bf AE})\nn
\\&=&n_a(\tetrahedron_{\bf B})\Sigma^{ab}(\triangle_{\bf AC})+n_a(\tetrahedron_{\bf B})\Sigma^{ab}(\triangle_{\bf AD})+n_a(\tetrahedron_{\bf C})\Sigma^{ab}(\triangle_{\bf AB})\nn
\\& & + n_a(\tetrahedron_{\bf C})\Sigma^{ab}(\triangle_{\bf AD})+n_a(\tetrahedron_{\bf D})\Sigma^{ab}(\triangle_{\bf AB})+n_a(\tetrahedron_{\bf D})\Sigma^{ab}(\triangle_{\bf AC})
\eea
and similar ones coming from $n_a(\tetrahedron_{\bf E})\Sigma^{ab}(\triangle_{\bf BE})=0$ and $n_a(\tetrahedron_{\bf E})\Sigma^{ab}(\triangle_{\bf CE})=0$. These are precisely the analogue of the continuum constraints (\ref{sumconst}).

Similarly, in the case of quadratic simplicity constraints, one can use 3d closure to eliminate $\Sigma^{ab}(\triangle_{\bf AE}), \Sigma^{ab}(\triangle_{\bf BE}), \Sigma^{ab}(\triangle_{\bf CE}), \Sigma^{ab}(\triangle_{\bf DE})$. Then one observes that additional quadratic cross-simplicity constraints give expressions such as
\bea
0& = &\epsilon_{abcd}\Sigma^{ab}(\triangle_{\bf AE})\Sigma^{cd}(\triangle_{\bf BE})\nn
\\&=&\epsilon_{abcd}\Sigma^{ab}(\triangle_{\bf AC})\Sigma^{cd}(\triangle_{\bf BD})+\epsilon_{abcd}\Sigma^{ab}(\triangle_{\bf AD})\Sigma^{cd}(\triangle_{\bf BC})
\eea
which are equivalent to the desired (two) volume constraints.  

All of this is an exercise in solving a system of linear equations for which there might be a more simple and elegant description, but the upshot is the following. The sufficient set of linear volume constraints (\ref{sumconst}) does indeed follow from the linear simplicity constraints and the closure constraints, once one also imposes a four-dimensional closure constraint on the normals to tetrahedra that seems very natural in light of their geometric interpretation. Just as in the formulation in terms of quadratic simplicity constraints \cite{consistent}, the volume constraints can be viewed as secondary constraints that imply conservation of the simplicity constraints in time, or put differently, the volume constraints follow if the simplicity constraints hold everywhere. Once more this strenghtens the relationship between the discrete linear volume constraints we have identified and the edge simplicity constraints of \cite{biancajimmy}.

\sect{Lagrangian and Hamiltonian Formulation}
Let us briefly outline the Lagrangian formulation of 4D gravity resulting from our linear constraints added to BF theory.
One adds the linear simplicity and volume constraints to the action of BF theory using Lagrange multipliers:
\ben
S = \int d^4 x\;\left(\frac{1}{4}\epsilon^{\mu\nu\rho\sigma}\Sigma^{ab}_{\mu\nu}R_{ab\rho\sigma}[\omega]+\frac{\gamma}{8}\epsilon^{\mu\nu\rho\sigma}\epsilon_{abcd}\Sigma^{ab}_{\mu\nu}R^{cd}_{\rho\sigma}[\omega]+\Xi_b^{\mu\nu\sigma\tau\upsilon}n_{a\sigma\tau\upsilon}\Sigma^{ab}_{\mu\nu}\right)\,,
\label{newaction}
\een
where the Lagrange multiplier field $\Xi_b^{\mu\nu\sigma\tau\upsilon}$ satisfies $\Xi_b^{\mu\nu\sigma\tau\upsilon}\equiv \Xi_b^{[\mu\nu][\sigma\tau\upsilon]}$, and
\ben
\epsilon_{\mu\tau\upsilon}\Xi_b^{\mu\nu\nu\tau\upsilon}=0\quad\mbox{(no sum over }\nu)\,.
\een
Indeed, varying with respect to $\Xi_b$ then gives back the constraints
\ben
n_{a\sigma\tau\upsilon}\Sigma^{ab}_{\mu\nu} = \begin{cases}
0, & \{\mu,\nu\}\subset\{\sigma,\tau,\upsilon\}, \cr \epsilon_{\mu\tau\upsilon}f_\nu^b, & \nu=\sigma\;(\mbox{for some }f_\nu^b)\,.
\end{cases}
\een
Note that the second line corresponds to the set of constraints (\ref{linearvol}) and not to the summed version (\ref{sumconst}), and that it is clearly sufficient for the geometric interpretation of $\Sigma^{ab}$. The field equation from varying with respect to the connection $\omega$ is the usual
\ben
\nabla^{(\omega)}_{[\mu} \Sigma^{ab}_{\nu\rho]}=0\,,
\een
where $\nabla$ is the covariant derivative for the connection $\omega^{ab}$. The remaining equations involve the Lagrange multipliers, as would be expected:
\ben
\frac{1}{4}\epsilon^{\mu\nu\rho\sigma}R_{ab\rho\sigma}[\omega]+\frac{\gamma}{8}\epsilon^{\mu\nu\rho\sigma}\epsilon_{abcd}R^{cd}_{\rho\sigma}[\omega]+\Xi_{[b}^{\mu\nu\sigma\tau\upsilon}n_{a]\sigma\tau\upsilon}=0\,,\qquad \Xi_b^{\mu\nu\sigma\tau\upsilon}\Sigma^{ab}_{\mu\nu}=0\,.
\een
We have seen that the constraints imply that $\Sigma^{ab}=\pm e^a\wedge e^b$, and when substituting this back into the action one will recover general relativity, modulo the possible sign ambiguity we have already discussed.

We leave a complete Hamiltonian analysis of this theory to future work. However, we note a feature of the theory that follows directly from the use of linear constraints, and from the introduction of the additional variables $n_a$. 

As in unconstrained BF theory the initial dynamical variables will be the spatial part of the connection $\omega^{ab}_k$ and its conjugate momentum $P_{ab}^k\equiv\frac{1}{2}\epsilon^{ijk}\Sigma_{abij}$. We also saw that the equation of motion $\nabla^{(\omega)}_{[\mu} \Sigma^{ab}_{\nu\rho]}=0$ is unaffected by the constraints. Hence there will be Gauss constraints of the form\index{constraints!Gauss}
\ben
\mathcal{G}^{cd}\equiv\partial_i P^{cdi}+{\omega^c}_{ei}P^{edi}+{\omega^d}_{ei}P^{cei}
\label{gauss}
\een
on the canonical momenta. Their role is to generate $G$ gauge transformations. 

Looking at the action (\ref{newaction}), one would already require that (\ref{gauss}) should be modified to generate gauge transformations on the normals $n_{a\sigma\tau\upsilon}$; (\ref{newaction}) is only invariant under gauge transformations if the three-forms $n_{a\sigma\tau\upsilon}$ are transformed. The need for such a modification is also seen if one computes Poisson brackets between the linear simplicity and Gauss constraints. Define ``smeared" constraints
\ben
C[\Xi]:=\int\Xi_b^{ij,\sigma\tau\upsilon}n_{a\sigma\tau\upsilon}\epsilon_{ijk}P^{abk}\,,\quad\mathcal{G}[\Lambda]:=\int\Lambda^{cd}\mathcal{G}_{cd}\,.
\een
One then finds that
\bea
\{C[\Xi],\mathcal{G}[\Lambda]\} & = & -\int\frac{\delta C[\Xi]}{\delta P^{abk}}\frac{\delta \mathcal{G}[\Lambda]}{\delta \omega_{abk}}\nn
\\& = & - \int \Xi_B^{ij,\sigma\tau\upsilon}n_{a\sigma\tau\upsilon}\epsilon_{ijk}\left[\Lambda^{ad}{{P^b}_d}^k-\Lambda^{bd}{{P^a}_d}^k\right]\nn
\\& = & -C[\Lambda\cdot\Xi]-\int \Xi_B^{ij,\sigma\tau\upsilon}\Lambda^{ad}n_{a\sigma\tau\upsilon}\epsilon_{ijk}{{P^b}_d}^k\,,
\eea
where $(\Lambda\cdot\Xi)_d^{ij,\sigma\tau\upsilon}={\Lambda_d}^b \Xi_b^{ij,\sigma\tau\upsilon}$. The first term alone would imply that $\mathcal{G}[\Lambda]$ generates gauge transformations, but the second term is an unwanted extra piece. For $\mathcal{G}$ to be a generator of gauge transformations, it must be first class\index{constraints!first class} (\ie commute with other constraints up to linear combinations of constraints). We can remedy this by adding the variables $n_{a\mu\nu\rho}$ to the phase space, together with their conjugate momenta $\pi^{a\mu\nu\rho}$. Now we can define a new Gauss constraint
\ben
\mathcal{G}'^{cd}\equiv\mathcal{G}^{cd}-n^{[c}_{\mu\nu\rho}\pi^{d]\mu\nu\rho}\,.
\een
Then, computing the Poisson brackets of the new Gauss constraint with $C[\Xi]$, one finds
\ben
\{C[\Xi],\mathcal{G}'[\Lambda]\} = \{C[\Xi],\mathcal{G}[\Lambda]\} -\int \Xi_B^{ij,\sigma\tau\upsilon}\Lambda^{ca}n_{c\sigma\tau\upsilon}\epsilon_{ijk}P^{abk}=-C[\Lambda\cdot\Xi]\,,
\een
as desired. We have however increased the number of phase space variables at each point by 32.

A similar reformulation of the Gauss constraint, leading to a relaxation of the gauge invariance properties of spin network states, has been already suggested by the Hamiltonian analysis of the Plebanski theory \cite{henneaux}, and it has been advocated in the loop quantum gravity context in \cite{sergeietera, projected} as well as spin foam and group field theory context \cite{sergei, danieleAristide, danieleGFtsimpl}.

\sect{Summary and Outlook}
We have investigated a formulation of classical BF-Plebanski theory where the constraint $\Sigma^{ab}=\pm e^a\wedge e^b$, needed to reproduce general relativity in four dimensions, starting from topological BF theory, is imposed through constraints linear in the bivector field $\Sigma^{ab}$. 
The discrete counterpart of a part of these linear constraints (the ``simplicity constraints"), in fact, has proven very useful in the spin foam approach to quantum gravity \cite{eprlong,freidelkrasnov}. 

The corresponding continuum constraints have been easily identified, and can indeed be used to replace the quadratic ``diagonal'' and ``off-diagonal'' parts of the simplicity constraints appearing in the Plebanski formulation. As in the discrete case, one needed to introduce a new set of variables $n_a$ which are assumed to form a basis of three-forms at each point of spacetime, and are slightly stronger than the quadratic constraints: they eliminate two of the four sectors of solutions that are present for quadratic constraints.

In the second part of the analysis we found that the quadratic volume constraints of the Plebanski formulation, needed to complete the identification $\Sigma^{ab}=\pm e^a\wedge e^b$, can also be replaced by linear constraints, which again are stronger than their quadratic analogues. They do not require an additional non-degeneracy assumption on $\Sigma^{ab}$. However, a non-degeneracy assumption on the three-forms $n_a$ is still necessary, and only when this is imposed one can hope to eliminate all ``non-geometric'' degenerate configurations for $\Sigma^{ab}$, which are feared to dominate the quantum theory in the case of quadratic volume constraints. Also, while for quadratic volume constraints the variables $n_a$ merely specify normals to submanifolds $\{x^\mu=\const\}$ and hence can be independently rescaled arbitrarily at each point, for linear volume constraints they directly specify, up to an overall rescaling, the frame field encoding the metric geometry, \ie an orthonormal basis in the cotangent space at each spacetime point.

We have then analysed the discrete (simplicial) translation of the linear constraints we identified. In the context of spin foams, the quadratic volume constraints follow from imposing the (quadratic) diagonal and off-diagonal simplicity constraints everywhere together with closure constraints on the discrete variables $\Sigma^{ab}(\triangle)$. We have shown a similar property for the linear volume constraints. If (linear) diagonal and off-diagonal simplicity constraints and closure constraints for {\it both} bivector variables $\Sigma^{ab}(\triangle)$  {\it and} normals $n_a(\tetrahedron)$ are imposed everywhere, a sufficient set of linear volume constraints follows. This means that ``non-geometric'' bivector configurations cannot appear if the additional closure constraint on the normals holds, and the same normals are assumed to be non-degenerate.

We have not performed a complete Hamiltonian analysis of the resulting linear constrained BF action for gravity, but only noted that the use of linear simplicity and volume constraints immediately requires a modification of the usual Gauss constraint to generate a transformation of the normal 3-form variables $n_a$ alongside that of the $\Sigma$'s; a similar relaxation of the Gauss constraint, which translates at the spin foam and discrete gravity level into a closure constraint for simplices, and in the canonical quantum gravity context into a generalisation of spin network states, has been suggested on more than one occasion in the literature \cite{sergei, sergeietera, danieleAristide, projected}, even if its proper implementation at the quantum level has not been yet developed. On the classical level, therefore, a full Hamiltonian analysis of the constraints would be highly desirable. This would involve adding momenta for the components $\Sigma^{ab}_{0i}$, which are Lagrange multipliers in unconstrained BF theory, as well as those for the normals $n_{a}$ we have introduced, so that all variables can transform nontrivially under $G$ gauge transformations generated by a modified Gauss constraint, as shown.

Still at the classical level, but with obvious implications for the quantisation, one aspect of our construction that deserves further work is the relation between the discretised linear volume constraints we have found and the edge simplicity constraints used in \cite{biancajimmy}. As noted, the two sets of constraints appear to be very similar, and their role in the classical theory is the same, in particular, they remove (partly) the non-geometric configurations from the configuration space (or phase space) of the theory and appear as ``secondary'' in the sense specified above. So it natural to conjecture that one is simply a reformulation of the other. The implications for the quantum theory are not only due to the dominant role that non-geometric configurations may play in the quantum theory, if not removed, but also in the fact that  one discrete formulation of these constraints can actually be simpler to implement in a spin foam context than the other.

The possible use of our findings in the spin foam and group field theory context, and more generally in any quantisation based on the formulation of gravity as a constrained BF theory, are in fact most interesting. In particular, it seems to be important to explore how a closure constraint on normals could be implemented into existing spin foam models, given that we found it to be necessary for the full imposition of the geometric constraints on the variables of topological BF.  A convenient setting to do so could be the GFT formulation of \cite{danieleAristide}, since there the simplicial geometry and the contact with classical actions is brought to the forefront.

\part{Gauge Symmetries and $CP$ Violation}

\chapter{$SU(3)$ and its Quotients}
\label{su3quot}
\sect{Introduction}
The Lie group that will play a central role in this second part of the thesis is the special unitary group $SU(3)$. In the standard model of particle physics, this group plays two important roles: Firstly as the ({\bf colour}) gauge group of quantum chromodynamics (QCD)\index{quantum chromodynamics}, the version of Yang-Mills theory describing the strong interactions, secondly as the group of {\bf flavour} symmetry which relates different generations of quarks, famously introduced in Gell-Mann's ``Eightfold Way" \cite{gellmann}. We shall in the following focus on the electroweak sector\index{electroweak model} of the standard model \cite{electroweak, electroweak3, electroweak2}, where the gauge group is $SU(2)\times U(1)/\mathbb{Z}_2\simeq U(2)$ \cite{penrose}, and possible extensions of it, where one adds another $SU(2)$ group acting on ``right-handed" fields. We are interested in $CP$ violating processes which can change quark flavour, which are again described by an element of $SU(3)$, the {\bf Cabibbo-Kobayashi-Maskawa} (CKM) matrix. Ambiguities in the definition of the bases with respect to which this matrix is defined mean that all matrices obtained from a given element of $SU(3)$ by left or right multiplication by an element of the maximal Abelian subgroup $U(1)^2$ should be regarded as equivalent to the original matrix. Hence the space of CKM matrices is the double quotient\index{double quotient} $U(1)^2 \backslash SU(3) / U(1)^2$. In chapter \ref{natural} we will make statements about statistics of the Jarlskog invariant $J$, a measure of the magnitude of $CP$ violation, which assume a choice of measure on this quotient space. In this chapter we study the geometry of $SU(3)$ and its quotients to motivate different possible choices of measure.

We will also discuss measures on the space of Hermitian and complex $3\times 3$ matrices whose construction relies on the knowledge of measures on quotients of $SU(3)$, and which become important if one considers the space of quark mass matrices.

Let us first introduce coordinates on $SU(3)$ that we find convenient. The Lie algebra $\frak{su}(3)$ is generated by (i times) the Gell-Mann matrices\index{Gell-Mann matrices}
\bea
&&\lambda_1=\begin{pmatrix} 0&1 &0\\ 
                          1&0&0\\
                          0&0&0 \end{pmatrix}\,,\qquad
\lambda_2=\begin{pmatrix} 0&-\im &0\\
                          \im &0&0\\
                          0&0&0 \end{pmatrix}\,,\qquad
\lambda_4=\begin{pmatrix} 0&0&1\\
                          0&0&0\\
                          1&0&0 \end{pmatrix}\,,\nn
\\&&\lambda_5=\begin{pmatrix} 0&0&-\im\\
                          0&0 &0\\
                          \im&0&0 \end{pmatrix}\,,\qquad
\lambda_6=\begin{pmatrix} 0&0&0\\
                          0&0&1\\
                          0&1&0 \end{pmatrix}\,,\qquad
\lambda_7=\begin{pmatrix} 0&0&0\\
                          0&0&-\im\\
                          0&\im &0 \end{pmatrix}\,,\nn
\\&&\lambda_3 =\begin{pmatrix} 1&0&0\\
                             0&-1&0\\
                             0&0&0 \end{pmatrix}\,,\qquad
\lambda_8 = \frac{1}{\sqrt{3}} \begin{pmatrix} 1&0&0\\
                             0&1&0\\
                             0&0&-2 \end{pmatrix}\,.
\eea
To make the passage from $SU(3)$ to the single or double quotients straightforward, we note that any element of $SU(3)$ can be written as
\ben
U= A_L\, V\, A_R\,,
\label{su3decomp}
\een
where
\ben
A_L= e^{\frac{\im}{2}(3p-q)\lambda_3 + \frac{\im \sqrt{3}}{2}(p+q)\lambda_8}\,,
\quad A_R= e^{\im t \lambda_3 + \im \sqrt{3} r\lambda_8}\,,
\een
and\ben
V = e^{\im x \lambda_7}\, e^{-\im w \lambda_3}\, e^{\im y \lambda_5} \,e^{\im w \lambda_3}\, e^{\im z \lambda_2}\,,
\een
where the ranges of the Euler angles\index{Euler angles} $x, y, z$ and the complex phases $w, p, q, r, t$ are
\ben
0 \le x\,, y\,, z \le \frac{\pi}{2}\,, \quad 0 \le w\,,p\,,q\,,r\,,t < 2\pi\,.
\een
Explicitly, the matrix $V$ takes the form
\ben
V = \begin{pmatrix} \cos y \cos z & \cos y \sin z & e^{-{\rm i}w} \sin y \\ -\cos x \sin z - e^{{\rm i}w}\sin x \sin y \cos z & \cos x \cos z - e^{{\rm i}w} \sin x \sin y \sin z & \sin x \cos y \\ \sin x \sin z - e^{{\rm i}w} \cos x \sin y \cos z & - \sin x \cos z - e^{{\rm i} w} \cos x \sin y \sin z & \cos x \cos y\end{pmatrix} \,,
\label{ckmmatrix}
\een
whereas the diagonal matrices $A_L$ and $A_R$ are 
\ben
A_L = {\rm diag}(e^{2{\rm i}p},e^{{\rm i}(q-p)},e^{-{\rm i}(p+q)})\,,\quad A_R = {\rm diag}(e^{{\rm i}(r+t)},e^{{\rm i}(r-t)},e^{-2{\rm i}r})\,.
\een
It should be clear the orbits of the left action of the subgroup $U(1)^2$ are parametrised by $(p,q)$, while the orbits of the right action are parametrised by $(r,t)$. Therefore one may use $(x,y,z,w)$ to parametrise the double quotient $U(1)^2 \backslash SU(3)/U(1)^2$. Indeed, (\ref{ckmmatrix}) is commonly used in the literature as a parametrisation of the CKM matrix.

We may now obtain left-invariant forms on $SU(3)$ by computing the Maurer-Cartan form\index{Maurer-Cartan form}
\ben
U^{-1} dU = \im \,\lambda_a\,\sigma_a\,.
\een
The resulting expressions are somewhat involved, and we use the usual complex notation to slightly shorten the expressions (taking real and imaginary parts one recovers the $\sigma_i$ which are all real),
\bea
\sigma_1+\im \sigma_2 & = & \im e^{2\im t}\, \omega - e^{2\im t}\,\sin y\, \cos (2z)(\sin w\,dx+\cos w\, \sin (2x)\,dq) +\frac{1}{4}e^{2\im t}\,\sin(2z)\,\Omega\,,\nn
\\\sigma_3 & = & \frac{1}{4}\cos(2z)\,\Omega+ dt+\sin y \,\sin(2 z)\,(\sin w\,dx+\cos w\,\sin (2x)\,dq)\,,\nn
\\\sigma_4+\im\sigma_5 & = & \im e^{\im(3r+t+w)}\,\cos z\,dy-\cos y\,\sin z(\im e^{\im(3r + t)}\,dx+e^{\im(3r + t)}\sin(2x)\,dq)\nn
\\& & +e^{\im(3r + t + w)}\,\cos y\,\sin y\,\cos z\,(3\,dp-dw+\cos(2x)\,dq)\,,\nn
\\\sigma_6+\im\sigma_7 & = & \im e^{\im(3r-t+w)}\,\sin z\,dy+\cos y\,\cos z(\im e^{\im(3r -t)}\,dx+e^{\im(3r-t)}\sin (2x)\,dq)\nn
\\& & +e^{\im(3r-t+w)}\,\cos y\,\sin y\,\sin z(3\,dp-dw +\cos(2x)\,dq)\,,
\\\sigma_8 & = & \frac{\sqrt{3}}{4}\left[-dp+4\,dr+dw+2\,dq\,\cos(2x)\,\cos^2 y + (3\,dp-dw)\cos(2y)\right]\,,\nn
\eea
where we defined the 1-forms
\bea
\omega & = & dz + dx\,\cos w\,\sin y-2\,dq\,\cos x\,\sin w\,\sin x\, \sin y\,,\nn
\\\Omega & = & 3 \, dp + dw + dq\,\cos (2x)(\cos(2y) - 3)+(3\, dp - dw)\cos (2y)\,.
\eea
A general left-invariant metric\index{left-invariant metric} on the group will have the form
\ben
ds^2 = g^{ab}\sigma_a\sigma_b
\een
for some constant symmetric matrix $g^{ab}$. Depending on the choice of $g^{ab}$, the resulting metric will have a certain subgroup of $SU(3)$ as isometries acting by right multiplication on $SU(3)$. The possible cases are detailed in \cite{coqjad}. The most symmetric case is the bi-invariant metric, invariant under $SU(3)_{{\rm L}}\times SU(3)_{{\rm R}}$, which results from taking $g^{ab}$ proportional to $\delta^{ab}$.

Since we will be mainly interested in the measure, we note that for any choice of $g^{ab}$ the (left-invariant) {\it volume form} is, up to an (irrelevant) overall constant,
\bea
\mu^{SU(3)}_{{\rm l.inv.}} & \propto & \sigma_1\wedge\sigma_2\wedge\sigma_3\wedge\sigma_4\wedge\sigma_5\wedge\sigma_6\wedge\sigma_7\wedge\sigma_8
\label{volform}
\\& = & \frac{3\sqrt{3}}{2}\,\sin (2x)\,\sin y\,\cos^3 y\,\sin (2z)\,dx\wedge dy\wedge dz\wedge dw\wedge dp\wedge dq\wedge dr\wedge dt\,,\nn
\eea
which is independent of the complex phases $w,p,q,r$ and $t$. Clearly left invariance fixes the volume element up to an overall constant: Pick a group element $g$ (say the identity), then the space of 8-forms at $g$ is one-dimensional. The volume form at any other group element is then determined to be the push-forward of the volume form at $g$ under left translation. The volume form (\ref{volform}) will be the central quantity in many of the following calculations.

One could of course have started demanding right-invariance, computing the right-invariant forms from
\ben
 dU\,U^{-1} = \im \,\lambda_a\,\tau_a\,.
\een
If the left- and right-invariant measures are identical the group is called {\em unimodular}\index{unimodular}. Compact and semisimple groups are unimodular. Hence we can use left and right invariance as interchangeable conditions on a natural measure on $SU(3)$.

If $G$ is compact and semisimple, (the negative of) the Killing form\index{Killing form} is a bi-invariant Riemannian metric which is also an Einstein metric \cite{arvani}. It may therefore be regarded as the most natural choice of metric on such a group. For a (complex) matrix Lie group this metric is defined by
\ben
ds^2 = \Tr (dg\, dg^{\dagger})\,.
\een
It can also be viewed as the induced metric on the group $G$, interpreted as a submanifold of $\mathbb{C}^{n\times n}$ with flat metric. (The same considerations are of course true for a group of real matrices.)

We will use this metric as the most natural choice later on. It does give the same measure as any left-invariant metric, as we have noted. For completeness, we give its form explicitly for $SU(3)$\index{bi-invariant metric!on $SU(3)$}:
\bea
ds^2 & = & \frac{1}{2}\Tr\left(dU\, dU^{\dagger}\right) = \delta^{ab}\sigma_a\sigma_b\nn
\\& = & 3\,dp^2+ dq^2 + 3\, dr^2 + dt^2 + \frac{3}{2}(3\cos (2y) -1)\, dp\, dr\nn
\\& & + 3\, \cos^2 y\, (\cos(2 z)\, dp\, dt + \cos(2x)\, dq\, dr )\nn
\\& & + \frac{1}{2}\left[\cos (2x) \, \cos (2z)\, (\cos (2y) - 3) +4\, \sin (2x)\, \sin (2z)\, \sin y\, \cos w\right]\, dq\, dt\nn
\\& & - \sin^2 y\, (3\,dp -3\, dr + \cos(2x)\, dq -\cos(2z)\, dt)dw \nn
\\& & +2 \sin y\, \sin w\, (\sin (2z)\, dt\, dx - \sin (2x)\, dq\, dz)\nn
\\& & + dx^2 + dy^2 + dz^2 + \sin^2 y\, dw^2 + 2 \sin y\, \cos w\, dx\, dz\,.
\label{su3metric}
\eea

\sect{The Homogeneous Space $SU(3)/U(1)^2$}

The coset space $SU(3)/U(1)^2$ is a reductive geometry\index{reductive geometry} as explained in section \ref{lie}: The splitting of $\frak{su}(3)$ into an Abelian subalgebra $\frak{u}(1)^2$ and the orthogonal subspace is invariant under the adjoint action of $SU(3)$. We may hence identify an appropriate subset of generators of $SU(3)$ and use the corresponding one-forms to define metrics on the homogeneous space $SU(3)/U(1)^2$.

The one-forms $\sigma_3$ and $\sigma_8$ correspond to the Gell-Mann matrices generating the Abelian subgroup $U(1)^2$. We can construct left-invariant metrics on $SU(3)/U(1)^2$ from the remaining six left-invariant forms $\sigma_1,\sigma_2,\sigma_4,\ldots,\sigma_7$. The left group action of $SU(3)$ on the homogeneous space again guarantees the existence of a unique (up to a constant) left-invariant volume form, and we find that this unique volume form is
\bea
\mu^{SU(3)/U(1)^2}_{{\rm l.inv.}} & \propto & \sigma_1\wedge\sigma_2\wedge\sigma_4\wedge\sigma_5\wedge\sigma_6\wedge\sigma_7
\label{cosetvolform}
\\& = & \frac{3\sqrt{3}}{2}\,\sin (2x)\,\sin y\,\cos^3 y\,\sin (2z)\,dx\wedge dy\wedge dz\wedge dw\wedge dp\wedge dq\,,\nn
\eea
where we note that, due to the $SU(3)$ decomposition (\ref{su3decomp}), we may parametrise the quotient $SU(3)/U(1)^2$ by the $SU(3)$ coordinates $(x,y,z,w,p,q)$. This is of course the same volume form as the one obtained by starting with the left-invariant volume form on $SU(3)$ and integrating over the coordinates $r$ and $t$, or by just considering a slice $\{r=\const,t=\const\}$ in $SU(3)$ with the measure (\ref{volform}). (We shall see in the next section that this need not be true if one reduces to the double quotient $U(1)^2\backslash SU(3)/U(1)^2$.) 

There is an additional possible source of non-uniqueness of the measure, namely a possible ambiguity of the definition of the same manifold as different homogeneous spaces $X=G/H=G'/H'$. In practice, this does not seem to lead to different results for the measure.

It should be noted that the homogeneous space considered here, where the subgroup is the maximal Abelian subgroup, the ``maximal torus", of a compact group, is what is known as a {\it flag manifold}\index{flag manifold}. Such a manifold is naturally a K\"ahler manifold. For any $G$-invariant complex structure on a flag manifold, there is a unique $G$-invariant K\"ahler-Einstein metric \cite{arvani}. This metric may then be regarded as the most natural choice on such a homogeneous space. An explicit construction for such metrics on spaces $SU(M)/U(1)^{M-1}$, where $U(1)^{M-1}$ is the maximal torus of $SU(M)$, is given in \cite{picken}. The case $M=6$ might be of interest for neutrino mixing; here the analogue of the CKM matrix is the {\bf Maki-Nakagawa-Sakata matrix} \cite{mns}, usually taken to be an element of the single quotient $U(1)^2\backslash SU(3)$. In the see-saw mechanism one adds very heavy right-handed neutrinos, and the most general mixing matrix would be an element of $U(1)^5 \backslash SU(6)$. The construction of \cite{picken} would be the starting point of an analysis of this quotient, which is beyond the scope of this thesis.

\sect{Metrics on $U(1)^2 \backslash SU(3)/U(1)^2$}
\label{biquotmet}
The double quotient $U(1)^2\backslash SU(3)/U(1)^2$ does not have a natural $SU(3)$ group action, because the left and right actions of $SU(3)$ on itself do not commute with the Abelian subgroup $U(1)^2$ and hence do not induce a group action on the double quotient: If $g$ and $g'$ are in the same equivalence class, $[g]=[g']$, then $g=hg'h'$ for some $h,h'\in U(1)^2$. For a group action to be defined on the double quotient, one would need $[j\cdot g]=[j\cdot g']$, $j\cdot g=k(j\cdot g')k'=k(j\cdot h^{-1}gh'^{-1})k'$ for some $k,k'\in U(1)^2$. This can only be done for those $j$ that commute with $h^{-1}$.

It is therefore less straightforward than in the previous cases to find a geometrically motivated measure. The simplest way of reducing the left-invariant measure on the group $SU(3)$ or the homogeneous space $SU(3)/U(1)^2$ to the double quotient would be to just integrate (\ref{cosetvolform}) over the coordinates $p$ and $q$, which is trivial since the volume form (\ref{cosetvolform}) is independent of $p$ and $q$. One would therefore essentially use the same measure for the double quotient.

Alternatively, one might start from a metric on either the group or the homogeneous space, reduce it in some way to the double quotient, and then determine the associated volume form, but this introduces two possible sources of ambiguity; the choice of metric and the choice of method of reduction. We will see that indeed the measure on the double quotient is vastly non-unique.

For example, one may be tempted to identify the double quotient with the submanifold $p=q=r=t=0$ of $SU(3)$, take a symmetric metric on $SU(3)$, say the bi-invariant metric (\ref{su3metric}), and set $p=q=r=t=0$. This approach was indeed followed by Ozsv\'ath and Sch\"ucking \cite{OzsvathSchucking}, and leads to an appealingly simple form of the metric
\ben
ds^2 = dx^2 + dy^2 +dz^2 + 2\sin y \cos w\, dx\, dz + \sin^2 y \, dw^2,
\label{schuck}
\een
but seems not to be justified geometrically: It corresponds to a gauge choice, a choice of section in the orbits of $U(1)^2\times U(1)^2$ over the double quotient, and introduces symmetries (three commuting Killing vectors) into the metric which should not be present. It seems much better motivated geometrically to do a Kaluza-Klein reduction\index{Kaluza-Klein reduction} of a metric, \ie to project it orthogonally to the orbits of $U(1)^2$. One writes a given metric on $SU(3)$ (say) as 
\ben
ds^2 = h_{ij}(x)\, (dy^i + {A^i}_\mu(x) dx^\mu)(dy^j + {A^j}_\nu(x) dx^\nu) + \tilde{g}_{\mu\nu}(x) dx^\mu\, dx^\nu\,,
\een
and identifies $\tilde{g}_{\mu\nu}$ as a metric on the quotient space parametrised by $(x^{\mu})$.

To understand what is going wrong in (\ref{schuck}), we compare with the simpler case of $SU(2)$. The bi-invariant metric
on $SU(2)$ is
\ben
ds^2 = (d\psi+\cos\theta\, d\phi)^2 + d\theta^2 + \sin^2\theta\, d\phi^2\,,
\een
where $\frac{\partial}{\partial\phi}$ generates the left action of $U(1)$ and $\frac{\partial}{\partial\psi}$ generates
the right action of $U(1)$.   Projecting the metric orthogonally to the orbits of right translations, {\it \`a la} Kaluza-Klein, gives the round metric
\ben
ds^2 = d\theta^2 + \sin^2\theta\, d\phi^2
\label{s2met}
\een
on $SU(2)/U(1)=S^2$.  By contrast, simply setting $d\psi=0$ (the analogue of the construction of Ozsv\'ath and Sch\"ucking) instead gives the flat metric 
\ben
ds^2 = d\theta^2 + d\phi^2\,.
\label{flatmet}
\een
The round metric (\ref{s2met}) is invariant under $SO(3)$.  The flat metric (\ref{flatmet}) appears to be invariant under the Euclidean group, with $\frac{\partial}{\partial\theta}$ and $\frac{\partial}{\partial\phi}$ having the appearance of
translations, but these are only local symmetries since $\phi$ is a periodic coordinate and $\theta$ lies in an interval.

The example of $SU(2)$ also illustrates the difference between taking the left-invariant measure on the homogeneous space and a measure obtained by Kaluza-Klein reduction of a metric to a biquotient. The biquotient $U(1)\backslash SU(2)/U(1)=U(1)\backslash S^2$ is just an interval. Its metric becomes, after performing another Kaluza-Klein reduction of (\ref{s2met}),
\ben
ds^2=d\theta^2\,.
\een
The measure would be $d\theta$, and not $d\theta\sin\theta$ as obtained by integrating a function $f(\theta)$ over the coordinate $\phi$. It is apparent from this simple example that there are inequivalent ways of calculating integrals of a function on a right quotient that is invariant under the left group action; namely, one can {\it either} reduce the metric to obtain a measure on the double quotient {\it or} take the measure on the single quotient and integrate out the left phases.

Performing a Kaluza-Klein reduction\index{Kaluza-Klein reduction} of the bi-invariant metric (\ref{su3metric}) we find a rather complicated metric on the double quotient, whose determinant gives the measure
\bea
\sqrt{\det \tilde{g}} & \propto &\sin 2x\, \sin 2z\, \sin y\, \cos^2 y\cdot \Big((\sin^2 2x+ \sin^2 2z)\sin^2y  \nn
\\& &+\frac{1}{8} (5\cos 2y-3)\sin^2 2x \,\sin^2 2z+\frac{1}{2} \sin 4x\, \sin 4z\, \sin^3 y\, \cos w\nn
\\& & +\frac{1}{8} (3\cos 2y -5)\sin^2 2x\, \sin^2 2z\, \sin^2 y\, \cos^2 w\Big)^{-\frac{1}{2}}\,.
\label{kkmeas}
\eea

One might think that, starting from a more general (``squashed") left-invariant metric on the homogeneous space $SU(3)/U(1)^2$, there might be some which after Kaluza-Klein reduction give rise to a simpler expression for the measure on the biquotient, but we have only found more complicated ones.

\sect{Hermitian and Complex Matrices}
\label{hermcomp}
In chapter \ref{natural} we will require measures on the space of either all Hermitian or all complex $3\times 3$ matrices which will appear as the space of quark mass matrices in the standard model of particle physics or certain extensions of it. These will involve the left- (or right-)invariant measure on the homogeneous space $SU(3)/U(1)^2$, as we will see shortly. Let us start with the space of $3\times 3$ Hermitian matrices. It has a natural group action by $U(3)$ by conjugation, corresponding to a change of basis. We would like to determine a metric invariant under this action.

Any Hermitian matrix can be diagonalised, \ie written in terms of a real diagonal and a unitary matrix:
\ben
M_{{\rm H}}=U^{\dagger}DU\,,
\label{diagonalise}
\een
It should be clear that $U$ is only defined up to left multiplication by elements of $U(1)^2$; $U$ should be regarded as an element of the homogeneous space $U(1)^2\backslash SU(3) \simeq SU(3)/U(1)^2$.

The most symmetric measure on the space of Hermitian matrices is induced by the bi-invariant metric
\ben
ds^2=\Tr(dM_{{\rm H}}\, dM_{{\rm H}})=\Tr\left(dD\, dD\right)+2\Tr\left(\left(dU\,U^{\dagger}\,D\right)^2-\left(dU\,U^{\dagger}\right)^2 D^2 \right)
\een
which is clearly invariant under the action of $U(3)$ by conjugation. Define right-invariant one-forms $\tau_a$ on $SU(3)$ by
\ben
dU\,U^{\dagger}=\im\,\lambda_a\, \tau_a\,,
\een
this becomes\index{bi-invariant metric!for Hermitian matrices}\footnote{Compare with the 
corresponding result for real matrices given in \cite{Giulini}}
 [with $D\equiv{\rm diag}(D_1,D_2,D_3)$]
\bea
ds^2 & = & \Tr\left(dD\cdot dD\right)-2\tau_a\tau_b\Tr
\left(\lambda_a[D,\lambda_b]D\right)\nn
\\ & = & dD_1^2+dD_2^2+dD_3^2+2\left\{(D_1-D_2)^2(\tau_1^2+\tau_2^2)+(D_1-D_3)^2(\tau_4^2+\tau_5^2)\right.\nn
\\ & & \left.+(D_2-D_3)^2(\tau_6^2+\tau_7^2)\right\}\,.
\eea
The corresponding volume form is
\ben
(D_1-D_2)^2 (D_1-D_3)^2 (D_2-D_3)^2\,dD_1\wedge dD_2\wedge dD_3\wedge \tau_1\wedge \tau_2\wedge \tau_4\wedge \tau_5\wedge \tau_6\wedge \tau_7\,.
\een
We have explained above that any left- or right-invariant measure on the coset $U(1)^2\backslash SU(3)$ must be given by (\ref{cosetvolform}) and hence there is no need to compute the $\tau_a$ explicitly. We obtain a Riemannian measure
\bea
DM_{{\rm H}}&:=&(D_1-D_2)^2 (D_1-D_3)^2 (D_2-D_3)^2 \sin (2x)\,\cos^3 y\,\sin y\,\sin (2z)\,dD_1\, dD_2\, dD_3\nn
\\& &\times\, dx\,dy\,dz\,dw\,dr\,dt
\label{hermit}
\eea
on the space of Hermitian $3\times 3$ matrices. Note that we use coordinates $(x,y,z,w,r,t)$ on $U(1)^2\backslash SU(3)$ according to the decomposition (\ref{su3decomp}). The measure (\ref{hermit}) favours strongly non-coinciding eigenvalues.

From (\ref{diagonalise}), it is apparent that each Hermitian matrix with three distinct eigenvalues is associated with six different elements of $\R^3 \times U(1)^2\backslash SU(3)$, related by the action of the discrete group $\frak{S}_3$:
\ben
M_{{\rm H}}=U^{\dagger}DU=(U^{\dagger}P^{-1})PDP^{-1}(PU)=:\tilde{U}^{\dagger}\tilde{D}\tilde{U}\,, \quad P\in \frak{S}_3\,,
\een
where $\frak{S}_3$ is the symmetric group of degree 3 (the dihedral group\index{dihedral group $\frak{S}_3$} of order 6, sometimes denoted by $D_3$ or $D_6$) which permutes the canonical basis vectors of $\R^3$. The set of matrices with coinciding eigenvalues has zero measure and hence can be ignored in the present discussion. 

Thus we need to consider the space $\R^3 \times (U(1)^2 \times \frak{S}_3)\backslash SU(3)$ instead\footnote{Note that the space of Hermitian matrices is not topologically identified with this space; we have discarded a subset of Hermitian matrices which has zero measure for our construction.}, restricting the coordinates on the quotient $U(1)^2\backslash SU(3)$ to an appropriate range to pick one of the six matrices related by the $\frak{S}_3$ action. We can use the fact that the $\frak{S}_3$ action permutes the rows of an $SU(3)$ matrix to demand that the elements of the third column (cf. (\ref{ckmmatrix}); note that absolute values will not depend on rephasing, \ie right multiplication by an element of $U(1)^2$) satisfy the relation
\ben
|\sin y| \le |\sin x \cos y| \le |\cos x \cos y|\,,
\een
which restricts the coordinates $x$ and $y$ to
\ben
0\le y \le \arctan (\sin x)\,,\quad 0\le x\le \frac{\pi}{4}\,.
\label{coordrestr}
\een

Using the right-invariant measure (\ref{cosetvolform}) on $U(1)^2\backslash SU(3)$, we see that this region has precisely one-sixth of the total volume:
\ben
\frac{\int\limits_0^{\pi/2} dz \int\limits_0^{\pi/4} dx \int\limits_0^{\arctan(\sin x)} dy\,\sin 2x\,\cos^3 y\,\sin y\,\sin 2z}{\int\limits_0^{\pi/2} dz \int\limits_0^{\pi/2} dx \int\limits_0^{\pi/2} dy\,\sin 2x\,\cos^3 y\,\sin y\,\sin 2z}=\frac{1}{6}\,.
\een

Notice that while the total volume of $(U(1)^2 \times \frak{S}_3)\backslash SU(3)$ will be finite, the integrals over the $\R^3$ part will diverge; the eigenvalues of arbitrary Hermitian matrices can of course be arbitrarily large. One would have to introduce some kind of cutoff to make integrals over the space of Hermitian matrices convergent, as we will do later on.

The space of $3\times 3$ complex matrices can be discussed along similar lines. Here, one has a left and right action by independent elements of $U(3)$. We should therefore determine a metric invariant under both of these group actions. An arbitrary complex matrix can be represented as
\ben
M_{{\rm c}}=U_L^{\dagger}DU_R\,,
\een
where $U_L,U_R$ are unitary and $D={\rm diag }(D_1,D_2,D_3)$ is real diagonal. $U_L$ and $U_R$ are only defined up to simultaneous left multiplication by a diagonal unitary matrix
\ben
U_L\rightarrow AU_L\,,\quad U_R\rightarrow AU_R\,,\quad A\in U(1)^3
\een
which so that one has the correct number of real parameters, only 18 instead of 21. Since here the signs of the elements $D_i$ can be changed by a different choice of $U_L$ and $U_R$, and one can permute the $D_i$ as for Hermitian matrices, there are additional discrete ambiguities given by elements of the group $\frak{S}_3\times\mathbb{Z}_2^3$. We will see shortly that the measure, similar to the measure for Hermitian matrices, vanishes whenever elements of $D$ coincide up to sign, so we can restrict to matrices with $D_1^2\neq D_2^2\neq D_3^2\neq D_1^2$. Hence we can identify this subspace of the space of $3\times 3$ complex matrices with $\mathbb{R}^3\times (U(1)^3\times\frak{S}_3\times \mathbb{Z}_2^3)\backslash(U(3)\times U(3))\simeq\mathbb{R}_+^3\times (U(1)^3\times\frak{S}_3)\backslash(U(3)\times U(3))$. 

The metric we will use to determine a measure is
\ben
ds^2 = \Tr (dM_{{\rm c}}\, dM^{\dagger}_{{\rm c}})\,,
\een
which clearly is invariant under $M\rightarrow OMO'$ for $O,O'\in U(3)$. To evaluate this, use
\bea
dM_{{\rm c}} & = & -U_L^{\dagger}\, dU_L\, U_L^{\dagger}\, D\, U_R + U_L^{\dagger}\, dD\, U_R + U_L^{\dagger}\, D\, dU_R\,,\nn
\\ dM_{{\rm c}}^{\dagger} & = & -U_R^{\dagger}\, dU_R\, U_R^{\dagger}\, D\, U_L + U_R^{\dagger}\, dD\, U_L + U_R^{\dagger}\, D \,dU_L
\eea
and $[D,dD]=0$ to obtain
\bea
\Tr (dM_{{\rm c}}\, dM_{{\rm c}}^{\dagger}) & = &  \Tr(dD\, dD) - \Tr(D^2(dU_L\, U_L^{\dagger})^2) - \Tr(D^2(dU_R\, U_R^{\dagger})^2)\nn
\\& & + 2\, \Tr (D\, dU_L\, U_L^{\dagger}\, D\, dU_R\, U_R^{\dagger})\,.
\eea
We introduce right-invariant one-forms
\ben
dU_L U_L^{\dagger}=\im\, \lambda_a\, \tau^a_L\,,\quad dU_R U_R^{\dagger}=\im\, \lambda_b\, \tau^b_R\,,
\een
where $\lambda_1,\ldots,\lambda_8$ are the Gell-Mann matrices, and
\ben
\lambda_9=\sqrt{\frac{2}{3}}\left(\begin{matrix} 1 & 0 & 0 \cr 0 & 1 & 0 \cr 0 & 0 & 1 \end{matrix}\right)\,,
\een
so that ${\rm i}\lambda_a$ are a basis for the Lie algebra $\frak{u}(3)$. Then
\ben
\Tr (dM_{{\rm c}}\, dM_{{\rm c}}^{\dagger}) = \Tr(dD\, dD) +\Tr(D^2\,\lambda_{(a}\lambda_{b)})(\tau^a_L\tau^b_L+\tau^a_R\tau^b_R) - \Tr (D\, \lambda_a\, D\, \lambda_b)(\tau^a_L\tau^b_R+\tau^a_R\tau^b_L)\,.
\label{traces}
\een
The only nonvanishing traces in (\ref{traces}) are
\bea
&&\Tr(D^2\lambda_1\lambda_1)=\Tr(D^2\lambda_2\lambda_2)=\Tr(D^2\lambda_3\lambda_3)=\Tr(D\lambda_3 D\lambda_3)=D_1^2+D_2^2\,,\nn
\\&&\Tr(D^2\lambda_{(3}\lambda_{8)})=\Tr(D\lambda_{(3}D\lambda_{8)})=\frac{1}{\sqrt{3}}(D_1^2-D_2^2)\,,\nn
\\&&\Tr(D^2\lambda_{(3}\lambda_{9)})=\Tr(D\lambda_{(3}D\lambda_{9)})=\sqrt{\frac{2}{3}}(D_1^2-D_2^2)\,,\nn
\\&&\Tr(D^2\lambda_{(8}\lambda_{9)})=\Tr(D\lambda_{(8}D\lambda_{9)})=\frac{\sqrt{2}}{3}(D_1^2+D_2^2-2D_3^2)\,,\nn
\\&&\Tr(D^2\lambda_4\lambda_4)=\Tr(D^2\lambda_5\lambda_5)=D_1^2+D_3^2\,,\nn
\\&&\Tr(D^2\lambda_6\lambda_6)=\Tr(D^2\lambda_7\lambda_7)=D_2^2+D_3^2\,,\nn
\\&&\Tr(D^2\lambda_8\lambda_8)=\Tr(D\lambda_8 D\lambda_8)=\frac{1}{3}(D_1^2+D_2^2+4D_3^2)\,,\nn
\\&&\Tr(D^2\lambda_9\lambda_9)=\Tr(D\lambda_9 D\lambda_9)=\frac{2}{3}(D_1^2+D_2^2+D_3^2)\,,\nn
\\&&\Tr(D\lambda_1 D\lambda_1)=\Tr(D\lambda_2 D\lambda_2)=2D_1 D_2\,,\nn
\\&&\Tr(D^2\lambda_4\lambda_4)=\Tr(D^2\lambda_5\lambda_5)=2 D_1 D_3\,,\nn
\\&&\Tr(D^2\lambda_6\lambda_6)=\Tr(D^2\lambda_7\lambda_7)=2 D_2 D_3\,.
\eea
The metric can be written in the form\index{bi-invariant metric!for complex matrices}
\bea
ds^2 & = & dD_1^2+dD_2^2+dD_3^2 + \frac{1}{2}(D_1-D_2)^2(\tau_L^1+\tau_R^1)^2 + \frac{1}{2}(D_1+D_2)^2(\tau_L^1-\tau_R^1)^2\nonumber
\\ & & + \frac{1}{2}\left[(D_1-D_2)^2(\tau_L^2+\tau_R^2)^2 + (D_1+D_2)^2(\tau_L^2-\tau_R^2)^2\right.\nn
\\ & & \left.+ (D_1-D_3)^2(\tau_L^4+\tau_R^4)^2 + (D_1+D_3)^2(\tau_L^4-\tau_R^4)^2\right]\nonumber
\\ & & + \frac{1}{2}\left[(D_1-D_3)^2(\tau_L^5+\tau_R^5)^2 + (D_1+D_3)^2(\tau_L^5-\tau_R^5)^2\right.\nn
\\ & & \left.+ (D_2-D_3)^2(\tau_L^6+\tau_R^6)^2 + (D_2+D_3)^2(\tau_L^6-\tau_R^6)^2\right]\nonumber
\\ & & + \frac{1}{2}(D_2-D_3)^2(\tau_L^7+\tau_R^7)^2 + \frac{1}{2}(D_2+D_3)^2(\tau_L^7-\tau_R^7)^2\nonumber
\\ & & + 2 D_1^2 \left(\frac{1}{\sqrt{2}}(\tau_L^3 - \tau_R^3) + \frac{1}{\sqrt{6}}(\tau_L^8 - \tau_R^8)+\frac{1}{\sqrt{3}}(\tau_L^9 - \tau_R^9)\right)^2 \nonumber
\\ & & + 2 D_2^2 \left(-\frac{1}{\sqrt{2}}(\tau_L^3 - \tau_R^3) + \frac{1}{\sqrt{6}}(\tau_L^8 - \tau_R^8)+\frac{1}{\sqrt{3}}(\tau_L^9 - \tau_R^9)\right)^2\nn
\\ & & + 2 D_3^2 \left(\sqrt{\frac{2}{3}}(\tau_L^8 - \tau_R^8)+\frac{1}{\sqrt{3}}(\tau_L^9 - \tau_R^9)\right)^2.
\eea
The fact that the metric only depends on $\tau_L^3-\tau_R^3$ etc., and not on $\tau_L^3+\tau_R^3$ etc., again reflects the $U(1)^3$ that has to be factored out. One finds that the volume form is proportional to 
\ben
(D_1^2-D_2^2)^2(D_1^2-D_3^2)^2(D_2^2-D_3^2)^2 |D_1 D_2 D_3|\,dD_1\wedge dD_2\wedge dD_3\wedge \tau_L^1\wedge \tau_R^1\wedge\ldots\wedge(\tau_L^8-\tau_R^8)\wedge(\tau_L^9-\tau_R^9)\,.
\een
Note that this expression only depends on the absolute values of $D_i$, as expected. Since the range of the $D_i$ is infinite, integration over these coordinates will give an infinity, as was the case for Hermitian matrices. Note the main difference to the measure (\ref{hermit}) in the different powers of $D_i$.

Now assume that all quantities we are interested in are independent of the parameters on $U_R$, which will be the case in the discussion of chapter \ref{natural}. Then we can integrate over these coordinates, obtaining a constant which is irrelevant in the averaging process. 

We are then left with integrating over the space of possible matrices $U_L$, the coset $U(1)^3 \backslash U(3) = U(1)^2 \backslash SU(3)$, and the volume form is proportional to
\ben
(D_1^2-D_2^2)^2(D_1^2-D_3^2)^2(D_2^2-D_3^2)^2 |D_1 D_2 D_3|\,dD_1\wedge dD_2\wedge dD_3\wedge \tau_L^1\wedge \tau_L^2\wedge \tau_L^4 \wedge \tau_L^5\wedge\tau_L^6\wedge\tau_L^7\,.
\een
In terms of the coordinates on $U(1)^2\backslash SU(3)$ introduced in (\ref{su3decomp}), the wedge product of right-invariant forms gives the usual measure on $SU(3)$, so that we finally get
\ben
DM_{{\rm c}} = \prod_{i<j}(D_i^2-D_j^2)^2 |D_1 D_2 D_3|\,\sin 2x \cos^3 y \sin y \sin 2z\,dD_1\, dD_2\, dD_3\,dx\,dy\,dz\,dw\,dr\,dt.
\label{su3meas}
\een

The discrete $\frak{S}_3$ symmetry can be taken into account just as in the previous case: We integrate only over one sixth of the homogeneous space $U(1)^2\backslash SU(3)$, corresponding to
\ben
0\le y\le \arctan(\sin x)\,,\quad 0\le x\le\frac{\pi}{4}\,.
\een
This restriction amounts to removing unitary matrices that permute the elements of $D$ and hence to fixing an ordering.

\chapter{Naturalness of $CP$ Violation}
\epigraph{\em{``Well, I don't like to get involved in these philosophical issues very much."}}{Murray Gell-Mann (attributed)}
\label{natural}

\sect{Introduction and Motivation}

In this chapter we will apply the mathematical theory of measures on quotients of $SU(3)$, outlined in the previous chapter, to the space of CKM matrices in the electroweak sector of the standard model in order to make statements about naturalness of the observed magnitude of $CP$ violation in electroweak processes. We will see that the observed value is rather small when compared to the maximal possible magnitude and aim to reach a conclusion if this should be viewed as a fine-tuning problem. The task of making mathematically precise statements is simplified by the fact that $SU(3)$ is compact, and so normalisation of measures on $SU(3)$ or a quotient of it will normally not be problematic.

In modern theoretical physics one often tries to make statements about ``naturalness" or ``fine-tuning"\index{fine-tuning} of the observed values of fundamental parameters, where fine-tuning of a parameter is interpreted as an indication for incompleteness of the theory. Popular examples of fine-tuning problems include the quark mass hierarchy and the cosmological constant problem in particle physics. Since statements about naturalness are fundamentally of a statistical character, to make them mathematically precise one has to assume a well-motivated probability distribution on the parameter space relevant for the theory. A fine-tuning problem then indicates that the probability distribution one has used should be modified by introducing new physical considerations. As an example, as long as observation was consistent with a vanishing cosmological constant, it seemed reasonable to assume that a postulated symmetry would constrain it to vanish. With more recent observations indicating that it must be taken to be very small and positive, there seems to be an issue of fine-tuning\index{cosmological constant problem}\footnote{For an alternative interesting but presumably non-mainstream viewpoint, see \cite{rovellilambda}}. We note that much of the motivation to extend the standard model of particle physics is driven by such considerations, that it is by no means necessary to contemplate a ``Multiverse'' where all possible values of a given parameter are actually realised. We also need not consider anthropic arguments, which for the problem at hand would not give a satisfactory explanation of very weak $CP$ violation; strong $CP$ violation might be anthropically preferred since baryogenesis necessarily requires $CP$ violation, and the $CP$ violation in the electroweak sector does not seem sufficient \cite{baryogen}.

Of course we can only ever observe and make measurements in a single Universe, and if a parameter takes a value that appears unlikely maybe this just means that an unlikely possibility is realised in our Universe. One is merely doing statistics. But one should recall {\bf Bayes' theorem} \cite{bayes}\index{Bayes' theorem}
\ben
P(A|B)=\frac{P(B|A)P(A)}{P(B)}=\frac{P(B|A)P(A)}{\sum_{A_i} P(B|A_i)P(A_i)},
\een
where one can take $A$ as a hypothesis and $B$ as an observation. In order to calculate the probability that a hypothesis follows from a given observation, one either needs {\em a priori} probabilities for both $A$ and $B$ and the conditional probability of $B$ assuming $A$ (which should be calculable), or {\em a priori} probabilities for a complete set of hypotheses $\{A_i\}$, together with all conditional probabilities of $B$ given these hypotheses. These {\em a priori} probabilities are referred to as {\bf priors}\index{priors}. In the second case the sum also has to be made explicit, and if it is an integral one needs to specify a measure.

In cosmology, the role of priors is particularly important as there is no possibility to confirm observations in different experimental situations, and $P(B)$ does not seem accessible. Here $A$ would be a statement about our Universe. While it may be appropriate to consider $P(A)$ as independent of $A$ (Laplace's Principle of Indifference), a measure on the set of all universes has to be determined \cite{garytalk}.

The situation considered in this chapter is similar, as ``experiments'' would involve different universes with different values of the fundamental physical parameters. For instance, one could try to use anthropic arguments, taking $B$ to be the observation that human beings have evolved in our Universe, to claim that the values of (in this case) $CP$ violating CKM matrix parameters must take values close to the observed values (hypothesis $A$). The conditional probability $P(B|A)$ could in principle be calculated using baryogenesis etc., but then one still needs a measure on the space of possible values that $A$ can take. Again, one is forced to pick a measure on the parameter space. We shall therefore investigate predictions for the magnitude of $CP$ violation, measured by the Jarlskog invariant $J$\index{Jarlskog invariant}, comparing different choices for the measure on the parameter space that we regard as natural. The observation by Kobayashi and Maskawa \cite{km} that $CP$ violation is only possible for at least three quark families has led to a Nobel prize, but the issue of possible fine-tuning in the magnitude of $CP$ violation is much less understood.

It is true that, if one is talking about parameters in quantum field theory, they should really not be regarded as constants, but have an evolution with energy scale given by renormalisation\index{renormalisation} group equations. However, since fine-tuning means a discrepancy of several orders of magnitude, it may well be that a fine-tuning problem is present at all energy scales. This is true for the quark mass hierarchy \cite{massref}, and also for the case of $CP$ violation: Recent numerical studies \cite{cpviolref} indicate that $J^2$ does not run strongly with energy scale, but that the value at extremely high energies $(\sim 10^{15} {\rm GeV})$ is merely about twice the value at low energies. It is then meaningful to talk about ``naturalness" of the value of such a parameter.

 Since any statements one tries to make depend very directly on the choice of measure, it is helpful if geometric considerations allow for a natural choice of probability distribution. We saw in the previous chapter that, if the parameter space is a homogeneous space $G/H$ for $G$ a compact Lie group\index{Lie group} and $H$ a closed subgroup, the natural requirement on the measure determining the probability distribution is invariance under the left action of $G$, which leads to a unique measure (up to normalisation). Probabilities for a given function on $G/H$ to take certain values are then well-defined. 

We also saw that if, as in the case of the space of CKM matrices, one has a double quotient $H\backslash G/H$, the problem of determining a natural probability distribution is more involved than for a homogeneous space $G/H$. We have discussed several possible choices that we compare here. We will find that while there is no clearly preferred choice of measure, there always seems to be fine-tuning in the observed value for $J$, unless additional input is used.

In a second part we shall take a different approach, taking the observed values for the quark masses into account by considering not the space of CKM matrices, but the space of mass matrices, as the fundamental parameter space. This is motivated by the observation that the mass matrices are directly linked to the Yukawa couplings and the Higgs vacuum expectation value, whereas the CKM matrix is only a derived quantity. The observed values for the quark masses will be taken as given, and a probability distribution be constructed that can reproduce these values. The choice made for this distribution will be as simple as possible in the following sense: The natural group action on the space of Hermitian mass matrices is the action of the unitary group $U(3)$ by conjugation, $M\rightarrow UMU^{\dagger}$. There is essentially a unique measure invariant under this action. This measure is then modified by introducing the simplest possible function that would allow for a modification of the expectation values for quark masses fitting observation. No further assumptions will be needed. We will then find that this simple choice for the measure gives an expectation value for $J$ that is remarkably close to the observed value. Hence we will conclude that once one assumes the quark masses as given, one does not face an additional fine-tuning problem with $J$. This statement, while not new, is hence made precise using a geometrically motivated measure on the parameter space.

In the standard model one can restrict, without loss of generality, to Hermitian mass matrices, which is why we construct a measure on the space of $3\times 3$ Hermitian matrices. However, in left-right symmetric extensions of the standard model, such as Pati-Salam\index{Pati-Salam model}, such an assumption can no longer be made and the mass matrices have to be regarded as arbitrary complex matrices\footnote{We thank Ben Allanach for pointing this out.}. This presumably has an effect on the statements one makes about naturalness of $J$; a similarly well-motivated measure on the space of all $3\times 3$ complex matrices might lead to very different results. We will therefore in a further calculation redo the same analysis for general complex matrices. We will again use the most symmetric measure on the space of mass matrices, here the space of general complex matrices, and modify it in the simplest possible way to incorporate the observed quark mass hierarchy\index{mass hierarchy}. While this is a choice that could of course be made very differently, it is a simple choice that uses as few assumptions as possible, and that has worked very well for the case of Hermitian mass matrices. What we will find is that the resulting probability distribution is different, and the result is different too: The observed value for $J$ now appears to be unnaturally large, since $CP$ violation should be more heavily suppressed by the quark mass hierarchy. One faces a fine-tuning\index{fine-tuning} problem, and needs additional assumptions to modify the measure appropriately.

We should point out that the only real input we use from the physical theory (standard model or a left-right symmetric extension of it) is, apart from the very definition of $CP$-violating parameters, how the theory restricts the type of mass matrices that appear. In particular, for any extension of the standard model (such as left-right symmetric models where parity is the left-right symmetry\footnote{We thank an anonymous referee for Phys. Rev. D for clarification on this point.}) that also has Hermitian mass matrices we would not see any modification in the results. Nevertheless, our calculations provide another example of how symmetries, in this case the presence or absence of a left-right symmetry, influence physical predictions of a given theory.

The structure of the remaining part of the chapter is as follows: In section \ref{cpviolation}, we review $CP$ violation in the electroweak sector and detail why mass matrices may be assumed to be Hermitian in the standard model, but not if one has an extended left-right symmetry. In section \ref{jstats}, we compute expectation values for the Jarlskog invariant $J$ using several of the measures on the space of CKM matrices that we have introduced. We also perform a more detailed analysis of how unlikely the observed value for $J$ appears in these distributions. In section \ref{gaussian} we focus on the space of mass matrices instead, detailing how the observed values for the quark masses can be incorporated into a measure and how this completely changes predictions about likely values for $J$. We briefly mention how our analysis relates to the case of neutrinos and close with a summary and discussion of our findings.

\sect{$CP$ Violation in the Standard Model and \mbox{Beyond}}
\label{cpviolation}

We summarise how $CP$ violation arises, first in the standard model and then in the more general case of left-right symmetric extensions, essentially following \cite{jarlskog}. In the standard model, the quark fields appear as left-handed $SU(2)$ doublets and right-handed $SU(2)$ singlets:
\ben
\left(\begin{matrix}q_{jL}\\ q'_{jL}\end{matrix}\right),\quad q_{jR}\,,\quad q'_{jR}\,,\qquad j=1,2,\ldots,N\,.
\een
Here $N$ is the number of quark families, which is arbitrary in the standard model, and normally taken to be three. The fields are written in a flavour basis which can be considered unphysical, since flavour eigenstates do not correspond to mass eigenstates.

The coupling of the Higgs doublet $H$ to quarks is given by
\ben
\mathcal{L}_{{\rm Higgs}}=\sum_{j,k=1}^N \left( Y_{jk} \overline{(q,q')_{jL}}H^C q_{kR} + Y_{jk}' \overline{(q,q')_{jL}}H q'_{kR}+{\rm h.c.}\right),
\een
where the Higgs\index{Higgs} doublet $H$ and its $C$-conjugate $H^C$ can be written as
\ben
H=\frac{1}{\sqrt{2}}\left(\begin{matrix}\phi_1 + {\rm i} \phi_2 \\ \phi_0 + {\rm i}\phi_3\end{matrix}\right),\quad H^C=\frac{1}{\sqrt{2}}\left(\begin{matrix}\phi_0 - {\rm i} \phi_3 \\ -\phi_1 + {\rm i}\phi_2\end{matrix}\right),
\een
and $Y_{jk}$ and $Y'_{jk}$ are complex (Yukawa) couplings. $H$ transforms under $SU(2)$ as a doublet and the term $\mathcal{L}_{{\rm Higgs}}$ is invariant under $SU(2)$. This symmetry is then spontaneously broken by the Higgs potential $V(H)$ which gives rise to a vacuum expectation value $v$ for $\phi_0$. The remaining components $\phi_j$ are `eaten' by the $W$ and $Z$ bosons, which become massive, and the only remaining terms involve $\phi'_0=\phi_0-v$, giving masses to the quark fields:
\ben
\mathcal{L}_{{\rm Higgs}}\stackrel{{\rm SSB}}{\longrightarrow}-\sum_{j,k=1}^N \left( m_{jk} \overline{q_{jL}} q_{kR} + m_{jk}' \overline{q'_{jL}} q'_{kR}+{\rm h.c.}\right)\left(1+\frac{1}{v}\phi'_0\right)\,.
\label{ssb}
\een
The new {\bf mass matrices}\index{quark mass matrix} $m$ and $m'$ are (up to a factor of $\sqrt{2}$) given by the original Yukawa couplings, multiplied by $v$. Thus, it seems appropriate to regard either the set of Yukawa couplings together with the Higgs vacuum expectation value or the collection of elements of $m$ and $m'$ as fundamental parameters of the theory. 

This part of the Lagrangian is formally $CP$ invariant if and only if $m$ and $m'$, which are so far arbitrary complex matrices, are real. Since this condition is not satisfied in Nature, one has formal $CP$ violation. However, as remarked before, the Lagrangian has been written in the unphysical flavour basis. One can, for general $m$ and $m'$, pass to a different basis, namely the basis of mass eigenstates, by diagonalising the mass matrices with unitary matrices:
\ben
m=U_L^{\dagger}\Delta U_R\,,\quad m'=(U'_L)^{\dagger}\Delta'U_R'\,,
\label{massdiag}
\een
thus the basis of mass eigenstates is related to the previously considered basis by
\ben
q_{L {\rm phys}}=U_L q_L\,, \quad q_{R {\rm phys}}=U_R q_R\,,\quad {\rm etc.}
\een

It is always possible to choose the unitary matrices so that $\Delta$ and $\Delta'$ are real, and in this new basis this part of the Lagrangian is invariant under $C$ and $P$ separately, and hence also under $CP$. However, the electroweak Lagrangian also contains charged current terms mixing up- and down-type quarks, coupled to the $W$ boson fields via (we are now using the basis of mass eigenstates)
\ben
X_C:=(W_{\mu}^1-iW_{\mu}^2)J_c^{\mu}+{\rm h.c.}\,,\quad J_c^{\mu}:= \overline{(u,c,t)_L}\gamma^{\mu}V\left(\begin{matrix}d_L\\ s_L \\ b_L \end{matrix}\right)\,,
\label{chargedc}
\een
where $V:=U_L (U_L')^{\dagger}$ is the {\bf Cabibbo-Kobayashi-Maskawa} (CKM) matrix\index{CKM matrix}. In the basis of mass eigenstates, this term $X_C$ is not invariant under $CP$ unless $V$ is real. Since we consider the mass eigenstates as physical, $CP$ is violated through these charged current terms.

An important observation made in \cite{jarlskogframpton} is that one can redefine the right-handed quark bases by arbitrary unitary transformations,
\ben
U_R\rightarrow OU_R\,,\quad U'_R\rightarrow O'U_R'\,,
\label{redefine}
\een
obtaining a new basis which is to be regarded as equally physical. This is due to the absence of charged current terms involving right-handed quarks, since they are singlets under $SU(2)$. It is therefore no loss of generality to set $U_R=U_L$ and $U'_R=U'_L$ in (\ref{massdiag}), and to assume that $m$ and $m'$ are Hermitian. 

A natural way to extend the standard model is to assume the existence of a second $SU(2)$ symmetry which acts on the right-handed quarks, as in the Pati-Salam model\index{Pati-Salam model} \cite{patisalam}. In such extensions, one adds a term (\ref{chargedc}) for right-handed quarks to the Lagrangian. This has the important consequence that a general transformation (\ref{redefine}) for arbitrary unitary transformations $O$ and $O'$ can no longer be regarded as giving an equivalent quark basis, since it modifies this new charged current term. The mass matrices cannot in general be taken to be Hermitian, but are arbitrary complex matrices. 

There are now also two possibly $CP$ violating terms, and two CKM matrices. We focus on $V=U_L (U_L')^{\dagger}$ which involves the processes that are actually observed \cite{jarlsprl} and disregard $V_R=U_R(U_R')^{\dagger}$ in the following.

Mathematically, $V$ is an element of $SU(N)$, but since the phases of the quark fields are arbitrary (even for non-Hermitian mass matrices), $V$ is only defined up to left or right multiplication by a diagonal element of $SU(N)$, \ie an element of the maximal torus $U(1)^{N-1}$. The space of CKM matrices is therefore the double quotient $U(1)^{N-1}\backslash SU(N)/U(1)^{N-1}$, characterised by $(N-1)^2$ parameters, out of which $\frac{1}{2}N(N-1)$ may be taken to be real (Euler) angles and the remaining $\frac{1}{2}(N-1)(N-2)$ appear as complex phases. It follows that the matrix $V$ can be taken to be real for $N=2$, and so in the formalism explained here one needs at least three quark families to have a possibility of $CP$ violation. Kobayashi and Maskawa were awarded the 2008 Nobel Prize for using this observation to predict the existence of a third quark family \cite{km}. We set $N=3$ in what follows.

The mathematical theory of observable measures of $CP$ violation in the standard model was developed by Jarlskog \cite{jarlskog}. She showed \cite{jarlsprl} that all necessary and sufficient conditions for $CP$ violation can be summarised as the following condition on the commutator of the Hermitian matrices $m,m'$:
\ben
\det C := \det \left(-{\rm i}\left[m,m'\right]\right)\neq 0\,.
\een
One finds that explicitly
\ben
\det C=-2J(m_t-m_c)(m_c-m_u)(m_u-m_t)(m_b-m_s)(m_s-m_d)(m_d-m_b)\,,
\een
where $J:=\frak{Im}(V_{11}V_{22}V_{12}^*V_{21}^*)$ is the {\bf Jarlskog invariant}\index{Jarlskog invariant} which is invariant under left or right multiplication of $V$ by a diagonal matrix, \ie an element of $U(1)^2$. The geometrical interpretation of the quantity $J$ is given by the so-called unitarity triangles\index{unitarity triangle}. These express the requirement on $V$ to be unitary, so that for instance
\ben
(VV^{\dagger})_{12}=V_{11}V^*_{21}+V_{12}V^*_{22}+V_{13}V^*_{23}=0\,.
\een
In the complex plane, the three complex numbers that sum to zero form the sides of a triangle. The absolute value $|J|$ is twice the area of this triangle. Since there are different unitarity triangles corresponding to different elements of $VV^{\dagger}$, all with the same area, there are several ways of expressing $J$ in terms of the elements of $V$. A general formula is given by \cite{jarlsprl}
\ben
J\sum_{\gamma, l}\epsilon_{\alpha\beta\gamma}\epsilon_{jkl}=\frak{Im}(V_{\alpha j}V_{\beta k}V_{\alpha k}^*V_{\beta j}^*)\,.
\een
The quantities describing the CKM matrix which are invariant under rephasing of the quark fields are $J$ and the absolute values $|V_{\alpha j}|$.

In the general case where $m$ and $m'$ are not assumed to be Hermitian, the corresponding quantity would be
\bea
\det {\bf C} & := & \det \left(-{\rm i}\left[m m^{\dagger},m' m'^{\dagger}\right]\right)
\\& = & -2J(m_t^2-m_c^2)(m_c^2-m_u^2)(m_u^2-m_t^2)(m_b^2-m_s^2)(m_s^2-m_d^2)(m_d^2-m_b^2)\,,\nn
\eea
which of course leads to the same conditions on the mass matrices as before (given that we took all masses to be positive before). In the literature, the use of ${\bf C}$ is perhaps more common than the use of $C$, and one may well argue that this second measure of $CP$ violation should be considered more fundamental as its value does not depend on the arbitrary signs of the mass terms in the Lagrangian.

Recall from chapter \ref{su3quot} the parametrisation of the CKM matrix (\ref{ckmmatrix})\index{CKM matrix}
\ben
V = \left( \begin{matrix} \cos y \cos z & \cos y \sin z & e^{-{\rm i}w} \sin y \\ -\cos x \sin z - e^{{\rm i}w}\sin x \sin y \cos z & \cos x \cos z - e^{{\rm i}w} \sin x \sin y \sin z & \sin x \cos y \\ \sin x \sin z - e^{{\rm i}w} \cos x \sin y \cos z & - \sin x \cos z - e^{{\rm i} w} \cos x \sin y \sin z & \cos x \cos y\end{matrix} \right)\,,
\een
where the ranges of the Euler angles\index{Euler angles} $x, y, z$ and the complex phase $w$ are
\ben
0 \le x, y, z \le \frac{\pi}{2}\,, \quad 0 \le w < 2\pi\,.
\een

In this parametrisation, the Jarlskog invariant $J$\index{Jarlskog invariant} is given by
\ben
J = \frac{1}{4} \sin (2x)\, \sin (2z)\, \sin y\, \cos^2 y\, \sin w\,.
\label{jdef}
\een

It appears that the observed value for $J$ is very small, since the maximal value would be $\frac{1}{6\sqrt{3}}\approx 0.1$, whereas in Nature \cite{particledb}
\ben
J=3.05_{-0.20}^{+0.19}\times 10^{-5}\,.
\label{jobserved}
\een
This discrepancy in 3.5 orders of magnitude is the fine-tuning problem we are investigating in this chapter.

In a general discussion where the values of the quark masses are not fixed, $J$ is not an appropriate measure of $CP$ violation, since even with nonvanishing $J$ one could have $CP$ conservation if, for example, the up and charm quark masses were coinciding. It was suggested in \cite{jarlskog2} to use an appropriately normalised form of $\det C$, namely
\ben
a_{CP}= 3\sqrt{6}\frac{\det C}{(\Tr\,C^2)^{3/2}}
\een
for three quark families as the unique basis independent measure of $CP$ violation. This is a dimensionless number which takes values between $-1$ and $+1$, and is again observed to be very close to zero. When written out in terms of the CKM matrix parameters and quark masses, it is a rather complicated expression which is therefore not extremely useful in practical computations. In the present analysis, we assume the quark masses as known and regard $J$ as the measure of $CP$ violation. 

\sect{Statistics on the Space of CKM Matrices}
\label{jstats}

Let us compute expectation values for $J$ for different choices of measure on the double quotient $U(1)^2\backslash SU(3)/U(1)^2$, parametrised by coordinates $(x,y,z,w)$. Obviously, given a metric $g$, the expectation value of any function $f(x,y,z,w)$ is
\ben
\langle f\rangle = \frac{\int dx\,dy\,dz\,dw\,\sqrt{\det g} \,f(x,y,z,w)}{\int dx\,dy\,dz\,dw\,\sqrt{\det g}}\,.
\een

As we mentioned in chapter \ref{su3quot}, one possible viewpoint of a given function on $U(1)^2\backslash SU(3)/U(1)^2$ is to see it as function on the homogeneous space\index{homogeneous space} $SU(3)/U(1)^2$ which is independent of the coordinates parametrising the left $U(1)^2$ ($p$ and $q$ in our conventions). Since we also saw that there is a unique left-invariant measure on the homogeneous space, this gives a unique definition of expectation values of such functions.

We computed this left-invariant measure on $SU(3)/U(1)^2$ to be
\ben
\sqrt{g} = N \sin (2x)\, \sin (2z)\, \sin y\, \cos^3 y\,
\label{homogenmeas}
\een
for some constant $N$.

Note that there is no need to pass from the double quotient to the homogeneous space if one regards the former as the fundamental domain of definition of a given function. One might equally well regard the left-invariance on the homogeneous space that has led to a unique measure as a spurious symmetry requirement.

Remembering that the CKM matrix is defined as $V=U(U')^{\dagger}$ in terms of two unitary matrices $U$ and $U'$ (denoted by $U_L$ and $U_L'$ for the general case of non-Hermitian mass matrices above), one might think about defining the Jarlskog invariant $J$ in terms of coordinates on the Cartesian product of $SU(3)\times SU(3)$, the spaces of $U$ and $U'$.\footnote{Ambiguities in the definition of $U$ and $U'$ mean that they should be regarded as living in quotient spaces such as $U(1)^2\backslash SU(3)$, but since our left-invariant measures do not depend on the left and right $U(1)^2$ parameters such subtleties may be ignored.} The natural measure on $SU(3)\times SU(3)$ is, in terms of left-invariant forms $\sigma_i$ and $\sigma_i'$,
\be
\mu = N\, \sigma_1\wedge \sigma_2\wedge \sigma_3 \wedge \sigma_4\wedge\sigma_5\wedge
   \sigma_6\wedge \sigma_7 \wedge \sigma_8 \wedge \sigma'_1\wedge \sigma'_2\wedge \sigma'_3 \wedge \sigma'_4\wedge\sigma'_5\wedge
   \sigma'_6\wedge \sigma'_7\wedge \sigma'_8\,.
\label{prodmeasure}
\ee

Since it is only $V=U{U'}^{\dagger}$ that enters into the $CP$ violating parameters, one could consider $U$ and $V$ as independent variables, \ie write $U'=V^{\dagger}U$ for some matrix $V$. Then the Maurer-Cartan form\index{Maurer-Cartan form} on the second $SU(3)$ is
\ben
\im\, \lambda_a \sigma_a' \equiv {U'}^{\dagger}dU'= U^{\dagger} dU - U^{\dagger} (dV \, V^{\dagger}) U \, ,
\een
which gives $\sigma_a'=\sigma_a - h_{ab}\tau_b$, where $\tau_b$ are right-invariant forms on $SU(3)$ in terms of $V$ coordinates and $h_{ab}$ only depends on the $U$ coordinates. The measure (\ref{prodmeasure}), expressed in terms of $V$ and $U$ coordinates, is thus a product of a function of the $U$ coordinates and the natural measure in $V$ coordinates (left- and right-invariant forms on $SU(3)$ give the same measure). Integration over the $U$ coordinates then just gives an irrelevant constant, and one is left with the measure (\ref{homogenmeas}) on the space of $V$ matrices. This justifies the use of (\ref{homogenmeas}) instead of the more complicated constructions obtained by reducing to the double quotient, and we will regard (\ref{homogenmeas}) as the most natural choice of measure on the parameter space.

For the measure (\ref{homogenmeas}) the evaluation of the necessary integrals is very simple and we find that all odd powers of $J$ average to zero, and 
\ben
\langle J^2 \rangle = \frac{1}{720} \approx 1.389 \times 10^{-3}\,,\qquad\langle J^4 \rangle =  \frac{1}{201600} \approx 4.960 \times 10^{-5}\,.
\een
Thus we find that $\Delta J$ for the Jarlskog invariant is given by
\ben
\Delta J = \frac{1}{12\sqrt 5} \approx 0.0373\,,
\een
which is about three orders of magnitude larger than the experimental value (\ref{jobserved}).

We argued in section \ref{biquotmet} that a geometrically motivated measure on the double quotient space arises from Kaluza-Klein reduction\index{Kaluza-Klein reduction} of a left-invariant metric on $SU(3)$ or $SU(3)/U(1)^2$, and obtained the measure (\ref{kkmeas}) from the bi-invariant metric on $SU(3)$. It is interesting how expectation values for powers of $J$ compare to the previous calculations if this measure is used. Unfortunately, the expression for the measure is too complicated to allow us to perform the integrations analytically.  Using numerical integration, we find that
\ben
\langle J^2\rangle \approx 1.1161\times 10^{-3}\,,\qquad\langle J^4 \rangle \approx 3.750 \times 10^{-6}\,,
\een
with the odd powers of $J$ again averaging to zero.
Thus we find
\ben
\Delta J \approx \sqrt{\langle J^2 \rangle} \approx 0.03341\,,
\een
which is very close to the previous result.

Naively, one might have thought that since $J$ is independent of all the $U(1)$ phases, the results would be the same whether one averaged over the space $U(1)^2\backslash SU(3)/U(1)^2$, or else the flag manifold\index{flag manifold} $SU(3)/U(1)^2$. Of course we know that this is not correct, since, as we have seen, the measure for the biquotient takes a different form than the measure one would obtain from just integrating over the left $U(1)^2$ parameters. Nevertheless, this does not seem to have a great impact on expectation values for $J$.

We briefly mentioned in chapter \ref{su3quot} that one could construct more general measures on the double quotient by Kaluza-Klein reduction of more general left-invariant metrics on $SU(3)/U(1)^2$. While this makes the expression for the measure even more complicated, the effect on expectation values of $J$ seems somewhat limited, unless one takes extreme values for the ``squashing parameters". The general conclusion, namely that there appears to be a fine-tuning problem, is unchanged.\footnote{These calculations, mainly carried out by Chris Pope, are not included in detail in this thesis, but can be found in \cite{cpviolation}.}

Given that we argued that the measure used by Ozsv\'ath and Sch\"ucking (\ref{schuck}) was not appropriate from a geometric viewpoint, we should compute expectation values for this measure also. We find that again $\langle J\rangle =0$ and
\ben
\langle J^2 \rangle = 35\times 2^{-16} \approx  5.341\, \times 10^{-4} \,,\qquad  \langle J^4\rangle = 27027\times 2^{-34}\approx  1.573 \,\times 10^{-6} \,,
\een
and that the standard deviation is 
\ben
\Delta J^2 = \sqrt{\langle J^4 \rangle - \langle J^2 \rangle ^2 } = \sqrt{22127}\times 2^{-17} \approx 1.135\, \times 10^{-3}\,, 
\een
and
\ben
\Delta J= \sqrt{ \langle J^2 \rangle  }=\fft{\sqrt{35}}{256}\approx 0.02311\,.
\een
Again, the results are rather similar to the previous cases.

Assuming a uniform distribution over the angles, and hence treating the double coset as a flat four-dimensional manifold so that the measure is simply $\sqrt{\det g}=1$, would give
\be
\langle J \rangle =0,\quad \langle J^2 \rangle = \fft{1}{2048}
  \approx 4.883\times 10^{-4}\,,
\qquad \langle J^4\rangle =189\times 2^{-27} \approx 1.408\times 10^{-6}\,.
\ee
Hence, this simplest possible choice gives
\ben
\Delta J \approx 0.02210 \,.
\een
We have to conclude that, although we argued that some choices of measure on the space of CKM matrices are better motivated geometrically than others, if one is only interested in predictions about the magnitude of $J$ there seems to be little difference between all of the choices we have examined. In all cases one would expect the observed value of $J$ to be at least of order $10^{-2}$, which leaves three orders of magnitude between prediction and observation.

\subsection{Fine-tuning of $J$}
\label{fine}

In the previous subsection we saw that different measures on the space of mixing angles all seem to lead to expectation values for $J$ which are about three orders of magnitude larger than the observed value. The value for $J$ that we observe hence appears to be finely tuned. In this section we shall do a closer, mainly numerical, analysis of the fine-tuning involved. We compare results obtained by taking the $SU(3)$-invariant (descending from the left-invariant measure on $SU(3)/U(1)^2$) and Kaluza-Klein (obtained from Kaluza-Klein reduction of the bi-invariant metric on $SU(3)$) measures, which seem natural from a geometric perspective, with a uniform distribution which is just the simplest possible choice.

\begin{figure}[htp]
\centering
\includegraphics[scale=0.9]{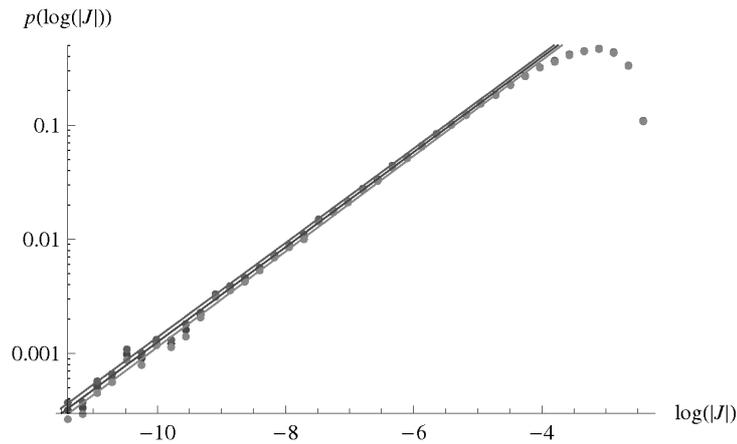}
\caption{{\small Probability distribution for $\log|J|$ using the $SU(3)$-invariant flag measure.}}
\label{fig1}
\end{figure}

\begin{figure}[htp]
\centering
\includegraphics[scale=0.9]{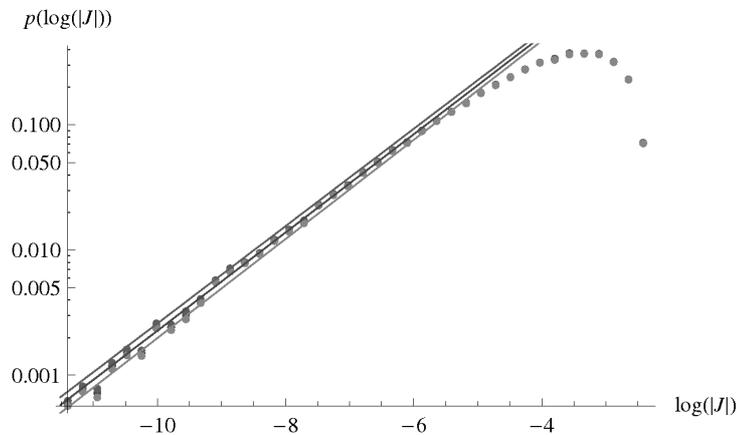}
\caption{{\small Probability distribution for $\log|J|$ using the Kaluza-Klein measure.}}
\label{fig2}
\end{figure}

\begin{figure}[htp]
\centering
\includegraphics[scale=0.9]{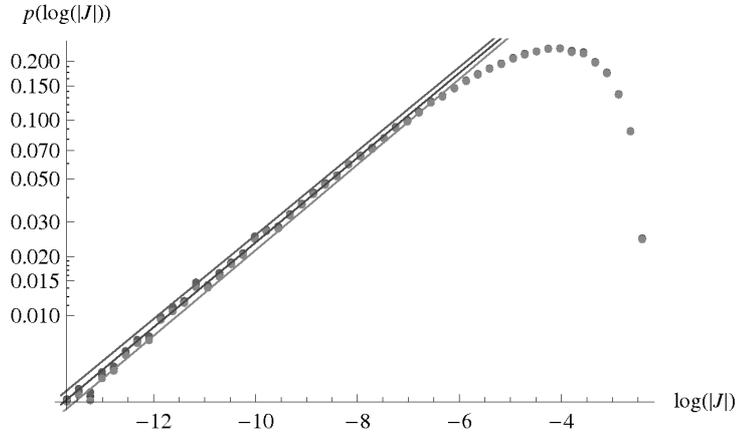}
\caption{{\small Probability distribution for $\log|J|$ using a uniform distribution.}}
\label{fig3}
\end{figure}

The observed value for the Jarlskog invariant $J$ is
\ben
J\approx 10^{-4.51}\approx e^{-10.39}\,.
\een
In order to obtain a probability distribution for $J$ we have used {\sc Mathematica} to numerically compute integrals of the form
\ben
\int dx\,dy\,dz\,dw\,\sqrt{\det g}\,\theta(a-|J|)\;\theta(|J|-b)\equiv P(b \le|J|\le a)\cdot \int dx\,dy\,dz\,dw\,\sqrt{\det g}
\een
using Monte Carlo methods. The $SU(3)$-invariant measure and the Kaluza-Klein measure disfavour small values of $J$ more strongly than a uniform distribution would. For example, we obtain
\ben
P_{{\rm flag}}(|J|\le 10^{-4})\approx 0.25 \%\,,\quad P_{{\rm KK}}(|J|\le 10^{-4})\approx 0.44 \%\,.
\een
Taking a uniform distribution $\sqrt{\det g}\equiv 1$, we get
\ben
P_{{\rm unif}}(|J|\le 10^{-4})\approx 7\%\,.
\een
In figures \ref{fig1} to \ref{fig3}, we have plotted the probability distribution of $|J|$ logarithmically, together with a fit to a power law $p(|J|)\propto |J|^{\lambda}$ assumed to be valid for small $|J|$.
\\The degree of fine-tuning required to reproduce a very small $J$ is considerably higher if one uses the measure induced by a $SU(3)$-invariant flag metric or the Kaluza-Klein metric, maybe contrary to what one might expect. Values of $J$ close to its maximal value of $\frac{1}{6\sqrt{3}}\approx 0.0962$ are disfavoured in both cases. This motivated the use of a logarithmic scale for $|J|$. 
\\In all three cases the numerical results for small $|J|$ are well approximated by a power law of the form $p(|J|)=\alpha\cdot |J|^{\lambda}$ for the probability density of $|J|$. The logarithmic graphs show $p(\log|J|)\propto |J|^{\lambda+1}$. For the $SU(3)$-invariant flag measure (Fig. \ref{fig1}), the best fit to the data in the region below $|J|=10^{-2.3}$ or $\log|J|=-5.3$ is
\ben
\lambda_{{\rm flag}}=-0.042(\pm 0.006)\,,\quad\alpha_{{\rm flag}}=18.1(\pm 0.7)\,;
\een
for the Kaluza-Klein measure (Fig. \ref{fig2}) we fitted the data in the region below $|J|=10^{-2.7}$ or $\log|J|= -6.2$ and obtained
\ben
\lambda_{{\rm KK}}=-0.097(\pm 0.008)\,,\quad\alpha_{{\rm KK}}=18.9(\pm 1.0)\,;
\een
finally for the uniform measure (Fig. \ref{fig3}), the best fit to the data in the region below $|J|=10^{-3.4}$ or $\log|J|= -7.8$ is
\ben
\lambda_{{\rm unif}}=-0.500(\pm 0.005)\,,\quad\alpha_{{\rm unif}}=3.51(\pm 0.15)\,.
\een

\subsection{Wolfenstein Parametrisation}
\label{wolfenstein}
A different parametrisation of the CKM matrix which is frequently used in the particle physics literature was introduced by Wolfenstein and is based on the experimentally observed hierarchy
\ben
y\ll x\ll z\ll 1
\een
in the mixing angles. One rewrites \cite{Wolfenstein}\index{Wolfenstein parametrisation} 
\ben
\sin z=\lambda\,,\;\sin x=A\lambda^2\,,\;\sin y e^{-\im w}=A\lambda^3 (\rho-\im\eta)
\een
and treats $\lambda$ as a small parameter while $A,\rho$, and $\eta$ are supposed to be parameters of order unity. In the modern literature one also frequently uses $\bar\rho,\bar\eta$ instead of $\rho$ and $\eta$ because then the combination $\bar\rho + i\bar\eta$ is independent of the phase convention in the 
CKM matrix \cite{particledb}. These parameters are defined by
\bea
&&\rho=\sqrt{\frac{1-A^2 \lambda^4}{1-\lambda^2}}\frac{\bar\rho-A^2\lambda^4(\bar\rho^2+\bar\eta^2)}{(1-A^2\lambda^4 \bar\rho)^2+A^4\lambda^8\bar\eta^2}\,,\nn
\\&& \eta=\sqrt{\frac{1-A^2 \lambda^4}{1-\lambda^2}}\frac{\bar\eta}{(1-A^2\lambda^4 \bar\rho)^2+A^4\lambda^8\bar\eta^2}\,.
\eea
The experimental values for $\lambda, A, \bar\rho, \bar\eta$ are \cite{particledb}\footnote{Note that only even powers of $\lambda$ appear in all expansions, so that it is $\lambda^2\approx 0.05$ which is the small parameter.}
\ben
\lambda=0.2272\pm 0.0010,\; A=0.818_{-0.017}^{+0.007},\; 
\bar\rho=0.221_{-0.028}^{+0.064},\;\bar\eta=0.340_{-0.045}^{+0.017}\,.
\een
One viewpoint on the Wolfenstein parametrisation is that it is adapted to the values for the CKM matrix entries that we observe and has no deeper significance; but often the viewpoint is expressed that this parametrisation expresses some kind of ``natural hierarchy" in the mixing angles coming from physics beyond the standard model (see \eg \cite{biggi}). Treating the other parameters as ``naturally of order unity" reduces our calculations to a one-dimensional problem as everything is only expanded in terms of $\lambda$. We find that the $SU(3)$-invariant measure on the flag manifold is now, to leading order in $\lambda$,
\ben
\left|\frac{\partial(x,y,z,w)}{\partial(\lambda,A,\bar\rho,\bar\eta)}
\right|\sqrt{g}\propto A^3 \lambda^{11}
\left(1+\lambda^2+O(\lambda^4)\right)\,,
\een
and the Jarlskog invariant $J$ is
\bea
J &= &\frac{A^2 \bar\eta \lambda^6 (1-A^2\lambda^4) \left(1-\lambda^2 - 2A^2\bar\rho\lambda^4 - A^2 (\bar\eta^2 + (\bar\rho-2) \bar\rho)\lambda^6  + A^4 (\bar\eta^2 + \bar\rho^2)\lambda^8 \right)}{(1-\lambda^2)\left(1 - 2 A^2 \bar\rho \lambda^4 +  A^4 (\bar\eta^2 + \bar\rho^2)\lambda^8\right)^2}\nn
\\&=&A^2 \bar\eta \lambda^6+O(\lambda^{10})\,.
\eea
Inverting this expression to leading order gives the probability distribution for $J$
\ben
p(J)\propto \frac{J}{A\bar\eta^2}\left(1+\left(\frac{J}{A^2\bar\eta}\right)^{1/3}+O(J^{2/3})\right)\,,
\een
which is incompatible with the numerical results. Trying to improve this approximate result by letting $A,\bar\rho$ and $\bar\eta$ take all possible values leads to inconsistencies since the expansion in powers of $J$ contains poles of arbitrary order in $A$. From our present viewpoint, where no mechanism for fixing these parameters close to one is known, the Wolfenstein parametrisation seems rather misleading when discussing geometric probability.

\sect{Statistics on the Space of Mass Matrices}
\label{gaussian}

In the previous sections we have focussed on $U(1)^2 \backslash SU(3) / U(1)^2$, the space of CKM matrices, as the space of $CP$ violating parameters. Since $SU(3)$ is compact, this space has finite volume for a natural measure. But the CKM matrix is derived from the quark mass matrices, which could be viewed as more fundamental and more directly determined by physics beyond the standard model. In this section, we try to obtain statistics of the Jarlskog invariant $J$ from a random distribution on the space of mass matrices; in the standard model, this is the space of $3\times 3$ Hermitian matrices, whereas in certain extensions with left-right symmetry one has to take all $3\times 3$ complex matrices into account. 

In this section, as suggested in \cite{jarlsprl}, we define a dimensionless mass matrix $M$ by $M=m/\Lambda$, where $m$ is the quark mass matrix in the notation of section \ref{cpviolation}, and $\Lambda$ is a scale which may be chosen for convenience. A natural choice would be $\Lambda=m_t$ for the up-type or $\Lambda'=m_b$ for the down-type quarks, but since we in principle allow arbitrary values for the quark masses we leave $\Lambda$ arbitrary.

We now need to pick a measure on the space of mass matrices in order to do statistics. Let us summarise the main results of section \ref{hermcomp}: For a given Hermitian matrix $M$, one defines a unitary matrix $U$ and a real diagonal matrix $D$ by
\ben
M_{{\rm H}} = U^{\dagger}DU\,,
\een
and the natural measure, invariant under conjugation by $U(3)$, on the space of such matrices was found to be
\bea
DM_{{\rm H}} & := & (D_1-D_2)^2(D_1-D_3)^2(D_2-D_3)^2\sin(2x)\,\cos^3 y\,\sin y\,\sin(2z)\nn
\\& & \times dD_1\,dD_2\,dD_3\,dx\,dy\,dz\,dw\,dr\,dt\,.
\label{hermmat}
\eea
We noted that to avoid overcounting one should integrate over $\mathbb{R}^3$ times a quotient of $SU(3)$, more precisely $(U(1)^2\times\frak{S}_3)\backslash SU(3)$.

Let us contrast this with the space of all complex matrices; here we used the representation
\ben
M_{{\rm c}} = U_L^{\dagger}DU_R\,,
\een
and identified the relevant space one has to integrate over as $\mathbb{R}_+^3\times (U(1)^3\times\frak{S}_3)\backslash (U(3)\times U(3))$. We then argued that if we are only interested in functions that depend on $U_L$, but not on $U_R$, such as the Jarlskog invariant $J$, we can integrate over the space of $U_R$. Then the resulting measure on $\mathbb{R}_+^3\times (U(1)^2\times\frak{S}_3)\backslash SU(3)$ is
\bea
DM_{{\rm c}} & := & \prod_{i<j}(D_i^2-D_j^2)^2|D_1 D_2 D_3|\sin(2x)\,\cos^3 y\,\sin y\,\sin(2z)\nn
\\& & \times dD_1\,dD_2\,dD_3\,dx\,dy\,dz\,dw\,dr\,dt\,,
\label{complmat}
\eea
which obviously differs from (\ref{hermmat}) by different powers of the real numbers $D_i$ that had the role of eigenvalues for Hermitian matrices. Since the measure (\ref{complmat}) is invariant under $D_i\rightarrow -D_i$, we will integrate over all of $\mathbb{R}^3$ for simplicity. This only leads to an extra factor 8 which drops out in expectation values.

It is clear that in both cases the integrals over $\mathbb{R}^3$ with the given measures will diverge. We could introduce a cutoff for the quark masses, but then any expectation values for quark masses would strongly contradict observation, as there is no way to explain the observed mass hierarchy.

To both make the integrals converge and introduce a possibility for introducing the quark masses as additional parameters into our distribution, we introduce a function decaying sufficiently fast for large $|D_i|$ into the measure. A natural and simple choice is a Gaussian.

Since there are actually two integrations over the space of mass matrices, corresponding to two mass matrices $M$ and $M'$ for the up-type and down-type quarks, we will use the measure\footnote{$DM$ is to replaced by $DM_{{\rm H}}$ and $DM_{{\rm c}}$ for Hermitian and complex matrices respectively, and from now on we write $U$ instead of $U_L$ in the case of complex matrices as well, since $U_R$ has disappeared from our calculations.}
\bea
&&DM\,DM'\,\exp(-\Tr(MM^{\dagger}A)-\Tr(M'(M')^{\dagger}A'))\nn
\\&=&DM\,DM'\,\exp(-\Tr(U^{\dagger}D^2 U A)-\Tr((U')^{\dagger}(D')^2 U' A'))\,,
\label{fullmeas}
\eea
which for complex matrices is still invariant under the two right $U(3)$ actions on the complex matrices $M$ and $M'$, but the invariance under the left $U(3)$ actions is broken to the subgroups commuting with $A$ and $A'$ respectively. We will assume $A$ and $A'$ to be Hermitian with non-negative eigenvalues, satisfying $[A,A']=0$ and use the redefinitions $U\rightarrow U W$, $U'\rightarrow U'W$, which leave $J$ invariant, to diagonalise $A$ and $A'$. Hence we can assume $A$ and $A'$ to be diagonal with positive entries in the following, and the subgroup commuting with $A$ and $A'$ is $U(1)^3$. For Hermitian matrices, (\ref{fullmeas}) becomes
\ben
DM_{{\rm H}}\,DM_{{\rm H}}'\,\exp(-\Tr(M_{{\rm H}}^2 A))\exp(-\Tr((M_{{\rm H}}')^2 A))\,,
\een
where the Gaussian breaks the symmetry of the measure from invariance under $(U(3)\times U(3))$ to invariance under $(U(1)^3\times U(1)^3)$, again the subgroup commuting with $A$ and $A'$. 

We will find that this symmetry breaking\index{symmetry breaking} is necessary to obtain a distribution that can reproduce the observed quark masses. Our proposal is to fit the diagonal matrices $A$ and $A'$ to the observed quark masses and to use the resulting probability distribution for statistics of $J$.

An expectation value of a quantity such as $J^2$ is then given by\footnote{In all cases considered, all odd powers of $J$ again have expectation value zero.} 
\bea
\langle J^2 \rangle & = & N \int DM \, DM'\, e^{-\Tr(MM^{\dagger} A)-\Tr(M'M'^{\dagger} A')} \,J^2(M,M')
\label{jintegral}
\\ & = & N \int_{\R^6} dD \,dD' \int_{((U(1)^2\times\frak{S}_3)\backslash SU(3))^2} DU \, DU'\, e^{-\Tr(D^2 UAU^{\dagger})-\Tr((D')^2 U'A'{U'}^{\dagger})}\,J^2\,.\nn
\eea
Here $DU$ and $DU'$ are the measures on $(U(1)^2\times \frak{S}_3)\backslash SU(3)$, given by (\ref{volform}),
\ben
DU:= \sin(2x)\,\sin y\,\cos^3 y\,\sin(2z)\,,
\een
and
\ben
dD:=\begin{cases}(D_1-D_2)^2 (D_1-D_3)^2 (D_2-D_3)^2 \,dD_1\,dD_2\, dD_3\,, & M \mbox{ Hermitian}\,,\cr (D_1^2-D_2^2)^2 (D_1^2-D_3^2)^2 (D_2^2-D_3^2)^2|D_1 D_2 D_3| \,dD_1\,dD_2\, dD_3\,, & M \mbox{ complex}\,,\end{cases}
\label{dmeasure}
\een
and the normalisation factor $N$ is defined by
\ben
\frac{1}{N}:=\int_{\R^6} dD \,dD' \int_{((U(1)^2\times\frak{S}_3)\backslash SU(3))^2} DU \, DU'\, e^{-\Tr(D^2UAU^{\dagger})-\Tr((D')^2U'A'{U'}^{\dagger})}\,.
\een
From $J=\Im \bigl (V_{11} \,V_{22}\, V_{12}^* \, V_{21}^* \bigr )$ and $V=U{U'}^{\dagger}$, we have
\ben
J(U,U')=\sum_{a,b,c,d=1}^3\Im\bigl( U_{1a}U_{2b}U_{1c}^*U_{2d}^*{U'}^*_{1a}{U'}^*_{2b}{U'}_{2c}{U'}_{1d}\bigr)\,.
\een
At this point, it is perhaps instructive to note that choosing $A$ proportional to the identity would split the integral (\ref{jintegral}) into a product of an integral over $\mathbb{R}^6$, \ie the quark masses, which just gives a constant, and an integral of $J^2$ over $((U(1)^2\times\frak{S}_3)\backslash SU(3))^2$. Since all even powers of $J$ are invariant under the $\frak{S}_3$ action on $U$ and $U'$, this can be replaced by an integral over $(U(1)^2\backslash SU(3))^2$ if averages are concerned. By the arguments presented in section \ref{jstats}, a change of coordinates reduces this to a single integration over a copy of $SU(3)/U(1)^2$, and one recovers the results of section \ref{jstats} for expectation values of powers of $J$. The introduction of more general diagonal matrices $A$ and $A'$ means that the maximal invariance of the measure $DM\,DM'$ is broken down to the action of a subgroup. 

It should be clear from (\ref{jintegral}) that multiplying $A$ (or $A'$) by a constant is the same as rescaling the eigenvalues $D_i$ (or $D_i'$) and so amounts to a rescaling of $\Lambda$ (or $\Lambda'$). We can therefore, without any loss of generality, choose
\ben
A=\left(\begin{matrix}
1 & 0 & 0 \\ 0 & 1/\mu_c^2 & 0 \\ 0 & 0 & 1/\mu_u^2
\end{matrix}\right),\quad
A'=\left(\begin{matrix}
1 & 0 & 0 \\ 0 & 1/\mu_s^2 & 0 \\ 0 & 0 & 1/\mu_d^2
\end{matrix}\right)\, ,
\label{amatrices}
\een
where $\mu_c, \mu_u, \mu_s$, and $\mu_d$ are dimensionless parameters that we are free to choose so as to reproduce the observed quark masses as expectation values. (In the case of an exponential $\exp(-\Tr(D^2 A))$, these would of course be equal to the respective quark masses, expressed in units where $\Lambda=m_t$ and $\Lambda'=m_b$.) Because of experimental uncertainties in the up and quark masses, one can modify this distribution to reproduce different values for these masses.

\subsection{Hermitian Mass Matrices}

We shall now present the result of evaluating the integral (\ref{jintegral}), first for Hermitian and then for complex mass matrices, and therefore for the two different choices of measure (\ref{dmeasure}). We will have to use various approximations which we try to justify in the following, and the final result is obtained independently by numerical integration and by an analytical approximation, where we find relatively good agreement between the two.

The need for approximations in the integral (\ref{jintegral}) is immediate since the expression for $J$ in terms of coordinates on $((U(1)^2\times \frak{S}_3)\backslash SU(3))^2$ is too complicated to be given explicitly. However, since
\ben
\Tr(D^2 U A U^{\dagger})=\sum_a D_a^2 \sum_c A_c |U_{ac}|^2=: \sum_a D_a^2 \xi_a\,, \quad \Tr((D')^2 U' A' {U'}^{\dagger})=: \sum_a (D'_a)^2 \xi'_a
\een
with
\bea
&& \xi_1= A_1 \cos^2 y \cos^2 z + A_2 \cos^2 y \sin^2 z + A_3 \sin^2 y\,,\nn
\\&& \xi'_1= A'_1 \cos^2 y' \cos^2 z' + A'_2 \cos^2 y' \sin^2 z' + A'_3 \sin^2 y'\,,
\eea
and we assume $A_3\gg 1$ and $A_3'\gg 1$, the integrand is negligibly small unless $y\approx 0$ and $y'\approx 0$. We use this to approximate the integrals over $y$ and $y'$ (recall the restriction of the coordinates on $U(1)^2\backslash SU(3)$ given by (\ref{coordrestr}) which restricts from $U(1)^2\backslash SU(3)$ to $(U(1)^2\times\frak{S}_3)\backslash SU(3)$):
\bea
& & \int\limits_{0}^{\arctan(\sin x)} dy \int\limits_{0}^{\arctan(\sin x')} dy' \cos^3 y \sin y \cos^3 y' \sin y' e^{-\Tr(D^2 UAU^{\dagger})-\Tr((D')^2 U'A'{U'}^{\dagger})}\,J^2(U,U') \nn
\\ & \approx & \int\limits_{0}^{\infty} dy \int\limits_{0}^{\infty} dy' \, y\, y'\, e^{-A_3 y^2 -A_3' (y')^2}\cdot\left(e^{-\Tr(D^2 UAU^{\dagger})-\Tr((D')^2 U'A'{U'}^{\dagger})}\,J^2(U,U')\right)\big|_{y=y'=0} \nn
\\ & \approx & \frac{1}{4A_3 A_3'}\left(e^{-\Tr(D^2 UAU^{\dagger})-\Tr((D')^2 U'A'{U'}^{\dagger})}\,J^2(U,U')\right)\big|_{y=y'=0}\,.
\eea

It turns out that this is independent of $w$ and $w'$. Constant prefactors such as $1/4A_3 A'_3$ appearing in both numerator and denominator can be dropped, and so we have
\ben
\langle J^2 \rangle \approx \frac{\int_{\R^6} dD dD' \int d^4 x d^4 x'\,s_{2x}s_{2z}s_{2x'}s_{2z'} \left(e^{-\Tr(D^2 UAU^{\dagger})-\Tr((D')^2 U'A'{U'}^{\dagger})}\,J^2(U,U')\right)\big|_{y=y'=0}}{\int_{\R^6} dD dD' \int d^4 x d^4 x'\,s_{2x}s_{2z}s_{2x'}s_{2z'} \left(e^{-\Tr(D^2 UAU^{\dagger})-\Tr((D')^2 U'A'{U'}^{\dagger})}\right)\big|_{y=y'=0}}\,,
\label{newjint}
\een
where $s_{2x}=\sin 2x$ etc. and
\ben
\int d^4 x\equiv \int\limits_0^{\pi/4}dx\int\limits_0^{\pi/2}dz \int\limits_0^{2\pi}dr\int\limits_0^{2\pi}dt
\een
and similarly for $\int d^4 x'$.

Now we can integrate over both copies of $\R^3$ in (\ref{newjint}), using
\bea
& &\int\limits_{-\infty}^{\infty}dD_1\int\limits_{-\infty}^{\infty}dD_2\int\limits_{-\infty}^{\infty}dD_3\,(D_1-D_2)^2(D_1-D_3)^2(D_2-D_3)^2 e^{-\xi_1 D_1^2-\xi_2D_2^2-\xi_3D_3^2}\nn
\\& = &\frac{3\pi^{3/2}}{8 \xi_1^{5/2}\xi_2^{5/2}\xi_3^{5/2}}\left(\xi_1^2(\xi_2+\xi_3)+\xi_2^2(\xi_1+\xi_3)+\xi_3^2(\xi_2+\xi_1)-2\xi_1\xi_2\xi_3\right)\,.
\eea

The explicit expression for $J$ at $y=y'=0$ is
\bea
J(U,U')\big|_{y=y'=0} & = &\frac{1}{4}s_{2x} s_{2x'} \left\{c^2_{z'} s^3_{z} s_{z'}\sin(3 \hat{r} + \hat{t})  + c^3_{z} c_{z'} s^2_{z'}\sin(3 \hat{r} - \hat{t}) \right.\label{jexp}
\\& &  - c^2_{z} s_{z} s_{z'} (c^2_{z'} \left[\sin(3 \hat{r} + \hat{t}) + \sin(3\hat{r} -3\hat{t})\right] - s^2_{z'}\sin(3\hat{r} + \hat{t}) )\nn
\\& & \left. + c_{z} c_{z'} s^2_{z} (c^2_{z'} \sin(3\hat{r} - \hat{t}) -  s^2_{z'}\left[\sin(3 \hat{r} + 3\hat{t}) + \sin(3 \hat{r} - \hat{t})\right])\right\}\nn
\eea
where $s_x=\sin x, c_{z'}=\cos z'$, etc., $\hat r= r-r'$, and $\hat t=t-t'$. Integrating (\ref{jexp}) over $r,r',t$, and $t'$ indeed gives zero, which is why we choose to use $J^2$.

We need to determine the parameters appearing in the matrices $A$ and $A'$ in (\ref{amatrices}) by fitting them to the expectation values for quark masses that we want to reproduce. We first observe that expectation values for squared mass matrices take the relatively simple form
\ben
\langle D_1^2 \rangle \approx \frac{\int_{\R^3} dD\, D_1^2 \int\limits_0^{\pi/4} dx \int\limits_0^{\pi/2} dz\,\sin 2x \sin 2z  \left(e^{-\Tr(D^2 UAU^{\dagger})}\right)\big|_{y=0}}{\int_{\R^3} dD\,  \int\limits_0^{\pi/4} dx \int\limits_0^{\pi/2} dz\,\sin 2x \sin 2z  \left(e^{-\Tr(D^2 UAU^{\dagger})}\right)\big|_{y=0}}\,.
\label{d1int}
\een
The denominator of (\ref{d1int}) is explicitly
\bea
I_D&:=&\int\limits_0^{\pi/4} dx \int\limits_0^{\pi/2} dz\,\sin 2x \sin 2z \,\frac{3\pi^{3/2}}{8 \xi_1^{5/2}\xi_2^{5/2}\xi_3^{5/2}}\times\nn
\\&&\left(\xi_1^2(\xi_2+\xi_3)+\xi_2^2(\xi_1+\xi_3)+\xi_3^2(\xi_2+\xi_1)-2\xi_1\xi_2\xi_3\right)\,,
\label{denom}
\eea
where within our approximation the quantities $\xi_a$ are taken at $y=0$,
\bea
& \xi_1= A_1 \cos^2 z + A_2 \sin^2 z\,,\quad \xi_2= A_1 \cos^2 x \sin^2 z + A_2 \cos^2 x \cos^2 z + A_3 \sin^2 x\,,\nn
\\& \xi_3= A_1 \sin^2 x \sin^2 z + A_2 \sin^2 x \cos^2 z + A_3 \cos^2 x\,,
\label{xidef}
\eea
with $A_3 \gg A_2 \gg A_1$. We notice that all $\xi_a$ are nonzero for all values of $x$ and $z$. Furthermore, the integral is dominated by very small $x$ and $z$ (we cannot have $x=\frac{\pi}{2}$), and we can approximate $I_D$ well by only keeping the terms of leading order in $x$ and $z$ in the trigonometric functions, and 
\ben
\xi_1^2(\xi_2+\xi_3)+\xi_2^2(\xi_1+\xi_3)+\xi_3^2(\xi_2+\xi_1)-2\xi_1\xi_2\xi_3\approx A_3^3 x^2 + A_3^2 A_2\,,
\een
which are the leading terms (as we shall see, the first of these is effectively also of order $A_3^2 A_2$):
\bea
I_D & \approx &\frac{3\pi^{3/2}}{8}\int\limits_0^{\pi/4} dx \int\limits_0^{\pi/2} dz\,4 x z \,(A_3^3 x^2 + A_3^2 A_2)(A_1+A_2 z^2)^{-5/2}(A_2+A_3 x^2)^{-5/2}A_3^{-5/2}\nn
\\ & \approx &\frac{3\pi^{3/2}}{8}\int\limits_0^{\infty} dX \int\limits_0^{\infty} dZ\,(A_3^3 X + A_3^2 A_2)(A_1+A_2 Z)^{-5/2}(A_2+A_3 X)^{-5/2}A_3^{-5/2}\nn
\\ & = &\frac{3\pi^{3/2}}{8}\int\limits_0^{\infty} dX (A_3^3 X + A_3^2 A_2)(A_2+A_3 X)^{-5/2}A_3^{-5/2}\cdot\frac{2}{3A_1^{3/2}A_2}\nn
\\ & = &\frac{3\pi^{3/2}}{8}\frac{2}{3A_1^{3/2}A_2}\left(\frac{2}{3} A_2^{-1/2}A_3^{-3/2}+\int\limits_0^{\infty} dX \,\frac{2}{3A_3^{1/2}}(A_2+A_3 X)^{-3/2}\right)\nn
\\ & = & \frac{\pi^{3/2}}{4 A_1^{3/2}A_2}\left(\frac{2}{3} A_2^{-1/2}A_3^{-3/2}+\frac{4}{3A_3^{3/2}A_2^{1/2}}\right) = \frac{\pi^{3/2}}{2 A_1^{3/2}A_2^{3/2}A_3^{3/2}}\,.
\eea
Similarly, we find for the numerator of (\ref{d1int})
\bea
I_D\langle D_1^2 \rangle & \approx & \frac{15\pi^{3/2}}{16}\int\limits_0^{\infty} dX \int\limits_0^{\infty} dZ\,(A_3^3 X + A_3^2 A_2)(A_1+A_2 Z)^{-7/2}(A_2+A_3 X)^{-5/2}A_3^{-5/2}\nn
\\ & = & \frac{15\pi^{3/2}}{16}\int\limits_0^{\infty} dX \,(A_3^3 X + A_3^2 A_2)(A_2+A_3 X)^{-5/2}A_3^{-5/2}\cdot\frac{2}{5A_2 A_1^{5/2}} \nn
\\ & = & \frac{3\pi^{3/2}}{4 A_1^{5/2}A_2^{3/2}A_3^{3/2}}\,,
\eea
hence
\ben
\langle D_1^2 \rangle \approx \frac{3}{2A_1}\,.
\een
Redoing the same calculation for $D_2$ and $D_3$ gives
\ben
\langle D_2^2 \rangle \approx \frac{1}{2A_2}\,,\quad \langle D_3^2 \rangle \approx \frac{1}{2A_3}\,.
\label{relfactor}
\een
There is a relative factor of 3 which has to be taken into account when determining $A$ and $A'$.

Because of the dependence of masses on the energy scale in quantum field theory, described by the renormalisation\index{renormalisation} group, there is some ambiguity in what is meant by the ``quark masses" we want to reproduce. Following \cite{rosner}, for example, we take all the quark masses evolved to the scale of the $Z$ boson mass. These are given in \cite{massref}:
\bea
(m_u,m_c,m_t)=(1.27_{-0.42}^{+0.50}\;{\rm MeV},\; 0.619 \pm 0.084\;{\rm GeV},\;171.7 \pm 3.0\;{\rm GeV})\,; \nn
\\ (m_d,m_s,m_b)=(2.90_{-1.19}^{+1.24}\;{\rm MeV},\; 55_{-15}^{+16}\;{\rm MeV},\;2.89 \pm 0.09\;{\rm GeV})\,.
\label{qmass}
\eea
We use the central values
\bea
&& (m_u,m_c,m_t):=(1.27\;{\rm MeV},\; 0.619\;{\rm GeV},\;171.7\;{\rm GeV})\,;\nn
\\&& (m_d,m_s,m_b):=(2.9\;{\rm MeV},\; 55\;{\rm MeV},\;2.89\;{\rm GeV})\,.
\label{cvalues}
\eea

The mass scales $\Lambda$ and $\Lambda'$ that were so far arbitrary are fixed by setting $\langle D_1^2 \rangle=(m_t/\Lambda)^2$ and $\langle (D'_1)^2 \rangle=(m_b/\Lambda')^2$. By comparing the results obtained by numerical integration with the values we want to reproduce, we can then fix the parameters $\mu_c,\mu_u,\mu_s$ and $\mu_d$.

In the case of the positively charged top, charm and up quarks, which exhibit a more extreme quark mass hierarchy, we find that numerical calculations (using {\sc Mathematica}) reproduce the results we have obtained analytically very well (see Table \ref{results}). For the negatively charged quarks, we find numerically that we have to use relative factors different from 3 to reproduce the observed masses. Comparing the numerical results with (\ref{cvalues}), we fix the parameters appearing in $A$ and $A'$ to
\bea
\mu_c^2 = 3\left(\frac{m_c}{m_t}\right)^2\approx 3.90\times 10^{-5}\,,\quad \mu_u^2 = 3\left(\frac{m_u}{m_t}\right)^2\approx 1.64\times 10^{-10}\,,\nn
\\ \mu_s^2 = \frac{3}{2}\left(\frac{m_s}{m_b}\right)^2\approx 5.43\times 10^{-4}\,,\quad \mu_d^2 = \frac{12}{5}\left(\frac{m_d}{m_b}\right)^2\approx 2.42\times 10^{-6}\,.
\label{muvalues}
\eea

As a brief side remark, we see that the dominant contributions to these integrals come from the regions
\ben
y\approx\sqrt{\frac{1}{A_3}}\,,\quad y'\approx\sqrt{\frac{1}{A'_3}}\,,\quad x\approx\sqrt{\frac{A_2}{A_3}}\,,\quad z\approx\sqrt{\frac{A_1}{A_2}}\,,\quad x'\approx\sqrt{\frac{A'_2}{A'_3}}\,,\quad z'\approx\sqrt{\frac{A'_1}{A'_2}}\,,
\label{regions}
\een
and these values are all small compared to one. This observation allows us to give rough estimates for magnitudes of individual elements of the CKM matrix.

In the standard convention the ordering of the quark families is $(u,c,t)$ and not $(t,c,u)$ as used in (\ref{amatrices}), which means that in our parametrisation,
\ben
|(U{U'}^{\dagger})_{13}|=|V_{td}|\,,\quad |(U{U'}^{\dagger})_{12}|=|V_{ts}|\,,\quad |(U{U'}^{\dagger})_{23}|=|V_{cd}|\,.
\een
Since all of the numbers in (\ref{regions}) are small, we only keep leading terms in the angles on $U$ and $U'$:
\bea
|(U{U'}^{\dagger})_{13}|\approx|x'(z'-z)-e^{iw'}y'+\ldots|\approx x'z'\approx \mu_d \approx 2 \times 10^{-3}\,, \nn
\\ |(U{U'}^{\dagger})_{12}|\approx z'\approx \mu_s\approx 0.02 \,,\quad |(U{U'}^{\dagger})_{23}|\approx x'\approx \frac{\mu_d}{\mu_s}\approx 0.07\,. 
\eea
Experimental values are \cite{particledb}
\ben
|V_{td}|=(8.14_{-0.64}^{+0.32}) \times 10^{-3}\,,\quad |V_{ts}|=(41.61_{-0.78}^{+0.12}) \times 10^{-3}\,,\quad |V_{cd}|=0.2271_{-0.0010}^{+0.0010}\,.
\een
Our rough estimates reproduce the right ordering of the three parameters and are accurate to factors of order a few. A more careful analysis would involve computing expectation values for these parameters in the distribution we have assumed.

We return to the task of computing the expectation value of $J^2$. In order to obtain an analytical expression, we use the fact that the main contribution to the integral (\ref{newjint}) will come from small $z$ to only take the term in (\ref{jexp}) that is nonzero at $z=0$. Averaging over $r,t,r'$,and $t'$ gives a factor of 1/2, as one might have expected, and therefore we use
\ben
J^2_{{\rm small}\;z}:=\frac{1}{2}\sin^2 x\cos^2 x\sin^2 x'\cos^2 x'\cos^2 z'\sin^4 z'
\een
for our calculations. The numerator of (\ref{newjint}) is the product (using again that only small $z$ contributes)
\bea
& &\frac{9\pi^3}{32}\times \int\limits_0^{\pi/2} dz \, \frac{2z}{(A_1 + A_2 z^2)^{5/2}} \times \int\limits_0^{\pi/2} dz' \, \frac{\sin 2z' \cos^2 z'\sin^4 z'}{(A_1 \cos^2 z'+ A_2 \sin^2 z')^{5/2}}
\label{longintegral}
\\ & \times &\int\limits_0^{\pi/4} dx \,\frac{\sin 2x\,\sin^2 x\,\cos^2 x(A_3^3 \cos^2 x \sin^2 x + A_3^2 A_2(\cos^6 x + 2 \cos^2 x \sin^2 x + \sin^6 x))}{(A_2 \cos^2 x + A_3 \sin^2 x)^{5/2} (A_3 \cos^2 x + A_2 \sin^2 x)^{5/2}}\nn
\eea
times an integral identical to the second line of (\ref{longintegral}), except that $A_2$ and $A_3$ are replaced by $A_2'$ and $A_3'$.
The first two factors are $\frac{2}{3A_1^{3/2}A_2}$ and $\frac{4}{3\sqrt{A_2'}(\sqrt{A_1'}+\sqrt{A_2'})^4}$, respectively; for the second line of (\ref{longintegral}) we change variables to $X=\cos^2 x$ to obtain
\ben
\int\limits_{1/2}^{1} dX \,\frac{X(1-X)(A_3^3 X(1-X) + A_3^2 A_2(X^2 - X + 1))}{(A_2 X + A_3 (1-X))^{5/2} (A_3 X + A_2 (1-X))^{5/2}}\approx \frac{1}{A_3^2}\left(\arctan\left(\frac{1}{2}\sqrt{\frac{A_3}{A_2}}\right)-2\sqrt{\frac{A_2}{A_3}}\right)\,,
\een
where we are dropping corrections of order $\frac{A_2}{A_3}$. Putting everything together, we obtain
\bea
\langle J^2_{{\rm small}\;z}\rangle & \approx &\frac{(A_1')^{3/2}A_2'\sqrt{A_2}}{\sqrt{A_3 A_3'}(\sqrt{A'_1}+\sqrt{A'_2})^4}\left(\arctan\sqrt{\frac{A_3}{4A_2}}-\sqrt{\frac{4A_2}{A_3}}\right)\nn
\\&&\times\left(\arctan\sqrt{\frac{A'_3}{4A'_2}}-\sqrt{\frac{4A'_2}{A'_3}}\right)
\label{japprox}
\\ & = & \frac{\frac{4}{\sqrt{15}}m_u\,m_d\,m_b}{m_c \,m_s^2\left(1+\sqrt{\frac{2}{3}}\frac{m_b}{m_s}\right)^4}\left(\arctan\frac{m_c}{2m_u}-\frac{2m_u}{m_c}\right)\nn
\\&&\times\left(\arctan\sqrt{\frac{5}{32}}\frac{m_s}{m_d}-\sqrt{\frac{32}{5}}\frac{m_d}{m_s}\right)\,,\nn
\eea
where the numerical factors appearing in the last line come from the different factors chosen in (\ref{muvalues}). Note that the top quark mass does not appear in this approximate result.

We also use numerical integration to check the validity of our analytical approximations. In these calculations we use both the simplified expression $J^2_{{\rm small}\;z}$ and the expression for $J$ given in (\ref{jexp}). We find that for the first quantity, the numerically evaluated expectation value $\langle J^2_{{\rm small}\;z}\rangle$  is about 7/6 of (\ref{japprox}), and the numerical result for $\langle J^2 \rangle$ (taken at $y=y'=0$) is
\ben
\langle J^2 \rangle \approx 5.28\times 10^{-9}\,,
\een
which gives
\ben
\Delta J = \sqrt{\langle J^2 \rangle} \approx 7.27 \times 10^{-5}
\een
which is much closer to the observed value than any of the previously obtained results. Assuming a Gaussian distribution for $J$ which is peaked at zero, the probability of finding a small $J$, in the sense of Sec. \ref{fine}, is now
\ben
P_{{\rm mass}}(|J|\le 10^{-4})\approx 83\%\,,
\een 
whereas the probability of finding a $J$ which is even smaller than the observed value is
\ben
P_{{\rm mass}}(|J|\le 3\times 10^{-5})\approx 32\%\,.
\een 
The observed value for $J$ can no longer be viewed as being finely tuned if the distribution used in our calculations is assumed.

\begin{table}[htp]
\caption{{\small Analytical and numerical results for integrals of interest.}}
\begin{center}
\begin{tabular}{c | c | c | c | c }
 Quantity & $I_D$ (over $x, z$) & $I_D\langle D_1^2\rangle$  & $I_D\langle D_2^2\rangle$  & $I_D\langle D_3^2\rangle$ 
\\\hline
 Analytical result & $ 1.43 \times 10^{-21}$ & $ 2.14 \times 10^{-21}$  & $ 2.78 \times 10^{-26}$  & $ 1.17 \times 10^{-31}$ 
\\\hline
 Numerical result & $ 1.43 \times 10^{-21}$ & $ 2.14 \times 10^{-21}$  & $ 2.78 \times 10^{-26}$  & $ 1.18 \times 10^{-31}$ 
\\\hline\hline
 Quantity & $I'_D$ (over $x', z'$) & $I'_D\langle (D'_1)^2\rangle$  & $I'_D\langle (D'_2)^2\rangle$  & $I'_D\langle (D'_3)^2\rangle$ 
\\\hline
 Analytical result & $ 1.32 \times 10^{-13}$ & $ 1.99 \times 10^{-13}$  & $ 3.60 \times 10^{-17}$  & $ 1.60 \times 10^{-19}$ 
\\\hline
 Numerical result & $ 1.32 \times 10^{-13}$ & $ 1.98 \times 10^{-13}$  & $ 7.28 \times 10^{-17}$  & $ 1.99 \times 10^{-19}$ 
\\\hline\hline
 Quantity & $\tilde{I}_D$ (over $x,z,x', z'$) & $\langle J^2_{{\rm small}\;z}\rangle$  & $\langle J^2\rangle$
\\\hline
 Analytical result & $ 1.89 \times 10^{-34}$ & $ 3.22 \times 10^{-9}$  & ---
\\\hline
 Numerical result & $ 1.88 \times 10^{-34}$ & $ 3.73 \times 10^{-9}$  & $ 5.28 \times 10^{-9}$ 
\\\hline\hline
\end{tabular}
\end{center}
\label{results}
\end{table}

To test the sensitivity of our results to changes in the parameters, we take values at the upper or lower limit in (\ref{qmass}) and try to find the highest and lowest values for $\langle J^2\rangle$. We find that setting 
\bea
&& (m_u,m_c,m_t):=(0.85\;{\rm MeV},\; 0.535\;{\rm GeV},\;174.7\;{\rm GeV})\,;\nn
\\&& (m_d,m_s,m_b):=(1.71\;{\rm MeV},\; 40\;{\rm MeV},\;2.98\;{\rm GeV})
\eea
gives
\ben
\langle J^2 \rangle \approx 1.86 \times 10^{-9}
\een
and
\ben
\Delta J = \sqrt{\langle J^2 \rangle} \approx 4.31 \times 10^{-5}\,,
\een
whereas setting
\bea
&&(m_u,m_c,m_t):=(1.77\;{\rm MeV},\; 0.535\;{\rm GeV},\;168.7\;{\rm GeV})\,;\nn
\\&& (m_d,m_s,m_b):=(4.14\;{\rm MeV},\; 71\;{\rm MeV},\;2.8\;{\rm GeV})
\eea
gives
\ben
\langle J^2 \rangle \approx 1.52 \times 10^{-8}
\een
and
\ben
\Delta J = \sqrt{\langle J^2 \rangle} \approx 1.23 \times 10^{-4}\,.
\een
While the result for $\Delta J$ depends on the choice of parameters, we see that the general conclusion, which is a comparison of orders of magnitude, is unchanged by reasonable modifications of the parameters (\ie values for the quark masses). Even the greatest possible value for $\Delta J$ is significantly lower than any of the values obtained in previous sections.

In this subsection, we have established that assuming the observed hierarchy in quark masses in a Gaussian distribution over the space of mass  matrices gives expectation values for $J^2$ which are small enough to regard the observed value as ``natural'' and not finely tuned. This statistical observation seems to open up the possibility that the same mechanism that is responsible for the apparently unlikely hierarchy in quark masses might also explain why the observed value for $|J|$ is so small.

A more detailed analysis including a probability density for $|J|$ for this distribution is left to future work, since the numerical methods used here do not give sufficiently accurate results.
  
\subsection{Complex Mass Matrices}
\label{nonherm}

The corresponding calculation for general complex mass matrices differs from the one for Hermitian mass matrices detailed in the previous subsection by a different measure over $\mathbb{R}^6$, the space of possible quark masses. Hence after the integration over $\mathbb{R}^6$ has been performed, the resulting integrands will be different functions of the quantities $\xi_a$ introduced in the previous section. The steps of calculating an expectation value for $J^2$ will be the same as in the previous subsection, and we will use similar analytical approximations as well. It suffices therefore to highlight those steps where the resulting formulae differ from the previous ones.

In the previous subsection we concluded that a natural measure on the space of Hermitian mass matrices that involves the observed quark masses reproduces the observed magnitude of $CP$ violation very well. The quark mass hierarchy had the effect of significantly lowering the expectation value of $J^2$. Looking at the different measures (\ref{dmeasure}), we would expect this effect much stronger if general complex mass matrices are considered. This expectation will be confirmed: There now appears to be fine-tuning in the opposite way, \ie one would expect $CP$ violation to be much smaller than is observed.

Using the same approximation of small $y$ and $y'$ as previously, we are again left with computing an integral of the form
\ben
\langle J^2 \rangle \approx \frac{\int_{\R^6} dD dD' \int d^4 x d^4 x'\,s_{2x}s_{2z}s_{2x'}s_{2z'} \left(e^{-\Tr(D^2 UAU^{\dagger})-\Tr((D')^2 U'A'{U'}^{\dagger})}\,J^2(U,U')\right)\big|_{y=y'=0}}{\int_{\R^6} dD dD' \int d^4 x d^4 x'\,s_{2x}s_{2z}s_{2x'}s_{2z'}  \left(e^{-\Tr(D^2 UAU^{\dagger})-\Tr((D')^2 U'A'{U'}^{\dagger})}\right)\big|_{y=y'=0}}\,,
\label{newjint2}
\een
where
\ben
\int d^4 x\equiv \int\limits_0^{\pi/4}dx\int\limits_0^{\pi/2}dz \int\limits_0^{2\pi}dr\int\limits_0^{2\pi}dt
\een
and similarly for $\int d^4 x'$. The symbol $dD$ of course now has a different meaning. Again we can integrate over both copies of $\R^3$ in (\ref{newjint2}), using
\bea
f_{\xi_1\xi_2\xi_3} & := &\int\limits_{-\infty}^{\infty}dD_1\,dD_2\,dD_3\,(D_1^2-D_2^2)^2(D_1^2-D_3^2)^2(D_2^2-D_3^2)^2 |D_1 D_2 D_3| e^{-\xi_1 D_1^2-\xi_2D_2^2-\xi_3D_3^2}\nn
\\& = &\frac{24}{\xi_1^{5}\xi_2^{5}\xi_3^{5}}\left(2(\xi_1^2\xi_2^2+\xi_1^2\xi_3^2+\xi_2^2\xi_3^2)(\xi_1^2+\xi_2^2+\xi_3^2-\xi_1\xi_2-\xi_1\xi_3-\xi_2\xi_3)\right.
\\& & \left.-3\xi_1\xi_2\xi_3(\xi_1^3+\xi_2^3+\xi_3^3-\xi_2^2\xi_3-\xi_3^2\xi_1-\xi_1^2\xi_2-\xi_3^2\xi_2-\xi_1^2\xi_3-\xi_2^2\xi_1)-8\xi_1^2\xi_2^2\xi_3^2\right)\,.\nn
\eea

To fix the parameters appearing in the matrices $A$ and $A'$, we again use expectation values for squared quark masses:
\ben
\langle D_1^2 \rangle \approx \frac{\int_{\R^3} dD\, D_1^2 \int\limits_0^{\pi/4} dx \int\limits_0^{\pi/2} dz\,\sin 2x \sin 2z  \left(e^{-\Tr(D^2 UAU^{\dagger})}\right)\big|_{y=0}}{\int_{\R^3} dD\,  \int\limits_0^{\pi/4} dx \int\limits_0^{\pi/2} dz\,\sin 2x \sin 2z  \left(e^{-\Tr(D^2 UAU^{\dagger})}\right)\big|_{y=0}}\,.
\label{d12int}
\een
The denominator of (\ref{d12int}) is explicitly
\ben
I_D:=\int\limits_0^{\pi/4} dx \int\limits_0^{\pi/2} dz\,\sin 2x \sin 2z \,f_{\xi_1\xi_2\xi_3}\,,
\label{denom2}
\een
where $\xi_1,\xi_2,\xi_3$ at $y=0$ were defined in (\ref{xidef}). Again, we can approximate $I_D$ well by only keeping the terms of leading order in $x$ and $z$ in the trigonometric functions, and approximating $f_{\xi_1\xi_2\xi_3}$ by the leading term
\ben
f_{\xi_1\xi_2\xi_3} \approx \frac{24}{\xi_1^5\xi_2^5\xi_3^5}\times 2\xi_3^4\xi_2^2\approx \frac{48}{(A_1+A_2 z^2)^5 (A_2 + A_3 x^2)^3 A_3}\,,
\een
which leads to
\bea
I_D & \approx & \frac{48}{A_3} \int\limits_0^{\pi/4} dx \int\limits_0^{\pi/2} dz\,4 x z \,(A_1+A_2 z^2)^{-5} (A_2 + A_3 x^2)^{-3}\nn
\\ & \approx &\frac{48}{A_3} \int\limits_0^{\infty} dX \int\limits_0^{\infty} dZ\,(A_1+A_2 Z)^{-5} (A_2 + A_3 X)^{-3}\nn
\\ & = &\frac{48}{A_3}\cdot\frac{1}{4A_2 A_1^4}\cdot\frac{1}{2A_3 A_2^2} = \frac{6}{A_1^4 A_2^3 A_3^2}\,.
\eea
The result is very well reproduced by numerical calculations. Similarly,
\bea
I_D\langle D_1^2 \rangle & \approx & \frac{240}{A_3} \int\limits_0^{\infty} dX \int\limits_0^{\infty} dZ\,(A_1+A_2 Z)^{-6} (A_2 + A_3 X)^{-3}\nn
\\ & = & \frac{240}{A_3}\cdot\frac{1}{5A_2 A_1^5}\cdot\frac{1}{2A_3 A_2^2} = \frac{24}{A_1^5 A_2^3 A_3^2}\,,
\eea
hence
\ben
\langle D_1^2 \rangle \approx \frac{4}{A_1}\,.
\een
Redoing the same calculation for $D_2$ and $D_3$ gives
\ben
\langle D_2^2 \rangle \approx \frac{2}{A_2}\,,\quad \langle D_3^2 \rangle \approx \frac{1}{A_3}\,.
\een
The relative factors, which are now slightly different from the ones obtained in (\ref{relfactor}), have to be taken into account when determining $A$ and $A'$.

Recall that we used the following central values for the quark masses:
\bena
&&(m_u,m_c,m_t):=(1.27\;{\rm MeV},\; 0.619\;{\rm GeV},\;171.7\;{\rm GeV})\,\nn
\\&& (m_d,m_s,m_b):=(2.9\;{\rm MeV},\; 55\;{\rm MeV},\;2.89\;{\rm GeV})\,.
\label{cvalues2}
\eea

We now proceed as before, using numerical calculations to check our analytical approximations. As before, the analytical approximations are better reproduced for the positively charged top, charm and up quarks, which exhibit a more extreme quark mass hierarchy, than for the negatively charged quarks. We use the numerical results to fix the parameters in $A$ and $A'$ to 
\bea
\mu_c^2 = 2\left(\frac{m_c}{m_t}\right)^2\approx 2.60\times 10^{-5}\,,\quad \mu_u^2 = 4\left(\frac{m_u}{m_t}\right)^2\approx 2.19\times 10^{-10}\,,\nn
\\ \mu_s^2 = \left(\frac{m_s}{m_b}\right)^2\approx 3.62\times 10^{-4}\,,\quad \mu_d^2 = 4\left(\frac{m_d}{m_b}\right)^2\approx 4.03\times 10^{-6}\,.
\label{muvalues2}
\eea

In order to obtain an analytical expression for expectation value of $J^2$, we again take the approximation of small $z$
\ben
J^2_{{\rm small}\;z}:=\frac{1}{2}\sin^2 x\cos^2 x\sin^2 x'\cos^2 x'\cos^2 z'\sin^4 z'
\een
for our calculations. Within this approximation for $J$, still taking $f_{\xi_1\xi_2\xi_3} \approx 48\xi_1^{-5}\xi_2^{-3}\xi_3^{-1}$, the numerator of (\ref{newjint2}) is the product (using again that only small $z$ contributes)
\bea
& &1152\times \int\limits_0^{\pi/2} dz \, \frac{2z}{(A_1 + A_2 z^2)^{5/2}} \times \int\limits_0^{\pi/2} dz' \, \frac{\sin 2z' \cos^2 z'\sin^4 z'}{(A_1 \cos^2 z'+ A_2 \sin^2 z')^5} \nn
\\ & \times &\int\limits_0^{\pi/4} dx \,\frac{\sin 2x\,\sin^2 x\,\cos^2 x}{(A_2 \cos^2 x + A_3 \sin^2 x)^3 (A_3 \cos^2 x + A_2 \sin^2 x)}
\eea
times the integral in the second line with all quantities replaced by ``primed" ones. The first two factors are $\frac{1}{4A_1^4 A_2}$ and $\frac{1}{12(A_1')^2(A_2')^3}$, respectively; for the other two (which have the same form) we change variables to $X=\cos^2 x$ to obtain
\ben
\int\limits_{1/2}^{1} dX \,\frac{X(1-X)}{(A_2 X + A_3 (1-X))^3 (A_3 X + A_2 (1-X))}= \frac{1}{2A_3^3 A_2}\left(1-\frac{2A_2}{A_3}+O\left(\left(A_2/A_3\right)^2\right)\right)\,;
\een
putting everything together, we obtain
\ben
\langle J^2_{{\rm small}\;z}\rangle \approx \frac{1}{6}\frac{A_2 (A_1')^2}{A_3 A_2' A_3'} = \frac{4}{3}\frac{m_s^2 m_u^2 m_d^2}{m_b^4 m_c^2}\,,
\label{japprox2}
\een
where the numerical factors appearing in the last line come from the different factors chosen in (\ref{muvalues2}). Just as before, the top quark mass does not appear in this approximation. This compares with the scaling behaviour obtained in the previous section,
\ben
\langle J^2_{{\rm small}\;z}\rangle \sim \frac{m_s^2 m_u m_d}{m_b^3 m_c}\,.
\een

For numerical calculations we use both the simplified expression $J^2_{{\rm small}\;z}$ and the expression for $J$ given in (\ref{jexp}). We find that for the first quantity, the numerically evaluated expectation value, $\langle J^2_{{\rm small}\;z}\rangle\approx 1.89\times 10^{-15}$, is about $94\%$ of (\ref{japprox2}), and the numerical result for $\langle J^2 \rangle$ (taken at $y=y'=0$) is
\ben
\langle J^2 \rangle \approx 2.07\times 10^{-15}\,,
\een
which gives
\ben
\Delta J = \sqrt{\langle J^2 \rangle} \approx 4.55 \times 10^{-8}
\een
which is now almost three orders of magnitude {\it smaller} than the observed value. Assuming a Gaussian distribution peaked at zero, we now get
\ben
P(|J|\le 10^{-7})\approx 97\%\,,
\een 
When the measure presented here is used, there seems to be extreme fine-tuning in $J$, but now we would say that one observes unnaturally {\it large} $CP$ violation! This result may look surprising, given that the maximal value for $J$ is around 0.1 and the observed value just $3\times 10^{-5}$, but it shows how strongly the quark mass hierarchy suppresses large values of $J$ in our distribution.

\sect{Extension to Neutrinos}
\label{neutrinos}

In the standard model, neutrinos are taken to be massless. There are no Yukawa couplings to the Higgs field, and no possibility of $CP$ violation. However, since the phenomenon of neutrino oscillations has been confirmed experimentally, neutrinos must be assumed to have mass. Curiously, a mixing matrix for leptons, now referred to as the {\bf Maki-Nakagawa-Sakata (MNS) matrix}\index{MNS matrix}, was already proposed in 1962 \cite{mns}. It appears in a charged current that has the same form as (\ref{chargedc}) for baryons, with massive leptons taking the place of up-type quarks and neutrinos taking the place of down-type quarks. With compelling evidence for neutrino oscillations \cite{particledb} and hence massive neutrinos, such an interaction is now assumed to exist. By diagonalising the leptonic mass terms one will again encounter a mixing matrix $M$ \cite{King}.

The mathematical problem is therefore the same as before; the only outstanding issue is the ambiguity in the definition of this matrix due to possible rephasing of the lepton fields. There are two different possible types of mass terms for fermions. In the most commonly used notation, fermion fields are given by complex 4-component (Dirac) spinors\index{spinor}, consisting of two 2-component (Weyl) spinors:
\ben
\psi=\left( \begin{matrix} \psi_1 \\ \psi_2 \end{matrix} \right),\quad \bar{\psi}:=\psi^{\dagger}\gamma^0=(\psi_2^{\dagger} \quad \psi_1^{\dagger}).
\een
Normally, a mass term contains $\psi$ and its Dirac conjugate $\bar\psi$, thus coupling $\psi_1$ and $\psi_2$:
\ben
\mathcal{L}_{{\rm mass}}=m\bar\psi\psi=m(\psi_2^{\dagger}\psi_1+\psi_1^{\dagger}\psi_2).
\een
In this case, the four complex components of $\psi$ represent a particle together with its antiparticle, which have the same mass but are otherwise distinct. It is however also possible to have Majorana\index{spinor!Majorana} mass terms, which are of the form
\ben
\mathcal{L}_{{\rm Majorana}}=m_1\psi_1^{\dagger}\psi_1+m_2\psi_2^{\dagger}\psi_2.
\een
Here $m_1$ and $m_2$ are not necessarily equal. In the case $m_1\neq m_2$, the 4-spinor represents two different particles, which are each equal to their own antiparticle. It is then appropriate to take real and imaginary parts and rewrite the theory in terms of two real 4-component (Majorana) spinor fields.\footnote{There is of course no need for a theory formulated in terms of Majorana spinors to have an even number of fields, since the Dirac spinor should not be regarded as fundamental. The presentation only started from the Dirac spinor field because it is most commonly encountered in the literature.}

These fields, being real, have no rephasing freedom, and the matrix relating the flavour and mass eigenbases for neutrinos is in $SO(3)$. Therefore, for an MNS matrix which couples three Majorana neutrinos to the massive leptons, known to be Dirac spinors, there is only the freedom of rephasing on one side, and this matrix would be an element of the single quotient $SU(3)/U(1)^2$. Of course, all intermediate cases are also conceivable, where some neutrinos have Majorana and others have Dirac masses. The most general mixing matrix is hence of the form $U(1)^k \backslash SU(N) / U(1)^{N-1}$, for some $k$ between 0 and $N-1$.

There are at present no strong constraints on the magnitude of $CP$ violation from observation. The results presented in this chapter may possibly put constraints on it, if there is some experimental information about neutrino masses, or conversely make near-coincident neutrino masses appear more or less likely, if the magnitude of $CP$ violation is known.

\subsection{Neutrino Mixing Matrix}

\label{neutrinomix}

If we assume that the neutrinos are Majorana spinors, the lepton mixing matrix belongs to $U(1)^2\backslash SU(3)$, since only phasing of the lepton charge eigenstates $(\nu_e,\nu_\mu,\nu_\tau)$, but not the mass eigenstates $(\nu_1,\nu_2,\nu_3)$ is possible. The former are defined by
\ben
\begin{pmatrix} \nu_ e  \cr \nu_ \mu \cr \nu_\tau \end{pmatrix} = M \begin{pmatrix} \nu _1  \cr \nu_ 2 \cr \nu_3 \end{pmatrix}\,.
\een
In other words, $|\nu_\alpha\rangle$ is the neutrino state created in the decay $W^+\rightarrow l_{\alpha}^+ + \nu$, where $l_{\alpha}$ is an antilepton of flavour $\alpha=e,\mu,\tau$ \cite{particledb}. Such a state is in general a superposition of mass eigenstates which will have different propagation amplitudes, leading to nonzero probabilities for neutrino oscillation $\nu_{\alpha}\rightarrow\nu_{\beta}$.

One conventionally fixes the phases so that $M$ takes the form
\ben
M= \begin{pmatrix}
 1&0&0 \cr 0 & c_{23} & s_{23} \cr 0 & -s_{23} & c_{23}\end{pmatrix}  
\begin{pmatrix} c _{13} &0& s_{13} e^{-\im \delta} \cr 0 & 1  & 0 \cr -s_{13}
  e^{\im\delta}
& 0 & c_{13}\end{pmatrix}  
\begin{pmatrix} c_{12} &s_{12} &0\cr -s_{12}  & c_{12} & 0 \cr 0 & 
0 & 1\end{pmatrix}
\begin{pmatrix} e^{\im\alpha_1 /2 } &0&0\cr 0 & e^{\im \alpha_2 /2 }   & 0  \cr 
0 & 0 & 1\end{pmatrix} \,,
\een
where the three angles $\theta_{12}$, $\theta_{13}$, and $\theta_{23}$ 
lie in the first quadrant.

The Jarlskog invariant for the neutrino mixing matrix is defined as in (\ref{jdef}) but with $(x,y,z,w)$ now parametrising the MNS matrix $M$. Note, in particular, that it is independent of the phases $\alpha_1$ and $\alpha_2$.

Experimentally\index{experiment}, parameters of the neutrino mixing matrix are not completely known. According to \cite{particledb}, 
\ben
\sin^2 (2\theta_{12}) = 0.87\pm 0.03\,,\qquad 0.92 < \sin^2 (2\theta_{23}) \le 1\,,\qquad \sin^2 (2\theta_{13}) < 0.19\,,
\een
and there is no experimental information about the Dirac angle $\delta$.
Thus, we can certainly deduce that there is an upper bound on the
Jarlskog invariant for the neutrino mixing matrix, given by
\ben
|J|< 0.049\,. \label{Jneutbound}
\een

For six different neutrino mass eigenstates, as in the see-saw mechanism, which are assumed to all couple to charged leptons, a general mixing matrix would be an element of $U(1)^5 \backslash SU(6)$, since one would diagonalise a $6\times 6$ Hermitian matrix. The present viewpoint is however that such additional neutrino fields would be ``sterile" and not couple to either $W$ or $Z$ bosons \cite{particledb}.

\subsection{Statistics of $J$}
\label{neutrinosec}

   We have seen that the parameter space for the neutrino mixing matrix is the six-dimensional single quotient $U(1)^2\backslash SU(3)$, and that the Jarlskog invariant for the neutrino mixing matrix takes the same form (\ref{jdef}) as it does for the CKM matrix. Therefore, all results obtained in section \ref{jstats} apply equally to the case of neutrinos.

  For completeness, we quote the results obtained in section \ref{jstats} for a right-invariant measure on the homogeneous space $U(1)^2\backslash SU(3)$,
\ben
\langle J^2 \rangle = \frac{1}{720} \approx 1.389 \times 10^{-3}\,,\qquad \langle J^4 \rangle =  \frac{1}{201600} \approx 4.960 \times 10^{-5}\,,
\een
and
\ben
\Delta J = \fft1{12\sqrt 5} \approx 0.0373\,.
\een
This can be compared with the experimental bound given in (\ref{Jneutbound}).

One should also repeat the calculations of section \ref{gaussian}, assuming particular values for the neutrino masses (and including the known masses for the charged leptons). A strong hierarchy\index{mass hierarchy} in the neutrino masses would then presumably again lead to ``naturally'' small $CP$ violation from the corresponding mixing matrix. Alternatively, an experimental observation of small $CP$ violation for neutrinos would perhaps be an indication of a mass hierarchy in neutrinos. At present, neither the magnitude of $CP$ violation nor any values of neutrino masses have been measured sufficiently accurately to allow predictions.

\sect{Summary and Outlook}

In this chapter, we analysed the problem of finding a natural measure on a space of coupling constants, which in our case was the space of CKM matrices, the double quotient\index{double quotient} $U(1)^2 \backslash SU(3) / U(1)^2$. We saw that the measure on this double quotient is nonunique, and we analysed several possible choices of measure on the double quotient. One class of measures was given by squashed Kaluza-Klein measures, induced by a Kaluza-Klein reduction of a left-invariant metric on the flag manifold. Alternatively, one could take the unique measure on $SU(3) / U(1)^2$ and simply integrate over the left angles. The measure used by Ozsv\'ath and Sch\"ucking seemed not to be very well motivated from a geometric perspective.

When calculating expectation values for $J$, we found that all of the measures we considered led to rather similar results. In each case, the observed value was about three orders of magnitude below what one would normally expect; the observed value appears to be finely tuned. The same applied to the Ozsv\'ath-Sch\"ucking measure, an extremely squashed Kaluza-Klein measure, or a flat measure, which is just the simplest choice and not justified geometrically.

In section \ref{gaussian}, we adopted the different viewpoint that the CKM matrix should not be viewed as separate from the quark masses, but that it is really the mass matrices which are ``chosen" by a yet unknown physical mechanism. We took the observed values for the quark masses as an input and chose the simplest distribution which was able to reproduce these observed values, while inducing a different measure on the space of CKM matrices. While this is a choice we made, and all results depend on this choice, our measure is the combination of a maximally symmetric measure, invariant under a redefinition of a Hermitian matrix by conjugation by a unitary matrix, and a Gaussian incorporating the observed values of the quark masses. We would have to make additional rather strong assumptions to motivate a different choice of measure that would differ appreciably from this simple construction. It may well be that such assumptions are justified by the underlying mechanism determining the mass matrices, but we do not know of such a mechanism yet. Assuming such a distribution, together with the assumption that the mass matrices can be taken as Hermitian as is possible in the standard model, we found that the observed value of $J$ now appears very natural and not finely tuned at all. In this statistical approach, regarding the Yukawa couplings determining the mass matrices as randomly chosen seems more appropriate than separating quark masses and mixing angles.

In order to test the dependence of this nice result on the assumption of Hermitian mass matrices, we then repeated the calculation under the assumption that there is a left-right symmetry which implies that the quark mass matrices can not in general be taken to be Hermitian. We constructed the analogous probability distribution on the space of $3\times 3$ complex matrices, again fitting four free parameters to expectation values for quark masses that can reproduce the observed values. We saw that using such a probability distribution the conclusion one reaches is rather different: One would now expect $J$ to be about three orders of magnitude smaller than the observed value. Hence, there is a fine-tuning problem; without further assumptions, a fundamental theory leading to a left-right symmetric electroweak sector at low energy should generically be expected to reproduce very weak $CP$ violation. Invoking the principle of Occam's razor,\index{Occam's razor} ``{\sc entia non sunt multiplicanda praeter necessitatem}," we would like to conclude that, only looking at possible explanations for the magnitude of $CP$ violation in the electroweak sector, the standard model should be preferred to left-right symmetric extensions such as Pati-Salam\index{Pati-Salam model}: In the latter one needs additional assumptions on the fundamental parameters that resolve the issue of observing ``unnaturally large" $CP$ violation, that are not necessary in the standard model, or any extension of it that allows a restriction to Hermitian mass matrices.

In the general context of this thesis, these calculations provide an example of gauge symmetries, described by one or two copies of $SU(2)$, whose presence or absence has a direct effect on measurable quantities, although we have followed a purely statistical approach rather than proposing dynamical mechanisms that could lead to certain values for these quantities.

Although we have tried to argue that our results are independent of renormalisation\index{renormalisation} group flow since the relevant quantities do not run strongly with energy scale, there is another more subtle issue: The low-energy limit of a left-right symmetric extension with non-Hermitian mass matrices would still be the standard model, where mass matrices can be assumed to be Hermitian, leading us back to the measure we considered previously. It would be desirable to incorporate this dependency of the assumptions one has used to construct the measure on energy scale into the analysis, namely to use a measure which depends on energy scale also. A starting point would be a quantification of ``non-Hermiticity" that could then flow from zero at low energies to some non-zero value at high energies. At present these ideas are however somewhat vague, so that we will have to leave them to exploration in future work.

We have argued that our analysis also applies to the case of massive neutrinos, where the predictions will conceivably be tested by future experiments. We have seen that for Majorana spinors, the Maki-Nakagawa-Sakata matrix \cite{mns} which appears is naturally an element of the single quotient $U(1)^2\backslash SU(3)$. Since the right phases do not play any role in neutrino oscillations and the relevant $J$ is independent of these phases, the calculations are identical to the ones presented here, although with the appropriate values of the $\mu$ parameters appearing in $A$ and $A'$. Based on our results, one would expect a strong hierarchy in neutrino masses to lead to suppressed $CP$ violation.

In the seesaw mechanism one adds very heavy right-handed neutrinos, and the most general mixing matrix (making the additional, at present experimentally unjustified, assumption that the heavy neutrinos couple to charged leptons as well) would be an element of $U(1)^5 \backslash SU(6)$. This is naturally a K\"ahler manifold, and the measure induced by the K\"ahler metric can be obtained from the analysis in \cite{picken}. We leave a detailed treatment of this case, following our approach here, to future work.

Finally, one could analyse the effects of a fourth generation of quarks on $CP$ violation by repeating the calculations for $4\times 4$ Hermitian matrices. If this generalisation spoils the agreement with the observed $J$, one might obtain interesting lower bounds on the masses of a hypothetical extra generation of quarks.

\chapter*{Summary and Outlook}
\fancyhead[LO]{\scshape Summary and Outlook}
\epigraph{\em{``La recherche de la v\'erit\'e doit \^etre le but de notre activit\'e; c'est la seule fin qui soit digne d'elle.'' \cite{poincarevaleur}}}{Henri Poincar\'e}
\addcontentsline{toc}{chapter}{Summary and Outlook}

In this thesis we have investigated several examples within general relativity and particle physics where symmetries, and more precisely the geometry of symmetry groups and homogeneous spaces, played a fundamental role. In each of these cases, the study of geometric aspects of the underlying symmetries led to physical predictions, providing illuminating examples for the interplay between mathematical structures and the physical world that modern theoretical physics is based on.

\

The first part consisting of chapters \ref{defspacetime} to \ref{bflinear} was concerned with deformed and constrained symmetries in relativity.

In chapter \ref{defspacetime}, we studied exact solutions of general relativity in four dimensions from a group-theoretic viewpoint, focussing on their global symmetries (isometries). The isometry groups of these spacetimes act simply-transitively; the spacetimes are homogeneous spaces $G/H$, where $H$ was either trivial or one-dimensional. We applied the theory of deformations of Lie groups and algebras to these isometry groups, finding that some solutions are related by deformation of their isometry groups, and that these algebraic properties are partly related to physical properties; we have at least found indications that the Kaigorodov-Ozsv\'ath solution contains closed timelike curves (CTCs) just as the Petrov solution which it is a deformation of. This work, which focussed on a few concrete relatively straightforward examples, leaves much scope for extension. First, one could go to higher dimensions, since in four dimensions all homogeneous solutions are more or less known. Second, one could investigate what the generic relationship is between algebraic properties (related to isometry groups) and physical properties (such as causal behaviour); one might be able to relate a classification of group deformations to a classification of physical properties of certain classes of exact solutions. We leave this to further work.

In chapter \ref{defgenrel}, we considered a formulation of general relativity as a gauge theory of the de Sitter group, shifting our focus to local symmetries. This formulation, an extension of the MacDowell-Mansouri formulation given by Stelle and West, replaces the perhaps more usual Poincar\'e group by the de Sitter group, and provides another example of group deformation. While the viewpoint on general relativity as some kind of gauge theory is not new, we argued how Einstein-Cartan-Stelle-West theory may provide a ``deformation" of general relativity analogous to the framework of deformed special relativity (DSR), and focussed on Minkowski space as an explicit example, thereby uncovering close relations between two apparently disconnected branches of the general relativity literature. We showed how the apparent ``noncommutativity of spacetime" given by noncommuting translations on the internal (``momentum") de Sitter space is manifest in torsion of the connection. We gave estimates for the torsion one would get in the simplest situation, and discussed worrying physical consequences if electromagnetic fields couple to torsion, such as breaking of charge conservation. While this means that the framework we discussed is presumably not viable as a physical theory, it provided an interesting alternative interpretation of DSR. 

Our study was an application of Cartan geometry which is a generalisation of (pseudo-)Riemannian geometry in that the ``tangent" space is replaced by an arbitrary homogeneous space $G/H$. It would be worth formulating other examples of such a situation in physics in this mathematical language. One example is the traditional formulation of loop quantum gravity (LQG) where the $\frak{so}(3,1)$-valued connection is reduced to an $\frak{so}(3)$-valued connection by partial gauge fixing. A recent study \cite{lqgprl} seems to generalise this gauge choice by essentially working in a Cartan geometric construction, but the precise relation is rather unclear to us.

In chapter \ref{bflinear} we studied yet another reformulation of general relativity, as a topological BF theory with constraints. Instead of deforming a symmetry, we were constraining it; the large symmetry on the phase space of the topological theory which implies the (local) equivalence of all solutions to the equations of motion could be constrained to be only the gauge symmetry of general relativity in four dimensions, where not all solutions are locally equivalent. Such a formulation, which is popular in covariant approaches to quantising gravity, has so far mainly been given in terms of quadratic constraints, which suffer from ambiguities in their solution whose relevance in a quantum theory is not entirely clear. It is therefore an improvement if, as we have shown, an alternative formulation in terms of linear constraints can be found. This formulation seems closely related to discrete constraints on representations that are used in spin foam models, but requires in its corresponding classical formulation the introduction of a new ``4D closure" constraint that has not been studied very much so far. An obvious extension of our work would be to study the effects of the presence of such a new constraint on spin foam models for quantum gravity. Linear constraints, whose solutions are formally given by projections, are also more closely related to the group field theory approach to quantum gravity, where constraints are usually enforced as projections \cite{iogft}.

\

The second part consisting of chapters \ref{su3quot} and \ref{natural} was a geometric study of gauge symmetries in the electroweak model and their role in $CP$ violation.

In chapter \ref{su3quot} we introduced a class of metrics on the group $SU(3)$, the homogeneous space $SU(3)/U(1)^2$ and the double quotient space $U(1)^2\backslash SU(3)/U(1)^2$. While for the first two cases the $SU(3)$ left group action provided a clear criterion (left-invariance) for preferred classes of metrics, the situation for the double quotient was less clear and several possible choices were introduced. We also computed bi-invariant metrics (with respect to the natural group actions) on the space of Hermitian and on the space of complex $3\times 3$ matrices.

In chapter \ref{natural} we saw how $CP$ violation in the electroweak sector is determined by the Jarlskog invariant $J$ which can be viewed as a function on $U(1)^2\backslash SU(3)/U(1)^2$, the space of Cabibbo-Kobayashi-Maskawa (CKM) matrices. We tried to give a mathematically precise answer to the question whether the observed magnitude of $J$, which appears to be rather small, should be considered natural or fine-tuned. Since finding this answer is equivalent to specifying a natural measure on the double quotient space, we analysed several possible choices of the measure induced by metrics presented in the previous chapter. While it seems not entirely clear which of these metrics should be preferred, we found that they all suggest fine-tuning in the magnitude of $J$. We then reformulated the problem by considering the space of quark mass matrices instead, using a measure which incorporates the observed quark masses and modifies the induced distribution on the space of CKM matrices. The effect of this modified distribution was the following: If the mass matrices can be taken to be Hermitian as in the standard model, we found that the input of observed quark masses is sufficient to reproduce the observed magnitude of $CP$ violation to a remarkably high accuracy. If on the other hand one takes general complex mass matrices, one finds an ``inverse" fine-tuning problem; the magnitude of $J$ should be expected to be much smaller than is observed. This distinction is physically relevant since in models with extended left-right symmetry, such as the Pati-Salam model, one has to consider general complex mass matrices. Our results, which rely on assuming the simplest and geometrically best motivated probability distributions, seem therefore to suggest that the standard model should be preferred by Occam's razor: One does not need additional fine-tuning assumptions to explain the magnitude of $CP$ violation. Hence, these calculations provide another example of how the modification of symmetries (the absence or presence of a left-right symmetry) leads to different physical (in this case statistical) predictions.

Similar calculations could be performed for the case of four quark generations, where the mass of a fourth quark would have an effect on the probability distributions we have considered. Another important extension of our results is to the case of neutrinos, where the formalism for $CP$ violation is rather similar and the mathematics essentially the same. Our results would suggest that a large mass hierarchy in neutrinos (together with the known hierarchy in lepton masses) makes highly suppressed $CP$ violation likely, whereas on the other hand large $CP$ violation might suggest that near-coincident neutrino masses appear more likely. Such predictions can obviously be tested in experiments such as those at LHC.\index{LHC}

\clearpage
\fancyhead[LO]{\scshape\leftmark}
\phantomsection

\listoffigures
\phantomsection
\addcontentsline{toc}{chapter}{List of Figures}

\listoftables
\phantomsection
\addcontentsline{toc}{chapter}{List of Tables}

\clearpage
\phantomsection
\addcontentsline{toc}{chapter}{Index}
\printindex

\end{document}